\documentclass[single]{uclathes}

\usepackage[Lenny]{fncychap} 

\usepackage{epsf} 
\usepackage[dvips]{color} 
\usepackage{multirow} 
\usepackage{rotating} 
\usepackage{graphicx} 
\usepackage{tabularx} 
\usepackage{pstricks} 
\usepackage{afterpage} 
\usepackage{amssymb} 
\usepackage{subfigure} 
\usepackage{supertabular}
\usepackage{rotating} 
\usepackage{fancyheadings}

\renewcommand{\chaptermark}[1]{\markboth{#1}{}}
\newcommand{\beq}{\begin{equation}} 
\newcommand{\eeq}{\end{equation}} 
\newcommand{\beqa}{\begin{eqnarray}} 
\newcommand{\eeqa}{\end{eqnarray}} 
\newcommand{\beqas}{\begin{equation}\begin{array}{lll}} 
\newcommand{\eeqas}{\end{array}\end{equation}} 
\newcommand{\mx}{\left[\begin{array}} 
\newcommand{\finmx}{\end{array}\right]} 
\newcommand{\mxp}{\left(\begin{array}} 
\newcommand{\finmxp}{\end{array}\right)} 
\newcommand{\casos}{\left\{\begin{array}} 
\newcommand{\fincasos}{\end{array}\right.} 
\newcommand{\rcasos}{\left.\begin{array}} 
\newcommand{\rfincasos}{\end{array}\right\}}

\def\lsim{\ \rlap{\raise 3pt \hbox{$<$}}{\lower 3pt \hbox{$\sim$}}\ }
\def\gsim{\ \rlap{\raise 3pt \hbox{$>$}}{\lower 3pt \hbox{$\sim$}}\ }
\def\ominus#1{\ \rlap{\raise 0pt \hbox{$#1$}}{\raise 1pt \hbox{$^{^{(-)}}$}}\ }

\renewcommand{\a}{\alpha}
\renewcommand{\b}{\beta}

\renewcommand{\t}{\theta}

\newcommand{\e}{\epsilon}
\newcommand{\s}{\sigma}

\newcommand{\cerenkov}{\v{Cerenkov}\ }

\newcommand{\lik}{{\cal L}}

\newcommand{\barr}[1]{\overline{#1}}

\newcommand{\anue}{{\overline{\nu}_e}}
\newcommand{\anux}{{\overline{\nu}_x}}
\newcommand{\nue}{{{\nu}_e}}
\newcommand{\nux}{{{\nu}_x}}
\newcommand{\anu}{{\overline{\nu}}}

\newcommand{\onubb}{$0\nu\beta\beta$}

\newcommand{\aver}[1]{\left<#1\right>}

\newcommand{\it}{\itshape}

\newcommand{\bf}{\bfseries}
\newcommand{\rm}{\mathrm}
\newcommand{\cal}{\mathcal}

\newcommand{\fder}[2]{\frac{d #1}{d #2}}

\newcommand{\mycaption}[2]{\caption[#1]{\baselineskip 10pt\small\it%
#2%
}}

\newcommand{\blankline}[1]{\multicolumn{#1}{c}{\strut}\\}

\newcommand{\titulo}[1]{\medskip{\bf\em #1}\medskip}

\newcommand{\D}{{\cal D}}
\newcommand{\M}{{\cal M}}

\newcommand{\msq}{m^2_\nu}

\newcommand{\PreserveBackSlash}[1]{\let\temp=\\#1\let\\=\temp}
\let\PBS=\PreserveBackSlash

\newcommand{\dmsq}[1]{\Delta m^{2}_{#1}}

\def\vvec#1{\rlap{\raise 0pt \hbox{$#1$}}{\raise 1.3ex \hbox{$\rightarrow$}}}

\newcommand{\derf}[2]{{d#1\over{d#2}}}

\newcommand{\degr}{^\circ}

\newcommand{\Aei}[1]{\left|U_{e#1}\right|^2}

\newcommand{\se}{{\barr{e}}}
\newcommand{\sm}{{\barr{\mu}}}
\newcommand{\st}{{\barr{\tau}}}
\newcommand{\sx}{{\barr{x}}}

\newcommand{\cita}[1]{[0]}

\newcommand{\Matrix}[1]{{\mathbf #1}}

\newcommand{\tt}{\ttfamily}

\titlesize{\LARGE}
\title{
Study of core collapse neutrino signals and constraints on neutrino
masses from a future Galactic Supernova
}
\author{Jorge I. Zuluaga}
\department{Instituto de F\'\i sica}
\degreemonth{January}
\degreeyear{2005}

\chair{Enrico Nardi}
\member{Luis Alberto Sanchez}
\member{Daniel Jaramillo}
\member{Alejandra Melfo}

\dedication{
\textsl{
Nothing is built on stone;\\
all is built on sand,\\
but we must build as if the
sand were stone.\\
}
{\bfseries Jorge Luis Borges}
}

\acknowledgments {

First I want to express my gratitude to all the institutions that made
possible the realization of this work.  The {\it Universidad de
Antioquia} that provided an initial financial support through the {\it
Fondo de Becas Doctorales de la Vicerrector\'\i a de Docencia} and to
COLCIENCIAS and its {\it Programa de Apoyo a la Comunidad Cient\'\i
fica Nacional a trav\'es de los Programas de Doctorado} for providing
the fellowship that allowed me to complete my Ph.D. studies.

I also want to thank to the {\it Posgrado de F\'\i sica} and the {\it
Grupo de Fenomenolog\'\i a de Interacciones Fundamentales} of the
Universidad de Antioquia for providing partial economical support for
my visits to institutes abroad.

Special thanks go to the Abdus Salam ICTP in Trieste (Italy) that I
had the pleasure to visit several times during the development of this
work, for the nice hospitality and for the economical support provided
for my travel and living expenses.  The Abdus Salam ICTP has always
been for me the ideal place to meet and interact with experienced
scientists of all nationalities, an this has been fundamental for
opening my mind.  A special acknowledgment goes to Professor Alexei
Smirnov for the comments and suggestions he made during a seminar I
gave in one of my visits.  I also want to acknowledge the precious
support that the staff of the Abdus Salam ICTP Computer Division gave
me that allowed me to optimize the use of the computational facilities
of the Center.

I am very grateful to the {\it Laboratori Nazionali di Frascati}
(INFN-LNF) in Frascati-Rome (Italy) where I had the opportunity to
spend a period of six months during my {\it Pasant\'\i a Doctoral}.
In particular it is a pleasure to acknowledge the INFN-LNF Theory
Group for the kind hospitality and for the partial economical support
they provided me during my stay in such a beautiful place.  I also
thank the {Laboratori Nazionali del Gran Sasso} (INFN-LNGS) for the
kind invitation to give a seminar and present the preliminary results
of this work.  Special thanks go to Professor Francesco Vissani for
his useful comments and quite valuable suggestions, as well as for the
visit that he personally organized for me to the amazing LNGS
underground installations.

I am specially indebted to my advisor, Professor Enrico Nardi, not
only for his constant guidance during all these years and for his
inspiring collaboration, but also for his continuous support and
advice and for his huge patience in proofreading the manuscript of
this thesis.

Many other people has contributed directly and indirectly to this
work.  Specially I want to thank to Juan Jose Quiros for loan me his
laptop to write the final version of this work in a more quit
environment and my friend Jorge Johnson who help me in the
organization of the computer codes in order to prepare a public
version of the same codes (SUNG).

To my friends, parents and brothers that always encouraged me despite
the circumstantial difficulties faced during these years.

And for their love and support in any possible way I owe my deepest
thanks to my family (Olga, Andres and Daniel, {\it mis mu\~necos}) who
always was there (and even follow me to Frascati) giving a profound
sense to all my efforts.  Words cannot express my gratitude and love
for them.  This work is dedicated to them.\\

\begin{flushright}
Jorge Zuluaga\\
January, 2005
\end{flushright}

}

\abstract{
We study the sensitivity to neutrino masses of a Galactic supernova
neutrino signal as could be measured with the detectors presently in
operation and with future large volume water \v{C}erencov and
scintillator detectors.  

The analysis uses the full statistics of neutrino events. The method
proposed uses the principles of Bayesian inference reasoning and has
shown a remarkably independence of astrophysical assumptions.

We show that, after accounting for the uncertainties in the detailed
astrophysical description of the neutrino signal and taking into
account the effects of neutrino oscillations in the supernova mantle,
detectors presently in operation can have enough sensitivity to reveal
a neutrino mass (or to set upper limits) at the level of 1 eV.  This
is sensibly better than present results from tritium $\beta$-decay
experiments, competitive with the most conservative limits from
neutrinoless double $\beta$-decay and less precise but remarkably less
dependent from prior assumptions than cosmological measurements.

Future megaton water \v{C}erencov detectors and large volume
scintillator detectors will allow for about a factor of two
improvement in the sensitivity; however, they will not be competitive
with the next generation of tritium $\beta$-decay and neutrinoless
double $\beta$-decay experiments.

Using the codes developed to perform the generation of synthetic
supernova signals and their analysis we created a computer package,
SUNG (SUpernova Neutrino Generation tool,
http://urania.udea.edu.co/sungweb), aimed to offer a general purpose
solution to perform calculations in supernova neutrino studies.
}

\nodepartment

\begin {document}
\makeintropages


\chapter{Introduction}
\label{ch:introduction}


Supernovae (SN) and neutrinos form one of the most interesting (and
strange) couples in Astrophysics and Fundamental Physics.  The first
ones are the most energetic explosions in the Universe after the
Big-Bang while the second ones are the lightest and most elusive
(massive) elementary particles.

This strange relation was probably first recognized in the early 1940s
\cite{Gamow:1940aa} and arises from the fact that neutrinos are
produced in huge quantities by the nuclear processes which dominate
the evolution of a dying massive star.

When the iron core of a massive star $M\gsim10\,\rm{M}_\odot$ reaches
a critical mass $M_c\simeq1.0\,\rm{M}_\odot\,(Y_e/0.4)^2$
\cite{1935MNRAS..95..207C} by accumulation of the residual material
produced in the ``melting'' of silicon, it becomes unstable and starts
to collapse.  Density and temperature of the very compressed material
rise rapidly during the fast implosion of the core.  When temperature
is high enough the heavy nuclei synthesized during the whole stellar
life start to be dissociated through photon processes:

\beq
\begin{array}{c}
^{56}\rm{Fe}+\gamma\rightarrow 13\,^4\rm{He} + 4\,n - 124 \rm{MeV}\,,\\
^{4}\rm{He}+\gamma\rightarrow 2\,p + 2\,n - 27.3 \rm{MeV}.\\
\end{array}
\label{eq:phodisint.}
\eeq

This endothermic photodisintegration processes start to drain energy
from the core and trigger the collapse.  Additionally electron capture
on proton produces electron neutrinos and reduces the electron
degeneracy pressure:

\beq
p + e \rightarrow n + \nue.
\label{eq:electron.capt.}
\eeq

\begin{figure}[t]
\begin{center}
\epsfxsize=140mm
\epsfbox{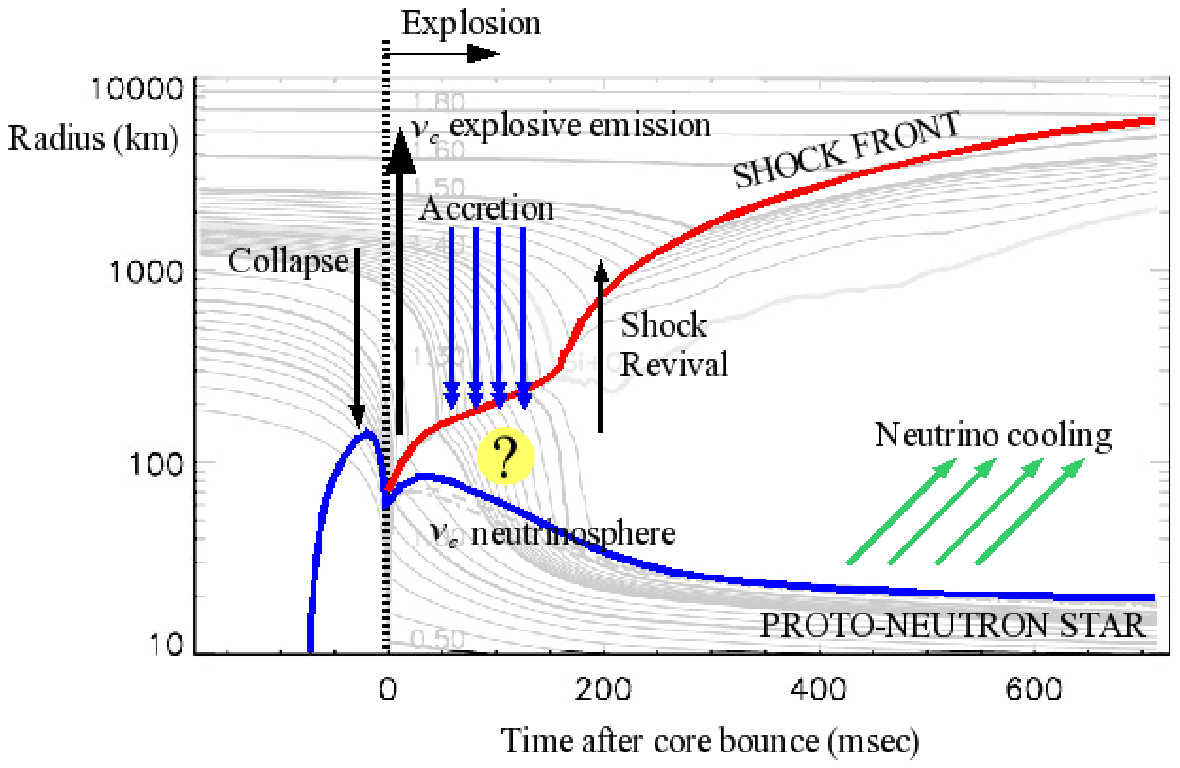}
\mycaption{Schematic illustration of the collapse of a massive star
core and the onset of the explosion}{
Schematic illustration of the collapse of a massive star core and the onset of
the explosion, superposed to a diagram of the evolution of mass-shells from a
supernova simulation \cite{Raffelt:2003en}.  The question-mark in the region
between the {\it neutrino-sphere} (neutrino last scattering surface) and the
stalled shock wave during the accretion phase indicates our present ignorance
about the exact mechanisms that concur to produce the revival of the
explosion.
}
\label{fig:schem.collapse}
\end{center}
\end{figure}

Collapse is not forever.  When the inner part of the stellar core
reaches nuclear densities $\rho>3-8\times 10^{14}$ g/cm$^3$ it
suddenly stops. Not all the core feels the interruption (collapse
velocities above the inner core are faster than acoustic waves) and
matter continues infalling.  In less than 1 msec a hydrodynamic shock
wave forms and starts to move outwards sweeping the falling matter and
reversing the collapse: the core bounces (see
fig.~\ref{fig:schem.collapse}).

%
%

At the very high densities of the collapsing stellar core neutrinos
are trapped in the first hundreds kilometers where production and
absorption processes are very fast.  After a few milliseconds from the
core bounce the shock wave reaches the region where neutrinos stream
out freely.  Electron neutrino luminosity suddenly increases by around
a factor of 10 in less than 1 msec and the fluxes of $\nu_{\mu,\tau}$,
$\anu_{\mu,\tau}$ are turned on (see sect.~\ref{sec:fluxes}).
Neutrino ``fireworks'' start.

In tens of seconds more than $10^{58}$ neutrinos of all flavors with
energies in the range 10-50 MeV, are emitted carrying out almost all
the gravitational binding energy released in the core collapse:

\beq
E_{B} \approx 3\times10^{53}\,\rm{erg}
\left(\frac{10\rm{km}}{R_\rm{NS}}\right)\,
\left(\frac{M_\rm{NS}}{1.4\,\rm{M}_\odot}\right)\,,
\label{eq:energy.released}
\eeq

where $R_\rm{NS}$ and $M_\rm{NS}$ is the final mass of the neutron
star formed in the process.

In the inner regions of the massive star the future of the shock wave
is uncertain.  More than thirty years of core collapse simulations
show that before leaving the core, the shock wave looses completely
its energy through photodisintegration of nuclear matter, and becomes
an almost static accretion front.  Then what produces the supernova
explosion?  No complete consensus still exists in the answer of this
important question \cite{Janka:2002ei,Buras:2003sn,Janka:2004jb}
although for sure, as was first realized twenty years ago
\cite{Bethe:1984ux}, neutrinos do play a central role.  The absorption
in the stalled accretion shock of even a small fraction of the huge
neutrino flux coming from the inner regions can be enough to revive
the shock wave and restart the explosion, hundred of milliseconds
after the initial core bounce.


This delayed explosion scenario is by now the preferred way to explain
to the ``supernova explosion enigma''.  However self consistent
hydrodynamic simulations have not yet been able to obtain a successful
explosion \cite{Janka:2002ei,Buras:2003sn,Janka:2004jb}.  The problem
does not seem to be the mechanism in itself, in the sense that it is
generally accepted that the delayed explosion occurs, and that
neutrinos play a central role.  More likely, other concomitant
processes (convection, rotation, magnetic fields, etc.) could give the
``last kick''.  However the processes at the moment are not completely
understood.

In summary, neutrinos are copiously produced in supernova and
(together with other not well understood concomitant processes) could
finally trigger the explosion itself.  However supernova neutrinos
can actually do more than producing the explosion.  The detection of
these neutrinos can open a door to understand the physics of
supernova by giving us an unique ``snapshot'' of the core collapse
processes in a way that no other signal (except perhaps the still
undetected gravitational radiation) can do.

Supernova neutrinos are also very interesting for particle physics.
On one hand they travel along huge distances, even in the astronomical
sense (10 kpc is a typical number which is $\sim 10^8$ times larger
than the Earth-Sun distance), and any small neutrino mass will produce
a tiny (but detectable) time delay among neutrinos of very different
energies.  On the other hand, its energy is in the right range to make
neutrino oscillations across the Earth ``visible'', allowing to
measure neutrino mixing parameters with unprecedented sensitivity.

In this work we will concentrate on the supernova neutrino potential
to give us information on the absolute scale of neutrino masses.

\medskip


During the past few years, atmospheric
\cite{Fukuda:1998mi,Fukuda:1998ah,Ambrosio:1998wu,
Ambrosio:2000ja,Sanchez:2003rb} and solar \cite{Ahmad:2001an,
Fukuda:1996sz, Cleveland:1998nv, Hampel:1998xg,Altmann:2000ft,
Fukuda:2001nj, Abdurashitov:2002nt,Smy:2003jf, Ahmed:2003kj} neutrino
experiments provided strong evidences for neutrino flavor oscillations
and therefore for non vanishing neutrino masses.  The KamLAND results
\cite{Eguchi:2002dm,Araki:2004mb} on the depletion of the $\anu_e$
flux from nuclear power plants in Japan, and the K2K indication of a
reduction in the $\nu_\mu$ flux from the KEK accelerator
\cite{Oyama:1998bd}, gave a final confirmation of this picture.

However, to date all the evidences for neutrino masses come from
oscillation experiments, that are only sensitive to squared mass
differences and cannot give any information on the absolute scale of
these masses.  The challenge of measuring the absolute value of
neutrino masses is presently being addressed by means of a remarkably
large number of different approaches, ranging from laboratory
experiments to a plethora of methods that relay on astrophysical and
cosmological considerations (for recent reviews see
\cite{Paes:2001nd,Bilenky:2002aw}).

From the study of the end-point of the electron spectrum in tritium
$\beta$-decay, laboratory experiments have set the limit $m_{\nu_e}< 2.2\,$eV
\cite{Bonn:2002jw,Lobashev:2001uu}. If neutrinos are Majorana particles, the
non observation of neutrino-less double $\beta$ decay can constrain a
particular combination of the three neutrino masses.  Interpretation of these
experimental results is affected by theoretical uncertainties related to
nuclear matrix elements calculations. This is reflected in some model
dependence of the corresponding limits, that lie in the range $m_\nu^\rm{eff}
< 0.2 - 1.3\,$eV \cite{Klapdor-Kleingrothaus:2000sn,
Klapdor-Kleingrothaus:2004wj, Aalseth:2002rf,Bilenky:2002aw}.


Tight bounds $\sum_i m_{\nu_i} < 0.6 - 1.8\,$eV have been recently set using
WMAP observations of cosmic microwave background anisotropies, galaxies
redshift surveys and other cosmological data (for a recent review see
\cite{Hannestad:2004nb} and references therein).  However, these limits become
much looser if the set of assumptions on which they rely is relaxed (see
\cite{Hannestad:2003xv,Elgaroy:2003yh,Crotty:2004gm} for discussions on this
point), and might even be completely evaded in exotic scenarios where
neutrinos can annihilate into hypothetical light bosons, implying a
suppression of their contribution to the cosmic matter density and negligible
effects on structure formation at large scales \cite{Beacom:2004yd}.


Zatsepin \cite{Zatsepin:1968aa} was probably the first to realize that
information on a neutrino mass could be provided by the detection of
neutrinos from a Supernova explosion.  Other early (independent)
contributions in studying the potential of supernova were put forth in
refs.~\cite{Pakvasa:1972gz,Piran:1981zz,1982Ap&SS..81..483S}.  The
basic idea relies on the time-of-flight (TOF) delay $\Delta
t_\rm{tof}$ that a neutrino of mass $m_\nu$ and energy $E_\nu$
traveling a distance $L$ would suffer with respect to a massless
particle:

\beq
\Delta t_\rm{tof} \simeq \frac{1}{2}\frac{m_\nu^2}{E_\nu^2}\,L.
\label{eq:tof.basic}
\eeq

Indeed, already in the past, the detection of about two dozens of
neutrinos from SN1987
\cite{Hirata:1988ad,VanDerVelde:1987hh,Alekseev:1988gp} allowed to set
upper limits on $m_\nu$.  Due to the low statistics, the model
independent bounds derived were only at the level of $m_{\overline
\nu_e}<30\,$eV \cite{Schramm:1987ra} while more stringent limits could
be obtained only under specific assumptions
\cite{Arnett:1987iz,Bahcall:1987nx,Spergel:1987ex,Abbott:1987bm} More
recently, a detailed reexamination of the SN1987 neutrino signal based
on a rigorous statistical analysis of the sparse data and on a
Bayesian treatment of prior informations on the SN explosion
mechanism, yielded the tighter bound $m_{\overline \nu_e}<5.7\,$eV
\cite{Loredo:2001rx}.

The first observation of neutrinos from a SN triggered in the years
following 1987 an intense research work aimed to refine the methods
for neutrino mass measurements, in view of a future explosion within
our Galaxy.  With respect to SN1987, the time delay of neutrinos from
a Galactic SN would be reduced by a factor of a few due to the shorter
SN-earth distance. However, the neutrino flux on Earth would increase
as the square of this factor and, most importantly, the large volumes
of the neutrino detectors presently in operation will yield a huge
gain in statistics.  In recent years several proposal have been put
forth to identify the best ways to measure the neutrino time-of-flight
delays, given the present experimental facilities.  Often, these
approaches rely on the identification of ``timing'' events that are
used as benchmarks for measuring the neutrino delays, as for example
the emission of gravitational waves in coincidence with the neutrino
burst \cite{Fargion:1981gg,Arnaud:2001gt}, the short duration $\nu_e$
neutronization peak that could allow to identify time smearing effects
\cite{Arnaud:2001gt}, the sudden steep raise of the neutrino
luminosity due to neutrino-sphere shock-wave breakout
\cite{Totani:1998nf}, the abrupt interruption of the neutrino flux due
to a further collapse of the star core into a black hole
\cite{Beacom:2000ng,Beacom:2000qy} ).  The more robust and less model
dependent limits achievable with these methods are at the level of
$m_{\nu}\lsim 3\,$eV.  Tighter limits are obtained only under specific
assumptions for the original profiles of the SN neutrino emission, or
for the astrophysical mechanisms that give rise to the benchmarks
events.


In this thesis we present a new approach to the measurement of the
absolute scale of neutrino mass from a future Galactic supernova
signal.  The new method proposed, which is formally based on the
principles of Bayesian reasoning, is rather independent on specific
astrophysical assumptions and makes use of the whole statistics of the
signal.  The method has been tested under a wide range of different
conditions for the neutrino emission, mixing schemes and for different
present and planned neutrino detectors.  The detailed results of such
tests are presented in this dissertation and allow us to conclude that
even with an approximate description of the neutrino flux and
spectrum, competitive mass limits can still be obtained from the
analysis of the next Galactic supernova signal.

We have organized this dissertation in four main chapters.  In
chapter~\ref{ch:fluxes} we review the information available about the
expected characteristics of the emitted neutrino fluxes and spectra.
We have put special attention on those characteristics that can be
identified as robust predictions regarding the supernova neutrino
emission.  We discuss in some detail how the physical processes
involved in the core collapse, shock wave emergence and revival
determine these robust features.  Chapter~\ref{ch:signal} is dedicated
to ``construct'' the detected signal according to the informations
acquired in chapter~\ref{ch:fluxes}.  The effect of neutrino
oscillations are also taken into account in this step.  Neutrino
flavor oscillations are determinant to model in a correct way an
observed supernova signal.  Although our aim will not be that of
studying oscillations signatures in supernova, the description of this
phenomenon is an obligatory aspect of any description of a neutrino
signal.  Using these informations we describe the procedure to
construct a neutrino signal, and we evaluate the expected main
properties of the signal under several mixing and detection
conditions.  Chapter~\ref{ch:mass} is devoted to explain with more
detail the different methods that have been devised to measure or
constrain a neutrino mass studying supernova neutrino.  More
importantly, we describe in this chapter with full detail the
statistical method that we have put forth to measure a neutrino mass
from the analysis of the whole statistics in a future supernova
signal.  The method has been tested with the procedures described in
the final chapter~\ref{ch:results}, where the results obtained from
such tests are presented in full detail.  A comprehensive comparison
of various results and an analysis of the conclusions that can be
drawn from our study is also contained in this last chapter.

In Appendix~\ref{ap:bayes} we describe the main ideas and
mathematical formalism of the Bayesian approach to data analysis.
Appendix~\ref{ap:sung} is devoted to present a simple computer tool
(publicly available) that has been designed specifically to generate
synthetic supernova signals.  Some of the underlying algorithms and
numerical techniques used in the development of our computer codes are
also presented in this appendix.

A complete list of symbols and abbreviations used along the text is
given in Appendix~\ref{ap:abbrev}.

\chapter{Neutrino fluxes and spectra}
\label{ch:fluxes}

Before proceeding to construct numerical simulations of a Supernova neutrino
signal it is necessary to establish which are the features of the neutrino
emitted signal for which the theory is reliable and which features are more
uncertain.

Although different Supernova simulations produce neutrino signals with
different sets of properties, there are a set of overall features common to
almost every result:

\begin{enumerate}

\item The neutrino luminosities and fluxes evolve with time in a rather well
  understood way.  In the first tens of milliseconds the luminosity and flux
  of all neutrino flavors increase rapidly.  After this, the luminosity and
  flux stabilize in their maxima and start to decay slowly for several hundred
  of milliseconds.  A slow decay on a time-scale of several seconds follows
  and finally the signal turns off.
  
\item The supernova core is the only place besides the early Universe where
  neutrinos are in thermal equilibrium.  The observational consequence of this
  fact is the emission of a quasi-thermal neutrino spectrum.
  
\item The average energies of different neutrino flavors are hierarchical,
   i.e.  muon- and tau-(anti)neutrinos (hereafter we will refer them globally
   as $\nu_x$) have a harder spectrum than the electron antineutrinos
   ($\anue$) which in turn are hotter than the electron neutrinos ($\nue$).

\end{enumerate}

Despite the wide range of detailed results obtained in supernova simulations
during the last three decades, we can be confident that a future supernova
signal will exhibit these properties.

In this chapter we will describe in some detail these robust features
of the neutrino emission from supernovae, its physical motivation and
how they are supported by self-consistent supernova simulations.  In
order to be prepared for the chapters where this information will be
used to generate and analyze realistic supernova signals, we have
attempted to translate some of the described properties into
analytical expressions.

A more comprehensive and complete description of the emitted neutrino
properties can be found in refs.
\cite{Mayle:1986ic,Burrows:1991kf,Keil:2003sw} and references therein.

\section{Evolution of Neutrino fluxes}
\label{sec:fluxes}

One of the best known features of neutrino emission from Supernovae is the way
the neutrino fluxes evolve with time (we are interested only in times after
the core-bounce).

Although the detailed time structure and some of the time-scales of the
neutrino fluxes differ from one simulation to another, it is possible to
recognize the existence of mainly three different phases in the evolution of
the rates.  Each of these phases are directly related to well known physical
processes in the collapsed core of the star.

The main phases of neutrino's fluxes evolution is illustrated in figure
\ref{fig:flux.evolution}.

\subsection{Fast rising - the shock breakout}
\label{subsec:shock.breakout}

Once the inner region of the collapsing stellar core reach supra-nuclear
densities a hydrodynamical shock wave forms and starts to travel out. After
less than few milliseconds the shock wave reach the region of the core where
neutrinos escape freely.  A sudden increase in the plasma temperature produced
by the energy deposited by the shock, together with the sudden transition from
coherent scattering on ``iron'' nuclei to scattering on nucleons produce a
huge release of electron neutrinos (the {\it prompt neutrino burst} or {\it
deleptonization burst}).

\begin{figure}[h]
\begin{center}
\epsfxsize=120mm
\epsfbox{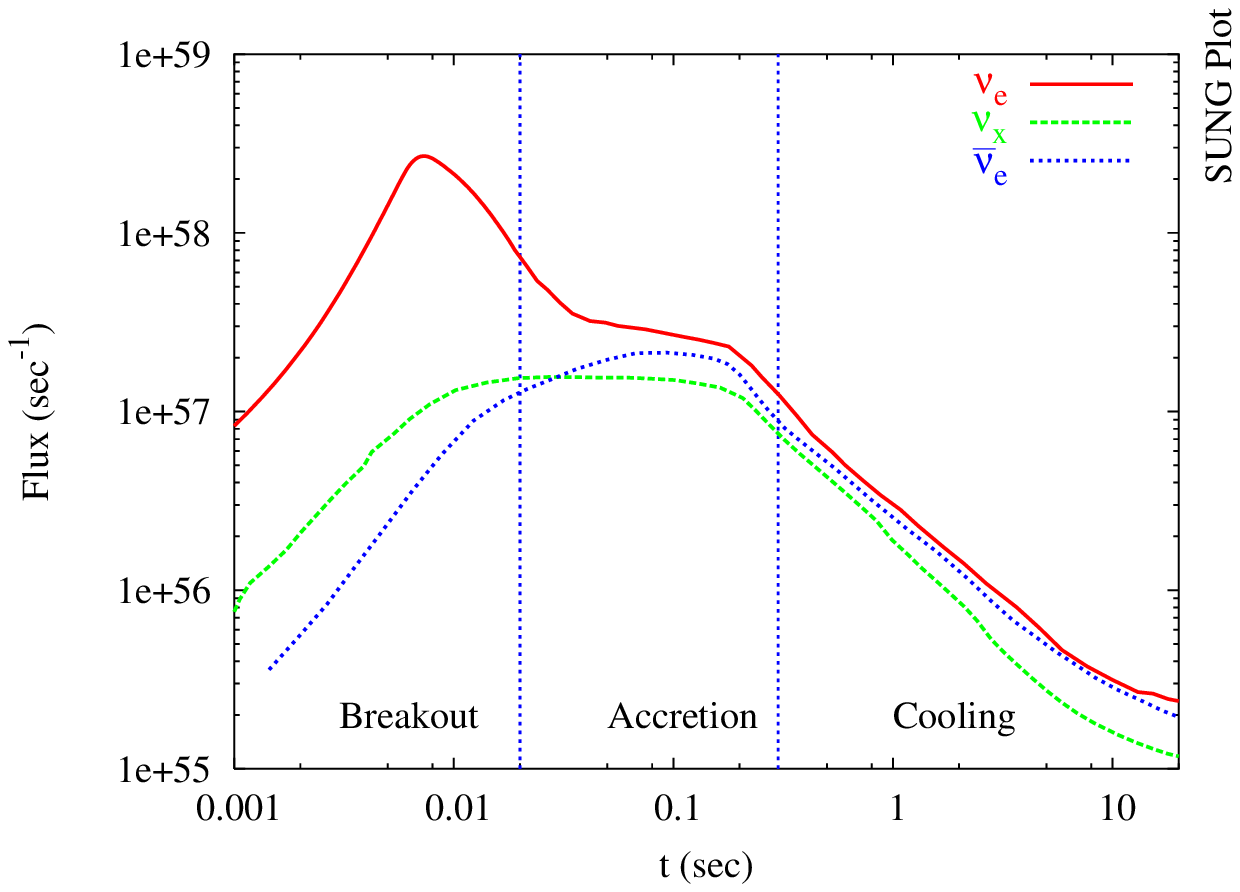}
\mycaption{Neutrino fluxes evolution}{
Neutrino fluxes evolution could be described with three main phases: shock
breakout (fast rising), accretion and mantle cooling (maximum and plateau) and
Kelvin-Helmholtz cooling (decay).  The fluxes were taken from simulations by
Totani et al. \cite{Totani:1998nf}.
}
\label{fig:flux.evolution}
\end{center}
\end{figure}

The shock breakout turn on the emission of $\nux$ in less than 1 msec
(neutrino-sphere crossing time of the shock wave \cite{Burrows:1991kf}).  After
this turn-on the heat from the shocked material above the neutrino-sphere and
below the shock wave produces a steadily increase in the flux that last from
20 to 50 msec (see figure \ref{fig:flux.evolution})

%

\subsection{Flux maximum - accretion, mantle cooling and explosion}
\label{subsec:acretion}

Neutrino emission and nuclear dissociation consume energy of the emergent
shock wave. Even before it leaves the iron core, the shock wave becomes an
almost stationary accretion front.  In these conditions a new source of
neutrinos appears.  The inner core (protoneutron star) emits its heat through
neutrinos in a steady fashion.  The same thing happens to the hot material
below the accretion front (core mantle).  However, as new material is accreted
in the stalled shock, it is dissociated and produce extra $\nue$ by electron
capture on protons.  The hot material in the shock front also produces
electron-positron pairs and the capture of positrons on neutrons, in the
neutron rich material of the mantle, creates an additional $\anue$ flux. The
emission of neutrinos during this phase is characterized by a slow decay in
the flux on a time-scale of one hundred of milliseconds.  

The emission of neutrinos by accretion is suddenly truncated when the shock
wave revives and the delayed explosion starts.

\subsection{Decay - Kelvin-Helmholtz cooling}
\label{subsec:cooling}

Once the revitalized shock wave emerges a more stable and long phase starts.
The gravitational binding energy of the newly formed protoneutron star
(\ref{eq:energy.released}) is radiated as neutrinos on time-scales comparable
with the diffusion time-scale ({\it Kelvin-Helmholtz cooling}).

This phase is characterized by a power law decrease in the neutrino flux and
the emission of an important fraction of the protoneutron star binding energy.
The Kelvin-Helmholtz cooling phase last for tens of seconds to minutes and it
ends when the material of the newly formed neutron star becomes transparent to
neutrinos.

\bigskip

As was mentioned before this basic description of the overall features
of the emitted neutrino flux is a robust prediction of supernova
models.  Its global features have been confirmed by the observation of
the neutrino signal from SN1987A.  The initial short burst of
neutrinos ($\sim 1$ sec) observed in the signal is consistent with an
accretion phase \cite{Loredo:2001rx} and the observed total duration
of the signal ($\sim 10$ sec) confirms the emission of neutrinos from
a cooling neutron star \cite{Spergel:1987ch}.

\subsection{Analytical flux model}
\label{subsec:analytical}

In order to quantify the overall features of the flux described above
it is possible to construct a simplified model for the neutrino
emission.  Such a model will be an useful tool to understand the
characteristics of fluxes obtained in simulations.  In section
\ref{subsec:flux.model}, we will start with this analytical approximation
to construct a test model to analyze a realistic neutrino signal.

We can recognize two main components in neutrino emission:

\begin{itemize}

\item {\it Cooling component}.  On a crude level of approximation neutrino emission
from the supernova core and mantle could be written in terms of the
Stefan-Boltzmann law (see \cite{Janka:1995cu} for a detailed discussion),

\beq
\fder{n}{t}^\rm{em} = 4\pi R_\rm{eff}(t)^2 \int F(E;T_\rm{eff}(t)) dE = C_c
R_\rm{eff}(t)^2 T_\rm{eff}(t)^4\,,
\label{eq:cooling.flux}
\eeq

where $R_\rm{eff}$ is the effective radius where neutrinos are emitted and a
quasi-thermal spectral distribution $F(E;T_\rm{eff}(t))$ is assumed, with
$T_\rm{eff}(t)$ as the spectral temperature.  $C_c$ is a constant in which we
have absorbed geometrical and physical terms.

\item {\it Accretion component}.  When accretion starts an additional
  contribution to the flux is turned on.  $\beta$-processes producing $\nue$
  and $\anue$ neutrinos in the accreting matter are in local thermal
  equilibrium (LTE) and thus neutrino spectra could be also described as thermal
  with a temperature equal to the accreted material temperature $T_a$.  The
  neutrino emission rate per unit of mass of accreted material will be given
  by \cite{Loredo:2001rx},

\beq
\frac{dn^\rm{em}}{dt}\sim M_a(t)\int E^2 F(E;T_a) dE = C_a(T_a) M_a(t).
\label{eq:accretion.flux}
\eeq

  The $E^2$ term arises from the energy dependence of the $\beta$-processes
  cross-section.  $C_a(T_a)$ is a parameter depending on the spectral
  temperature of the accreted matter that can be assumed constant in time.
  $M_a(t)$ is the mass of accreted material at time $t$.

\end{itemize}

When the cooling emission component (\ref{eq:cooling.flux}) and the accretion
component (\ref{eq:accretion.flux}) are summed up the total flux on $\nue$ and
$\anue$ is

\beq
\frac{dn^\rm{em}}{dt} = C_c R_\rm{eff}(t)^2 T_\rm{eff}(t)^4 + C_a(T_a) M_a(t).
\label{eq:analytical.flux}
\eeq

As can be seen the temporal evolution of neutrino flux is a function
of the temporal evolution of the effective emission radius
$R_\rm{eff}$, the effective thermal temperature $T_\rm{eff}$ and the
rate of matter accretion $M_a$.  Different phenomenological models
could be formulated to describe the behavior of this quantities.  For
the cooling component it is common to use a exponential cooling law,
i.e. $T_\rm{eff}=T_o e^{-t/4\tau}$ and assume an almost constant
effective neutrino-sphere radius.  This kind of models seems to fit
well the gross behavior of the SN1987A signal \cite{Spergel:1987ch}.
In a more complete and detailed reanalysis of SN1987A signal Loredo
and Lamb \cite{Loredo:2001rx} has recently found that a power-law
cooling model, $T(t)=T_o (1+t/4 \gamma \tau)^{-\gamma}$ which is
capable to reproduce a wide range of cooling behaviors (e.g. when
$\gamma\rightarrow \infty$ we obtain an exponential cooling) could fit
even better the signal \cite{Loredo:2001rx}.

For the accretion component a truncated power law decay for the mass of
accreted matter, $M_a(t)=M_o e^{-(t/t_a)^{n_a}}(1+t/t_b)^{-n_b}$, seems to
fit well supernova simulations results \cite{Loredo:2001rx}.

Using the preferred phenomenological models the final neutrino flux for times
posterior to the shock breakout (\ref{eq:analytical.flux}) is written as

\beq
\frac{dn^\rm{em}}{dt} = C\frac{1}{(1+t/\gamma\tau)^{\gamma}} +
\frac{e^{(-t/t_a)^{n_a}}}{(1+t/t_b)^{n_b}}\,,
\label{eq:flux.model}
\eeq

where $C\equiv C_c T_o$ and $A\equiv C_a(T_a)M_o$ and we have
reparametrized the cooling component in terms of $n_c=4\gamma$ and
$t_c=\gamma\tau$.  Note that this flux does not behaves properly for
$t\to 0$ since it is aimed to describe the emission of neutrinos for
times after the shock breakout.  In order to use this analytical
approximation to fit a signal we will modify
(\ref{eq:analytical.flux}) in section~\ref{subsec:flux.model} to
ensure the proper asymptotic behavior of this function.

\begin{figure}[h]
\begin{center}
\epsfxsize=120mm
\epsfbox{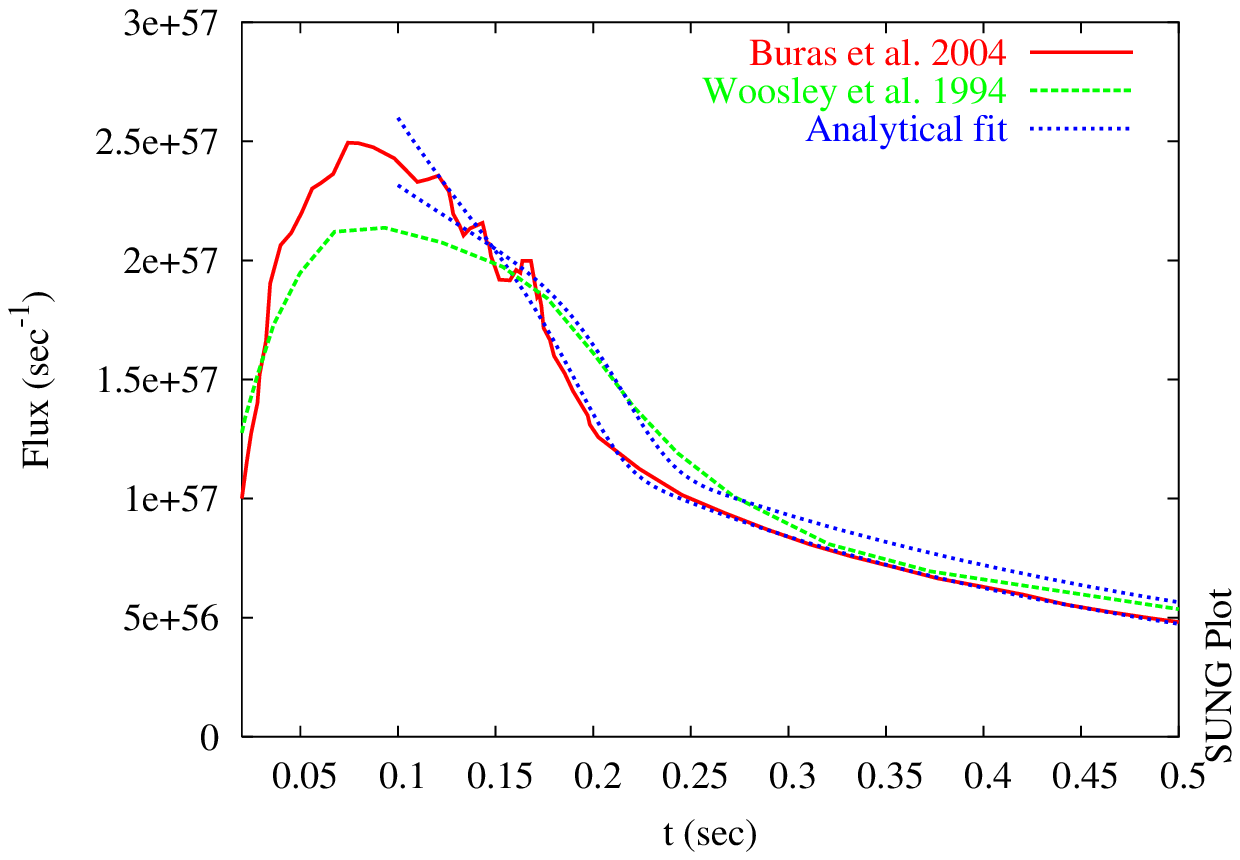}
\mycaption{Schematic analytical fits of $\anue$ fluxes from two supernova
simulations}{
Schematic analytical fits of $\anue$ fluxes from two supernova simulations
using the flux model (\ref{eq:flux.model}).
}
\label{fig:analytical.real}
\end{center}
\end{figure}

In figure \ref{fig:analytical.real} we compare the $\anue$ and $\anux$ fluxes
from supernova simulations with this analytical model.  As can be seen the
flux model (\ref{eq:flux.model}) reproduces well the behavior of the flux
in the accretion and cooling phases.

%

\begin{figure}[p]
\begin{center}
\epsfxsize=120mm
\epsfbox{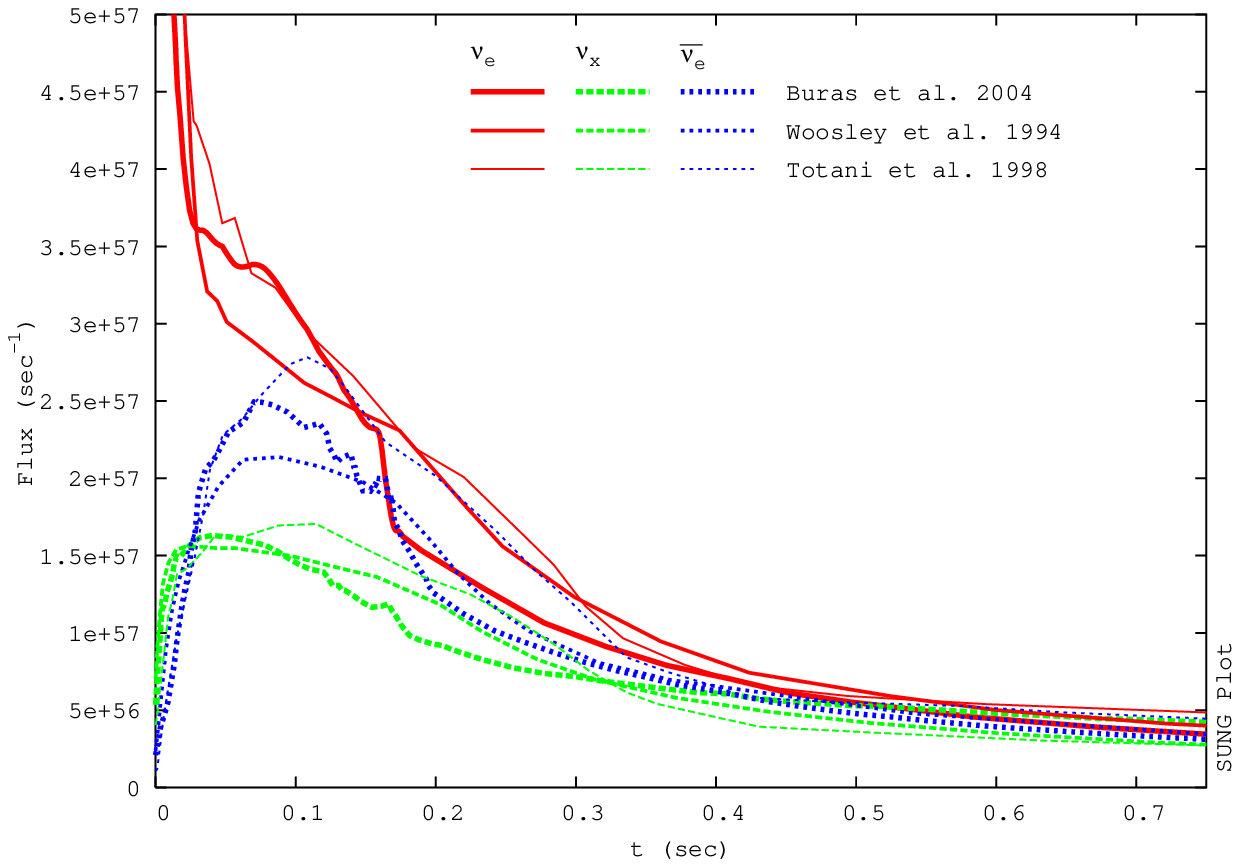}
\mycaption{Supernova neutrino fluxes from three different self-consistent
simulations}{
Supernova neutrino fluxes from three different self-consistent simulations:
Totani et al. \cite{Totani:1998nf}, Woosley et al. \cite{Woosley:1994ux} and
Buras et al. \cite{Buras:Private2004}.  The main overall features of the
fluxes described in the text could be recognized here.
}
\label{fig:supernova.flux}
\end{center}
\end{figure}

Finally in figure \ref{fig:supernova.flux} we present fluxes of all neutrino
flavors coming from three different supernova simulations.

%


\section{Neutrino Spectra}
\label{sec:spectra}

Neutrinos trapped inside the inner regions of a supernova core are maintained
in thermal equilibrium with the surrounding plasma through inelastic
scattering.  For this reason it is expected that they will be emitted with a
quasi-thermal spectrum characterized by a spectral temperature close to the
temperature of the plasma in the region where neutrinos decouple
energetically.  More than 30 years of analytical and numerical studies of
neutrino spectra formation in supernova have confirmed this basic picture.
However, the fine details of neutrino transport and emission near to the last
scattering surface are highly non trivial and the precise determination of the
spectral properties of the emitted neutrinos is not a simple task.

Let us examine here the main features expected for the neutrino
spectra and its origin.  A more comprehensive and complete discussion
could be found in ref. \cite{Keil:2003sw} and references therein.

\subsection{Electron neutrinos and antineutrinos}
\label{subsec:electron-neutrinos}

The production and transport of $\nue$ and $\anue$ is simple when
compared with the transport of $\nux$.  Due to the presence of
electron and positrons, $\nue$ and $\anue$ can undergo charge current
reactions ($\beta$-processes $ep\leftrightarrow n\nue$,
$e^+n\leftrightarrow p\anue$) that are almost absent for all other
flavors.  With a cross-section much larger than other scattering
processes, $\beta$-processes dominate completely the transport of
$\nue$ and $\anue$ inside the supernova core.  The continuous creation
and capture of neutrinos in these processes inside the core maintain
$\nue$ and $\anue$ in LTE with the surrounding medium.

\begin{figure}[h]
\begin{center}
\epsfxsize=120mm
\epsfbox{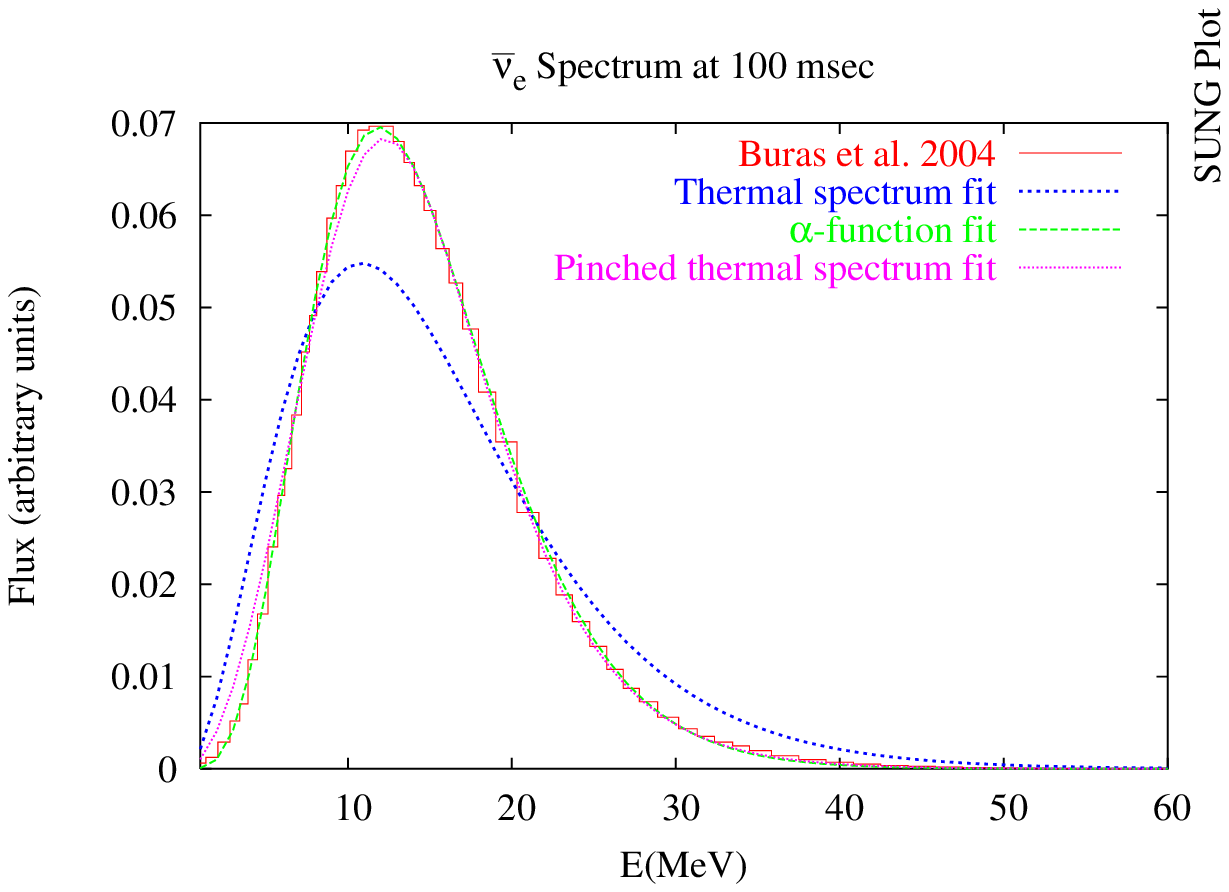}
\vskip-4mm
\mycaption{$\anue$ spectrum at 100 msec}{
$\anue$ spectrum at 100 msec (accretion and mantle cooling phase) taken from
the simulation of Buras et al. \cite{Buras:Private2004} (histogram).  The
dotted curve correspond to a Fermi-Dirac fit with zero chemical potential.
The pinching effect, i.e. the depletion of the distribution at high and low
energy, can be clearly recognized in the distorted Fermi-Dirac and
$\alpha$-distribution fits.  As can be seen, for this particular case the
$\alpha$-distribution fits slightly better the spectrum than the pinched
Fermi-Dirac.
}
\label{fig:pinching}
\end{center}
\end{figure}

Neutrinos diffuse outward until the $\beta$-processes become ineffective at
the neutrino-sphere.  Since the cross section of $\beta$-processes are energy
dependent the neutrino-sphere radius and therefore the characteristic thermal
temperature for $\nue$ and $\anue$, are different for neutrinos of different
energies.  High energy neutrinos will stream out from larger radius with lower
temperatures while low energy neutrinos will be radiated from a lower radius
and higher temperatures.  Therefore the energy distribution of the emitted
neutrinos will not be exactly thermal.  The emitted neutrino spectra will be
``pinched'', i.e. high energy and low energy tails of the spectrum will be
depleted due to the different thermal temperatures characteristic of the
radius where neutrinos of different energies are emitted (see
fig.~\ref{fig:pinching}).


%

\subsection{Non-electron neutrinos}
\label{subsec:nonelectron-neutrinos}

The emission of $\nux$ from the supernova requires a more complicated
description.  In this case different neutral current reactions play comparable
roles in the neutrino transport.

$\nux$ can undergo three kind of processes: creation reactions, energy
exchange scattering and isoenergetic scattering.

\begin{figure}[h]
\begin{center}
\epsfxsize=120mm
\epsfbox{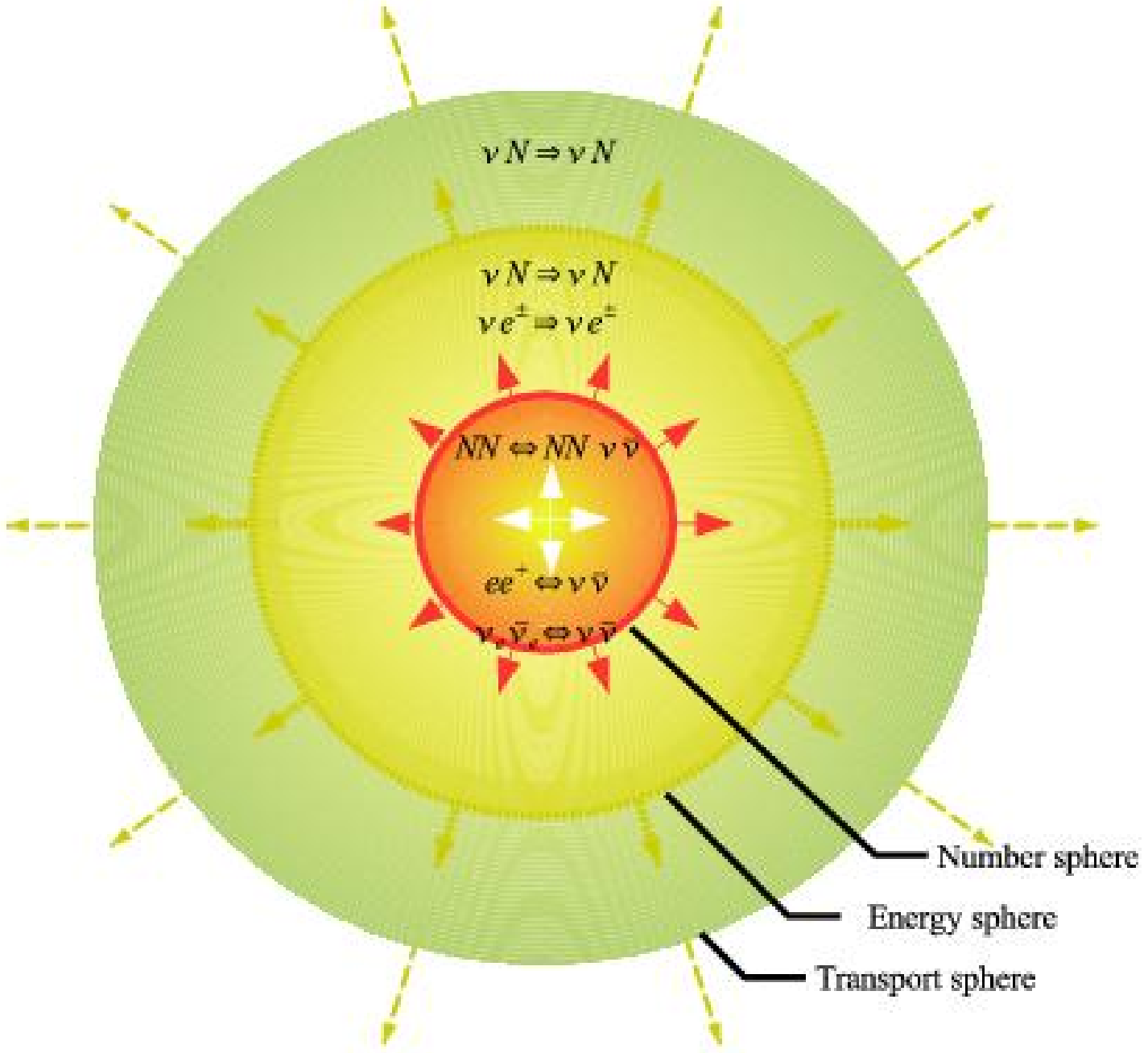}
\mycaption{Schematic representation of $\nux$ transport}{
Schematic representation of $\nux$ transport.  Neutrinos are produced and
annihilate through nucleon bremssthralung and pair processes.  When
creation/annihilation freeze-out at the number sphere the total flux is fixed
(number of arrows).  Neutrinos maintain a thermal contact with the medium
until they reach the energy sphere (style of lines indicates the neutrino
average energy).  Finally above the transport sphere (neutrino-sphere)
neutrinos freely stream out.  Its average energy at this layer (style of
arrows) is close to that at the energy sphere.
}
\label{fig:schem.transport}
\end{center}
\end{figure}

Deep inside the core neutrinos are created and annihilated through three main
processes: nuclear bremsstrahlung $NN\leftrightarrow NN\nu\nu$,
electron-positron annihilation $ee^+\leftrightarrow \nux\nux$ and
neutrino-neutrino annihilation $\nue\anue\leftrightarrow \nux\nux$.  There,
neutrinos are in LTE and trapped.  At a given radius creation/annihilation
processes freeze out and no more neutrinos are created.  This condition
defines what is called the {\it number sphere}.  The emergent neutrino total
flux is fixed in this layer.  Outside the number sphere neutrinos still
undergo energy exchange scattering on electrons, positrons and nucleons and
maintain some thermal contact with the plasma.  When density and temperature
is lower inelastic scattering on nucleons and $e^\pm$ becomes inefficient and
neutrinos decouple energetically from the surrounding medium.  This condition
defines the so-called {\it energy sphere}.  Above the energy sphere neutrinos
still interact strongly with the plasma through elastic scattering on
nucleons.  Finally at some radius elastic scattering becomes in turn
inefficient and neutrinos stream out from the neutrino-sphere or {\it transport
sphere}.  In figure~\ref{fig:schem.transport} we illustrate schematically this
process.

\afterpage{\clearpage} 
%

What will be the emergent energy distribution of $\nux$?  Naively one
could think that since thermal contact is maintained only inside the
energy-sphere neutrino average energy will be related with the plasma
temperature at that layer.  However the energy dependence of the
nucleon scattering outside the number-sphere acts as a filter
\cite{Raffelt:2001kv} lowering the average energy to 50-60\% its value
close to the energy sphere.

The complex interplay of all factors which determine the emergent $\nux$ flux
makes harder to predict exactly which will be the spectral properties of these
flavors.  We will come back on this issue in section \ref{subsec:hierarchy}.

\subsection{Spectral pinching}
\label{subsec:pinching}

The quasi-thermal nature of the neutrino spectra is a robust feature of
neutrino emission from a supernova.
 
There are two ways to quantify the non-thermal distortion (pinching effect) of
the neutrino spectra.  On one hand it is possible to use higher order momenta
of the neutrino energy distribution that contain informations about the high
energy tail.  Raffelt \cite{Raffelt:2001kv} first quantifies the pinching
effect by introducing a ``pinching parameter'' defined in terms of the ratio
of the second and first momentum

\beq
p\equiv \frac{1}{a}\frac{\aver{E^2}}{\aver{E}}
\label{eq:pinching.par}
\eeq

Where $a\simeq 1.3029$ is a weight constant equal to the ratio of the second
and first momentum of a Fermi-Dirac distribution with zero chemical
potential.  In the absence of any pinching and without degeneracy $p=1$.  A
pinched spectrum will be characterized by $p<1$ while an anti-pinching,
i.e. an enhancement of the spectrum tails, corresponds to $p>1$.

On the other hand has been common in the literature to fit the supernova
neutrino spectrum using a Fermi-Dirac distribution with an effective
degeneracy parameter $\eta$ that enters the distribution like an effective
chemical potential:

\beq
F_\rm{FD}(E;T,\eta)\propto \frac{E^2}{1+e^{E/T-\eta}}.
\label{eq:fit-fermi}
\eeq

The introduction of $\eta$ allows to fit the distorted tails of the
distribution.  Its value could therefore used as a measure of the spectrum
pinching.

Recently an alternative two-parameter distribution has been proposed
\cite{Raffelt:2003en,Keil:2002in}:

\beq
F_\a\left(E;\bar\epsilon,\alpha\right) = N(\bar\epsilon,\alpha)\> 
\left({E}/{\bar\epsilon}\right)^{\alpha}e^{-(\alpha+1)\,E/\bar\epsilon}\,,
\label{eq:alpha.dist}
\eeq

where the normalization constant is given by
$N(\bar\epsilon,\alpha)={(\alpha+1)^{\alpha+1}}/{\Gamma(\alpha+1)
\bar\epsilon}$.  It has been shown that this new distribution, denoted {\it
$\alpha$-distribution}, in \cite{Raffelt:2003en} fits slightly better the
neutrino spectra and has the nice property of allowing a simple analytical
estimation of the two spectral parameters $\bar\epsilon$ and $\alpha$ directly
in terms of the first and second momentum of the energy distribution.  Using
the well known relation $\alpha\,\Gamma(\alpha)=\Gamma(\alpha+1)$ it is easy
to verify that:

\beq
\bar\epsilon = \langle E \rangle\,; \qquad \qquad 
\frac{2+\alpha}{1+\alpha}= \frac{\langle E^2 \rangle}{\langle E \rangle^2}.
\label{eq:alpha.momenta}
\eeq

Using these relations, the pinching parameter $p$ could be expressed in terms
of $\alpha$ as:

\beq
a\,p=(2+\alpha)/(1+\alpha).
\label{eq:pinching-alpha}
\eeq

Therefore when an $\alpha$-distribution is used to fit the neutrino spectrum
the value of the $\alpha$ parameter can quantify the spectrum pinching.

Figure \ref{fig:pinching} illustrate the quality of fits of a numerical
neutrino spectrum with a Fermi-Dirac and an $\alpha$-distribution.

\begin{sidewaysfigure}
\begin{center}
\epsfxsize=180mm
\epsfbox{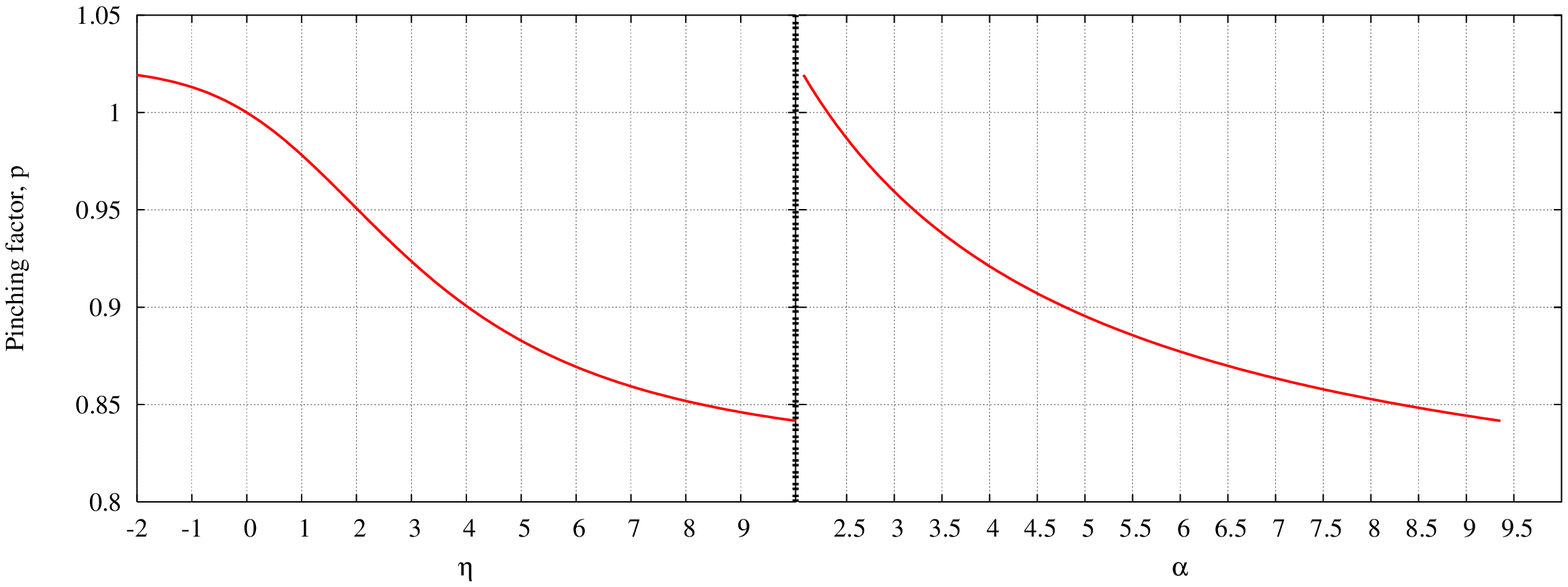}
\mycaption{Relation between pinching $p$, $\alpha$ and the effective
degeneracy parameter $\eta$}{
Relation between the pinching parameter $p$, $\alpha$ and the effective
degeneracy parameter $\eta$ used to quantify the thermal distortion of
supernova neutrino spectra.
}
\label{fig:eta.alpha.pinching}
\end{center}
\end{sidewaysfigure}
\afterpage{\clearpage} 
%

\begin{figure}[p]
\begin{center}
\epsfxsize=90mm
\epsfbox{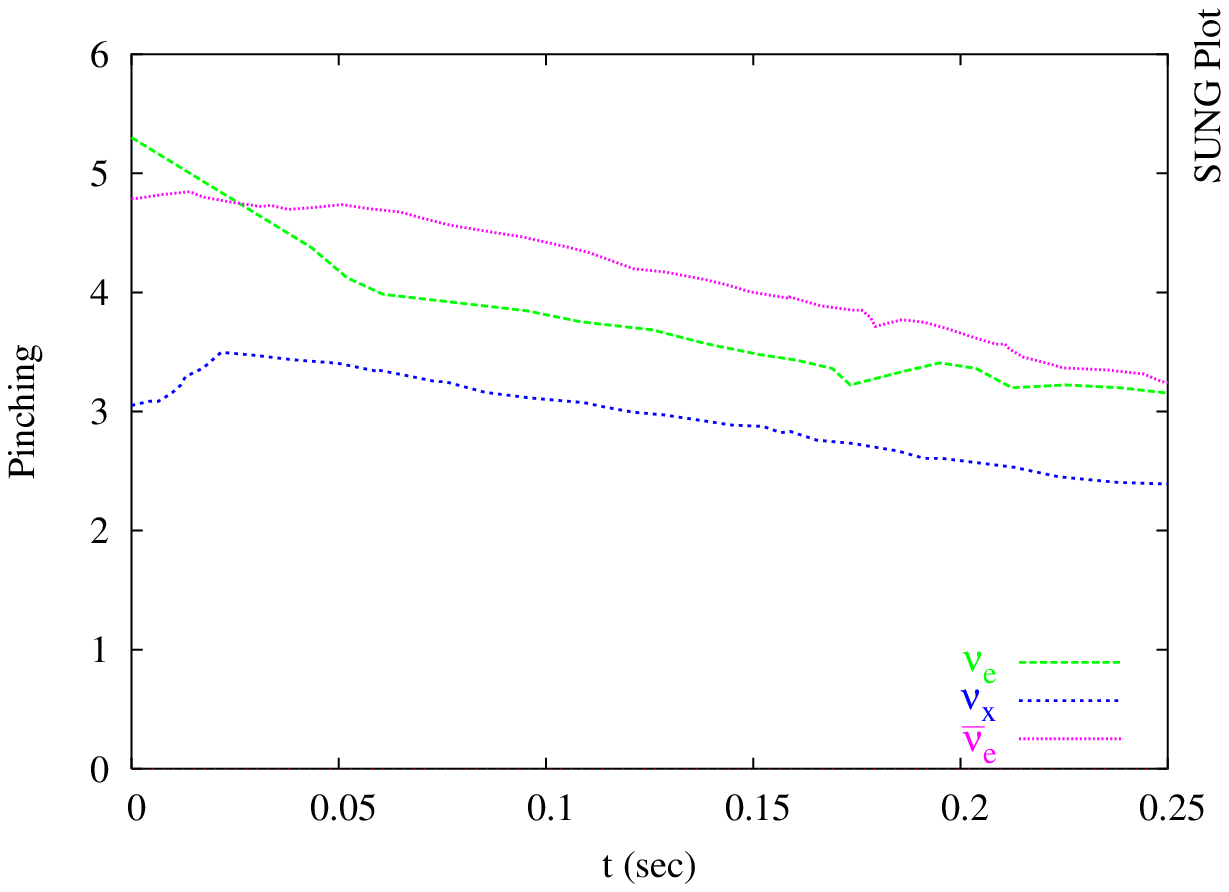}
\vskip-4mm
\mycaption{$\alpha$ parameter evolution}{
Evolution of the $\alpha$ parameter in the supernova simulation by Buras et
al. \cite{Buras:Private2004}.  The pinching gets reduced at later times when
neutrinos are emitted from more compact regions inside the protoneutron star.
}
\label{fig:real.pinching}
\end{center}
\end{figure}

The relation between $p$, $\eta$ and $\alpha$ is depicted in figure
\ref{fig:eta.alpha.pinching}.  Typical values for $\alpha$ from a
self-consistent supernova simulation are presented in figure
\ref{fig:real.pinching}.  As can be seen the spectral pinching evolves to
lower values as the protoneutron star and the core mantle settles and neutrino
emission is confined to more compact regions \cite{Raffelt:2003en}.

Typical ranges for the spectral pinching obtained in simulations are
\cite{Raffelt:1996wa,Raffelt:2003en}:

\beq
\begin{array}{ccccc}
p_\nue\simeq 0.88\div 0.92 &,& \eta_\nue \simeq 3\div5 &,&
\alpha_\nue\simeq 4.0\div5.6\\
p_\anue\simeq 0.90\div 0.95 &,& \eta_\anue\simeq 2\div4 &,&
\alpha_\anue\simeq 3.0\div5.0\\
p_\nux\simeq 0.95\div 1.00 &,& \eta_\nux\simeq 0\div2 &,&
  \alpha_\nux\simeq 2.3\div3.2
\end{array}
\label{eq:typical-pinching}
\eeq

%

\section{Hierarchy of average energies}
\label{subsec:hierarchy}

The last robust feature of the emission of neutrinos from supernova is the
hierarchy of the average energies of different neutrino flavors.

As was discussed in the previous section the spectral properties of
neutrinos emitted from the supernova core are determined by the
transport processes near the neutrino-spheres.  In the simple case of
$\nue$ and $\anue$ the average energy is determined by the temperature
of the plasma around the neutrino-spheres.  Although the cross section
for $\b$-processes (the main source of opacity for these flavors) has
the same energy dependence for $\nue$ and $\anue$, the asymmetry
between protons and neutrons in the supernova core produces a
significant difference in the opacity of antineutrinos and neutrinos.
Since neutrons are more abundant, $\nue$ decouple from matter at a
larger radius than $\anue$, and therefore has a lower average energy,
i.e. $\aver{E_\nue}<\aver{E_\anue}$.

\begin{sidewaysfigure}
\begin{center}
\epsfxsize=200mm
\epsfbox{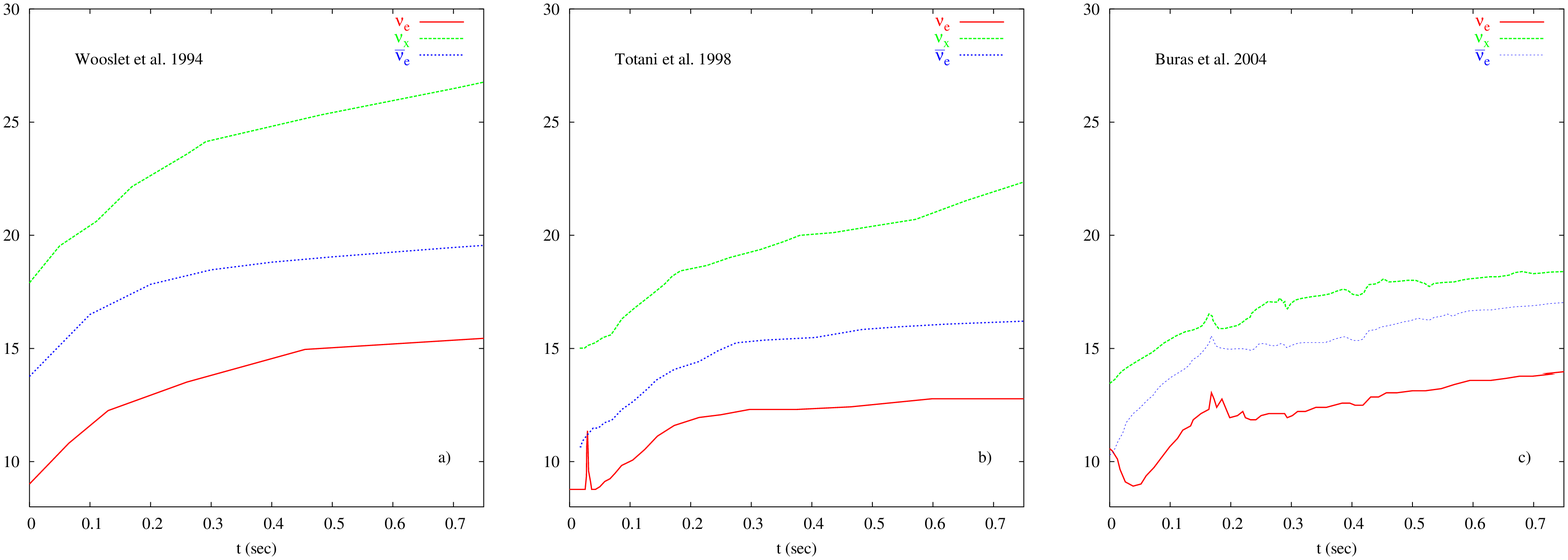}
\mycaption{Mean energy evolution from three different supernova simulations}{
Evolution of the mean energy from three different supernova simulations.  It
is apparent the big difference between the ``standard'' result of large
spectral differences in simulations where a simplified description of $\nux$
transport was used (panels a and b) compared with almost equal average
energies of $\nux$ and $\anue$ when a more precise and complete description of
$\nux$ interactions are included (panel c).
}
\label{fig:mean.energies}
\end{center}
\end{sidewaysfigure}

The size of the difference between the average energies of $\nue$ and
$\anue$ depends basically on the core structure (density, temperature
and lepton number profile) which determines the position of the
respective neutrino-spheres.  In supernova simulation typical values
ranges from 30\% \cite{Buras:Private2004,Burrows:1991kf} to 50\%
\cite{Woosley:1994ux,Totani:1998nf} (see figure
\ref{fig:mean.energies}).

\afterpage{\clearpage}
%

On the other hand the transport of $\nux$ is dominated by neutral current
reactions that have smaller rates than the $\b$-processes.  This implies that
$\nux$ decouples at a lower radius than $\anue$ and therefore it is expected
that $\aver{E_\nux}>\aver{E_\anue}>\aver{E_\nue}$.  However the determination
of how large are the differences between the average energies of $\nux$ and
$\anue$ is not easy.  If the average energies were only determined by the
spectral temperature at the energy-sphere, large differences would be expected.
However the filter effect of the elastic reactions outside the energy-sphere
for $\nux$ contribute reducing the average energy.  It has been shown in
ref. \cite{Raffelt:2003en} that the final difference will depend on the
details of the $\nux$ transport in that regions.

Most of the self-consistent simulations have made use of simplified
descriptions of $\nux$ transport mainly due to the fact that $\nux$ does not
play an important role in the supernova dynamics.  In these simulations the
resulting $\nux$ spectra has typical energies differences that range from 30\%
up to 100\%! (see figures \ref{fig:mean.energies}a,b).

Large spectral differences of this size have been exploited in several works
to study the potential information contained on a supernova signal.  Minakata
et al. \cite{Minakata:2000rx} have claimed that an inverse hierarchy of
neutrino masses could be rejected using the informations contained in the
SN1987A signal.  Lunardini and Smirnov \cite{Lunardini:2001pb} have derived
similar conclusions about the potential of SN1987A signal to give information
about neutrino oscillations parameters.  The potential of a future supernova
signal has also be studied in the presence of large spectral differences
\cite{Dighe:1999bi}.

Recently the inclusion of a detailed description of $\nux$
microphysics in self-consistent hydrodynamical supernova simulations
(see ref. \cite{Keil:2003sw} has suggested that the standard picture
of large spectral differences could be wrong.  Including all the
relevant reactions in the $\nux$ transport, and other effects that had
not been taken into account in previous simulations, the $\nux$
average energy results almost equal to that of $\anue$.  Although a
partial hierarchy could still be present (at least during part of the
emission) differences will not be larger than 20\%, with 10\% as a
typical value \cite{Keil:2003sw}.  Panel {\it c} in figure
\ref{fig:mean.energies} illustrates this point.

We conclude that, according to the most recent results, any attempt to
obtain informations on neutrino mixings and their mass hierarchy from
a Galactic supernova signal will have to take into account that the
differences between the neutrino energies will be probably not very
large.

\chapter{Simulation of a supernova neutrino signal}
\label{ch:signal}

Using what we have learned about neutrino emission in Supernova we can now
proceed to calculate the expected characteristics of a future supernova
signal.  One main part of this work regards using this description to create
artificial signals that will be later used to test the statistical method that
will be presented in the next chapter.

In this chapter we present the theoretical description of a supernova neutrino
signal and the general strategies that can be used to generate synthetic
realizations of such signals.  In section~\ref{sec:oscillations} we present
the most relevant ideas about supernova neutrino oscillations which will play
a central role in the determination of the final features of any neutrino
signal. Section~\ref{sec:signal.form.} introduces the formalism and describes
in some details how we model the neutrino detection process.  In
section~\ref{sec:sign.char} we compute the most relevant features of a future
supernova neutrino signal, assuming different emission models and oscillation
effects and analyze the interplay of the different aspects which are involved
in the determination of the signal characteristics.  Finally
section~\ref{sec:sign.generation} describes the particular strategies,
algorithms and numerical techniques used to create synthetic signals.

The material presented in this chapter is a collection of useful results
obtained from rather standard descriptions of neutrino propagation and
detection phenomenology.  We intend just to present synthetically the
theoretical results behind the computer codes used in our work.

\section{Oscillations of supernova neutrinos}
\label{sec:oscillations}

Neutrinos will undergo oscillations on their way from the supernova core to a
detector.  First, when they traverse the outer shells of the stellar core and
then when they arrive to the Earth and travel through our planet's mantle
and/or core before being detected.

In the last two decades many works has been devoted to study supernova
neutrino oscillation phenomenology.  Many of these efforts were devoted to
understand how neutrino oscillations affected the SN1987A signal
\cite{Minakata:2000rx,Lunardini:2001pb}.  More recently the effect of neutrino
oscillations on a supernova signal have been studied in the attempt to
determine how the distortions caused by oscillations could be used to estimate
and/or constrain the neutrino mixing parameters
\cite{Dighe:1999bi,Aglietta:2001jf,Takahashi:2001dc,Lunardini:2003eh,Dighe:2003jg}.

Here we will summarize the most relevant facts and predictions of these works.
More detailed discussions can be found fro example in
refs.~\cite{Dighe:1999bi,Lunardini:2001pb} and references therein.

\subsection{Neutrino Oscillations - Basic facts}
\label{subsec:oscil.basic}

Neutrinos are produced and detected as weak-interaction (flavor) eigenstates
$\nu_\a$ ($\a=e,\mu,\tau$).  We know that the flavor eigenstates do not
diagonalize the mass matrix, and after being produced they propagate as a
superposition of mass eigenstates $\nu_i$ ($i=1,2,3$).  The flavor and mass
neutrino basis are related through a unitary transformation:

\beq
\vvec{\nu_W}=U\vvec{\nu_M}\,,
\label{eq:basis.transform.}
\eeq

where we denote $\vvec{\nu_W}=(\nu_e\,\nu_\mu\,\nu_\tau)^\rm{T}$,
$\vvec{\nu_M}=(\nu_1\,\nu_2\,\nu_3)^\rm{T}$ and $U$ is the
Maki-Nakagawa-Sakata matrix (MNS).  $U$ (whose elements we will denote as
$U_{\a i}$) can be parameterized in terms of three mixing angles
$\theta_{12}$, $\theta_{23}$ and $\theta_{13}$, and assuming $\delta=0$ it
reads\cite{Maki:1962mu,Eidelman:2004wy}:

\beq
U=\left(
\begin{array}{ccc}
c_{12}c_{13} & s_{12}c_{13} & s_{13} \\
-c_{23}s_{12}-s_{23}s_{13}c_{12} &
c_{23}c_{12}-s_{23}s_{13}s_{12} & s_{23}c_{13} \\
s_{23}s_{12}-c_{23}s_{13}c_{12} &
-s_{23}c_{12}-c_{23}s_{13}s_{12} & c_{23}c_{13}
\end{array}\right)\,,
\label{eq:U.paramet.}
\eeq

where we have used the notation $c_{ij}=\cos \theta_{ij}$ and $s_{ij}=\sin
\theta_{ij}$.

The propagation of an ultra relativistic neutrino mass eigenstate $E\simeq
p(1+m_\nu^2/2E)$ obeys an approximate Schr\"odinger-like equation of
motion\cite{Kuo:1989qe}:

\beq
i\derf{\vvec{\nu_M}}{t} \simeq H_M \vvec{\nu_M}\,,
\label{eq:mass.eqofmot.}
\eeq

where the Hamiltonian in the mass basis is
$H_M\equiv(1/2E)\rm{diag}\,(m_1^2,m_2^2,m_3^2)$.  In the flavor basis the
propagation equation could be obtain from (\ref{eq:mass.eqofmot.}) applying the
transformation (\ref{eq:basis.transform.}),

\beqa
i\,\derf{(U^{\dagger}\vvec{\nu_W})}{t}&=&H_M\,(U^{\dagger}\vvec{\nu_W})\,,\\
i\,\derf{\vvec{\nu_W}}{t} &=& H_W\vvec{\nu_W}
\label{eq:flavor.eqofmot.}
\eeqa

with the Hamiltonian defined as $H_W=U\,H_M\,U^\dagger$.

The evolution of a mass eigenstate is given by $\nu_i(t) = \exp[-(m_i^2/2E)t]
\nu_i(0)$ and can be used to write down how a neutrino which is produced as a
definite flavor eigenstate $\nu_\a(0)=\sum{U_{\a i}\nu_i(0)}$ evolves:

\beq
\nu_\a(t) = \sum{U_{\alpha i}e^{-i\,{m_i^2\over{2E}}t}\nu_i(0)}.
\label{eq:flavor.evol.}
\eeq

We see that the neutrino flavor (given by the particular mix of mass
eigenstates) changes along the path, provided the neutrino masses are
different.  This is the reason underlying neutrino {\it vacuum oscillations}.

The projection of the neutrino state at $t$ on the flavor basis allows us to
find the general {\it conversion probability} in vacuum:

\beqa
P_{\a\b}\equiv P(\nu_{\alpha}\rightarrow\nu_{\beta}) &=&
\left|\big<\nu_{\beta}|\nu_\a(t)\big>\right|^2 \\ &=&\left|\sum{U_{\beta j}
U_{\alpha i}^* \big<\nu_j|\nu_i(0)\big>}\right|^2 \\ &=&\sum{U_{\b j}^*U_{\a
j}U_{\b i}U^*_{\a i}e^{-i\frac{\dmsq{ij}}{2E}t}}\,,
\label{eq:convprob.vac.}
\eeqa

where $\dmsq{ij}\equiv m_{i}^2 - m_{j}^2$ are the {\it squared mass
differences}.  The above equation tells us that a neutrino flavor oscillate
with a spatial period $\ell_\rm{osc}=4\pi E/\dmsq{ij}$ called the {\it
oscillation length}.

In matter, the effect of coherent neutrino interaction with the medium can be
accounted for by adding a potential term to the equation of motion
(\ref{eq:flavor.eqofmot.}),

\beqa
i\,\derf{\vvec{\nu_W}}{t} & = & \left[H_W + 
\left(\begin{array}{ccc}
  V & 0 & 0\\0 & 0 &0\\0 & 0 & 0
\end{array}\right)\right] \vvec{\nu_W}\,, \\
& = & H_W^\rm{eff} \vvec{\nu_W}
\label{eq:matter.flavor.eqofmot.}
\eeqa

where $V=\sqrt{2}G_F n_e$ is the effective electron neutrino potential ($-V$
for antineutrinos) that is due to coherent charge current scattering off
electrons.  In the last equation we have redefined the fields to absorb the
neutral current effective potential which is common to all flavors and
contributes an overall phase.

With the introduction of the interacting terms the vacuum mixing
matrix $U$ does not diagonalize the weak Hamiltonian and therefore the
vacuum mass eigenstates are no longer propagation eigenstates.  A new
matrix $U^m$ which diagonalize $H^\rm{eff}_W$ must be introduced, by
defining a new {\it instantaneous mass eigenstates} basis
$\vvec{\nu_W}=U^m \vvec{\nu_M^m}$ which in general is different from
point to point along the neutrino path if the density and therefore
$V$ changes.

In the two flavor case (which will be interesting for supernova mantle
oscillations described in section~\ref{subsec:oscil.mantle}) the effective
weak Hamiltonian $H^\rm{eff}_W$ is diagonalized by the in-matter mixing
matrix

\beq
U=\left(
\begin{array}{cc}
\cos\theta_m & \sin\theta_m \\
-\sin\theta_m & \cos\theta_m 
\end{array}
\right)\,,
\label{eq:U.matter}
\eeq

where $\theta_m$ is the mixing angle in matter and reads:

\beq
\sin 2\theta_m = \frac{\sin 2\theta}{\sqrt{\sin^2 2\theta+(cos 2\theta -
    2EV/\dmsq{})^2}}.
\label{eq:mixing.angle}
\eeq

The eigenvalues of $U^m H_W^\rm{eff}U^{m\dagger}$ give the squared effective
neutrino masses $\mu^2_{1,2}$,

\beqa
\mu^2_1&=&\frac{m_1^2+m_2^2}{2}+\frac{2EV}{2}-\frac{\dmsq{}}{2}\sqrt{\sin^2 2\theta+(\cos
    2\theta - 2EV/\dmsq{})^2}\,,\\
\mu^2_2&=&\frac{m_1^2+m_2^2}{2}+\frac{2EV}{2}+\frac{\dmsq{}}{2}\sqrt{\sin^2
    2\theta+(\cos 2\theta - 2EV/\dmsq{})^2}.
\eeqa

From this we get $\Delta \mu^2= \dmsq{}\sqrt{\sin^2 2\theta+(\cos 2\theta -
    2EV/\dmsq{})^2}$ and $\ell^m_\rm{osc}=4\pi E/\Delta\mu^2_{ij}$.

In summary, the effect of the different neutrino interaction with the
surrounding medium is that of producing a change in the mixing angle and
therefore in the oscillation rate $\ell^m_\rm{osc}$.  This change is maximal
when the {\it resonant condition} is fulfilled:

\beq
2EV = \dmsq{} \cos 2\theta.
\label{eq:reson.cond.}
\eeq

When the medium density changes another ``degree of freedom'' is added to the
oscillation dynamics of neutrinos. Starting from the equation of motion for
flavor eigenstates in matter (\ref{eq:matter.flavor.eqofmot.}) we can obtain
the instantaneous mass eigenstates $\vvec{\nu^m_M}$ by solving:

\beqa
i\derf{U^{m\dagger}\vvec{\nu_M^m}}{t}&=&H^\rm{eff}_W
U^{m\dagger}\vvec{\nu_M^m}\,,\\
i\derf{\vvec{\nu_M^m}}{t}&=&\left(\frac{1}{2E}\,\rm{diag}(\mu_1,\mu_2,\mu_3)-i
U^m\derf{U^{m\dagger}}{t}\right)\vvec{\nu_M^m}.
\label{eq:matter.mass.eqofmot.}
\eeqa

The variation of $U^m$ along the neutrino path makes the Hamiltonian in the
last equation not diagonal and therefore $\vvec{\nu_M^m}$\ are not longer
propagation eigenstates.  As a result not only neutrino flavor change due to
differences in the phase evolution of the propagation eigenstates as in
vacuum, also the states mix among themselves.

The mixing of instantaneous mass eigenstates is governed by the ratio between
the diagonal and off-diagonal terms of the Hamiltonian in
eq. (\ref{eq:matter.mass.eqofmot.}).  The off-diagonal terms which are
provided by $dU^{m\dagger}/dt$ depend on the rate of variation of the
in-matter mixing angle $d\theta_m/dt$ which is determined by the density
gradient.

Near the resonance region (\ref{eq:reson.cond.}) the diagonal terms of
the effective mass Hamiltonian are almost equal and cancel out by a
phase redefinition of the fields.  Therefore it is in that region
where the instantaneous mass eigenstate are more important.  If
density varies slowly inside the resonance region (and the
off-diagonal terms are close to zero) the instantaneous mass
eigenstates are almost propagation eigenstates and do not oscillate.
It is said that neutrinos perform an {\it adiabatic transition}.  But,
if the density gradient is steep enough, the instantaneous mass
eigenstates strongly oscillates and neutrinos could emerge from the
resonance layer as a different instantaneous mass {\it eigenstate}.
It is said that the neutrino state {\it jumps}.

Quantitatively, the adiabaticity condition can be expressed in terms of the
mixing parameters and the medium properties as \cite{Kuo:1989qe}:

\beq
\gamma \equiv \frac{\dmsq{}}{2E}\frac{\sin^2 2\theta}{\cos
  2\theta}\left|\derf{\log n_e}{t}\right|_\rm{res}^{-1}\,,
\label{eq:adiab.param.}
\eeq

where $\gamma$ is called the {\it adiabaticity} parameter.

The jumping probability of instantaneous mass eigenstates near the resonance
region is approximately given by the {\it Landau-Zener Formula}\footnote{In
principle the Landau-Zener formula is accurate under the assumption of a small
value of the adiabaticity parameter and a linear variation of the density {\it
inside} the resonance region.  We have verified that for typical supernova
density profiles and mixing parameters this formula works as well as the
double exponential formula used in other works \cite{Fogli:2001pm}.}:

\beq
P_j=\exp(-\pi\gamma/2).
\label{eq:lz.formula}
\eeq

The resonant conversion of neutrinos in matter through the above process is
conventionally called the {\it Mikheev-Smirnov-Wolfenstein effect (MSW)}
\cite{Wolfenstein:1977ue,Mikheev:1986wj}.

\subsection{Neutrino mixing parameters - experimental values and limits}
\label{subsec:oscil.param.}

As we saw in the previous section the dynamics of {\it active} neutrino
oscillation depends on 3 pairs of parameters: ($\dmsq{12}$,$\theta_{12}$),
($\dmsq{23}$,$\theta_{23}$), ($\dmsq{13}$,$\theta_{13}$).

The study of neutrino oscillations in solar, atmospheric, reactor and
accelerator experiments has allowed us to measure accurately some of these
parameters, and to put limits on other ones.  In the following, we present the
most recent values and limits obtained from a global analysis of all the
available experimental results \cite{Bahcall:2004ut}:

\begin{itemize}

\item $[\dmsq{12}=7.4-9.2\times10^{-5} eV^2$,\ $\theta_{12}=28\degr-37\degr]$
  (99.7\% CL).
  
  This is the so-called {\it LMA solution} to the solar neutrino problem.

  It is common to identify this pair of parameters as
  ($\dmsq{\odot},\theta_\odot$)

\item $[|\dmsq{23}|=1.5-3.9\times10^{-3} eV^2$,\
  $\theta_{23}=34\degr-45\degr]$ (99.7\% CL).

  The sign of $\dmsq{23}$ is still unknown and will depend on what is the mass
  hierarchy: {\bf normal hierarchy} (NH) $\dmsq{23}>0$ or {\bf inverted
  hierarchy} (IH) $\dmsq{23}<0$.

  It is common to identify this pair of parameters as
  ($\dmsq{\rm{atm}},\theta_\rm{atm}$).

\item $[\dmsq{13}\simeq\dmsq{23}$,\ $\theta_{13}<12.9\degr]$ (99.7\% CL).

\end{itemize}

\subsection{Neutrino Oscillations in the Supernova Mantle}
\label{subsec:oscil.mantle}

In the following sections we will apply these basic results to study neutrino
oscillations in the supernova mantle.

\subsection{Initial conditions}
\label{subsubsec:mantle.initial.cond.}

Inside the supernova core where neutrinos are produced, the effective neutrino
potential felt by the electron (anti)neutrinos overwhelms the off-diagonal
terms of the effective weak Hamiltonian $H^\rm{eff}_W$
(\ref{eq:matter.flavor.eqofmot.}) and flavor eigenstates are approximately
local mass eigenstate.  The correspondence between flavor and mass eigenstates
in the production region will depend on the mass hierarchy
\cite{Dighe:1999bi}:

\begin{itemize}

\item with {\bf normal hierarchy} we will have:

  \beq
  \begin{array}{ccc}
    \nue=\nu_{3m},&\nu\,'_\tau=\nu_{2m},&\nu\,'_\mu=\nu_{1m}\,,\\
    \anue=\nu_{1m},&\anu\,'_\mu=\nu_{2m},&\anu\,'_\tau=\nu_{3m}.\\
  \end{array}
  \label{eq:in.cond.NH}
  \eeq

\item for {\bf inverted hierarchy}:

  \beq
  \begin{array}{ccc}
    \nue=\nu_{3m},&\nu\,'_\mu=\nu_{2m},&\nu\,'_\tau=\nu_{1m}\,,\\
    \anue=\nu_{1m},&\anu\,'_\tau=\nu_{2m},&\anu\,'_\mu=\nu_{3m}.\\
  \end{array}
  \label{eq:in.cond.IH}
  \eeq

\end{itemize}

Where $\nu\,'_\mu$, $\nu\,'_\tau$, $\anu\,'_\mu$ and $\anu\,'_\tau$ are
redefined states which diagonalize the $(\nu_\mu,\nu_\tau)$ sub-matrix of
$H^\rm{eff}_W$ \cite{Dighe:1999bi}.  It must be noticed that such redefinition
does not change the supernova physics since for all practical purposes
$\nu_\mu$ and $\nu_\tau$ are phenomenological identical in supernova
conditions.

\subsubsection{Resonance regions}
\label{subsubsec:mantle.resonance}

Neutrinos emerging from the supernova core traverse matter with densities
ranging from $10^9$ g/cm$^3$ to almost 0.  In the process they will find the
required conditions for resonant conversion.  Two possible resonances are
allowed by the mixing parameters:

\begin{itemize}

\item a high density {\bf H-Resonance} corresponding to enhanced conversion
  into the heaviest mass eigenstate $\nu_3$ and determined by the condition:

  \beq
  2E\,V(\rho^\rm{res}_{H})=\dmsq{23}\cos 2\theta_{13}.
  \label{eq:reson.H}
  \eeq

\item and a low density {\bf L-Resonance} associated to transitions between
  the lightest eigenstates:

  \beq
  2E\,V(\rho^\rm{res}_L)=\dmsq{12}\cos 2\theta_{12}.
  \label{eq:reson.L}
  \eeq

\end{itemize}

Since $\dmsq{12}>0$ the {\it L-Resonance will be only in the neutrino channel}.

On the other hand the H-Resonance will affect neutrinos and antineutrinos
according to the mass hierarchy:

\begin{itemize}

\item for {\bf\em normal hierarchy} ($\dmsq{23}>0$) {\it H-Resonance will just
affect neutrinos},

\item while in the case of an {\bf\em inverted hierarchy} ($\dmsq{23}<0$) {\it
H-Resonance will occur in the antineutrino channel}.

\end{itemize}

\subsubsection{Factorization of Dynamics}
\label{subsubsec:mantle.fact.}

The hierarchical nature of the squared-mass differences
$\dmsq{23}:\dmsq{12}\simeq 100:1$ will produce a natural decoupling of the
conversion dynamics in the two resonance layers inside the supernova mantle.

At high densities (H-Resonance) the in-matter mixing of the lightest
eigenstates $U_{e2}^m$ will be suppressed.  Transitions will affect just the
heaviest eigenstate and the conversion description is reduced to the two
flavor case $(\nu_1,\nu_3)$.  At low densities (L-Resonance) the mixing of the
third eigenstate is near to its value in vacuum $U_{e3}^m\simeq U_{e3}\lsim
10^{-2}$ and almost constant.  Only the first and second eigenstate will mix
and again the problem is reduced to a two flavor case $(\nu_1,\nu_2)$.

Additionally as $U_{e3}$ is very small the total survival probabilities for
$\nue$ and $\anue$ can be factorized \cite{Kuo:1987qu,Mikheev:1988in}:

\beq 
P\approx P_H\times P_L.
\label{eq:surv.fact}
\eeq

Factorization of dynamics is a property which simplifies considerably the
computation of the total conversion probabilities in the supernova mantle
without a detailed numerical study of the neutrino propagation equation.

\subsubsection{Jumping probabilities}
\label{subsubsec:mantle.jump}

\begin{figure}[h]
\begin{center}
\epsfxsize=120mm
\epsfbox{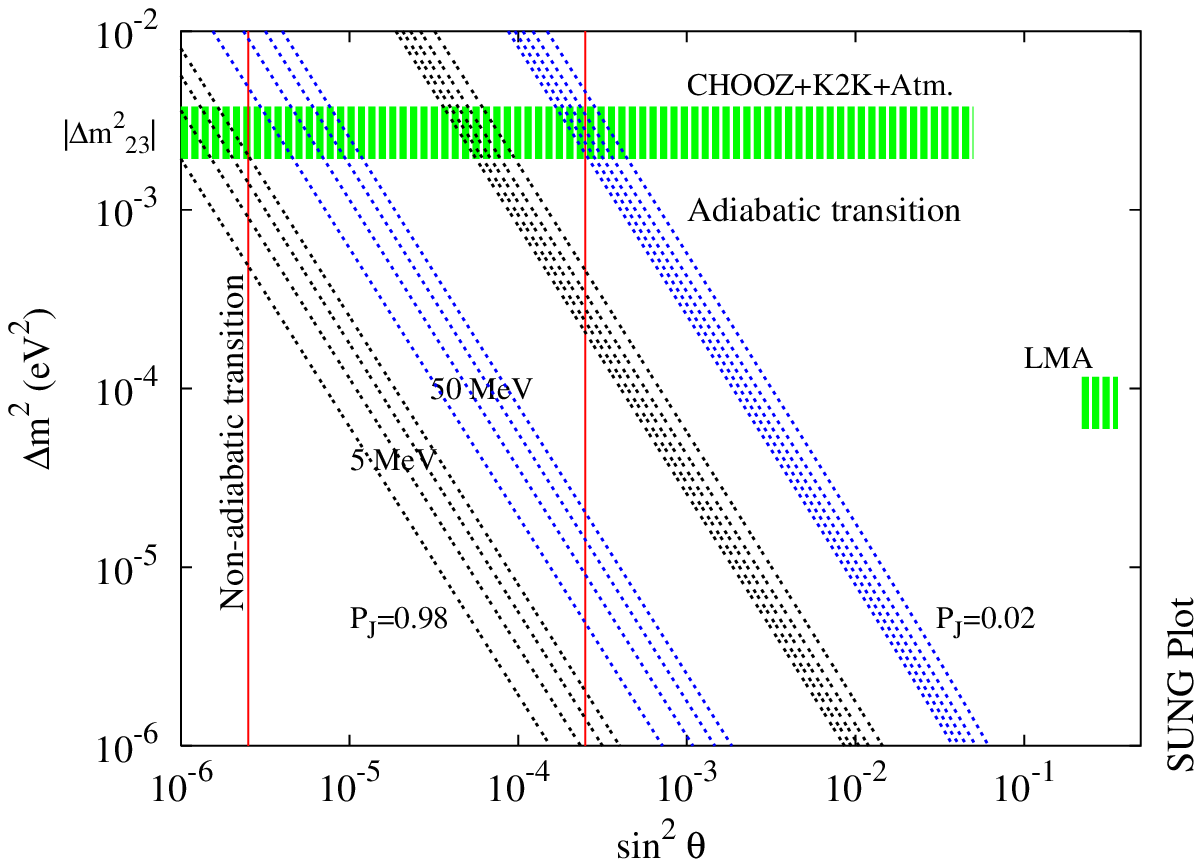}
\mycaption{Contours of equal neutrino jump probability across a resonance region}{
Contours of equal neutrino jumping probability across a resonance region with
different mixing parameters $(\dmsq{},\theta)$.  Dotted lines correspond to a
50 Mev neutrino while dashed lines are for 5 MeV.  Contours at the left range
from $P_J=0.02$ to $P_J=0.1$ and at right from $P_J=0.9$ to $P_J=0.98$.  It is
possible to recognize three regions where jumping probabilities have different
values: one were adiabatic transitions ($P_J\simeq 0$) take place
(upper-right), a second one for which there are non-adiabatic transitions
($P_J\simeq 1$) (lower-left) and an intermediate one where $P_J$ has
intermediate values for almost all neutrino energies.  Jump probabilities are
computed assuming the simple density profile
$\rho(r)Y_e=2\times10^4\,\rm{g/cm}^3(r/10^9\,\rm{cm})^3$ \cite{Dighe:1999bi}
}
\label{fig:jump.contours}
\end{center}
\end{figure}

We have computed generic jump probabilities using the Landau-Zener formula
(\ref{eq:lz.formula}) for typical conditions found in the supernova mantle
and generic mixing parameters. Contours of equal probability in the
$(\dmsq{},\sin^2\theta)$ plane for different neutrino energies, are depicted
in figure~\ref{fig:jump.contours}.

%

Observing the contours, we see that the neutrino transitions in the
L-Resonance, which are governed by the solar mixing parameters
$(\dmsq{12},\theta_{12})$ (LMA), is completely adiabatic, i.e. $P_L=0$.

The case for the H-Resonance is not as simple.  Since the neutrino transition
through that layer is governed by the unknown mixing angle $\theta_{13}$, the
adiabaticity will be also uncertain.

We can recognize two extreme cases \cite{Dighe:1999bi}:

\begin{itemize}

\item $\sin^2 \theta_{13}\gsim 10^{-4}$ ($\theta_{13}\gsim 1\degr$).  In this case
  for all the interesting neutrino energies the adiabaticity condition is
  fulfilled, i.e. $P_H\simeq0$.

\item $\sin^2 \theta_{13}\lsim 10^{-6}$ ($\theta_{13}\lsim 0.06\degr$).  In
  this case adiabaticity is strongly violated implying $P_H\sim 1$ for almost
  the entire range of supernova neutrino energies.

\end{itemize}

For $10^{-6}\lsim\sin^2 \theta_{13}\lsim 10^{-4}$ we will have an intermediate
behavior with jump probabilities ranging between 0.1 and 0.9.

\begin{figure}[h]
\begin{center}
\epsfxsize=120mm \epsfbox{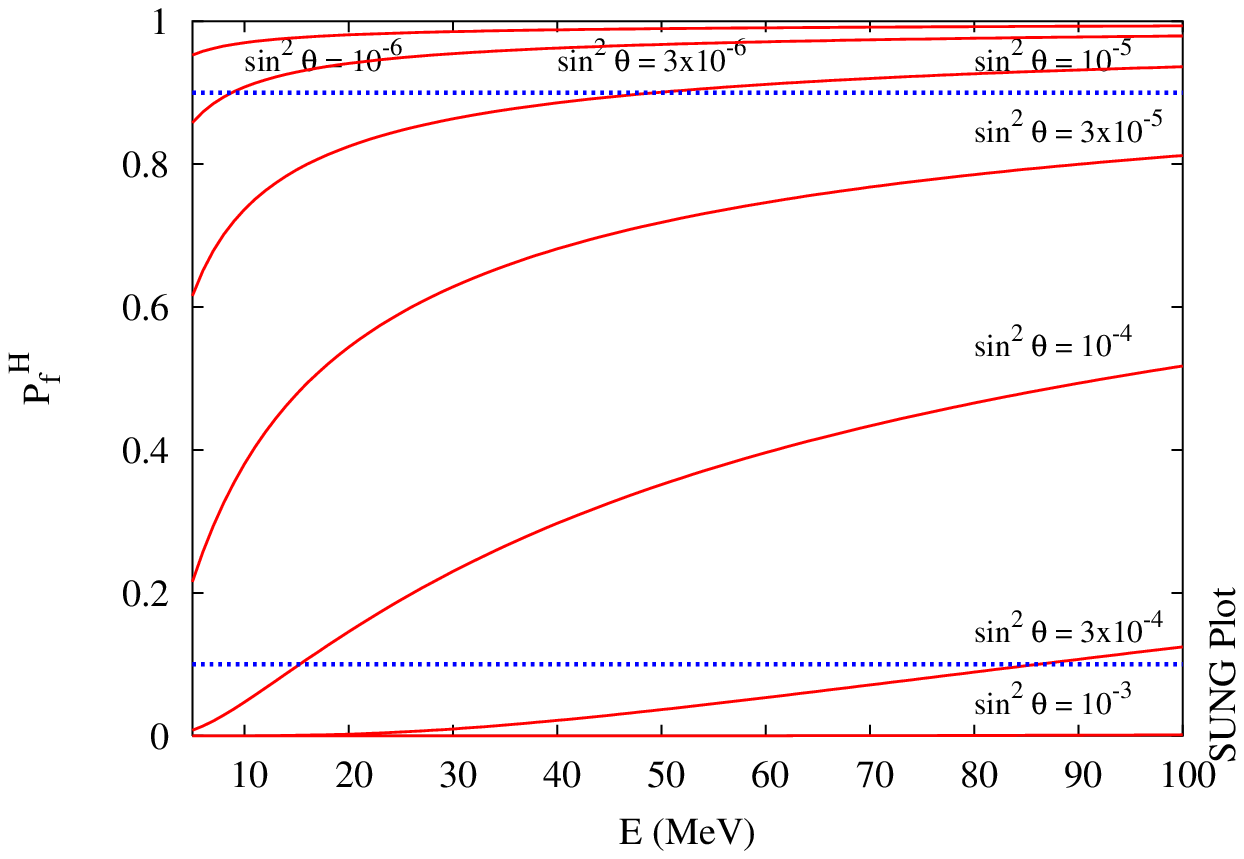} 
\mycaption{Jumping probability $P_H$ ($\barr{P}_H$ for inverted hierarchy)}{
Jumping probability $P_H$ ($\barr{P}_H$ for inverted hierarchy) as a function of
energy for different values of the mixing parameter $\sin^2\theta_{13}$.
}
\label{fig:jump.energy}
\end{center}
\end{figure}

The energy dependency of the jump probability in the resonance region is
another matter of concern.  We have plotted in figure~\ref{fig:jump.energy}
$P_H$ as a function of neutrino energy for different values of the mixing
parameter $\sin^2\theta_{13}$.  It can be seen that for the range of
interesting neutrino energies the jump probability in the extreme cases is
almost constant.  In the intermediate range of values of the mixing parameter,
$P_H$ will have only a mild variation in the energy range where most of the
neutrino events will be expected.

The above conclusions apply also to antineutrinos in the inverted mass
hierarchy case where the jumping probability $\barr{P}_H$ becomes identical to
$P_H$.

%

\subsubsection{Conversion probabilities}
\label{subsubsec:mantle.conv.prob.}

\begin{figure}[h]
\begin{center}
\epsfxsize=80mm
\epsfbox{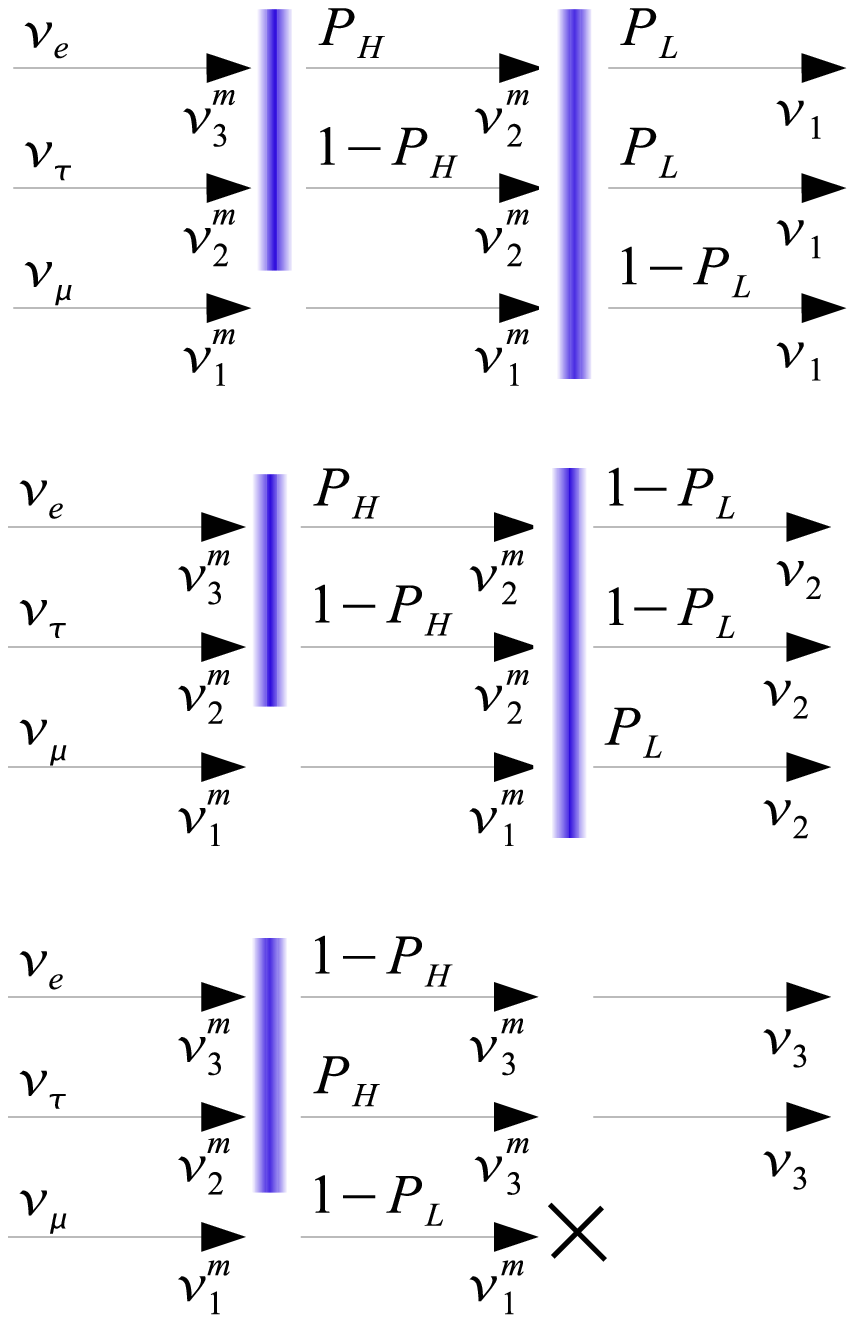}
\mycaption{Schematic illustration of how transition probabilities are computed}{
Schematic illustration of how transition probabilities are computed using
jumping probabilities and following the evolution of neutrino state along the
resonance regions.  The example illustrated is for $\nue$ transitions and
normal hierarchy.
}
\label{fig:schem.trans.prob.}
\end{center}
\end{figure}

Using the jumping probabilities and the fact that the dynamics in the
resonance region decouple we can compute the {\it transition
probability} ($\barr{p}_{{\a}i}$)$p_{{\a}i}$, i.e. the probability
that a given (anti)neutrino flavor $\a$ emerges from the supernova in
the $i^\rm{th}$ mass eigenstate.  Figure~\ref{fig:schem.trans.prob.}
illustrates schematically the way these probabilities are computed
following the evolution of the neutrino state from the emission region
to the star surface.

%

In table~\ref{tab:trans.prob.} we summarize the results of applying
rules as those presented in fig.~\ref{fig:schem.trans.prob.} to
compute transition probabilities for neutrinos and antineutrinos,
assuming normal and inverted hierarchies.

\begin{table}[h]
\mycaption{Transition probabilities}{Transition probabilities}
\begin{center}
\begin{tabular}{c|ccc||c|ccc}
\multicolumn{8}{c}{\bf Normal mass hierarchy}\\\hline\hline
\multicolumn{4}{c}{\bf Neutrinos} & \multicolumn{4}{c}{\bf Antineutrinos}\\
\multicolumn{1}{c}{\strut} & $\nu_1$ & $\nu_2$ & $\nu_3$ & \multicolumn{1}{c}{\strut} & $\anu_1$ & $\anu_2$ & $\anu_3$ \\
\cline{2-4}\cline{6-8}
$\nue$    & $P_H\,P_L$   & $P_H(1-P_L)$    & $(1-P_H)$    &$\anue$    & 1 & 0 & 0 \\
$\nu\,'_\mu$ & $(1-P_L)$    & $P_L$           & 0            &$\anu_\mu$ & 0 & 1 & 0 \\
$\nu\,'_\tau$& $(1-P_H)P_L$ & $(1-P_H)(1-P_L)$& $P_H$        &$\anu_\tau$& 0 & 0 & 1 \\
\blankline{8}
\multicolumn{8}{c}{\bf Inverted mass hierarchy} \\\hline\hline
\multicolumn{4}{c}{\bf Neutrinos} & \multicolumn{4}{c}{\bf Antineutrinos} \\
\multicolumn{1}{c}{\strut}  & $\nu_1$ & $\nu_2$ & $\nu_3$ & \multicolumn{1}{c}{\strut} & $\anu_1$ & $\anu_2$ & $\anu_3$ \\
\cline{2-4}\cline{6-8}
$\nue$    & 0 & $P_L$ & $(1-P_L)$ &$\anue$  & $\barr{P}_H$ & 0 & $(1-\barr{P}_H)$ \\
$\nu\,'_\mu$ & 0 & $(1-P_L)$ & $P_L$   &$\anu_\mu$ & 0 & 1 & 0 \\
$\nu\,'_\tau$& 1 & 0 & 0 &$\anu_\tau$& $(1-\barr{P}_H)$ & 0 & $\barr{P}_H$  \\
\end{tabular}
\end{center}
\label{tab:trans.prob.}
\end{table}

Transition probabilities allow for a straightforward computation of
the final conversion probabilities, i.e. the probability that a given
neutrino flavor $\a$ is detected as flavor $\b$.  In terms of $p_{\a
i}$ the conversion probabilities are simply given by

\beq
P_{\a\b}=\sum_{i}{p_{{\a}i} p_{i\b}}.
\label{eq:conv.prob.}
\eeq

Where $p_{i\b}$ is the probability that the mass eigenstate $i^\rm{th}$ be
detected as the neutrino of flavor $\b$.

In the absence of any other mixing effect neutrinos emitted from the supernova
arrive to the Earth surface as incoherent mass eigenstates and $p_{i\b}$ is
the amplitude of the projection on the flavor basis in vacuum:

\beq
p_{i\b}=|U_{{\b}i}|^2.
\eeq

We will focus on electron antineutrinos for reasons that we will give later.

As an example, the electron antineutrino conversion probabilities for normal
and inverted hierarchy read:

\beq
\left(
\begin{array}{c}
P_{\se\se}\\P_{\sm\se}\\P_{\st\se}
\end{array}
\right)=
\left\{
\begin{array}{cc}
\left(
\begin{array}{c}
\Aei{1}\\\Aei{2}\\\Aei{3}
\end{array}\right), & \rm{NH}\\
\left(
\begin{array}{c}
\barr{P}_H\Aei{1}+(1-\barr{P}_H)\Aei{3}\\\Aei{2}\\(1-\barr{P}_H)\Aei{1}+\barr{P}_H\Aei{3}
\end{array}\right), & \rm{IH}
\end{array}
\right.
\label{eq:surv.prob.}
\eeq

We can see that in the case of a normal hierarchy the observed $\anue$ will be
an admixture of neutrinos emitted originally as $\anue$
($P_{\se\se}=cos^2\theta_{12}\simeq 0.26$) and $\anu\,'_\mu$
($P_{\sm\se}=cos^2\theta_{12}\simeq 0.74$).  On the other hand in the inverted
hierarchy case the mixing will mainly depend on the adiabaticity of
antineutrino transitions in the H-Resonance.  If transitions are adiabatic,
$\barr{P}_H\simeq0$, the survival probability for $\anue$ will be suppressed
$P_{\se\se}\simeq\Aei{3}\lsim 10^{-2}$ and the observed $\anue$ events will be
$\anu_\mu$ and $\anu_\tau$ neutrinos converted in the supernova mantle into
$\anue$ ($P_{\sm\se}+P_{\st\se}\simeq 1$).

Similar considerations can be done for electron neutrinos (see
ref.\cite{Dighe:1999bi}).

\subsection{Neutrino oscillations in the Earth}
\label{subsec:oscil.earth}

If neutrinos travel inside the Earth oscillations will mix the arriving mass
eigenstates and the probabilities $p_{i\b}$ will be more complex.

Due to the hierarchical nature of the mass spectrum, antineutrino oscillations
in the Earth could be also reduced to a 2 flavor oscillation problem
\cite{Dighe:1999bi}.  In this case the probability that an arriving
antineutrino mass eigenstate be detected as an electron antineutrino after
crossing the Earth interior could be written as \cite{Fogli:2001pm}:

\beqas
\barr{p}_{1e} & \simeq & \cos^2 \theta_{13}\,(1-\barr{p}_{\oplus})\,, \\
\barr{p}_{2e} & \simeq & \cos^2 \theta_{13}\,\barr{p}_{\oplus}\,, \\
\barr{p}_{3e} & \simeq & \sin^2 \theta_{13}\,,
\label{eq:earth.trans.prob.}
\eeqas

where $\barr{p}_\oplus$ is the two-flavor oscillation probability after
crossing the Earth.  $\barr{p}_\oplus$ can be computed under simple
assumptions about the Earth interior.  For example, if neutrinos travel only
trough the Earth mantle and we assume a constant density, the oscillation
probability is given by \cite{Fogli:2001pm}:

\beq
\barr{p}_\oplus=\sin^2\theta_{12}+\sin^2
2\theta_m\sin(2\theta_m-2\theta_{12}) \sin^2
\left(\frac{\sin2\theta_{12}}{\sin2\theta_m}\frac{\dmsq{12}d}{4E}\right)\,,
\label{eq:earth.osc.prob.}
\eeq

where the in-matter mixing angle $\theta_m$ satisfies~\ref{eq:mixing.angle}.

\begin{figure}[h]
\begin{center}
\epsfxsize=120mm
\epsfbox{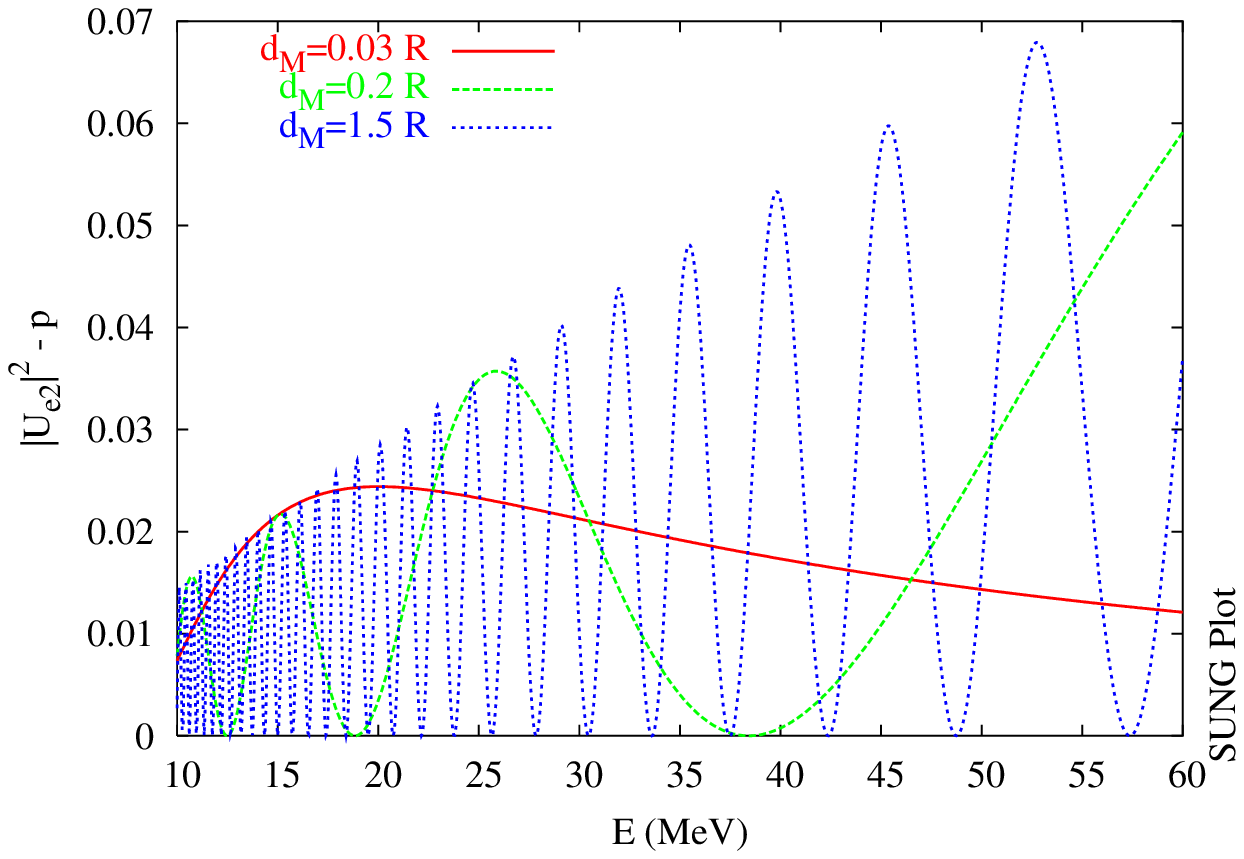}
\mycaption{Oscillation probability $p_\oplus$ in the Earth mantle}{
Oscillation probability $p_\oplus$ in the Earth mantle for different distances
$d_M$ traveled.  We used $\dmsq{12}=8\times10^{-5}$, $\theta_{12}=30\degr$,
$\barr{\rho}^\rm{mantle}_\oplus=4.5$ g/cm$^3$ and $Y_e=0.5$.
}
\label{fig:mantle.oscil.}
\end{center}
\end{figure}

Figure~\ref{fig:mantle.oscil.} depicts the oscillation probability
$\barr{p}_\oplus$ for antineutrinos as a function of the neutrino energy
assuming different distances $d$ traveled inside the Earth mantle.  We have
assumed standard values for $\barr{\rho}^\rm{mantle}_\oplus\simeq4.5$ g/cm$^3$
and $Y_e=0.5$, and the mixing parameter values of
sect.~\ref{subsec:oscil.param.}

%

A more complex problem is to describe neutrino oscillations when neutrinos
traverse the mantle and the Earth core.  In a dedicated study Akhmedov
\cite{Akhmedov:1998ui} obtained analytical relations assuming two layers of
constant density.  In that case the oscillation probability $p_\oplus$ for a
neutrino which travel distances $d_M+d_C+d_M$ could be parameterized in the
following way \cite{Fogli:2001pm}:

\beq
p_\oplus=\sin^2\theta_{12}+O_1^2 \cos 2\theta_{12} + O_1O_3\sin 2\theta_{12}\,,
\label{eq:earth.core.osc.prob.}
\eeq

where the ``oscillating'' coefficients $O_1$ and $O_3$ are the components of
the vector

\beq
\vvec{O}=2\sin\phi_M Y \vvec{t_M} + \sin\phi_C \vvec{t_C}
\label{eq:oscil.coef.}
\eeq

having defined $Y$,$\vec{t}$ and $\phi$ as

\beqa
\label{eq:earth.core.def.1}
Y & \equiv & \cos\phi_M\cos\phi_C - (\vvec{t_M}\cdot\vvec{t_C})\sin\phi_M\sin\phi_C\,,\\
\label{eq:earth.core.def.2}
\vvec{t} & \rightarrow & (\sin 2\theta_m,0,-\cos 2\theta_m)\,,\\
\label{eq:earth.core.def.3}
\phi & \equiv & \frac{\sin2\theta_{12}}{\sin2\theta_m}\frac{\dmsq{12}d}{4E}.
\eeqa

The subindices $M$ and $C$ for $\phi$ and $\vec{t}$ in (\ref{eq:oscil.coef.})
and (\ref{eq:earth.core.def.1}) refer to the mantle and core values of
$\theta_m$ and $d$.

The distances traveled by neutrinos inside the mantle and core are given in
terms of the Earth radius $R_\oplus=6371$ km, the core radius $r_C=3486$ km and
the zenith angle $z$:

\beqa
d_M & = & R\,\left(-\cos z-\sqrt{r_C^2/R_\oplus^2-\sin^2 z}\right)\\
d_C & = & 2 R\sqrt{r_C^2/R_\oplus^2-\sin^2 z}
\eeqa

\begin{figure}[h]
\begin{center}
\epsfxsize=120mm
\epsfbox{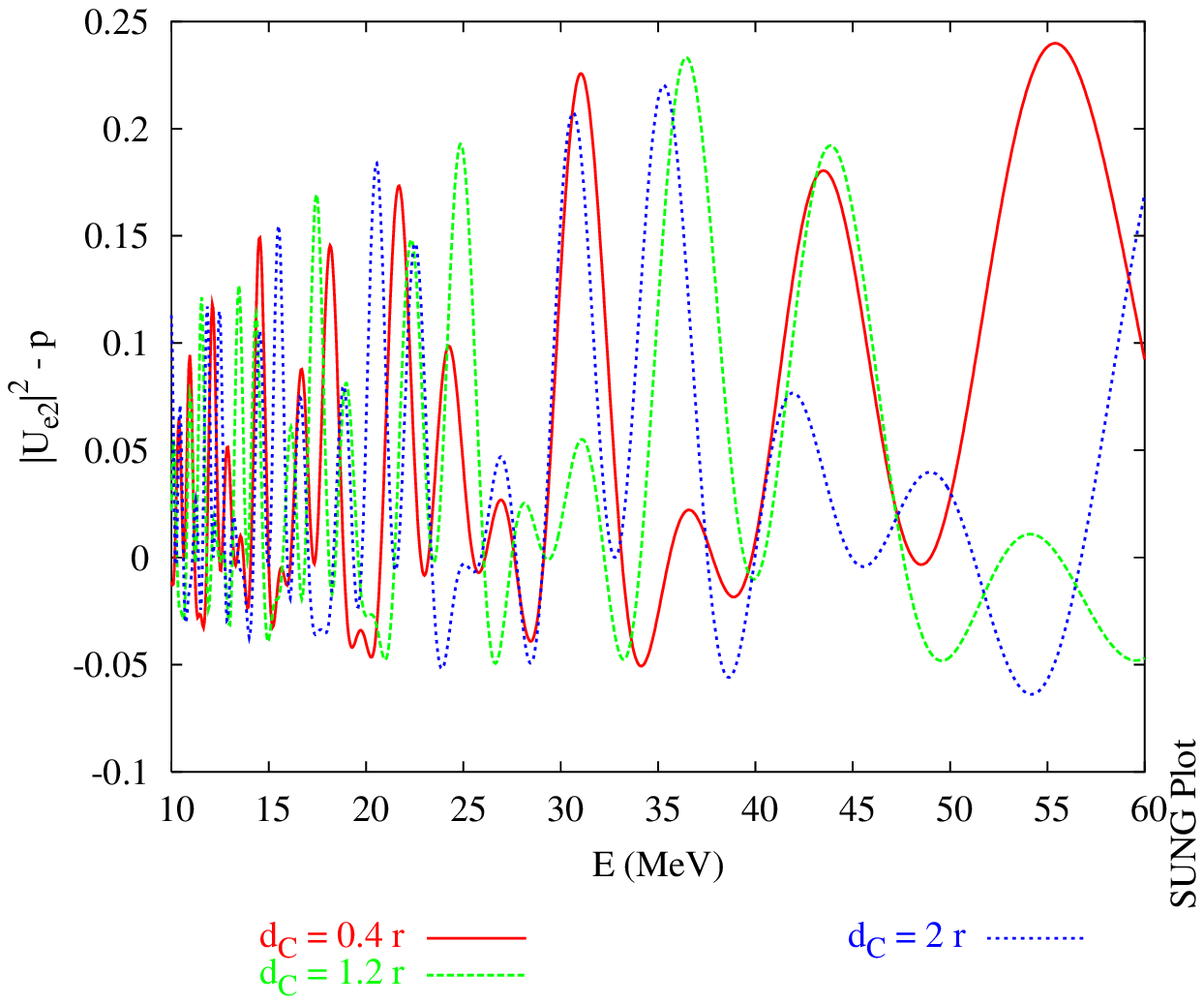}
\mycaption{Oscillation probability $p_\oplus$ when neutrino traverse the mantle and core}{
Oscillation probability $p_\oplus$ when neutrino traverse the mantle and core.
We used $\dmsq{12}=8\times10^{-5}$, $\theta_{12}=30\degr$,
$\barr{\rho}^\rm{mantle}_\oplus=4.5$ g/cm$^3$,
$\barr{\rho}^\rm{core}_\oplus=12$ g/cm$^3$ and $Y_e=0.5$.
}
\label{fig:mantle.core}
\end{center}
\end{figure}

In figure~\ref{fig:mantle.core} we show the probability $p_\oplus$ for
different energies and zenith angles.

%

\section{The neutrino signal formalism}
\label{sec:signal.form.}

The theoretical relation between neutrino emission in a supernova and the
detected signal involves three elements: i) the neutrino emission, ii)
neutrino oscillations and iii) the detection process.

With the conversion probabilities we can proceed to determine the way neutrino
oscillations mix the fluxes and spectra of the different neutrino flavors.

At the source, the total flux of neutrinos $\a$, $S_\a(E,t)$ can be written in
terms of the luminosity $L_\a(t)$, the neutrino average energy
$\barr{E_\a}(t)$ and the spectral distribution $F^\rm{em}_\a(E;t)$ as

\beq
S_\a(E,t)=\frac{L_\a(t)}{\barr{E_\a}(t)}\,F^\rm{em}_\a(E;t).
\label{eq:emitted.signal}
\eeq

At the detector these fluxes are mixed to construct the observed antineutrino
flux according to the conversion probabilities (\ref{eq:surv.prob.}):

\beq
L^2 S^\rm{det}_{\barr{e}}(E,t) = P_{\se\se} S_{\barr{\a}}(E,t) +
(P_{\sm\se}+P_{\st\se}) S_{\barr{x}}\,,
\label{eq:detected.signal.multiple}
\eeq

where $L$ is the supernova distance and we have used the fact that for all
purposes the $\anu_\mu$ and $\anu_\tau$ fluxes are identical
$S_{\barr{x}}\equiv S_{\barr{\mu}}\simeq S_{\barr{\tau}}$ (see
sect.~\ref{sec:fluxes}).

Using the unitarity property of the conversion probabilities $\sum_\a
P_{\barr{\a}\se}=1$ we can express the detected flux
(\ref{eq:detected.signal.multiple}) in terms of just one probability, namely
the electron antineutrino survival probability $\barr{p}\equiv
P_{\anue\anue}$:

\beq
L^2 S^\rm{det}_{\barr{e}}(E,t)=\barr{p}\,S_{\barr{e}} +
(1-\barr{p})\,S_{\barr{x}}.
\label{eq:detected.flux}
\eeq

From the detected flux of electron antineutrinos we can straightforwardly
compute the rate of events at a given neutrino detector:

\beq 
\frac{d^2n_\anue(E,t)}{dE\,dt} = N_T\, \int{dE'\,
S_{\barr{e}}^\rm{det}(E',t)\, \s(E')\, \epsilon(E')\, {\cal R}(E,E')}\,,
\label{eq:total.rate}
\eeq

where $\sigma(E)$ is the detection cross-section, $N_T$ is the number of
target particles in the fiducial volume, $\epsilon(E)$ is the detection
efficiency and ${\cal R}(E,E')$ is the energy resolution function that
accounts for the uncertainties in the measurement of the neutrino energies.

\subsection{Neutrino Detection}
\label{subsec:detect.process}

The statistical procedure that will be introduced in the next chapter is based
on two assumptions about the neutrino signal: it must have a large number of
events (several thousands) and both the energy and time measurements must be
available.  Both conditions are fulfilled for electron antineutrino events in
water \cerenkov detectors and in scintillator detectors.  This is the main
reason why we concentrate our attention on $\anue$.

Electron antineutrinos are detected in water and scintillator detectors
through the inverse $\beta$ process:

\beq
\anue+p \rightarrow n+e^+\;,(E_\rm{react}\simeq1.8 MeV).
\label{eq:anupn}
\eeq

\begin{figure}[h]
\begin{center}
\epsfysize=160mm
\epsfbox{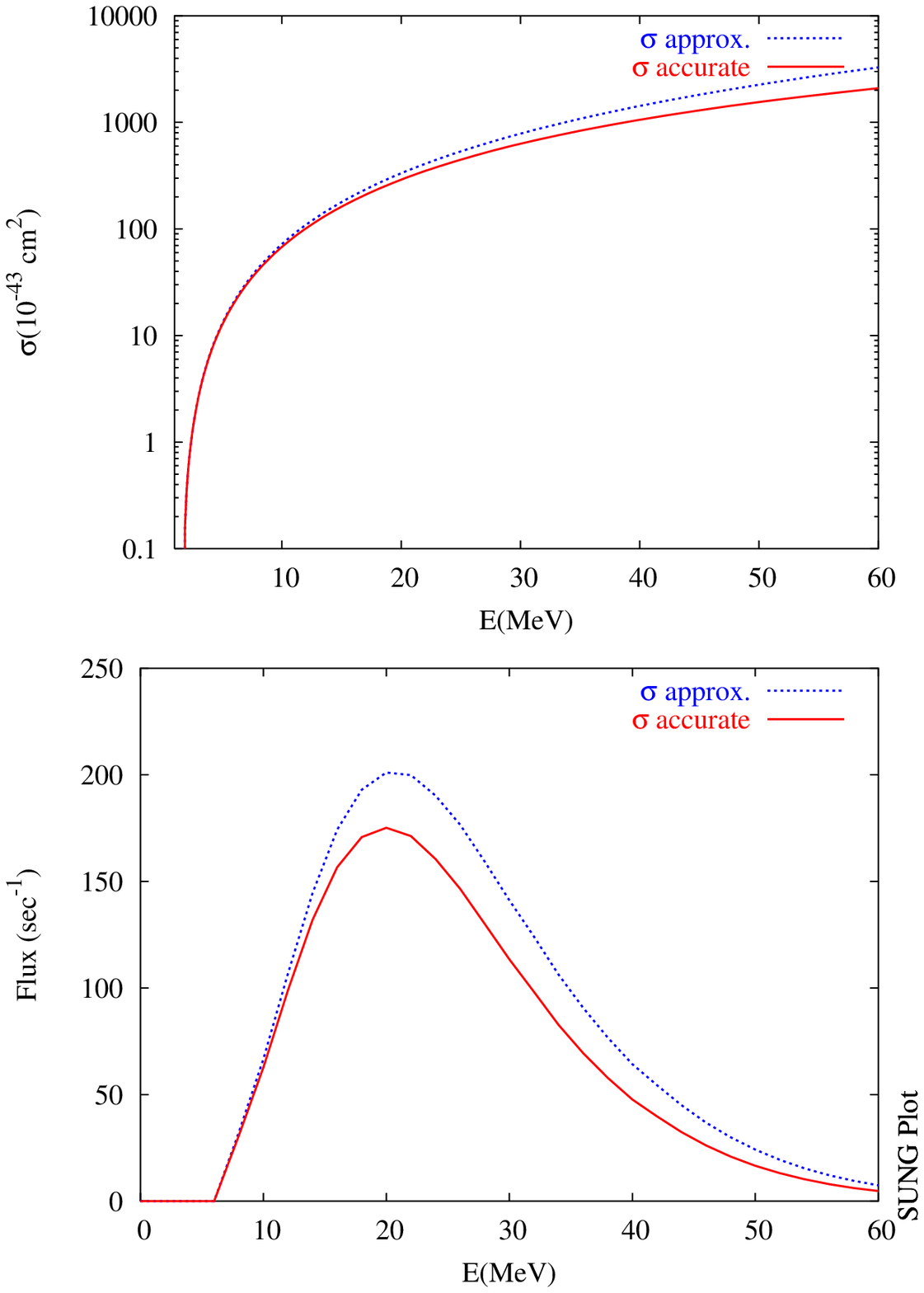}
\mycaption{Comparison between inverse $\beta$ process cross-sections}{
Upper panel: comparison between the approximate (dashed line) and the accurate
cross-section (\ref{eq:acc.xsection}).  Lower panel: time integrated spectrum
obtained when the approximate (dashed) and the accurate cross-section are
used.
}
\label{fig:xsection}
\end{center}
\end{figure}

An accurate parameterization of the cross-section for this process has been
recently provided by Strumia and Vissani\cite{Strumia:2003zx}:

\beq
\frac{\sigma_{\bar\nu}(\bar\nu_e p \to e^+n)}{ 10^{-43}\rm{cm}^2} = p_e E_e\,
E_{\bar\nu}^{-0.07056+0.02018\ln E_{\bar\nu}-0.001953\ln^3E_{\bar\nu} },
\label{eq:acc.xsection}
\eeq

where $E_e=E_{\bar\nu}-Q_{np}$ with $Q_{np}=m_n-m_p\approx 1.293\,$MeV and all
the energies are in MeV.  This parameterization is accurate on a wide range of
neutrino energies.  In figure~\ref{fig:xsection} we compare this cross-section
with the approximation $\sigma_{\bar\nu}(\bar\nu_e p \to e^+n)\simeq
9.52\times 10^{-43}\,\rm{cm}^2\;p_e E_e$ and the corresponding time-integrated
spectrum.  We see that using an accurate cross-section the neutrino flux is
reduced as much as 20\%.

\afterpage{\clearpage}
%

In heavy water (D$_2$O), $\anue$ can be detected through the charged current
reaction off deuterium:

\beq
\anue+d \rightarrow n+n+e^+\; (E_\rm{react}=4.03 MeV)
\label{eq:anudnn}
\eeq

SNO is the only operational heavy water detector and therefore, with fiducial
volume much smaller than Super Kamiokande, the number of neutrino events
detected through this process will not be enough large to apply efficiently
our method.

Other reactions (as described in details in ref.~\cite{Burrows:1991kf}),
including $\anue$ absorption in oxygen (water detectors) and carbon
(scintillator detectors), will produce typical number of events below 10\% of
the statistics from $\anue$ $\beta$ process.  Additionally, some of these
reactions have large energy thresholds or cannot provide energy and/or time
informations, a central condition for the application of any method.

In our simulations we have assumed 100\% detection efficiency above the energy
threshold for all the detectors.  This idealized situation will not change
much the signal characteristics.

The uncertainty in the energy measurement is an important aspect in our
simulations and their analysis.  We have approximated the resolution function
${\cal R}(E,E')$ with a Gaussian distribution with mean $E$ equal to the
measured energy, and as standard deviation $\Delta E$ equal to the energy
resolution, that can be parameterized as\cite{Lunardini:2001pb}:

\beq 
\frac{\Delta E}{\rm{MeV}} = a_E \sqrt{\frac{E}{\rm{MeV}}} + b_E
\,\frac{E}{\rm{MeV}}.
\label{eq:Euncertainty}
\eeq

The value of the adimensional coefficients $a_E$ and $b_E$ and other
characteristics of several operational detectors as well as of a few planned
detectors are presented in table~\ref{tab:detectors}.

\begin{table}[t]
\begin{tabular}{>{\bf}p{2cm}c>{\centering}p{1cm}p{1.5cm}>{\centering}p{1cm}c}
\hline\hline
\multicolumn{2}{c}{\bf Detector}
& $E_{\rm th}$ 
& $(a_E,b_E)$
& \bf Fiducial mass
& \bf $N^{\det}_\anue$ \\
 & & (MeV) & & kton & ($D = 10\;{\rm kpc}$) \\\hline\hline
\blankline{6}
%
\cerenkov & SK\cita{Nakahata:1998pz,Beacom:2003nk} &
5 & $(0.47,0)$ & $32$ & 5,900 - 9,990 \\
 & (H$_2$O) & & & & \\
 & SNO\cita{Virtue:2001mz,Aharmim:2004uf} & 4 & (0.35,0) &  &  \\
 & H$_2$O &  &  & 1.4 & 260 - 440\\
 & D$_2$O &  &  & 1.0 &  80 - 160 \\
Scintillator & KamLAND \cita{Iwamoto:2003aa} 
& 2.6 & (0,0.075) & 1.0 & 240 - 400 \\
 &  (N12+PC+PPO) & & & & \\
\hline
\blankline{6}
%
%
\cerenkov & HK\cita{Nakamura:2002aa}
& 5 & (0.5,0) & 540 & 100,000 - 170,000 \\
&  (H$_2$O) & & & & \\
 & UNO \cita{Jung:1999jq} 
& 5 & (0.5,0) & 650 & 120,000 - 203,000 \\
& (H$_2$O) & & & & \\
Scintillator
& LENA \cita{LENA}
& 2.6 & (0.1,0) & 30 & 7,500 - 12,600 \\
& (PXE) & & & & \\
\hline\hline
\end{tabular}
\mycaption{The relevant $\anue$ detection parameters for present and proposed
large volume detectors} {
The relevant $\anue$ detection parameters for present and proposed large
volume detectors. In the last column we give a range for the total number of
expected $\anue$ events from a Galactic SN at 10 kpc. The larger and smaller
numbers correspond respectively to a supernova model with large spectral
differences between $\anue$ and $\anux$ (Supernova model I in
chapter~\ref{ch:results}) and of a model where smaller spectral differences
are obtained (Supernova model II). As regards to neutrino oscillations, we
have assumed NH and $\sin^2\theta_{12}=0.26$ which give a mixing
$\anue:\anux\approx4:3$.  Only charged current reactions, that provide good
energy and time informations, have been considered.
}
\label{tab:detectors}
\end{table}

\section{Characteristics expected for the signal}
\label{sec:sign.char}

Starting from the detected neutrino rate (\ref{eq:rate}) we can compute the
properties of the detected signal.  The most important informations are: i)
the number of events, ii) the time-integrated spectrum and iii) the
energy-integrated time-profile.

Normally the analysis performed on a neutrino signal focus on one of these
properties.  In this section we will evaluate these quantities under different
conditions, in the attempt of understanding the interplay between the
different effects for the determination of the signal characteristics.

\subsection{Number of events}
\label{subsec:signal.number}

The number of events expected in a given energy and time interval is computed
integrating the total rate in the desired region:

\beq
N(\Delta E_{12},\Delta
t_{12})=\int_{E_1}^{E_2}{dE\int_{t_1}^{t_2}{dt\,\frac{d^2n\,(E,t)}{dE\,dt}}}.
\label{eq:signal.number}
\eeq

The counting of neutrino events provides very important global information on
the signal.  The total number of events, for example, is directly related to
the total energy released in the supernova.  In table~\ref{tab:detectors} we
present the total number of $\anue$ events for a future Galactic Supernova in
present and future detectors.

In our analysis the number of neutrino events in certain energy and time
intervals are clue properties of a signal.  For example, we are particularly
interested in low energy events because they are very sensitive to a neutrino
mass, and how many neutrinos of this type could be present in a signal becomes
a very important piece of information.

\begin{figure}[h]
\begin{center}
\epsfxsize=120mm
\epsfbox{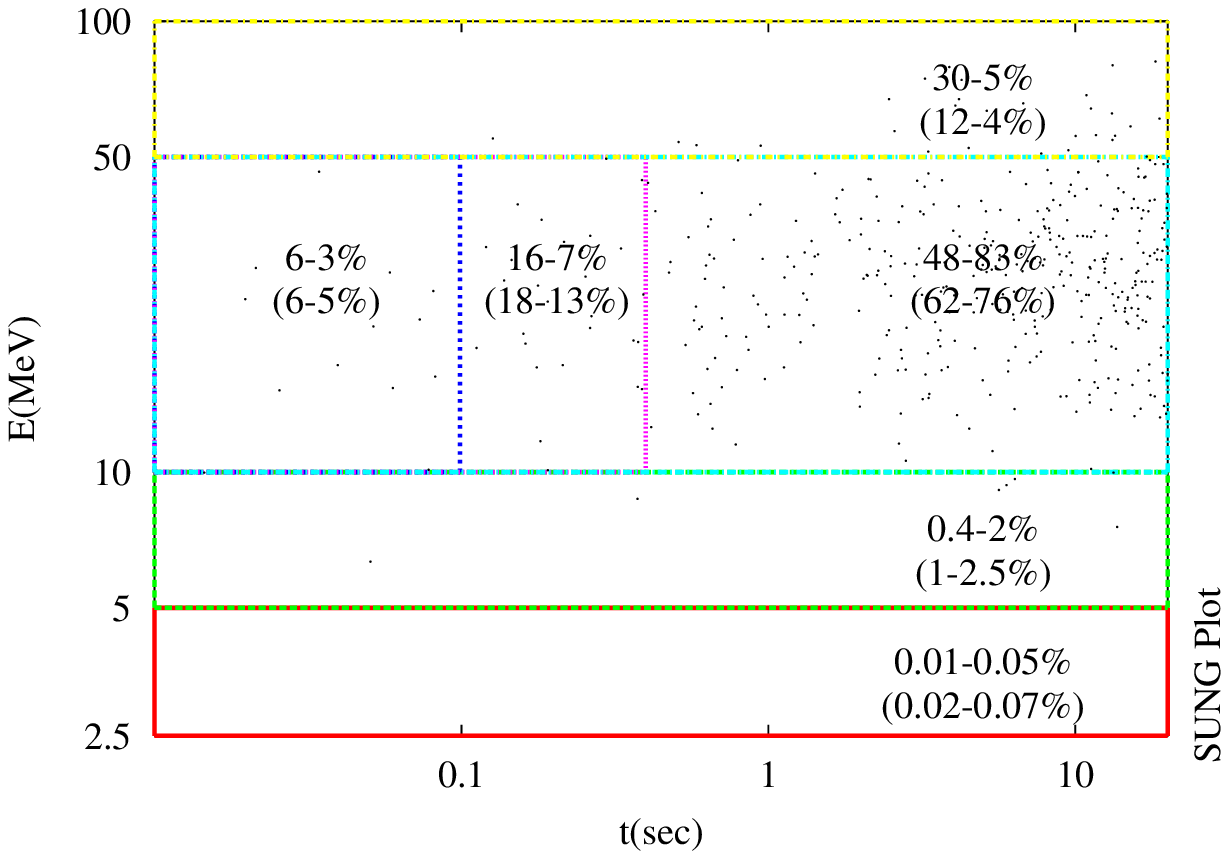}
\mycaption{Distribution of events in a supernova signal}{
Distribution of events in a supernova signal.  Fractions are computed using
(\ref{eq:signal.number}).  The ranges refer to results obtained with two
different supernova emission models.  The first figure correspond to Supernova
model I (see chapter~\ref{ch:results}) with large spectral differences; the
second one to Supernova model II with small spectral differences.  Two
different extreme mixing schemes were considered: a moderate mixing obtained
when NH is assumed (numbers in parenthesis), and a complete spectral swap that
can be produced with IH and $\sin^2\theta_{13}\gsim 10^{-3}$.
}
\label{fig:number.zones}
\end{center}
\end{figure}

In figure~\ref{fig:number.zones} we present the distribution of neutrinos
inside several interesting E-t regions.  To compute these numbers we used two
very different supernova emission models and two extreme cases for the
neutrino oscillation patterns.

%

We can see that almost in all cases the majority of events will have
intermediate energies and will arrive at late times ($t\gsim 1$).
Only few percent of the neutrinos will have energies below 10 MeV,
this means that in a signal with $10^4$ events, hundreds of neutrinos
of low energy will be available, enough to provide a good sensitivity
to the neutrino mass.  This suggests that it may be appropriate for
the purposes of mass related analysis to have energy thresholds below
the 10 MeV level.  Just few neutrinos with energies smaller than 5 MeV
are observed in almost all cases implying that an improvement in the
energy threshold below this level will not change too much the
potential to neutrino mass measurements.

\subsection{Time-integrated spectrum}
\label{subsec:time.integrated.spectrum}

One of the most interesting properties of a supernova signal is the
distribution of neutrino energies.  The energy spectrum of neutrinos
could bring us clues about the neutrino emission process and will
allow us to study the effects that oscillations in the supernova
mantle and Earth interior will have on the signal.

The time-integrated energy spectrum computed from the total detected rate
(\ref{eq:total.rate}) is given by:

\beq
f_E^\rm{det}(E)=\int{dt\,\frac{d^2n\,(E,t)}{dE\,dt}}.
\label{eq:integrated.spectrum}
\eeq

Substituting the detected flux $S^\rm{det}$
(\ref{eq:detected.signal.multiple}) in $d^2n/dt\,dE$ we can rewrite
$f_E$ in the following way

\beqas
f_E^\rm{det}(E) & = & \barr{p} F_\se(E) + (1-\barr{p}) F_\sx(E)\\
    & = & F_\se(E) - (1-\barr{p})[F_\se(E)-F_\sx(E)]\,,
\label{eq:integrated.spectrum.mod.}
\eeqas

where $F_\se(E)$ and $F_\sx(E)$ are the independent time-integrated energy
fluxes of $\anue$ and $\anux$ respectively.  In the last equation we can see
that neutrino mixing have a ``modulation'' effect on the neutrino spectrum
distortion.

\begin{figure}[h]
\begin{center}
\epsfxsize=100mm
\epsfbox{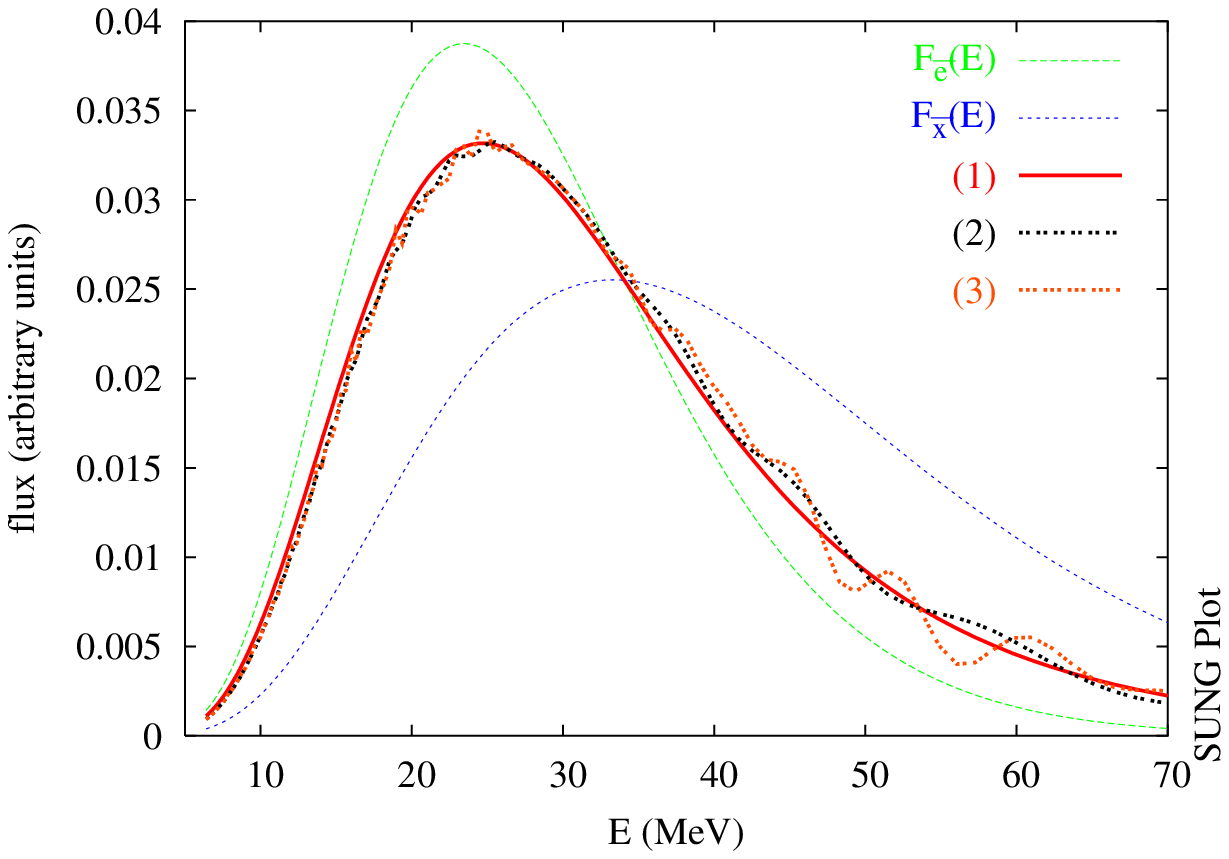}
\epsfxsize=100mm
\epsfbox{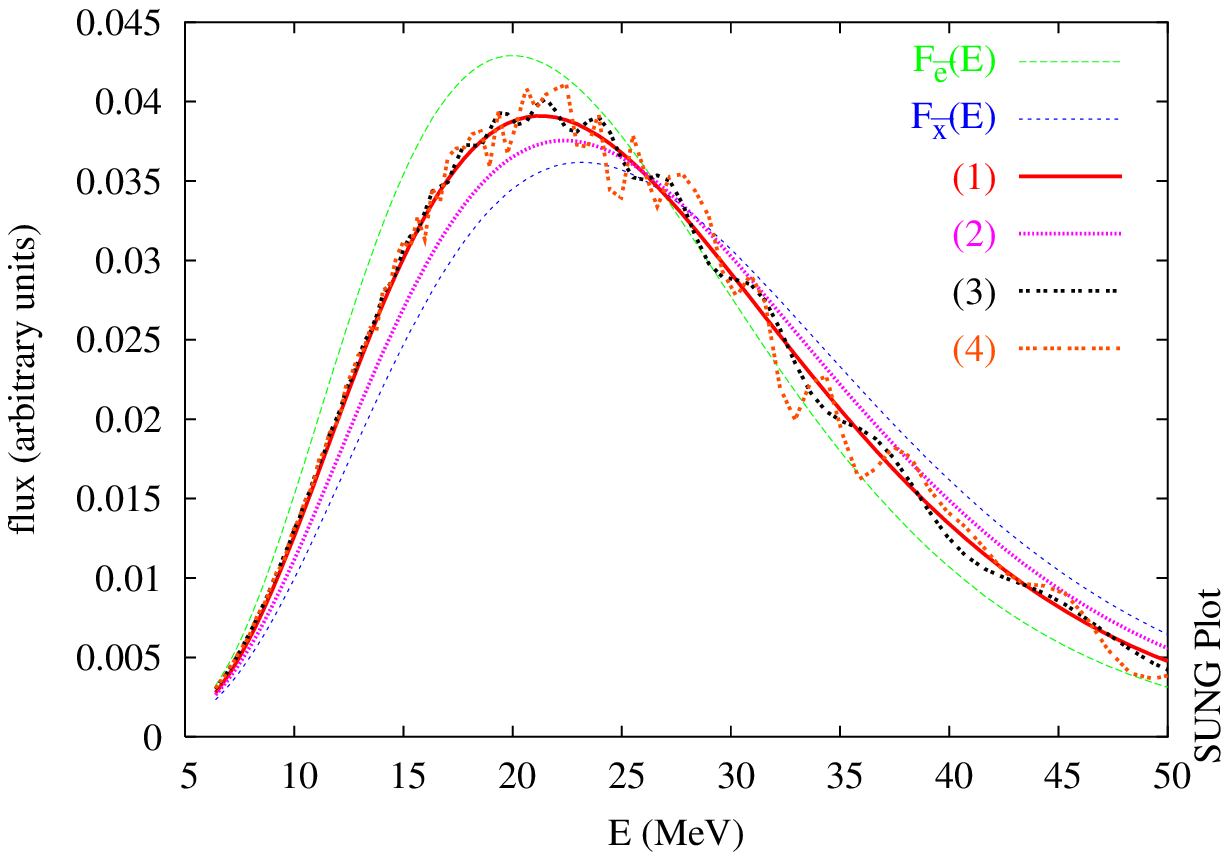}
%
\mycaption{Time integrated spectrum for two different supernova model and
oscillations schemes}{
Time integrated spectrum for two different supernova model and oscillations
schemes.  Upper panel: supernova model I: (1) NH, (2) IH,
$\sin^2\theta_{13}=10^{-2}$, (3) IH, $\sin^2\theta_{13}=10^{-2}$, Earth matter
effects, zenith angle $z=100$ (path just into the mantle) (4) same as (3) but
with $z=180\degr$ (mantle+core+mantle).  Lower panel: supernova model II: (1)
NH, (2) IH, $\sin^2\theta_{13}=10^{-2}$, Earth matter effects, $z=100$ (path
just into the mantle) (3) same as (2) but with $z=180\degr$
(mantle+core+mantle)
}
\label{fig:time.int.spectrum}
\end{center}
\end{figure}

In figure~\ref{fig:time.int.spectrum} we depict $f_E$ for two different
emission models, and assuming different neutrino oscillations schemes.  There
are two interesting facts in the spectral distortion observed in the figures.
The first one is that in the absence of Earth oscillations the distorting
effects introduced in the case of normal hierarchy can be mimicked with an
inverted hierarchy, if the mixing parameter $\sin^2\theta_{13}$ has a value in
the intermediate region of fig.~\ref{fig:jump.contours}.  The other observation is
that Earth matter effects will produce only a mild distortion.

%
\afterpage{\clearpage}
%

\subsection{Energy-integrated time profile}
\label{subsec:energy.integrated.profile}

The energy-integrated time profile is computed from the detected rate
(\ref{eq:total.rate}) as:

\beq
f_t^\rm{det}(t)=\int{dE\,\frac{d^2n\,(E,t)}{dE\,dt}}.
\label{eq:energy.int.timeprofile}
\eeq

$f_t$ provides global informations on how the neutrino flux changes in time,
regardless of the neutrino energies.

\begin{figure}[h]
\begin{center}
\epsfxsize=120mm
\epsfbox{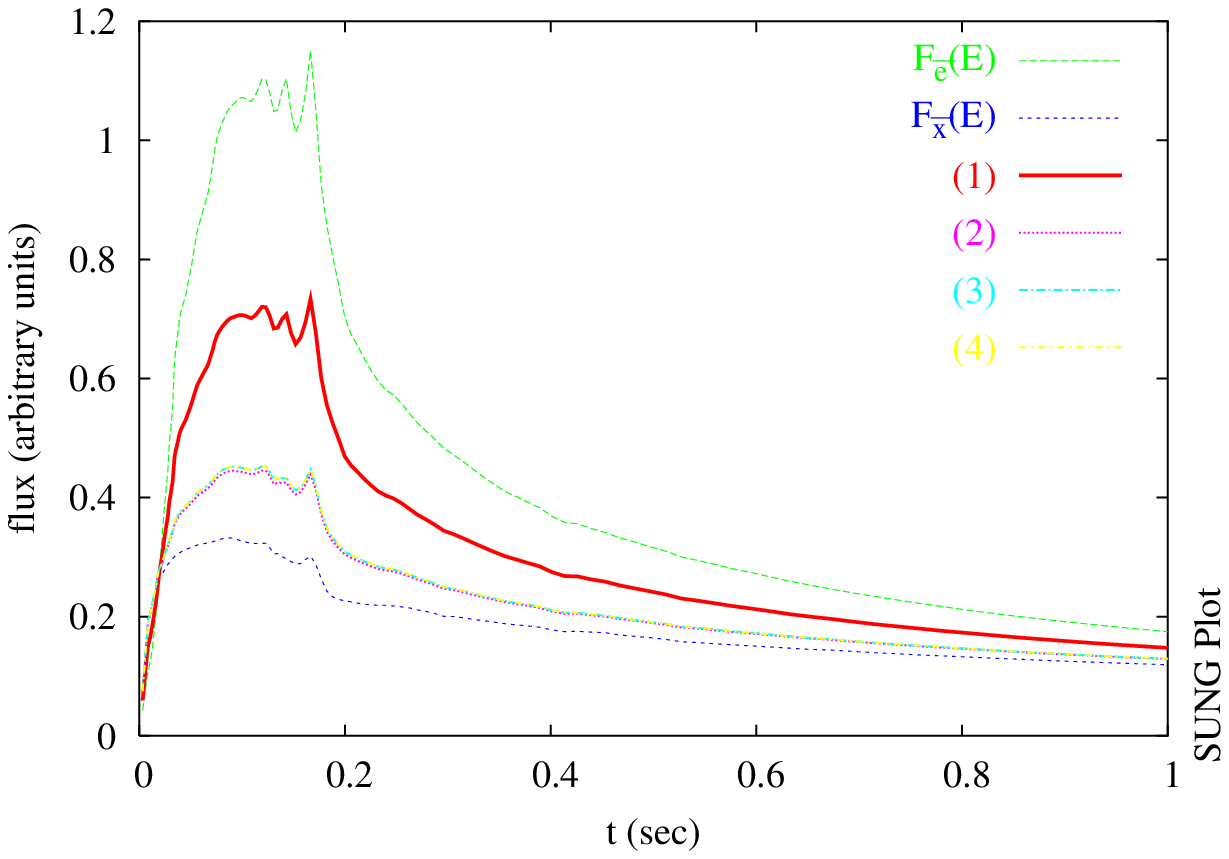}
\mycaption{Energy integrated time-profile}{
}
\label{fig:energy.int.profile}
\end{center}
\end{figure}

Figure~\ref{fig:energy.int.profile} depicts typical time-profiles obtained for
different oscillation schemes.

\afterpage{\clearpage}
%

We notice that besides a global modulation effect, the oscillatory distortion
of the flux that was clearly observed in the spectrum can not be discerned in
the time profile.  In particular, Earth matter effects become almost
undetectable.  This is easy to understand since being indeed small, Earth
effects are averaged out when the integration in the energy is performed.

\section{Generation of full statistics signals}
\label{sec:sign.generation}

Many of the techniques that have been considered to analyze a signal
from a future Galactic supernova use one or more of the properties
described in the previous section.  To obtain the number of events or
the integrated signal profiles it is not necessary to generate a
complete signal, and then testing these techniques is rather
straightforward.  On the other hand, if we want to perform a
statistical analysis of the whole statistics in the signal, as is done
in only a few of the proposed techniques, it becomes necessary to
generate a complete realization of the signal in all its details.

There are two different ways to generate a detailed synthetic supernova
neutrino signal.  The first one is an intuitive procedure where the various
physical processes are simulated.  Neutrinos of different flavors are
generated using the emitted fluxes $S_{\barr{\a}}(E,t)$.  Then the ``seed''
sample is filtered first according to the conversion probabilities, where
neutrinos change flavor depending on their energy, and after with the cross
section and efficiency of detection.  The final signal is obtained joining all
the resulting events.

Despite the apparent naturalness, this method has several defects.  Firstly,
although sampling the emitted fluxes seems to be rather simple, because they
are not affected by the distorting effects of oscillations, designing reliable
filters using the conversion probabilities, particularly in presence of an
energy dependence or a strong oscillatory behavior, is a non trivial matter.
On the other hand since the seed sample to which the filters are applied has a
finite number of neutrinos, statistical fluctuations could be amplified by
filtering, producing artificial enhancements or depletions of given parts of
the signal.

The second method consists in sampling directly the detected rate $d^2n/dtdE$
(\ref{eq:total.rate}):

\beq 
\frac{d^2n_\anue(E,t)}{dE\,dt} = N_T\,S_{\barr{e}}^\rm{det}(E,t)\,\s(E)\,\epsilon(E).
\label{eq:rate}
\eeq

The signal is constructed generating $N$ pairs of energies and arrival
times $(E,t)$ (with $N$ as given by eq. (\ref{eq:signal.number})).  If
a non zero neutrino mass is assumed, the time coordinate of each event
is shifted according to the time-of-flight delay (\ref{eq:tof.basic}).
To account for measurement uncertainties (which affect mainly the
measured energies) the energy of each event is reshuffled according to
a Gaussian distribution with mean equal to the generated value $E$,
and dispersion equal to the resolution of the detector
(\ref{eq:Euncertainty}).

The synthetic signals used in the MC analysis performed to test our method
were generated through this procedure.

\begin{figure}[h]
\begin{center}
\epsfxsize=160mm
\epsfbox{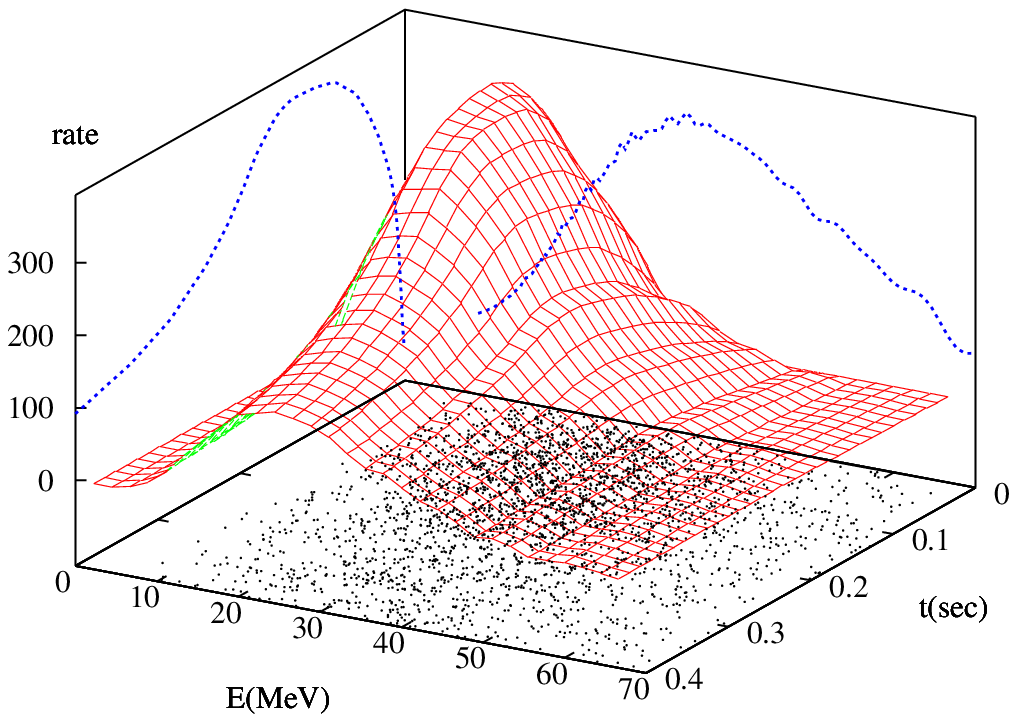}
%
\mycaption{Neutrino detected rate}{
Neutrino detected rate (surface), time-integrated energy spectrum (curve on
E-rate plane), energy-integrated flux (plane t-rate) and scatter plot (dots on
plane E-t) for a signal generated with Supernova model I assuming IH,
$\sin^2\theta_{13}=10^{-5}$ and $z=180\degr$ (mantle-core-mantle).  The
profiles have been normalized properly to fit into the plot area.
}
\label{fig:surface}
\end{center}
\end{figure}

Although the detected rate (\ref{eq:total.rate}) could be in principle
a complex multivariate function (see fig.~\ref{fig:surface}), this
method is free of the statistical flaws that could arise when the
first procedure is used.

In figure~\ref{fig:surface} we have summarized all the characteristics of a
synthetic signal.  We have plotted the surface defined by the total rate in
the most general case when Earth matter effects are present, in the E-t plane
we show a scatter plot of the events generated with such rate and on the
E-rate and t-rate planes we show the time-integrated spectrum and the
energy-integrated time profile.

%

Other details of supernova signal synthesis are presented in
Appendix~\ref{ap:sung} where we describe the computer tool that we have
developed to generate and analyze supernova neutrino signals.

\chapter{Mass limits with Supernova Neutrinos}
\label{ch:mass}

In this chapter we will describe a new statistical method to constraint
neutrino masses using a high statistics supernova neutrino signal.  The method
was described by the first time in a intuitive form in
ref.~\cite{Nardi:2003pr}.

We have organized this chapter as follows.  The first two sections are devoted
to present the main idea underlying almost all methods to measure or
constraint neutrino mass using supernova neutrinos. Then we describe
synthetically several of the recently proposed methods including those that
were used to constraint the neutrino mass using the SN1987A signal.  An
outline of the new method and the main assumptions behind it is presented in
section~\ref{sec:method.basic}.  Section~\ref{sec:method.formalism} presents a
rigorous description of the method and several of its mathematical and
numerical properties.  Details about the numerical computation of the
statistics and other quantities related to the method are presented in
sections~\ref{sec:method.likelihood} and~\ref{sec:method.margin.}.

\section{Time-of-flight delay of supernova neutrinos}
\label{sec:tof}

It was realized long time ago that valuable informations on the neutrino
masses could be provided by the detection of neutrinos from a Supernova
explosion \cite{Zatsepin:1968aa,Pakvasa:1972gz,
Piran:1981zz,1982Ap&SS..81..483S}.

The basic idea relies on the estimation or measurement of the time-of-flight
(TOF) delay $\Delta t_{\rm tof}$ that a relativistic neutrino of mass $m_\nu$
and energy $E_\nu$ traveling a distance $L$ would suffer with respect to a
massless particle:

\beqa
\nonumber
\frac{\Delta t_\rm{tof}}{L} & = & \frac{1}{v} - 1 \\ \nonumber
 & = & \frac{E}{\sqrt{E^2-m^2}} - 1\, =\, \frac{1}{2}\frac{m^2}{E^2} + {\cal
 O}\left(\frac{m^3}{E^3}\right)\\
\Delta t_\rm{tof} & \simeq & 5.1\,\rm{msec}\;\left(
\frac{L}{10\,\rm{kpc}}\right) \! \left(\frac{10\, \rm{MeV}}{E_\nu}\right)^2 \!\!
\!  \left(\frac{m_\nu}{1\,\rm{eV}}\right)^2\,.
\label{eq:tof.delay}
\eeqa

\begin{figure}[p]
\begin{center}
\epsfxsize=120mm
\epsfbox{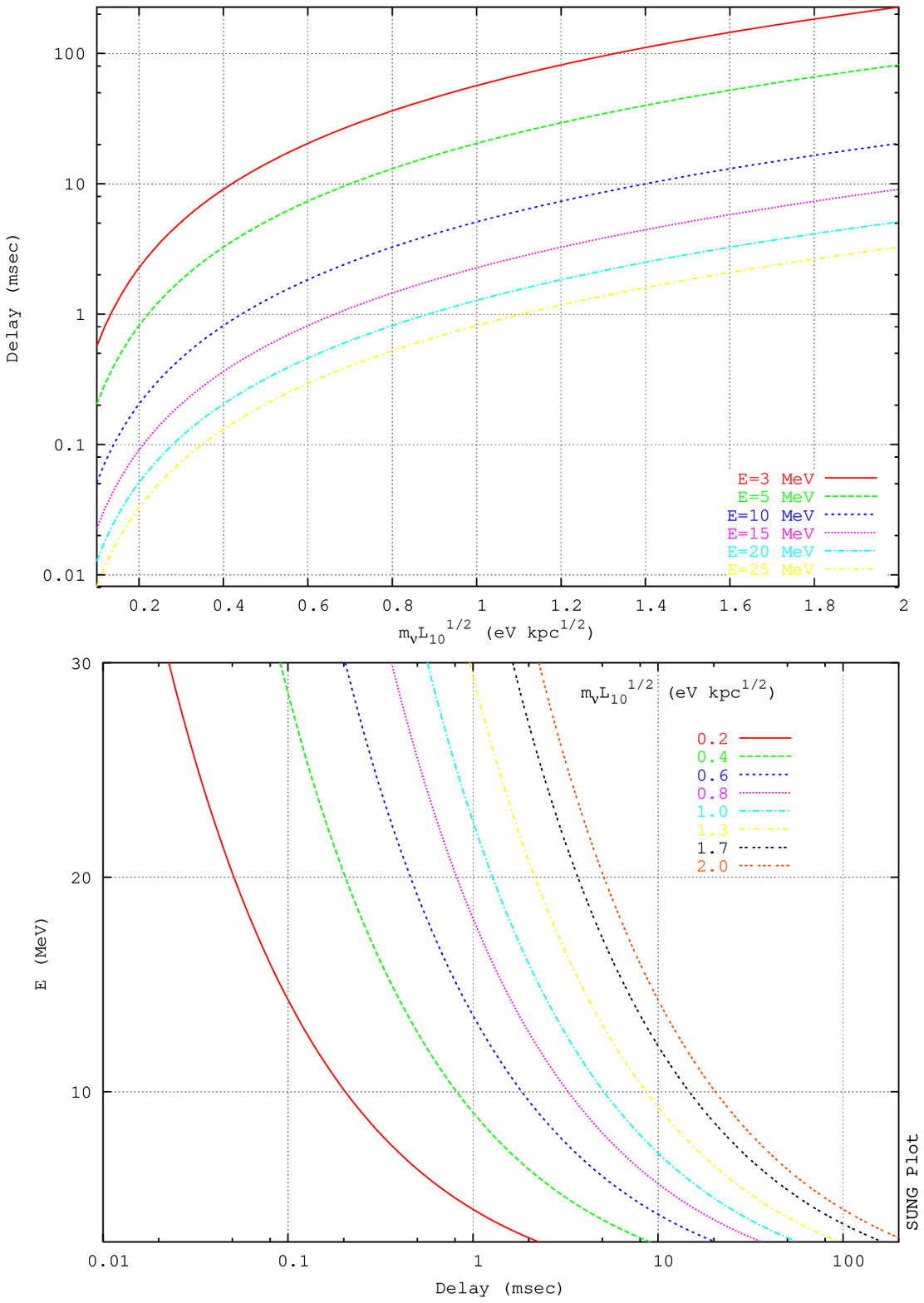}
\mycaption{TOF delay induced by a non-zero neutrino mass as a function of
neutrino mass and supernova distance}{
TOF delay as a function of the neutrino mass and the supernova distance (upper
panel) for different neutrino energies.  Lower panel depicts the continuous
dependence on the energy of the TOF delay for different values of the neutrino
mass.
}
\label{fig:delay}
\end{center}
\end{figure}

Figure~\ref{fig:delay} shows typical values of the TOF delay and its
dependence on neutrino mass and energy.

A natural question arises: what could be a ``benchmark'' in the signal that
could play the role of a ``massless particle'' to estimate the neutrino
delays?  Alternative answers to this question are at the root of different
approaches used to exploit this basic idea.  We will summarize several of them
in the next section.

\afterpage{\clearpage}
%

\section{Neutrino mass and the TOF delay}
\label{sec:delay.measurem.}

In order to measure or constrain a neutrino mass using the TOF delay is
necessary to identify some kind of ``timing'' information in the signal.  Many
different ideas have been proposed since the seminal work by Zatsepin
\cite{Zatsepin:1968aa}.  Some of them have been even already used to obtain
information on neutrino mass from SN1987A
signal~\cite{Schramm:1987ra,Arnett:1987iz,Bahcall:1987nx,Spergel:1987ex,Abbott:1987bm}.

We could classify all the methods in three general classes:

\begin{itemize}

\item Methods that use the change in the time spread of the signal due to
  delays between the arrival time of high and low energy neutrinos.

\item Methods that rely on the occurrence of specific timing events in
  coincidence with neutrino emission.

\item Methods that make use of a detailed statistical description of the
  signal.

\end{itemize}

\subsection{Time spread of the signal}
\label{subsec:timespread}

A supernova neutrino signal has a relatively short duration.  Most of the
neutrinos are emitted in just few seconds and although the detailed evolution
of the neutrino flux depends on the supernova model, the total duration of the
signal, which is determined by the time scale of neutrino diffusion in the
supernova core, can be predicted from basic physical principles.

Neutrinos of different energies emitted simultaneously from the supernova core
will arrive at different times.  This effect will increase the time spread of
the signal in an amount directly related to the value of neutrino mass and
supernova distance according to (\ref{eq:tof.delay}).  Therefore, knowing the
expected duration of the supernova neutrino burst and measuring the observed
time spread it is possible to constraint or measure the neutrino mass.

Using this kind of approach the total duration of the signal from SN1987A was
used to estimate model independent limits on the electron neutrino mass in the
range $20-30$ eV \cite{Schramm:1987ra}.  More stringent limits were obtained
using specific assumptions about the signal time structure
\cite{Arnett:1987iz,Bahcall:1987nx,Spergel:1987ex,Abbott:1987bm}.

More recently a method to measure the time spread of neutrinos emitted during
the shock breakout in a future Galactic supernova has been proposed
\cite{Totani:1998nf}.  The short duration emission of neutrinos in this phase,
determined by the time scale of the shock wave emergence in the supernova
core, could allow to set limits at the level of $3$ eV without make any
additional assumption about the time structure of the neutrino emission.

\subsection{Timing events}
\label{subsec:timing}

Another way to estimate a neutrino mass with a supernova signal is based on
``timing'' events that take place simultaneously with the supernova explosion,
and used as benchmarks to measure the neutrino delays.

Gravitational waves are expected to be emitted almost in coincidence
with the neutrino burst \cite{Fargion:1981gg,Arnaud:2001gt}.  If the
supernova is close enough to allow a clear detection of gravitational
waves, the peak in the emission of this radiation could be used to
measure the delayed arrival of neutrinos in the early phases of the
explosion, and constrain neutrino masses at the level of $1$ eV
\cite{Arnaud:2001gt}.

Another possibility is the early formation of a Black hole in the supernova
core that will abruptly truncate the neutrino emission.  With a non zero
neutrino mass the cutoff in the observed flux will not be turn off sharply,
and a measurement of the time spread of the neutrino signal could allow to put
limits on neutrino mass at the level of $1.8$ eV
\cite{Beacom:2000ng,Beacom:2000qy}.

\subsection{Detailed analysis of a signal}
\label{subsec:det.sig.anal.}

A non-zero neutrino mass will produce a global distortion in a supernova
signal that would be very difficult to be mimicked by astrophysical effects at
the source.  Using a detailed model of neutrino emission and performing a
complete statistical analysis, values of the neutrino mass could be
constrained by subtracting the effect of the masses and measuring the
agreement between the emission model and the signal.

A very accurate statistical analysis performed under these lines of reasoning
was done recently by Loredo and Lamb \cite{Loredo:2001rx}.  They used a
Maximum Likelihood analysis to fit several types of neutrino emission models
with the SN1987A data, and Bayesian reasoning to account for prior information
about those models and to perform statistical comparisons between them. They
obtained the limit $m_\anue<5.7 eV$ which is more stringent than any other
limit coming from the analysis of SN1987A signal, and close to laboratory
upper bounds on the electron neutrino mass.

The method proposed here belongs to this kind of approaches.  In the
following sections we present a detailed description of the method and
its properties.

In table~\ref{tab:supern.limits} we summarize the neutrino mass limits from a
supernova signal, for SN1987A and for a future Galactic supernova.

\begin{table}[t]
\raggedright
\centering
\small
\begin{center}
\renewcommand{\arraystretch}{0.9}
\setlength{\tabcolsep}{1mm}
\mycaption{Limits on the neutrino mass from supernova neutrinos}{
Limits on the neutrino mass from supernova neutrinos, SN1987A and from a
future Galactic supernova neutrino signal.
}
\vspace{12pt}
\begin{tabular}{l>{\PBS\raggedright}p{8cm}p{3cm}}
  \bf Limit & \bf Description & \bf References \\\hline\hline
\multicolumn{3}{c}{\bf SN1987A} \\ \hline
$m_{\anue}\lesssim 20-30$ eV & 
Model independent mass limits using the total duration of the signal.
& \cite{Schramm:1987ra} \\
$m_{\anue}\lesssim10-26$ eV & 
Model dependent mass limits with assumptions on the time structure of signal.
&  \cite{Arnett:1987iz,Bahcall:1987nx,Spergel:1987ex,Abbott:1987bm} \\
$m_{\anue}\lesssim5.7$ eV & 
Complete statistical analysis using different neutrino emission models and
Bayesian techniques.
& \cite{Loredo:2001rx} \\\hline\hline
\multicolumn{3}{c}{\bf Potential} \\ \hline
$m_{\anue}\lesssim3$ eV & 
Model independent limit using time spread after shock breakout
& \cite{Totani:1998nf} \\
$m_{\anue}\lesssim1.0$ eV & 
Gravitational waves as benchmark to measure neutrino delays in the early phase
of the emission
& \cite{Arnaud:2001gt} \\
$m_{\anue}\lesssim1.8$ eV & 
Time spread of the residual neutrino emission after the abrupt cutoff of the
flux by the formation of a Black hole
& \cite{Beacom:2000qy,Beacom:2000ng} \\\hline\hline
\end{tabular}
\label{tab:supern.limits}
\end{center}
\end{table}

\section{Basic description of method}
\label{sec:method.basic}

The method proposed has three basic characteristics:

\begin{enumerate}

\item It is based on a neutrino-by-neutrino kind of analysis and requires the
  determination of the time and energy of each event in the
  signal\footnote{$\anue$ events on scintillator and \cerenkov detectors are
  particularly well suited for this kind of analysis.  Additionally, since
  $\anue$ will provide the largest number of events on present and future
  detectors, we will focus on these events in the description of our method
  and in the tests performed to measure its sensitivity.}

\item The method uses the full statistics of the signal, i.e.  every neutrino
  event in the signal is used to construct the statistical estimator.

\item It can be applied independently of particular astrophysical assumptions
  about the characteristics of the neutrino emission (time evolution of the
  neutrino luminosity and spectral parameters) and does not rely on additional
  benchmarks events for timing the neutrinos TOF delays.

\end{enumerate}

The method relies on two basic assumptions:

\begin{enumerate}

\item The first and most important one is that inside the collapsing core
  neutrinos are kept in thermal equilibrium, by means of continuous
  interactions with the surrounding medium, and therefore are emitted with a
  quasi-thermal spectrum.  As was explained in chapter \ref{ch:fluxes} this is a
  solid prediction of almost all supernova models and simulations and has also
  been confirmed by the duration of the SN1987A signal that constitutes an
  evidence for efficient neutrino trapping within the high density core.

\item The second hypothesis is that the time scale for the variation of the
  characteristics of the neutrino spectrum is much larger than the time lags
  induced by a non-vanishing mass (say, much larger than 5 msec., see
  (\ref{fig:delay})).  Also this assumption is quite reasonable, since it is a
  robust prediction of all SN simulations
  \cite{Burrows:1991kf,Woosley:1994ux,Totani:1997vj,Raffelt:2003en} that
  sizeable changes in the spectral parameters occur on a time scale much
  larger than 5 msec (see fig.~\ref{fig:mean.energies}).

\end{enumerate}

According to the first assumption, a high statistics neutrino signal can be
considered as a `self timing' quantity, since the high energy part of the
signal, that suffers only negligible delays, could determine with a good
approximation the characteristics of the low energy tail, where the mass
induced lags are much larger.  The second assumption implies that the time
evolution of the spectral parameters as inferred from the detected sample will
reproduce with a good approximation the time evolution of the neutrino
spectrum at the source.  No additional timing events are needed, and each
neutrino, according to its specific energy, provides a piece of information
partly for fixing the correct timing and partly for measuring the time delays.

\begin{figure}[p]
\begin{center}
\epsfxsize=120mm
\epsfbox{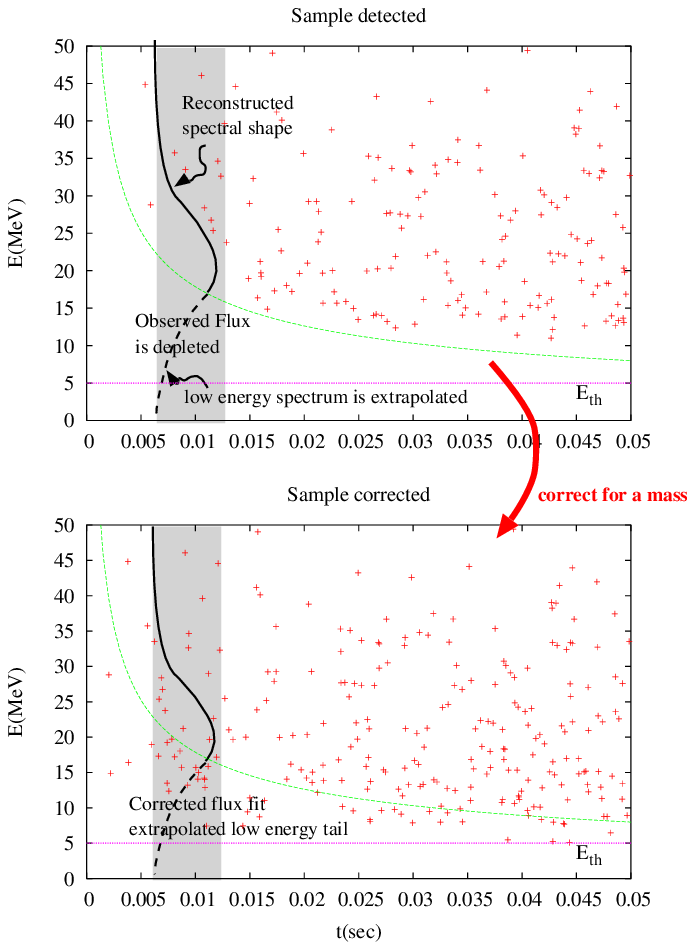}
\mycaption{Schematic explanation of the method strategy}{
Schematic explanation of the method strategy.  Using the observed neutrino
energy distribution at high energy the best-fit spectrum at a given time is
found (upper panel).  The neutrino mass can be measured by shifting the events
with a time delay according to the test mass and event energy, until the {\em
corrected} flux reproduces what is expected from the extrapolation to low
energies of the fitted spectra at each time (lower panel).
}
\label{fig:method.schem}
\end{center}
\end{figure}

Figure~\ref{fig:method.schem} illustrates schematically the basic strategy of
the method.

\afterpage{\clearpage}
%

\section{Formalism for the signal analysis} 
\label{sec:method.formalism}

In real time detectors, supernova electron antineutrinos are revealed through
the positrons they produce via charged current interactions, that provides
good energy informations as well.  Each $\anue$ event corresponds to a pair of
energy and time measurements $(E_i, t_i)$ together with their associated
errors.  In order to extract the maximum of information from a high statistics
SN neutrino signal, all the neutrino events have to be used in constructing a
suitable statistical distribution. The {\it Likelihood Function} is probably
the best option to perform the analysis of the whole statistics of a supernova
signal.

To compute a LF we need to specify a probability for each event in the signal
and therefore we need to construct or select a suitable analytical model to
describe the neutrino emission from supernova.  However, we are interested in
evaluating how much information about the neutrino mass is possible to extract
from a signal irrespective of our detailed knowledge on the neutrino emission
process.  Therefore we will need to {\it marginalize} the extra information
about the astrophysical description of the signal from the effect of a
mass. We have constructed our statistical method on the basis of Bayesian
principles of inference.  In Appendix~\ref{ap:bayes} we present a simple
introduction to Bayesian reasoning.  The definitions, results and terminology
used in this section to describe our method is presented there, and therefore
a first reading of that Appendix is recommended.

The power of Bayesian statistics to perform an analysis of supernova neutrino
data has been demonstrated with the analysis of the SN1987A signal by Loredo
and Lamb in ref.~\cite{Loredo:2001rx}.  Our method is similar in several
aspects to the Loredo's analysis but the fact that we will analyze signals
with thousands of events will require a somewhat different mathematical and
numerical approach.

\subsection{The Likelihood Function}
\label{subsec:likelihood}

We identify a signal as a set of pairs of energy and time measurements
$\D\equiv\{E_i,t_i\}$.  The Likelihood Function (LF) $\lik$, associated to a
particular parametric model of the emission $\M\equiv \M(\{\t\})$ and a given
neutrino mass $m_\nu^2$, could be {\it schematically} written as:

\beq
p(\D|m^2,\M)\equiv\lik(m^2,\{\t\};\D)=\prod_{i} f(E_i,t_i)\,,
\label{eq:lik.def.}
\eeq

where the index $i$ runs over the entire set of events in the signal.  Here,
$f(E,t)$ represents a {\it probability distribution function} (pdf) used to
evaluate the contribution to the likelihood of a single event.

Using the emission rate expression in eq. (\ref{eq:emitted.signal}) and the
total detected rate in eq. (\ref{eq:total.rate}), $f$ can be written as:

\beq
f(t,E;m^2,\{\t\}) = N^{-1}\;\phi(t) \times F(E;t)\times \sigma(E)\,,
\label{eq:event.prob.}
\eeq

where the normalization constant is $N=\int dt\int
dE\,\phi(t)\,F(E;t)\,\sigma(E)$.  In the above definition we identify three
components: the neutrino time profile flux $\phi(t)$, the energy spectrum
$F(E;t)$ which in general evolves in time, and the detection cross-section
$\sigma(E)$ which is a well known function of the neutrino energy.


\subsection{Posterior probabilities}
\label{subsec:posteriors}

In Bayesian inference the likelihood of a given model in the presence of some
experimental evidence must be always weighted with the prior probability that we
could give to the model itself and to the measurement process (see
Appendix~\ref{ap:bayes}).  Therefore, when using Bayesian principles the LF is
just the first step to obtain the {\it posterior} probability for the
theoretical model.

Using Bayes theorem (\ref{eq:bayes.theo.param.}) the {\it posterior
probability} of a signal model is given by:

\beqa
\nonumber
p(m^2,\{\t\}|\D) & \equiv & \frac{\lik(m^2,\{\t\};\D) p(m^2)
p(\{\t\})}{p(\D)}\,,\\
p(m^2,\{\t\}|\D) & = & {\cal N}^{-1}\;\lik(m^2,\{\t\};\D) p(m^2).
\label{eq:full.post.}
\eeqa

In the last equation the {\it evidence}, $p(\D)$, which does not depend on the
parameters of the model $\{\t\}$, has been absorbed in a normalization
constant ${\cal N}$.  Flat prior probabilities for the model parameters
$p(\{\t\})=\rm{const.}$ (as it is often the case) can also be absorbed in the
normalization.

The choice of some suitable form for the prior probability for $m^2_\nu$ is
more subtle.  We have used for simplicity a step function
$p(m^2)=\Theta(m^2)=1,(0)$ for $m^2\geq 0,(<0)$ to exclude unphysical values of
the neutrino mass.  It must be noticed however that this is not the better way
to describe our present knowledge about neutrino mass, since a prior
probability flat on $m^2_\nu$ implies an uneven prior probability on $m$.  We
will discuss later in sect.~\ref{sec:results.otherprop} how the choice of
other priors can affect the results of the analysis.


The posterior probability in eq. (\ref{eq:full.post.}) corresponds to the joint
probability of the emission model characterized by the set of parameters
$\{\t\}$ and a given neutrino mass.  However, here we are just interested on
what the signal could tell us about the mass and then we need the posterior
probability just for this quantity.  Bayesian statistics provides a natural
way to {\it marginalize} the {\it nuisance parameters} (the emission model
parameters $\{\t\}$) by integrating them out

\beqa
\nonumber
p(m^2_\nu|\D) & = & \int{d\{\theta\}\;p(m^2,\{\t\}|\D)}\,, \\
p(m^2_\nu|\D) & = & {\cal N}^{-1}\;\Theta(m^2)\,\int{d\{\theta\}\;
\lik(m^2_\nu,\{\theta\};\D)\;}.
\label{eq:marginal.mass}
\eeqa

Our final goal will be to compute $p(m^2_\nu|\D)$ and to derive from it all
the possible probabilistic informations contained in the signal.

All methods to constrain the neutrino mass using a supernova signal are
directly sensitive to $m^2_\nu$.  However it is common to search for a way to
express the limits on $m^2_\nu$ as limits on $m_\nu$.  In our case that
translation is natural since knowing the pdf for $m^2_\nu$, to write down the
pdf for $m_\nu$ requires only a variable transformation.  The posterior pdf
for $m_\nu$ reads:

\beq
p(m|\D) = 2|m| p(m^2|\D)
\label{eq:post.mass}
\eeq

\subsection{Best-fit value and neutrino mass limits}
\label{subsec:limits}

Once the posterior pdf for $m^2_\nu$ and $m_\nu$ are computed it is possible
to derive probabilistic statements about the neutrino mass.  There are three
pieces of information which can be obtained.

Firstly the LF alone could provide the best-fit values for $\msq$ and for the
parameters of the emission model (the ones that maximize the likelihood).
This {\it Maximum Likelihood} (ML) analysis gives us the most probable value
$m^2_\rm{fit}$ and its error, compatible with the neutrino signal, but
conditioned to the specific set of best-fit values $\{\t_\rm{fit}\}$ of the
model parameters.

For practical reasons in the ML analysis the logarithm of the likelihood
function, the {\it log-likelihood function} (log-LF) is used.  Using
eq. (\ref{eq:lik.def.}) the log-LF reads

\beqa
\nonumber
\log {\cal L} & = & \sum_{i} \log f(E_i,t_i;\{\t\})\\
& = & \sum_{i} \log\left[\phi(t_i) \times F(E_i;t_i)\times \sigma(E_i)\right]
- \log{\cal N}.
\label{eq:loglf}
\eeqa

It is the maximum of the posterior pdf $p(\msq|\D)$ which actually provide us
with the most probable neutrino mass, irrespective of the value of the model
parameters and taking into account all the prior information.

The most useful information about the mass is obtained computing the {\it
credible regions} (CR) that are the range where the neutrino mass lies with a
certain probability.

For our purposes we will use three CR that respectively give us the lower and
upper limits in $\msq$, and an upperbound on $m_\nu$.

{\bf Upper and lower limits for $\msq$}.  The upper and lower limit on $\msq$
is obtained requiring that

\beq
\int_{-\infty}^{m^2_{\rm up}} p(m^2_\nu|D)\;d\,m^2_\nu = CR\,,
\label{eq:muplimit}
\eeq

and 

\beq
\int_{m^2_{\rm low}}^{\infty} p(m^2_\nu|D)\;d\,m^2_\nu = CR.
\label{eq:mdwlimit}
\eeq

\begin{figure}[t]
\begin{center}
\epsfxsize=120mm
\epsfbox{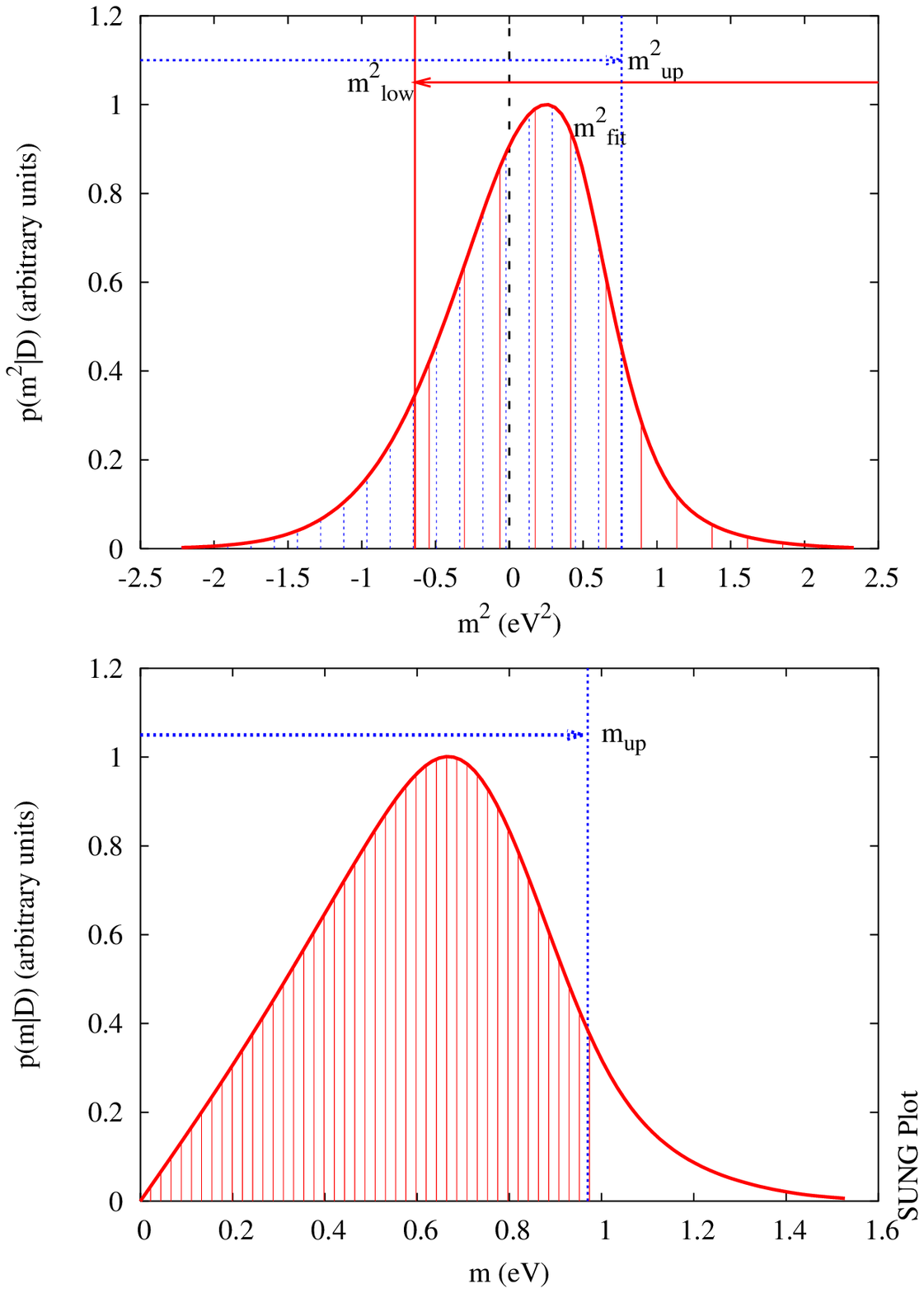}
\mycaption{Schematic illustration of the determination of mass limits from the
posterior probabilities}{
Schematic illustration of the determination of mass limits from the posterior
probabilities for $\msq$ (upper panel) and $m_\nu$ (lower panel).
}
\label{fig:mass.limits}
\end{center}
\end{figure}

Figure~\ref{fig:mass.limits} illustrate the definition of these limits.

{\bf Upper bound for $m$}.  If the neutrino mass is too small to be discerned
an upperbound on $m$ can be obtained from its posterior pdf
(\ref{eq:post.mass}) requiring

\beq
\int_{0}^{m_{\rm up}} p(m_\nu|D)\;d\,m_\nu = CL,
\label{eq:mupper}
\eeq

\afterpage{\clearpage}
%

\section{Construction of the Likelihood Function}
\label{sec:method.likelihood}

To compute the LF (\ref{eq:lik.def.}) two basic elements are required. First, we
need to deduce the time evolution of the spectral parameters from the sample
in order to describe the neutrino energy spectrum at any time.  Second, we
have to find a suitable class of parametric analytical function $\phi(t)$ that
could fit reasonably well the {\it detected} flux.

\subsection{Neutrino spectrum} 
\label{subsec:fluxmodel}

According to the first assumption in sect.~\ref{sec:method.basic} the spectrum
can be reasonably described by a quasi-thermal (analytical) distribution. If
for example a distorted Fermi-Dirac function~$\sim [1+\exp(E/T-\eta)]^{-1}$ is
used, as was done in ref.~\cite{Nardi:2003pr}, $F$ can be parameterized in
terms of a time dependent effective temperature $T$ and degeneracy parameter
$\eta$ describing the spectral distortions.  According to the second
assumption, the time dependence of the relevant spectral parameters can be
inferred directly from the data.

Here, we model the observed neutrino spectrum $F(E;t)$ by means of an
$\alpha$-distribution (\ref{eq:alpha.dist}),

\beq
\nonumber
F\left(E,\bar\epsilon(t),\alpha(t)\right) = N(\bar\epsilon,\alpha)\> 
\left({E}/{\bar\epsilon}\right)^{\alpha}e^{-(\alpha+1)\,E/\bar\epsilon}\,,
\eeq

where $\e(t)$ and $\a(t)$ describe the evolution in time of the spectrum.

The choice of this distribution to fit the observed neutrino spectrum instead
of the physically better motivated Fermi-Dirac, obeys similar reasons than
those presented in sect.~\ref{subsec:pinching}.  Starting from a discrete sample
of neutrinos, a Fermi-Dirac spectrum can be reconstructed only by carrying out
numerical fits to the energy momenta until the correct values of $T$ and
$\eta$ are determined through a minimization procedure. In contrast, the
$\alpha$-distribution can be straightforwardly determined through the simple
analytical relations (\ref{eq:alpha.momenta}) connecting the spectral parameters
and the two first momenta of the energy distribution.

Since the effect of the detection cross-section (see
sec.~\ref{subsec:detect.process}) modifies the measured energy distribution,
we need to find a way to reconstruct the energy momenta of the emitted
distribution from the set of observed energies.

If we call $F_e(E)$ the energy spectrum at the source (where for simplicity we
have dropped the temporal dependence) and $F_d(E)$ the (unknown) observed
neutrino spectrum, $F_d(E)=F_e(E)\times \sigma(E)$, the n$^th$-momentum of
$F_e$ is given by:

\beqa
\nonumber
\barr{E^n}&\equiv&\frac{\int_0^\infty{dE\;E^n\,F_e(E)}}{\int_0^\infty{dE\;F_e(E)}}\\
\nonumber
&=&\frac{\int_0^\infty{dE\;E^n\,F_d(E)/\sigma(E)}}{\int_0^\infty{dE\;F_d(E)/\sigma(E)}}.
\eeqa

The last relation allows us to define the new {\it cross-section weighted}
momenta:

\beq
\aver{E^n}\simeq\frac{\sum{E_i^n/\sigma(E_i)}}{\sum{1/\sigma(E_i)}}.
\label{eq:correct.momenta}
\eeq




This weighted momenta provide us an estimate of the momenta of the emitted
energy spectrum and allows us to find the spectral parameters around a given
time.  So, in order to obtain the spectral evolution, we slice the sample in
many time windows and find the values of $\alpha$ and $\epsilon$ inside them.

A couple of words about the procedure of signal slicing procedure are
necessary to explain in more details how the functions giving the spectral
parameters evolution are constructed.

Any slicing procedure faces the effects of statistical fluctuations.
In our case these fluctuations come from the limited statistics on
each window and from the fact that neutrinos from different times and
with different spectral temperatures at the source could be mixed up
inside a single window.

\begin{figure}[h]
\begin{center}
\epsfxsize=120mm 
\epsfbox{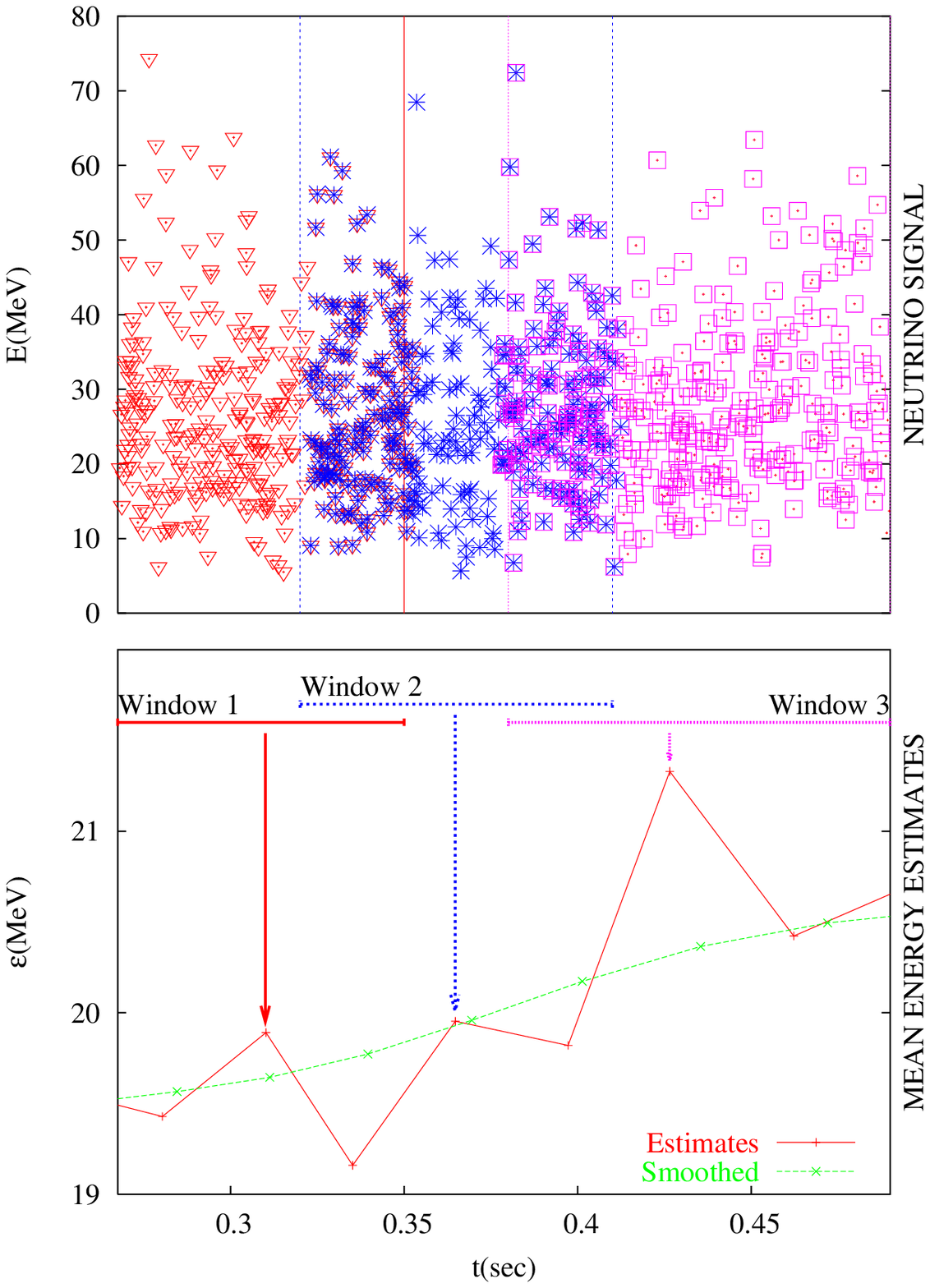} 
\mycaption{Schematic illustration of the spectral parameter estimation using
overlapped time windows}{
Schematic illustration of the spectral parameter estimation using overlapped
time windows.  The signal (upper panel) is sliced in overlapped time windows
with enough statistics (different points belong to three selected time
windows).  The emitted spectral mean energy $\epsilon$ is estimated using the
cross-section weighted momenta of the events contained on each window.  The
result is the polygonal line in the lower panel.  The final values used as
parameter $\epsilon$ in the analysis are the smoothed values of the dotted
line.
}
\label{fig:spec.estim.}
\end{center}
\end{figure}

To reduce statistical fluctuations in the spectral parameters estimation the
construction of the time windows and the processing of the resulting
information has been performed in the following way:

\begin{enumerate}

\item The width of each window is chosen large enough to contain a
  sufficient number of neutrinos (a few hundreds).  This reduces the
  statistical uncertainties coming from finite statistics effects.

\item Time windows are overlapped: the central value of each new window is
  determined as $t_{n+1}=t_n+\delta t$, with $\delta t \ll \Delta
  t_\rm{tof}$. Therefore many neutrinos of neighbor windows will be contained
  on each window and averaging the resulting fluctuations will produce smooth
  estimates for the spectral parameters (see fig.~\ref{fig:spec.estim.}).

\item Once the spectral parameters in each window have been estimated their
  values are smoothed to wash-out the statistical fluctuations and the effect
  of neighbors.

\item The time variation of $\alpha$ and $\epsilon$ is finally fitted with a
  non-linear least-square analysis.

\end{enumerate}

A schematic illustration of the above procedure is depicted in
fig.~\ref{fig:spec.estim.}.

\afterpage{\clearpage}
%

\begin{figure}[p]
\begin{center}
\epsfxsize=120mm 
\epsfbox{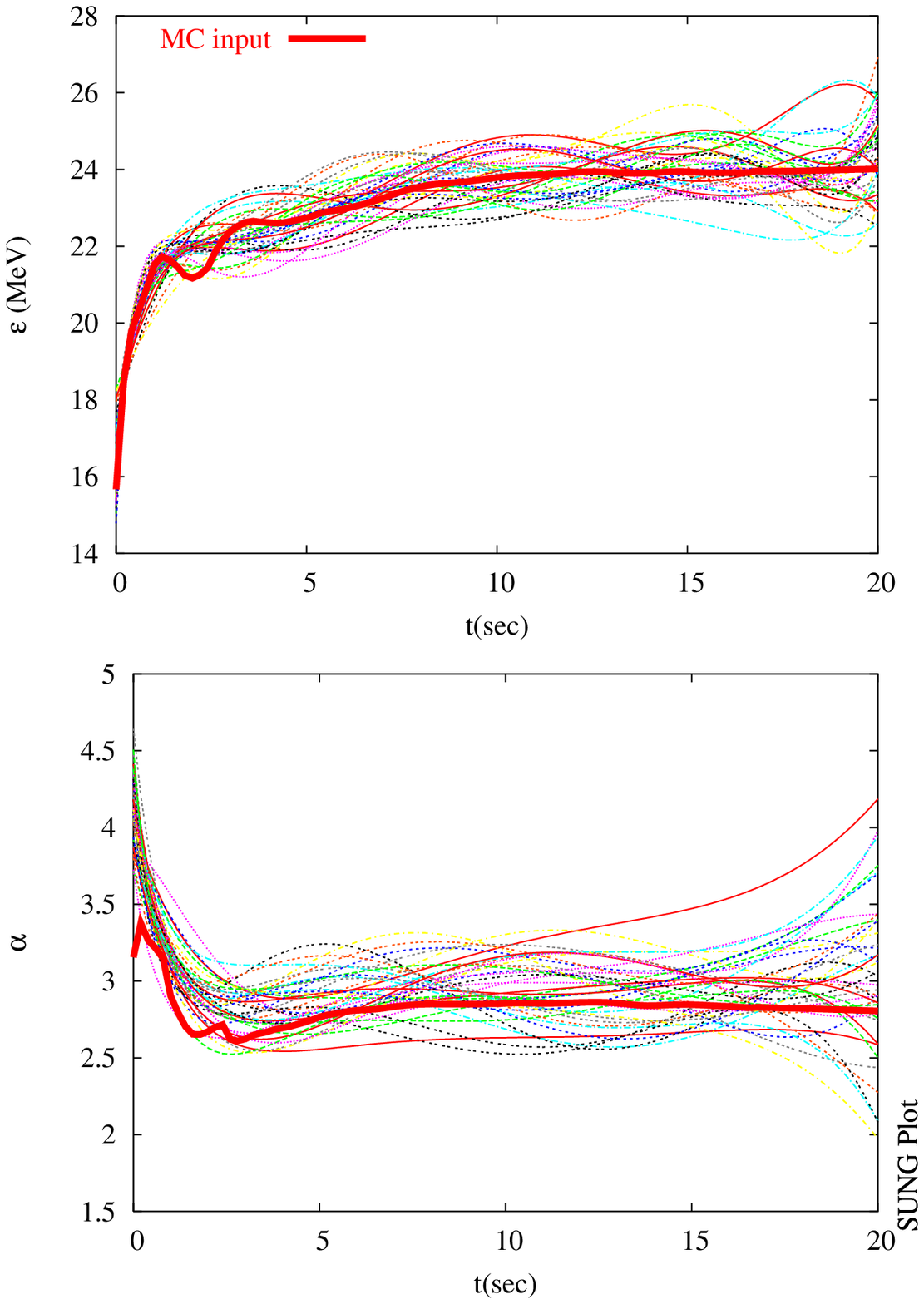} 
\mycaption{Results of the spectral parameters estimation for 40 synthetic
signals}{
Results of the spectral parameters estimation for 40 synthetic signals.  Upper
panel (lower panel): spectral mean energy $\epsilon$ (spectral pinching
$\alpha$), estimated from the sample with the procedure described in the text
(oscillatory curves).  Continuous line represent the expected value after the
mixing of $\anue$ and $\anux$ fluxes due to oscillations in the SN mantle.
The acceptable level of agreement confirms the validity of the procedure.
}
\label{fig:spec.result.}
\end{center}
\end{figure}

In figure~\ref{fig:spec.result.} we compare the estimation of the spectral
parameters with the described procedure, for a set of synthetic supernova
signals and the real values of those parameters used to generate the samples.

\afterpage{\clearpage}
%

\subsection{Detection cross-section}
\label{subsec:xsection}

Also for the analysis we have used the accurate parameterization presented in
sec.~\ref{subsec:detect.process}.  This is the same cross-section used to
generate the signals.

\subsection{Flux model} 
\label{subsec:flux.model}

Coming back to the problem of constructing the Likelihood function, and in
particular of choosing a specific time profile for the neutrino flux (namely
the model $\M$) we have proceeded according to the following requirements:

\begin{enumerate}

\item the analytical flux function must go to zero at the origin and at
  infinity

\item it must contain at least two time scales: the initial, fast rising phase
  of shock-wave breakout, and the later Kelvin-Helmholtz cooling phase.
  Another parameter could be required to describe the accretion phase,
  i.e. the transition point between these extreme phases.

\item it must contain the minimum possible number of free parameters to avoid
  degenerate directions in parameter space and to speed up the Maximum
  Likelihood analysis. Still, it must be sufficiently `adaptive' to fit in a
  satisfactory way the numerical flux profiles resulting from different SN
  simulations, as well as flavor mixed profiles as would result from neutrino
  oscillations (see section~\ref{sec:oscillations}).

\end{enumerate}

In order to test how much our results on the neutrino mass will depend
on the specific flux profile, we have performed several tests using
two different flux models constructed following the above criteria
(results are described in sect.~\ref{subsec:fluxmodel.fit}).

In the following paragraphs we will describe some of the main properties of
these two general fluxes.

\titulo{Flux model I: exponential rising and power-law decay}

The following model for the flux, in spite of being very simple, has all the
required behaviors, and moreover it showed a remarkable level of smoothness
and stability with respect to numerical ``extremization'' and multi-parameter
integrations:

\beq
\nonumber
\phi(t;\{\theta\}) = \frac{ e^{-(t_a/t)^{n_a}}}{[1 + (t/t_c)^{n_p}]^{n_c/n_p}} \sim
\casos{ll}
e^{-(t_a/t)^{n_a}} & (t \to 0). \\ 
(t_c/t)^{n_c} & (t \to \infty).
\fincasos
\label{eq:fluxmodel.simple}
\eeq

\begin{figure}[p]
\begin{center}
\epsfxsize=100mm
  \epsfbox{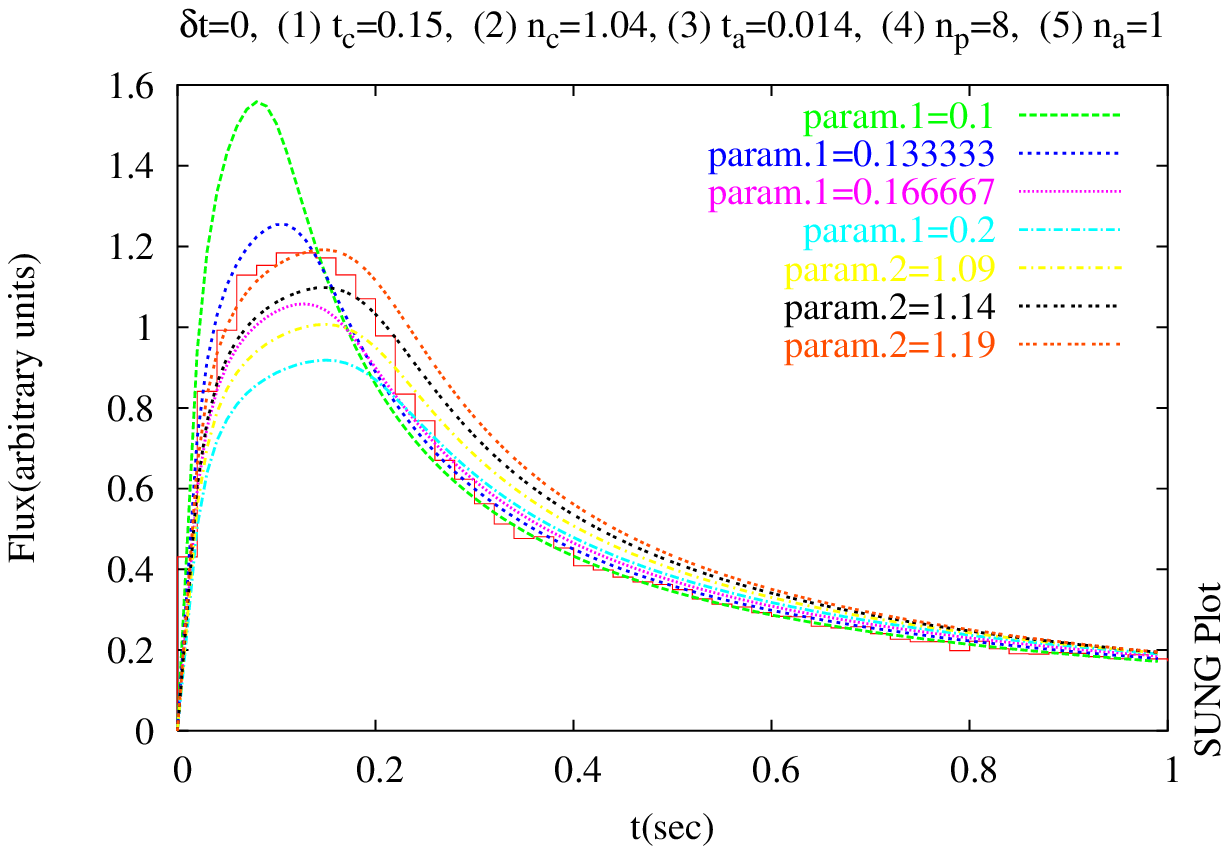}
\epsfxsize=100mm
  \epsfbox{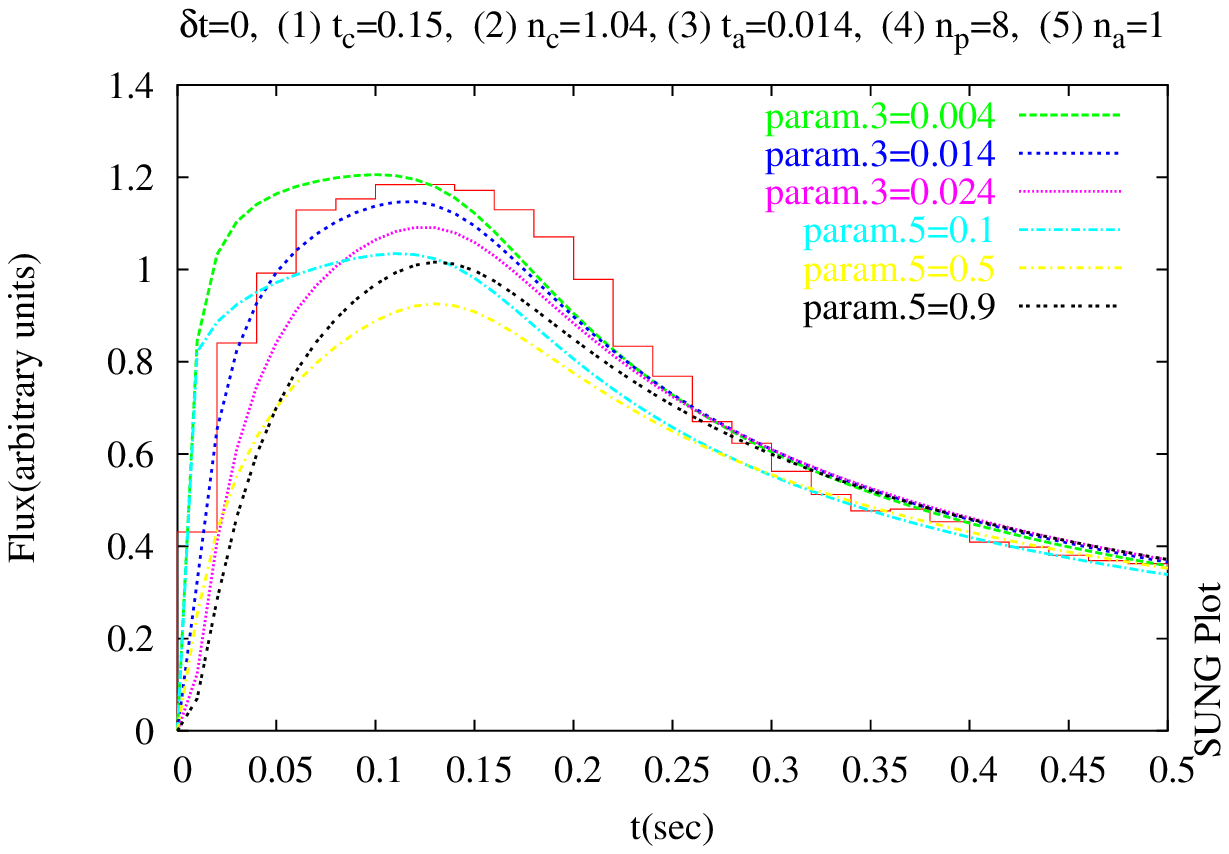}
\mycaption{Behavior of the flux model I}{
Flux model I (\ref{eq:fluxmodel.simple}) and the effect of each
parameter. Upper panel: the effects of parameters 1 (time scale $t_c$) and 2
(rate of decay $n_c$); as can be seen the effect of both parameters are very
different and little correlation between them is expected. Lower panel: the
effects of parameters 3 (time scale $t_a$) and 5 (rate of exponential rising
$n_a$); also in this case little correlation is expected.
}
\label{fig:fluxmodel.simple}
\end{center}
\end{figure}

This model has five free parameters that on the l.h.s of
(\ref{eq:fluxmodel.simple}) have been collectively denoted with $\{\t\}$: a
time scale $t_a$ for the initial exponentially fast rising phase, a second one
$t_c$ for the power law cooling phase, two exponents $n_a$ and $n_c$ that
control the detailed rates for these two phases, and one additional exponent
$n_p$ that mainly determines the width of the ``plateau'' between the two
phases (see figure~\ref{fig:fluxmodel.simple}).

\afterpage{\clearpage}
%

For the purposes of our numerical tests we have reduced the number of
parameters to three by fixing the values of exponents $n_a=1$ and $n_c=8$
found with a fit of realistic time profiles.  Although for each supernova model
it is necessary to find the best values of these exponents, the election of
these parameters will not affect too much the results of the analysis.

\titulo{Flux model II: Truncated accretion and power-law decay}

Here we describe a second analytical model for the evolution in time of the
emitted neutrino flux that is based on the phenomenological analysis presented
in sect.~\ref{subsec:analytical}. 

The analytical model in (\ref{eq:flux.model}) does not fulfill all the
requirements: for example, it does not include a description of the shock
breakout phase and therefore, does not go to zero when $t\to 0$.  To correct
this we have included to (\ref{eq:flux.model}) a multiplicative damping
factor.

Our phenomenologically motivated flux model reads:

\beqa
\phi(t;\{\t\}) & = & \left(1-e^{(-t/t_d)^{n_d}}\right) \left[ \frac{A\,
e^{-(t/t_{a})^{n_{a}}}}{ (1+t/t_b)^{n_b}} + \frac{C}{(1+t/t_c)^{n_c}}\right]\,,
\\ \nonumber
& \sim & 
\casos{ll}
(t/t_d)^{n_d} & (t \to 0). \\ 
(t_c/t)^{n_c} & (t \to \infty).
\fincasos
\label{eq:fluxmodel.phen.}
\eeqa

This model has 11 parameters: four time scales, $t_d$ for the initial fast
rising, $t_a$ and $t_b$ for the flux plateau and the early flux decay and
$t_c$ for the long term decay; four exponents, $n_a$ to $n_d$ which controls
the rate of each component; and two amplitudes $A$ and $C$ which determine the
relative contribution of the accretion and cooling terms to the total flux.

\begin{figure}[p]
\begin{center}
\subfigure[Parameters $A$(1) and $C$(2)]{
\epsfxsize=70mm
  \epsfbox{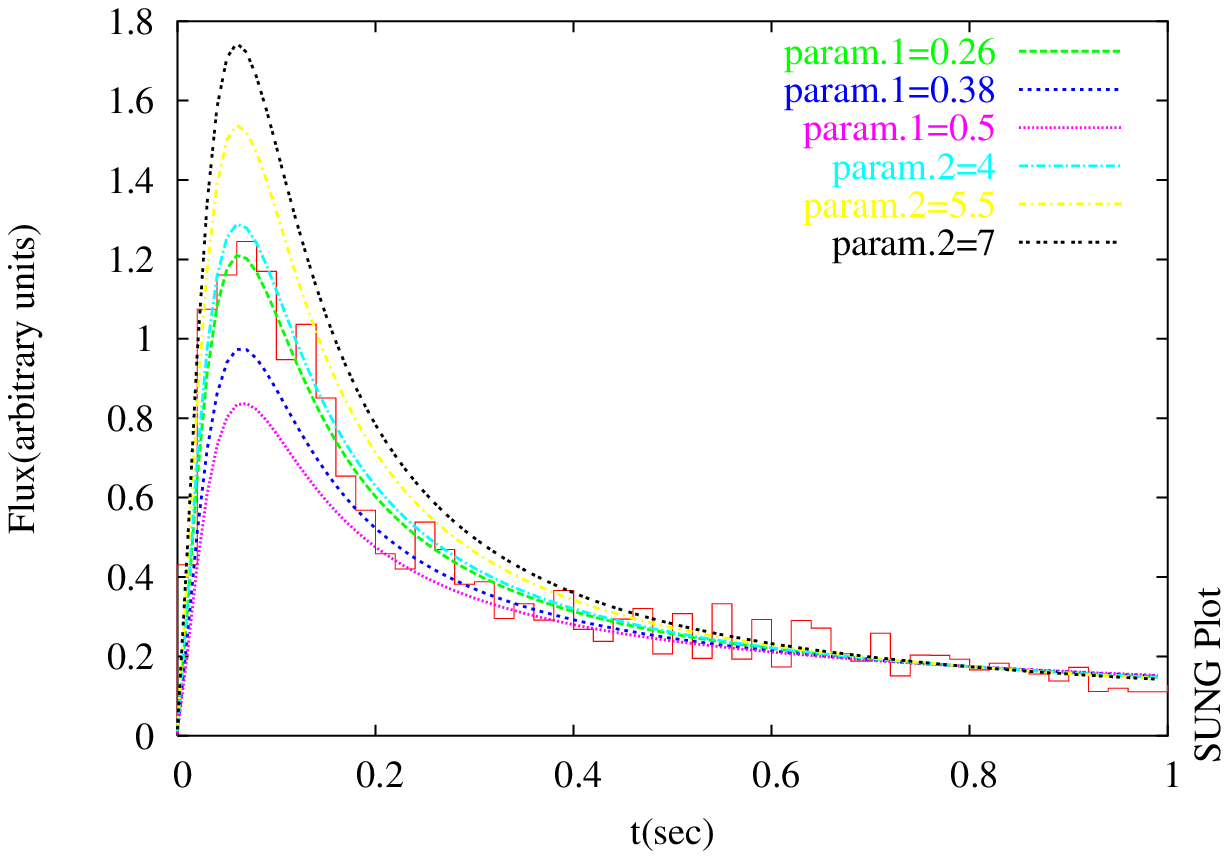}}
\subfigure[Parameters $t_a$(3) and $n_a$(7)]{
\epsfxsize=70mm
  \epsfbox{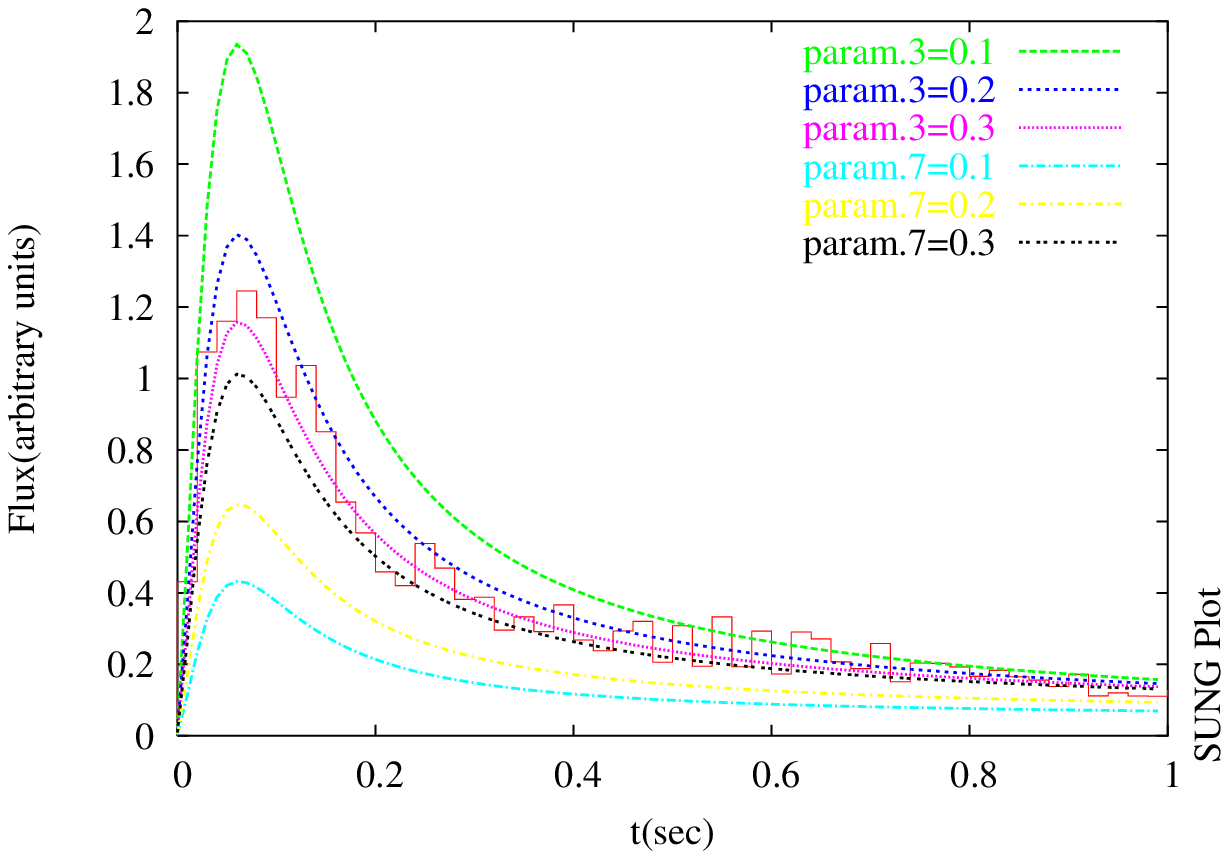}}
\subfigure[Parameters $t_c$(5) and $n_c$(9)]{
\epsfxsize=70mm
  \epsfbox{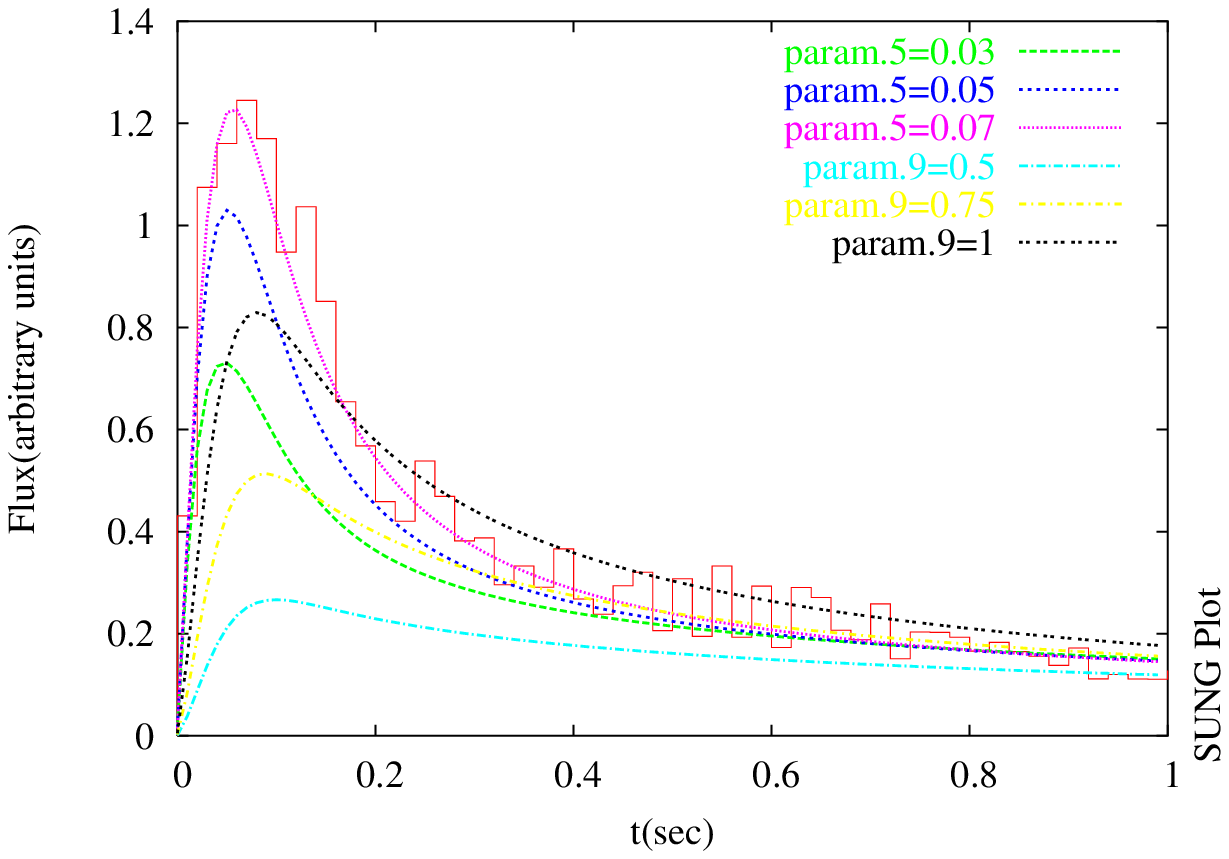}}
\subfigure[Parameters $t_d$(6) and $n_d$(10)]{
\epsfxsize=70mm
  \epsfbox{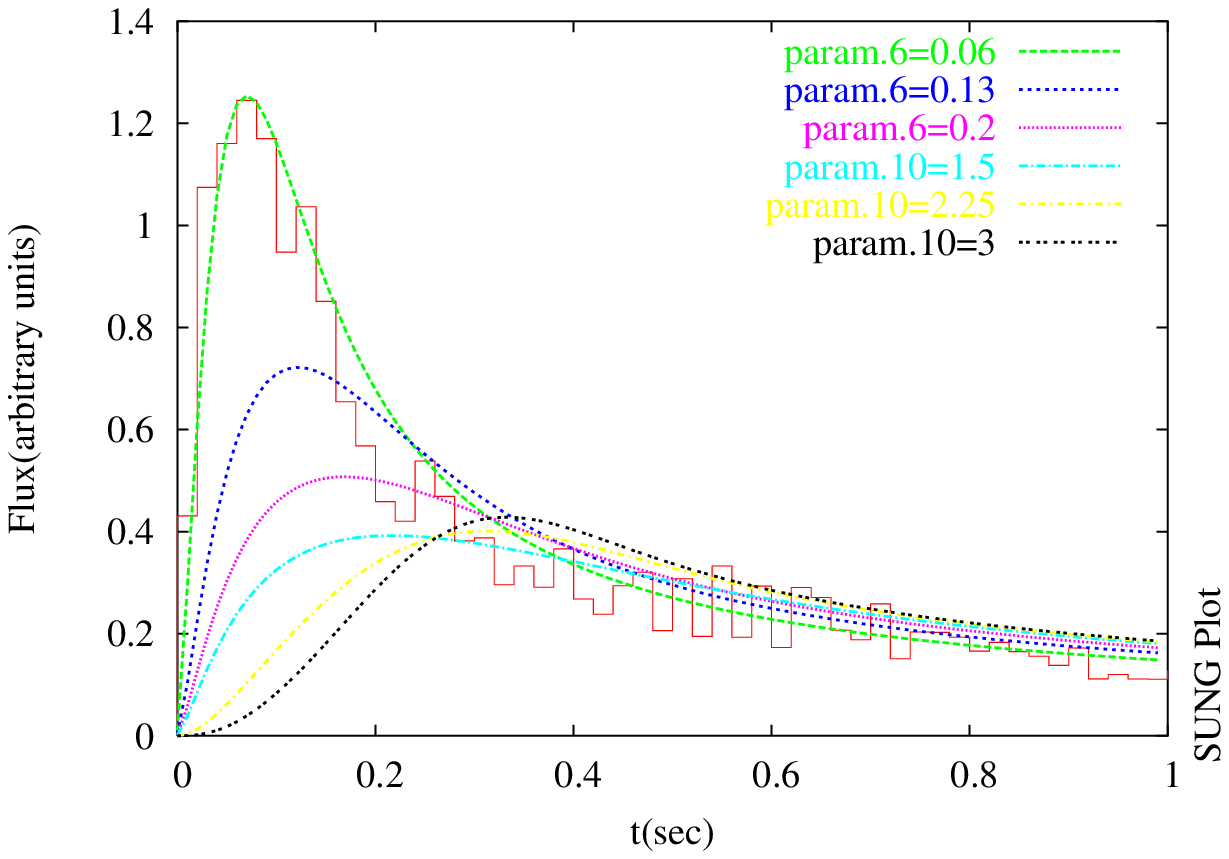}}
\mycaption{Behavior of the flux model II}{
Behavior of the flux model II in eq. (\ref{eq:fluxmodel.phen.}) under changes
in its parameters.  We start with the reference set of parameter values: (1)
$A=0.2$, (2) $C=1.5$, (3) $t_a=0.2$, (4) $t_b=300$, (5) $t_c=0.1$, (6)
$t_d=0.05$, (7) $n_a=0.3$, (8) $n_b=0.4$, (9) $n_c=2.0$ and (10) $n_d=1.5$.
In each panel a pair of parameters are studied. As can be seen in
panels~\ref{fig:fluxmodel.phen.}a,b,c the parameters $A$, $C$, $t_a$, $n_a$ and $t_c$
have similar overall effect on the flux shape.  To avoid correlations we have
left free just one of them ($t_a$).  On the other hand $n_c$, $t_d$ and $n_d$
determine different behaviors and therefore are all left free.
}
\label{fig:fluxmodel.phen.}
\end{center}
\end{figure}

As was done with the more simple flux model, we need to reduce the number of
free parameters to simplify the extremization procedures.  Studying the impact
of each parameter (see fig.~\ref{fig:fluxmodel.phen.}) we have determined
that enough flexibility could be still obtained free just three relevant
parameters: the cooling exponent $n_c$, and the time scales of fast rising
$t_d$ and early decay (truncated accretion) $t_a$.  The rest of parameters
could be fixed at the beginning performing a simple fit of the signal time
profiles with the flux model.

\afterpage{\clearpage}
%

In presence of neutrino oscillations the detected flux will be an admixture of
$\anue$ and $\nux$ fluxes.  Since the mechanism responsible for the emission
of the two species of neutrinos are different, a simple description of the
resulting flux is not possible.

\bigskip

Given that in the Likelihood analysis we will set the origin of times in
coincidence with the first neutrino detected, and this obviously cannot
correspond to the origin of time of a flux function which must satisfy
$\phi(0)=0$, an additional parameter $\delta t$ is needed to allow the
function to shift freely along the time axis according to $\phi(t)\to
\phi(t+\delta t)$.

\subsection{LF regularization}
\label{subsec:regularization}

Until now we have not discussed the role of the mass in the likelihood
function.

The neutrino mass determines the value of the time shift $\Delta t =
\Delta t(\msq,E,L)$ applied to the event probability
(\ref{eq:event.prob.}) in the definition of the LF
(\ref{eq:lik.def.}).  In that sense we could think about the neutrino
mass as another parameter of the pdf $f$.

\begin{figure}[h]
\begin{center}
\epsfxsize=120mm \epsfbox{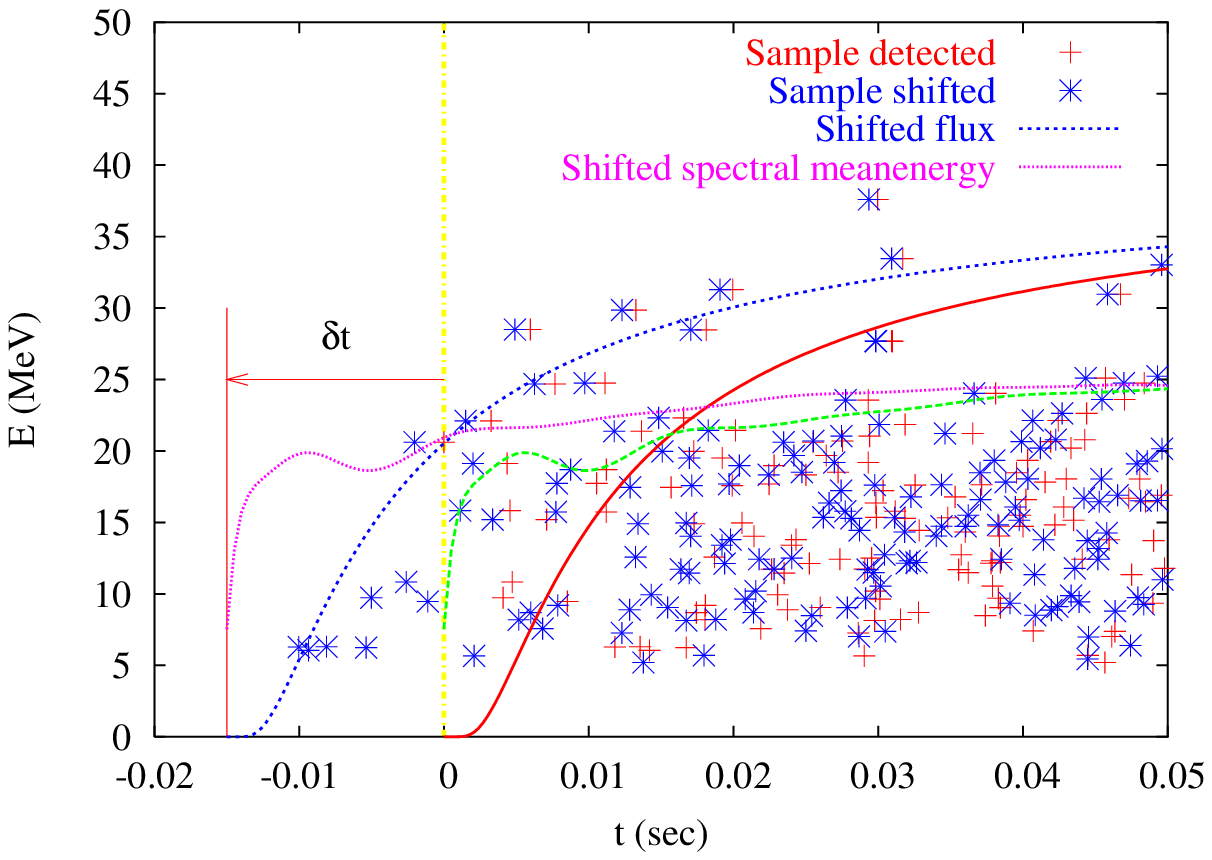}
\mycaption{Schematic illustration of the two-step process used to evaluate the
likelihood function of a given model}{
Schematic illustration of the two-step process used to evaluate the likelihood
function of a given model.  Given a test neutrino mass a new sample
(asterisks) is constructed shifting the times of all the events.  The LF is
computed by calculating the probability $f$ of the events in the shifted
sample, by applying a global shift to the flux model and spectral parameters
(curves).
}
\label{fig:shift.illust.}
\end{center}
\end{figure}

However, in order to make more transparent the way the neutrino mass enters
into the likelihood, we could treat the effect of this parameters in the
following completely equivalent way.  For each new value of the tested was we
could think as if the sample were changed: all neutrinos are shifted to
smaller times according to $\Delta t(\msq,E,L)$ and we define a new set of
pairs of energies and times $(E_i,t_i^\rm{sh})$ (see
fig.~\ref{fig:shift.illust.}).  The LF of the emission model for this
particular mass will be,

\beq
{\cal L}=\prod{f(E_i,t_i^\rm{sh};\{\t\})}.
\label{eq:lik.shift.}
\eeq 

According this procedure, it is natural that some neutrinos can end up with
negative values $t_i^\rm{sh}<0$.  This effect would induce a compensation
through a time shift of the flux model. The procedure is illustrated
schematically in figure~\ref{fig:shift.illust.}.

%

\begin{figure}[h]
\begin{center}
\epsfxsize=120mm
\epsfbox{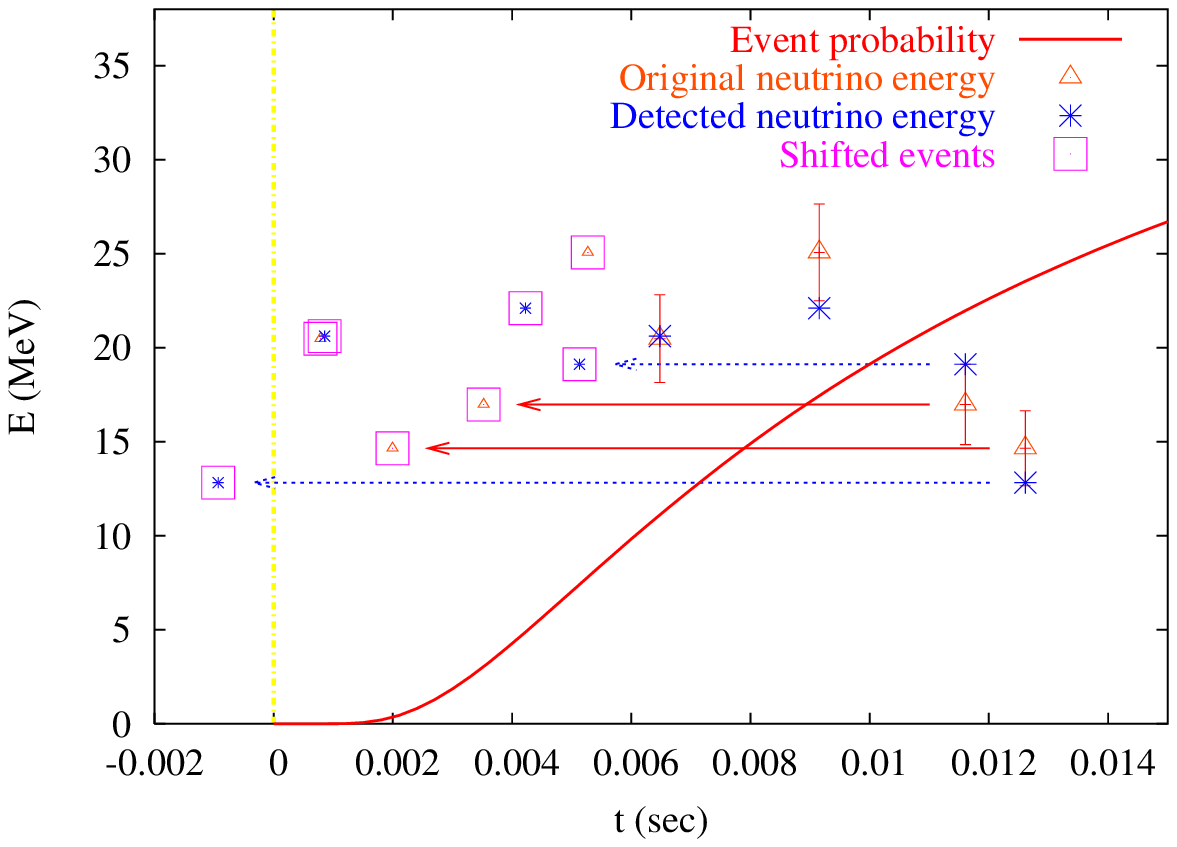}
\mycaption{Schematic illustration of the problem introduced because the detected neutrino
energies are not the true energies}{
Schematic illustration of the problem introduced because the detected neutrino
energies are not the true energies, after shift according to a test mass.  The
asterisks represent the detected events.  Triangles are at the original energy
of the event.  Bars indicate the size of the energy resolution of the detector
(Super Kamiokande resolution is used).  When shifted with a test mass the
position of the detected events are different from the original positions.
This is is a natural result of the energy dependence of this shifts.  However,
when the measured energy is smaller than the original one, as in the lowest
energy event, even if a mass equal to the real neutrino mass is used, the
resulting time shift could bring the event to a region where the log-LF
diverges.
}
\label{fig:reshuf.problem}
\end{center}
\end{figure}

Because of the finite energy resolution of any detector, the {\it measured}
neutrino energies that are used to evaluate the time shifts for neutrinos in
the sample do not correspond to the {\it true} energies that determine the
real neutrino delays.  Therefore, even when the correct value of the test mass
is used, the time-shifted neutrino sample will not correspond exactly to the
sample originally emitted.  Although completely natural (as well as
unavoidable), this behavior can produce a dangerous situation. When the energy
measurement yields a value {\it smaller} than the true energy, a neutrino
arrival time can be shifted to negatives values where the flux function, even
with a non-zero time shift, vanishes, implying that the log-Likelihood
diverges. This would imply rejecting the particular neutrino mass value for
which the divergence is produced, regardless of the fact that it could
actually be close to the true value.  Figure~\ref{fig:reshuf.problem}
illustrates this point.

\afterpage{\clearpage}
%

To correct this problem we have adopted the following procedure.  The
contribution $f$ to the Likelihood (\ref{eq:lik.shift.}) of a neutrino event
with measured energy $E_i\pm \Delta E_i$ for which, after subtracting the
delay $\delta t_i = m^2_\nu L/2E_i^2\,$, we obtain a negative value
$t_i^{sh}<0$ (or a value close to the origin of the flux function $t_i \sim
-\delta t$) is computed by convolving it with a Gaussian ${\cal
G}(t;t_i^\rm{sh},\sigma^t_i)$ centered in $t_i^\rm{sh}$ and with standard
deviation $\sigma_i = 2\,\delta t_i\, \Delta E_i/E_i\;$:

\beq 
\tilde{f}(E,t_i^\rm{sh};\{\t\}) = \int{dt\; f(E,t;\{\t\}) {\cal
G}(t;t_i^\rm{sh},\sigma_i)}.
\label{eq:regeventprob}
\eeq

Clearly this regularization of the divergent contributions to the Likelihood
is physically motivated by the fact that the origin of the problem is the
uncertainty in the energy measurements, that translates into an uncertainty in
the precise location in time of the neutrino events after the energy-dependent
shifts are applied.



\begin{figure}[h]
\begin{center}
\epsfxsize=120mm
\epsfbox{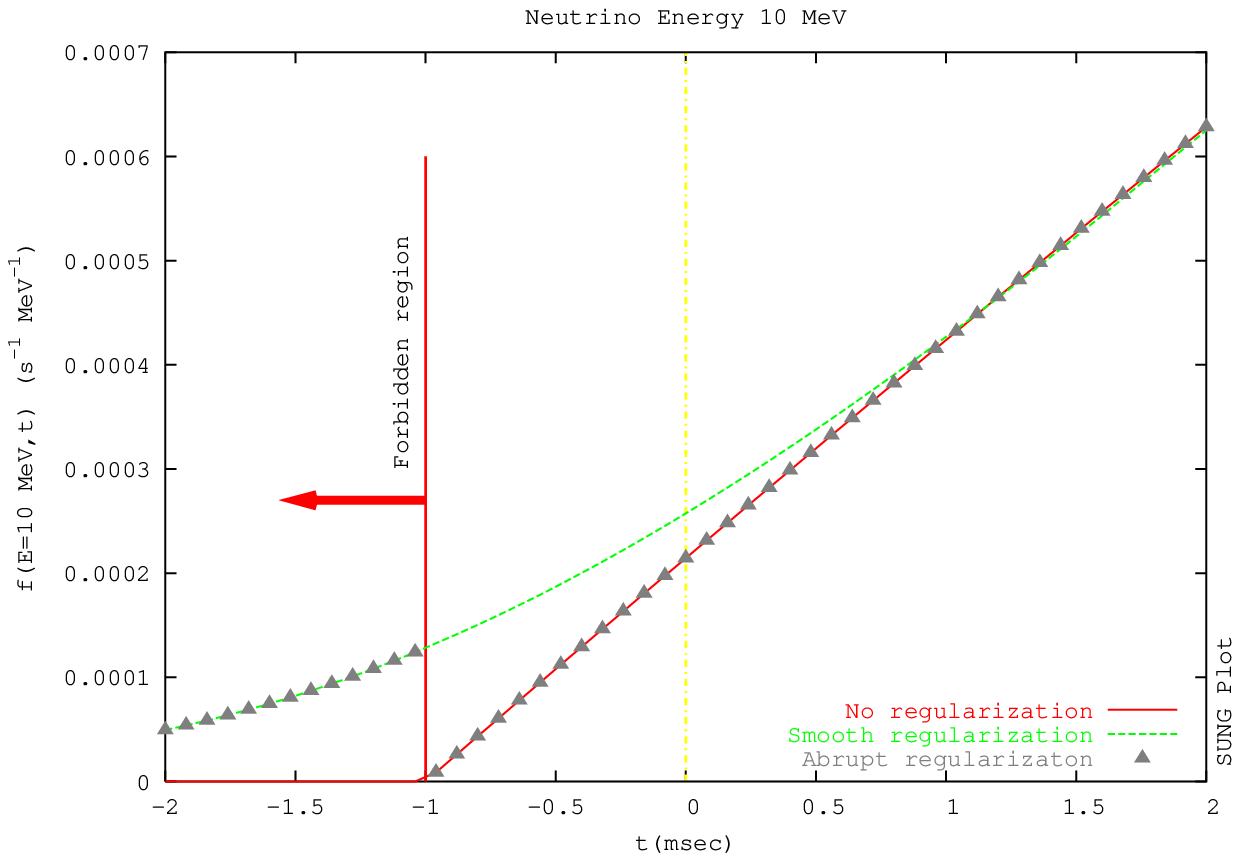}
\mycaption{Time profile of $\tilde f$ for a 10 MeV neutrino using three
different schemes of regularization}{
Time profile of $\tilde f$ for a 10 MeV neutrino using three different schemes
of regularization: no regularization at all (continuous line), regularization
of events just into the forbidden region $t^\rm{sh}<-\delta t$ (triangles) and
regularization of events inside and outside (but close) the forbidden region
(dashed line).  The last strategy intends to avoid discontinuous changes in
the likelihood produced when a single neutrino event leaves the region where
$f$ is regularized.
}
\label{fig:reg.smooth}
\end{center}
\end{figure}

The regularization procedure is one of the novel features of the method
proposed.  Although it is not free of numerical problems (e.g. the likelihood
can have discontinuities that must be corrected as illustrated in
fig.~\ref{fig:reg.smooth}), the most interesting property of this procedure is
that of allowing the inclusion in the likelihood of informations about the
resolution of the detector: the same event in different neutrino detectors
will have a different contribution to the LF, when shifted to the {\it
forbidden region} $t^\rm{sh}<-\delta t$), because the width of the resolution
function ${\cal G}$ in (\ref{eq:regeventprob}) will be also different.
Therefore detectors with a better energy resolution will be more sensitive to
changes in the position of the events and therefore in the tested neutrino
mass.

%

\section{Likelihood marginalization}
\label{sec:method.margin.}

The LF alone cannot give us the information that we want about the neutrino
mass without reference to the other parameters used to describe the signal.
Independent probabilistic information on $\msq$ is obtained by calculating the
{\it posterior} pdf $p(m^2_\nu|\D)$ that results from marginalizing the LF
with respect to the nuisance (flux) parameters.

In practice, the marginalization procedure can be a very expensive numerical
multidimensional integration of a particularly ``heavy'' function.  Each
evaluation of the LF for a typical signal from a future Galactic supernova
could require several thousands of evaluations of relatively complex functions
(flux model, spectrum and cross-section).  With all these conditions it is
clear that the CPU time required to carry out all the necessary operations
would be exceedingly large, specially considering that to draw any statistical
conclusion about the quality of our method, we need to analyze a large set of
neutrino samples, corresponding to different SN models, SN-earth distances and
also to different detectors.

\begin{figure}[h]
\begin{center}
\epsfxsize=140mm
\epsfbox{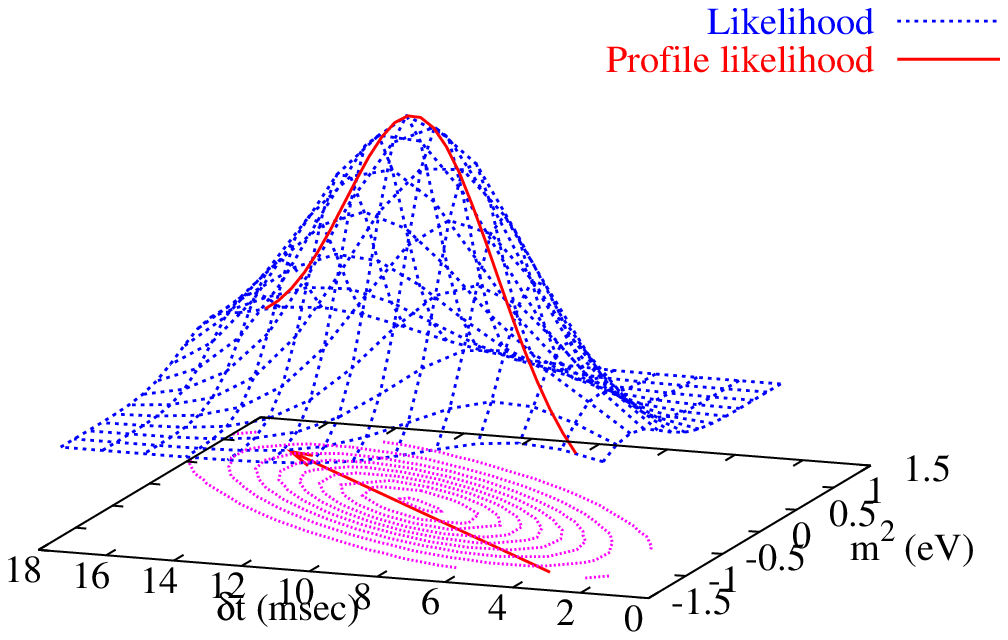}
\mycaption{Illustration of the profile likelihood computation for
$\msq$}{
Illustration of the profile likelihood computation for $\msq$.  
}
\label{fig:illust.margin.}
\end{center}
\end{figure}

A way to avoid this problem, is to approximate the marginal posterior
probability with the {\it profile likelihood} (PL) $\hat{\cal L}(m^2_\nu;D)$,
that corresponds to the trajectory in parameter space along which for each
given value of $m^2_\nu$ the Likelihood is maximized with respect to all the
other parameters (see fig.~\ref{fig:illust.margin.}).

%

\begin{figure}[p]
\epsfxsize=160mm
\epsfbox{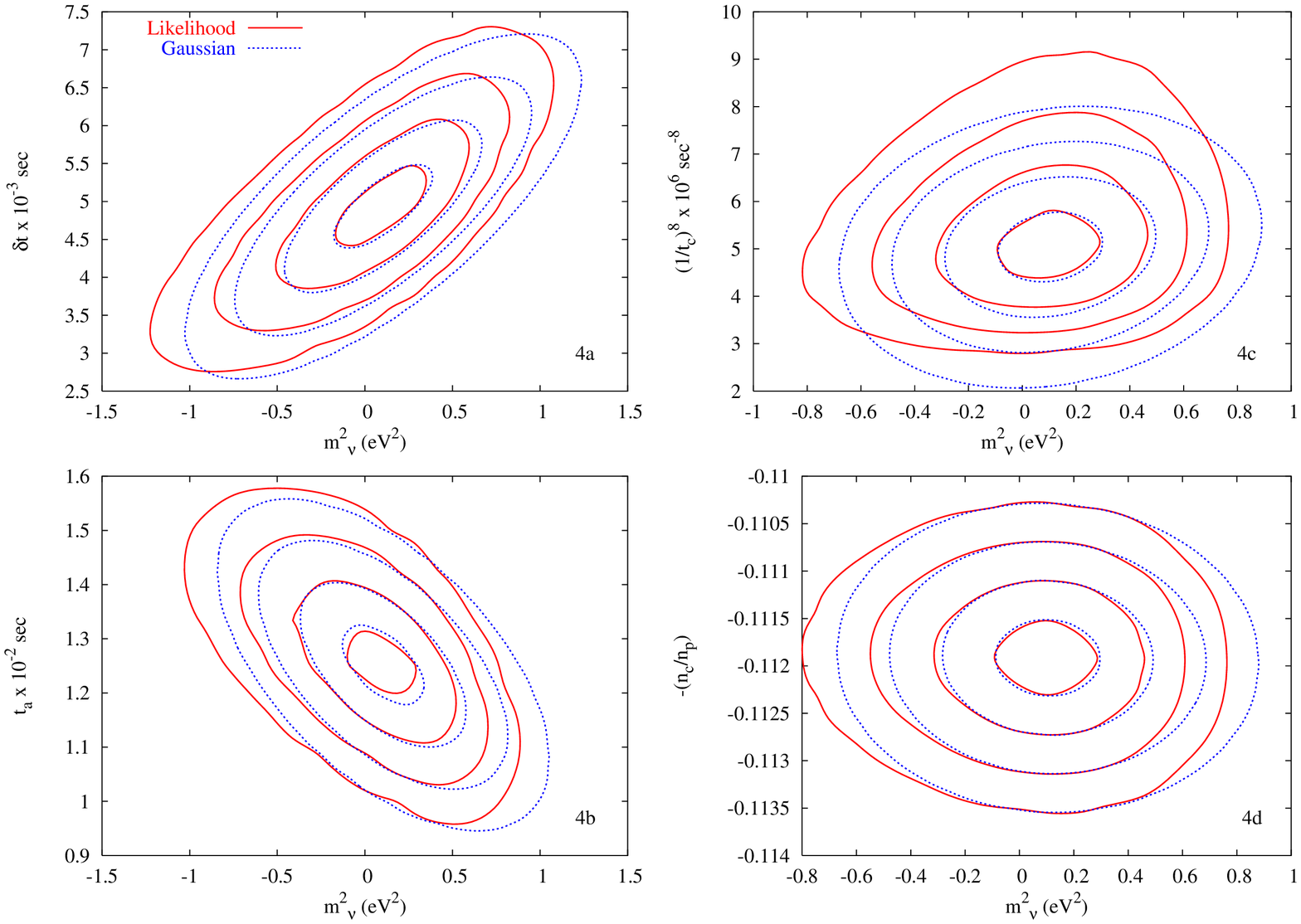}
\vskip-4mm
\mycaption{Contours of the log-LF compared with the contours of a
Gaussian distribution}{
Contours of the log-LF compared with the contours of a Gaussian distribution
of the same mean and covariance in four different two parameters spaces:
$m^2_\nu$ versus~\ref{fig:contours}a:~the time shift $\delta t$ of the flux
function; \ref{fig:contours}b:~the time scale of the fast initial rase $t_a$
($n_a=1$); \ref{fig:contours}c:~the time scale of the cooling phase $t_c$
($n_c=8$); \ref{fig:contours}d:~the ratio $n_c/n_p$ (see
eq.~(\ref{eq:fluxmodel.simple}).  For the construction of this contours we
have used a particular parameterization of the
flux-model~(\ref{eq:fluxmodel.simple}), the same used in
ref.~\cite{Nardi:2003pr}, in order to minimize correlations between
parameters.  The contours are plotted at $0.5,\,1.0,\,1.5$ and 2.0 $\sigma$.
}
\label{fig:contours}
\end{figure}

It can be shown that for a multivariate Gaussian the PL coincides with the
marginal posterior $p(m^2_\nu|D)$ (for details see appendix~\ref{ap:bayes}),
and therefore our results will be reliable to the extent the Likelihood
approximates well enough a normal distribution.  In fig.~\ref{fig:contours} we
compare different contours in parameter space for $\log {\cal
L}(m^2_\nu,\{\t\};D)$ with those of a corresponding multivariate normal
distribution, with the same mean and covariances than the likelihood.  We see
that within the region where the contributions to the integrations are large,
the behavior of the Likelihood is indeed approximately Gaussian.

\afterpage{\clearpage}
%

\chapter{Results and discussion}
\label{ch:results}

We present in this chapter the results of a series of tests performed to study
the behavior of the method and to estimate its sensitivity to a neutrino mass.

The tests were performed with a wide range of emission (SN model),
propagation (neutrino oscillations) and detection conditions, as
described in sect.~\ref{sec:results.tests}.  The general results of
the ML analysis are summarized in sect.~\ref{sec:results.general}.  To
quantify the sensitivity of the method we have measured two
statistical properties for each set of analyses.  The results of these
measurements are presented and analyzed in
sect.~\ref{sec:results.sens.quant.}.  Finally, we discuss in
sect.~\ref{sec:results.conclusions} the implications of the results
and its relevance in the context of current and future neutrino mass
searches.  A summary and the conclusions of this work are presented in
this section.

Part of the material presented in this chapter were published in a similar
form in the paper\cite{Nardi:2004zg}

\section{Description of tests}
\label{sec:results.tests}

In the absence of a real high statistics supernova signal, we can only
test our method applying it to realistic {\it synthetic} signals
generated with the methods described in chapter~\ref{ch:signal}.
However of course we cannot derive any general and reliable conclusion
from the analysis of a single signal and just one particular emission
model and supernova set of properties.

In order to have an overall understanding of what to expect if our method
could be applied to a real signal, we have carried out a Monte Carlo (MC)
analysis using many synthetic signals, that were generated under a wide range
of conditions for the emission (SN fluxes and spectra, SN distance),
propagation (oscillations and TOF delays) and neutrino detection (number of
events, energy resolution and energy thresholds).

We divide our MC analysis in a set of different {\it tests}.  For each test we
applied the method to a sample of $\sim40$ synthetic signals each one
generated with the same set of input conditions (we can imagine the samples
conforming one test, as a set of many realizations of the same supernova,
detected with the same detector).

The input conditions considered to construct the whole set of tests were:

\begin{itemize}

\item {\bf Neutrino emission}.  Two different supernova models were used:

  the first model ({\it supernova model I}) corresponds to the results of a
  simulation of the core collapse of a 20 $M_\odot$ star which was performed
  with the code developed by the Lawrence Livermore Laboratory group
  \cite{Mayle:1993uj}.  These results were used to study supernova r-processes
  \cite{Woosley:1994ux}.  Figure~\ref{fig:snmodel}a depicts the neutrino
  luminosity and mean energy evolution obtained in this simulation.  This
  model is characterized by large spectral differences between electron and
  non-electron neutrino species, which, as explained in
  sect.~\ref{subsec:hierarchy}, is a result of the approximate description of the
  $\nux$ transport in the SN core.

  Neutrino spectra and their evolution were not published in the original
  paper by Woosley et al. \cite{Woosley:1994ux}.  We have therefore used the
  spectral shapes taken from the dedicated study of Janka and Hillebrandt
  \cite{1989A&A...224...49J} and we have assumed for simplicity that the
  $\alpha$-parameter, which quantify the spectral distortions, remains
  constant during the emission. We have used the standard values
  $\alpha_\anue=3.5$ and $\alpha_\nux=3.0$ for this quantity.

%
\begin{figure}[p]
\begin{center}
\vskip-30mm
\subfigure[\it Supernova model I (adapted from ref.~\cite{Woosley:1994ux})]{
\epsfxsize=70mm
\epsfbox{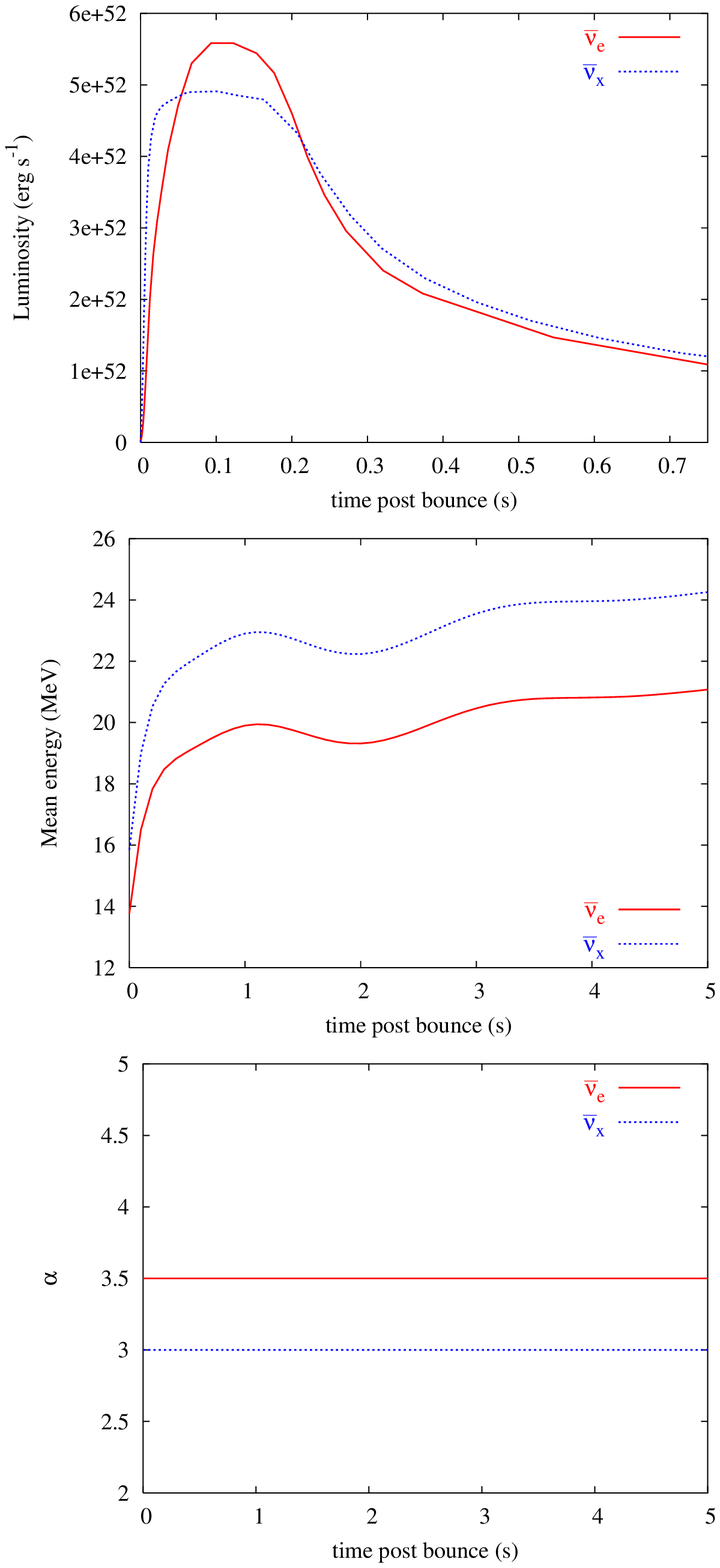}}
\subfigure[\it Supernova model II (adapted from ref.~\cite{Buras:Private2004})]{
\epsfxsize=70mm
\epsfbox{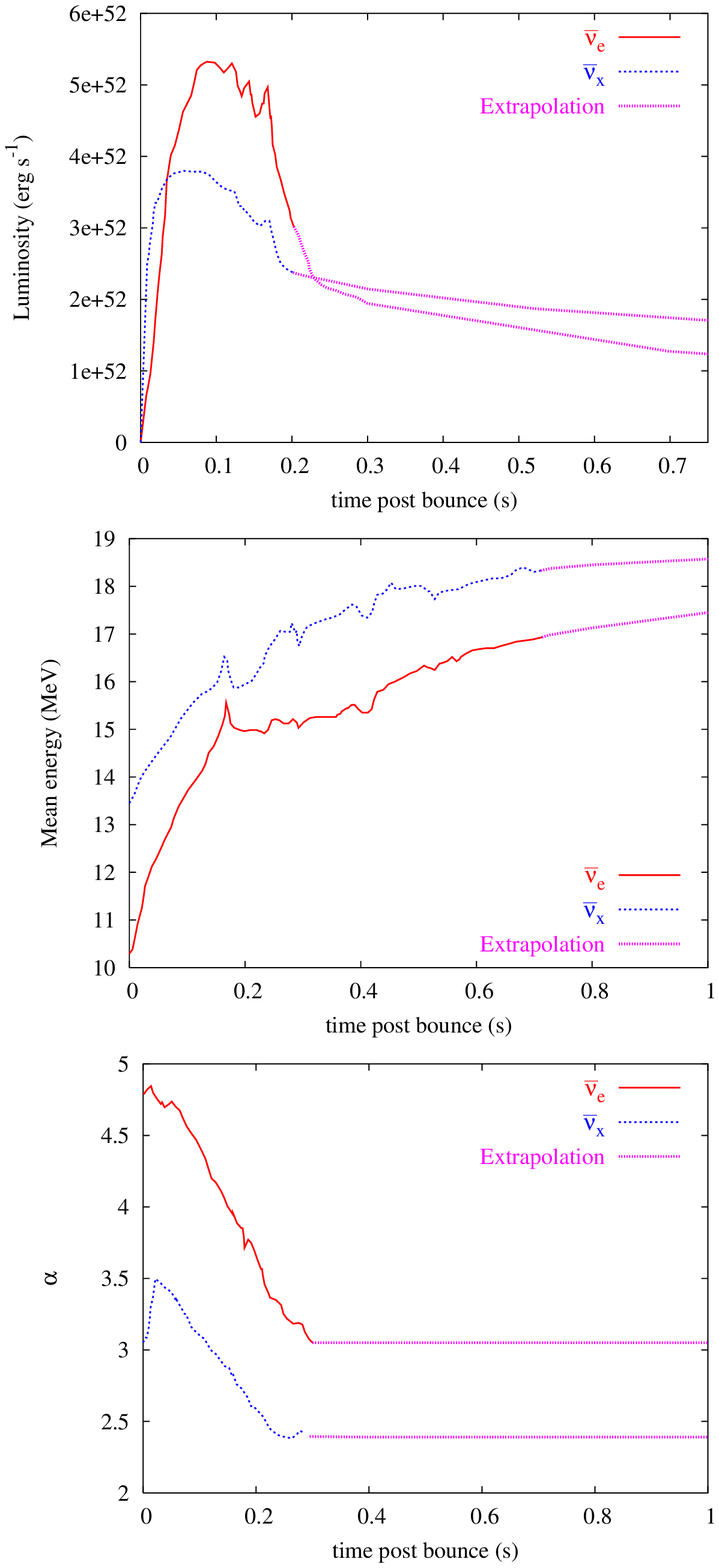}}
\mycaption{Luminosity, mean energy and $\alpha$-parameter for $\anue$ and
$\anux$ in the supernova models used in this work to generate the synthetic
signals}{
Luminosity, mean energy and $\alpha$-parameter for $\anue$ and $\anux$ in the
supernova models used in this work to generate the synthetic signals.
}
\label{fig:snmodel}
\end{center}
\end{figure} 
\afterpage{\clearpage}
%

  The second model ({\it supernova model II}) correspond to the recent
  state-of-the-art hydrodynamical simulation of the core collapse of
  15 $M_\odot$ progenitor star \cite{Raffelt:2003en,Buras:Private2004}
  carried out by means of the Garching group code
  \cite{Rampp:2000a,Rampp:2000ws,Rampp:2002bq}.  This simulation
  included a more complete treatment of neutrino opacities
  \cite{Buras:2002wt,Keil:2002in,Raffelt:2003en}, that resulted in
  smaller spectral differences between $\anue,\nue$ and $\nux$ (see
  fig.~\ref{fig:snmodel}a).

  The Garching group simulations were stopped after 750 msec, and the results
  were not completely reliable already after the firsts 300 msec
  \cite{Buras:Private2004}.  Since our study assumes a signal time duration of
  20 sec, after which the neutrino flux is assumed to become undetectable, we
  have done a conservative extrapolation of the signal to later times.  For
  the luminosities we have assumed a power law decay in agreement with general
  results of SN simulations
  \cite{Burrows:1991kf,Totani:1997vj,Woosley:1994ux,Raffelt:2003en} while for
  the mean energies we have assumed a mild decrease and a constant pinching
  after 750 msec.

  We have also varied SN-earth distance into the representative values 5, 10,
  and 15 Kpc.

\item {\bf Neutrino oscillations}.  To perform the MC analyses we used an
  oscillation scheme with a normal mass hierarchy and the LMA mixing
  parameters set (see sect.~\ref{subsec:oscil.param.}).  With this oscillation
  scheme the observed flux of $\anue$ correspond to an admixture of
  $\approx$74\% $\anue$ and $\approx$26\% $\anux$.

  The case of a complete swap $\anux\leftrightarrow\anue$ that can result from
  an inverted mass hierarchy and $|U_{e3}|^2\gsim10^{-3}$, can be considered
  equivalent to the non-oscillation case (studied in ref.~\cite{Nardi:2003pr})
  with a larger average energy for $\anue$. Indeed we can consider the results
  of the analysis performed with supernova model I as the harder spectrum
  version of that corresponding to supernova model II.

  We do not include Earth matter effects, since they will depend on the
  specific position in the sky of the SN relative to the Earth, on the
  specific location of each detector and on the time of the day.  However,
  given that even with a dedicated analysis it appears quite challenging to
  identify clearly these effects \cite{Dighe:2003jg}, we believe that this
  neglect is of no practical importance.

\item {\bf Detection}. As explained in the previous chapter the application of
  our method requires the knowledge of both energy and time of each neutrino
  event.  On the other hand the spectral estimation procedure described in
  sect.~\ref{sec:method.likelihood} requires a large statistics of
  neutrinos. Electron antineutrino detection will provide by far the largest
  number of events from a future Galactic supernova, and therefore our
  analysis will be restricted to this kind of neutrinos and to detection
  processes capable to measure the energies and arrival times with good
  precision.

  Today, the Super-Kamiokande (SK) detector, having the largest fiducial
  volume among all operative detectors, suited for the kind of analysis we
  propose.  Operative scintillator detectors are characterized by a lower
  threshold and/or better energy resolution, which in principle represents an
  advantage in the analysis.  However, their significatively smaller fiducial
  volume imply the possibility of detecting only a few hundred of $\anue$
  events (see tab.~\ref{tab:detectors}).  Such a low statistics is not well
  suited for the application of our method.  However, since KamLAND (KL) is
  located at the same site as SK (and equally affected by neutrino
  oscillations in the Earth) KL events can be the high statistics of SK,
  possibly yielding some sensitivity improvement.  We have performed this kind
  of joint analysis to study the impact of the addition of KL events to the SK
  signal.

  For the future large volume detectors we have chosen two of the most
  interesting proposals: Hyper-Kamiokande \cite{Nakamura:2003hk} (a megaton
  water \v{C}herenkov detector) and the multi-kiloton scintillator detector
  LENA (Low Energy Neutrino Astrophysics) \cite{Oberauer:2004ji}.

  The properties of all the detectors used in the MC analysis are summarized
  in tab.~\ref{tab:detectors} in sect.~\ref{sec:sign.char}.

\end{itemize} 

The following table defines the set of labels that will be used hereafter to
identify the different tests of the method and their particular input
conditions:

\begin{center}
%
%
\begin{supertabular}{>{\bf}p{2cm}>{\centering}p{3cm}c>{\centering}p{2cm}p{2cm}}
\bf Label & \bf Supernova model & \bf Detector(s) & \bf Distance (kpc) & \bf
Flux model \\ \hline\hline
TSNI-1 & model I & SK & 10  & model I \\
TSNI-2 & -- & -- & -- & model II \\
TSNI-3 & -- & -- & 5  & model I \\ 
TSNI-4 & -- & -- & 15  & -- \\ 
TSNI-5 & -- & SK + KL & 10  & --  \\ 
TSNI-6 & -- & LENA & -- & -- \\ 
TSNI-7 & -- & HK & -- & -- \\ 
TSNI-8 & -- & HK & -- & model II \\ \hline\hline
TSNII-1 & model II & SK & 10  & model II \\ 
TSNII-2 & -- & -- & -- & model I \\ 
TSNII-3 & -- & -- & 5  & model II \\ 
TSNII-4 & -- & -- & 15  & -- \\ 
TSNII-5 & -- & SK + KL & 10  & -- \\
TSNII-6 & -- & LENA & -- & -- \\ 
TSNII-7 & -- & HK & -- & -- \\\hline\hline
\end{supertabular}
\end{center}

In the last column of the table we have indicated what flux model (model I
(\ref{eq:fluxmodel.simple}), model II (\ref{eq:fluxmodel.phen.}) was used to
compute the signal pdf (\ref{eq:event.prob.}).

For the signals generated with supernova model II we have mainly used the more
flexible flux model II.  The analysis with such a flux in the case of
supernova model I does not show any relevant difference with respect to the
simple flux model I.

\section{Results of the maximum likelihood analysis}
\label{sec:results.general}

To illustrate some of the most interesting properties of the method and before
proceeding with the estimation of its sensitivity, we performed a set of MC
analyses using an hypothetical MC neutrino mass $m_\nu=1$ eV.

Several interesting facts about the behavior of the method could be observed
when the output of the ML analysis, namely the best-fit values of mass and
flux parameters, are studied.  In the following two sections we describe these
results.

\subsection{Best-fit values and limits on $\msq$}
\label{subsec:mass.fit}

\begin{figure}[t]
\begin{center}
\subfigure[TSNI-1]{
\epsfxsize=65mm
\epsfbox{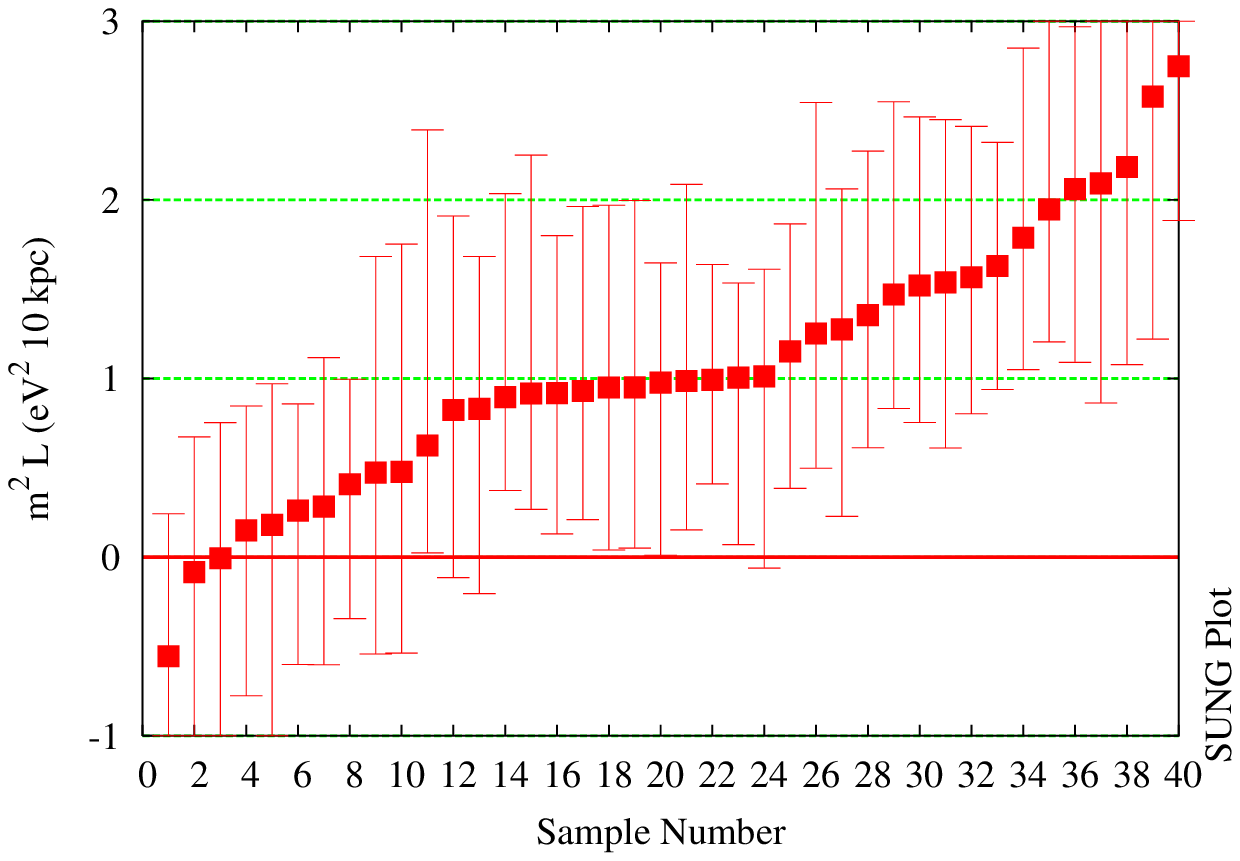}
}
\subfigure[TSNI-2]{
\epsfxsize=65mm
\epsfbox{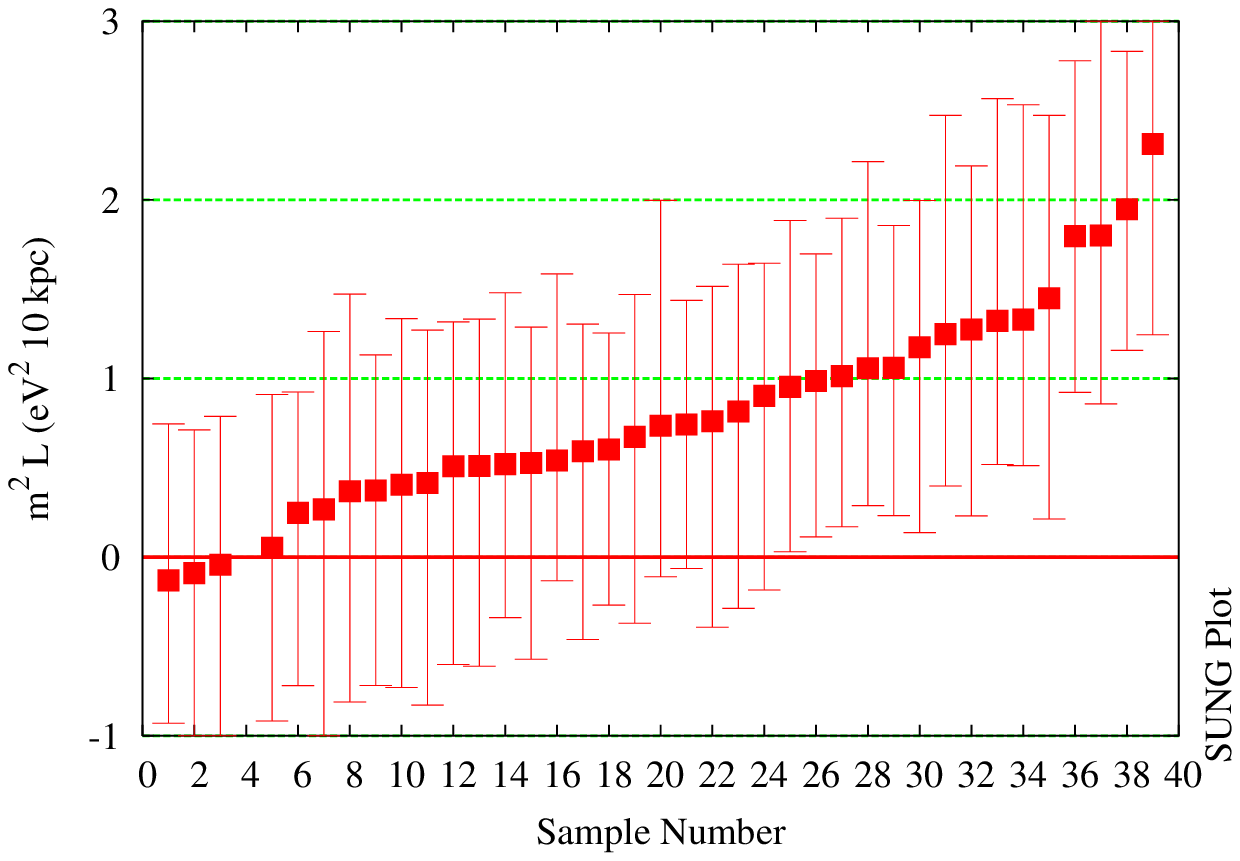}
}
\subfigure[TSNI-3]{
\epsfxsize=65mm
\epsfbox{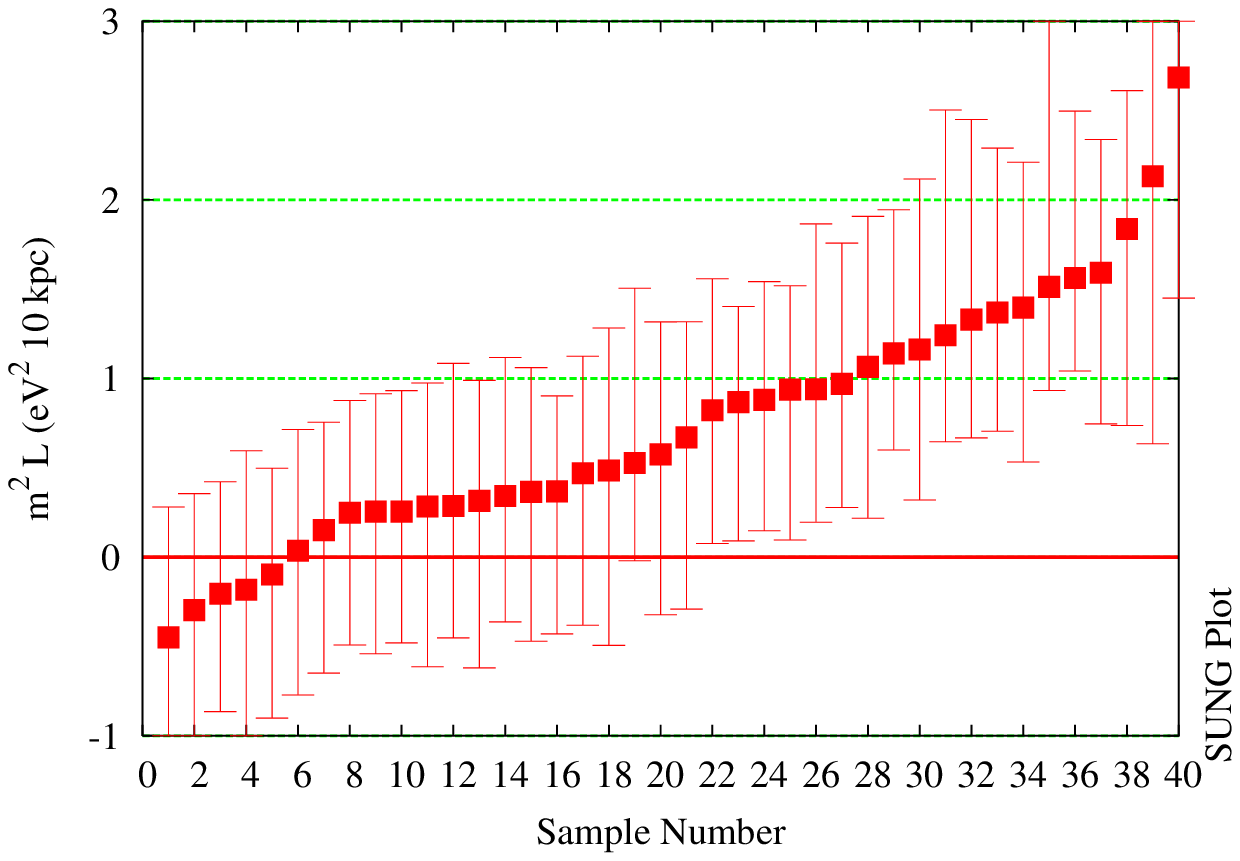}
}
\subfigure[TSNI-4]{
\epsfxsize=65mm
\epsfbox{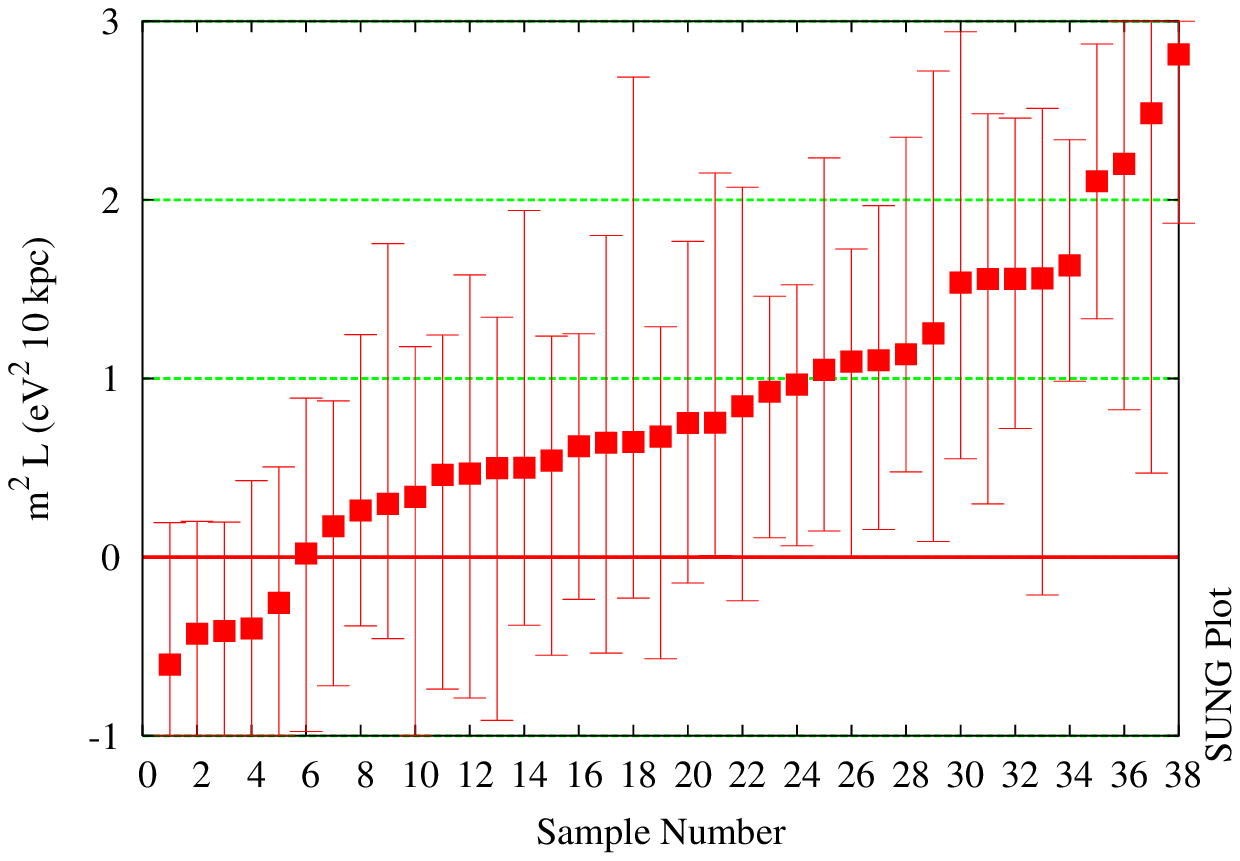}
}
\mycaption{First set of band plots for tests performed with supernova model
I}{
First set of band plots for tests performed with supernova model I and
$m^2_\rm{MC}=1.0\,\rm{eV}^2$.  The best-fit values for $\msq$
obtained from the ML analysis are represented by the filled squares.
The error bars extend from $m^2_\rm{low}$ to $m^2_\rm{up}$.
}
\label{fig:bandplots.SNI-1}
\end{center}
\end{figure}

\begin{figure}[t]
\begin{center}
\subfigure[TSNI-5]{
\epsfxsize=65mm
\epsfbox{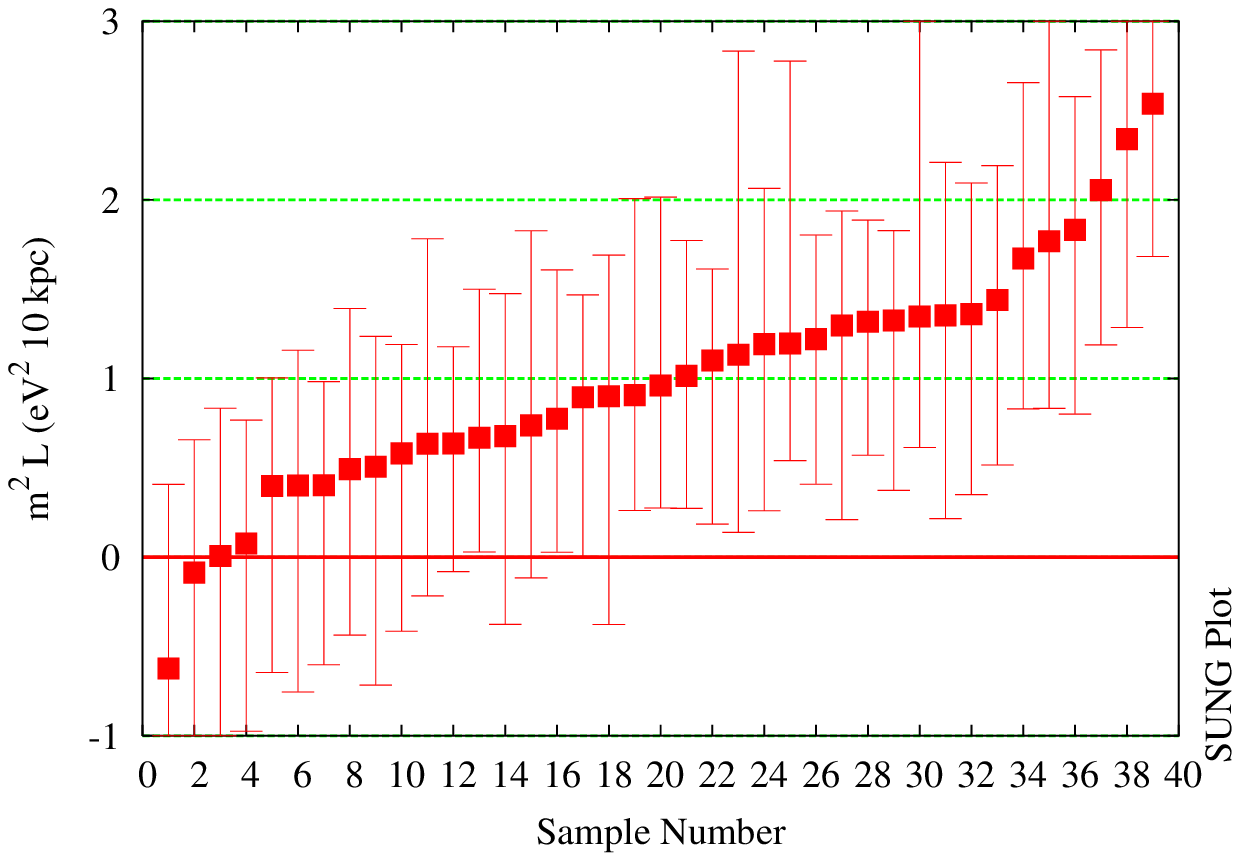}
}
\subfigure[TSNI-6]{
\epsfxsize=65mm
\epsfbox{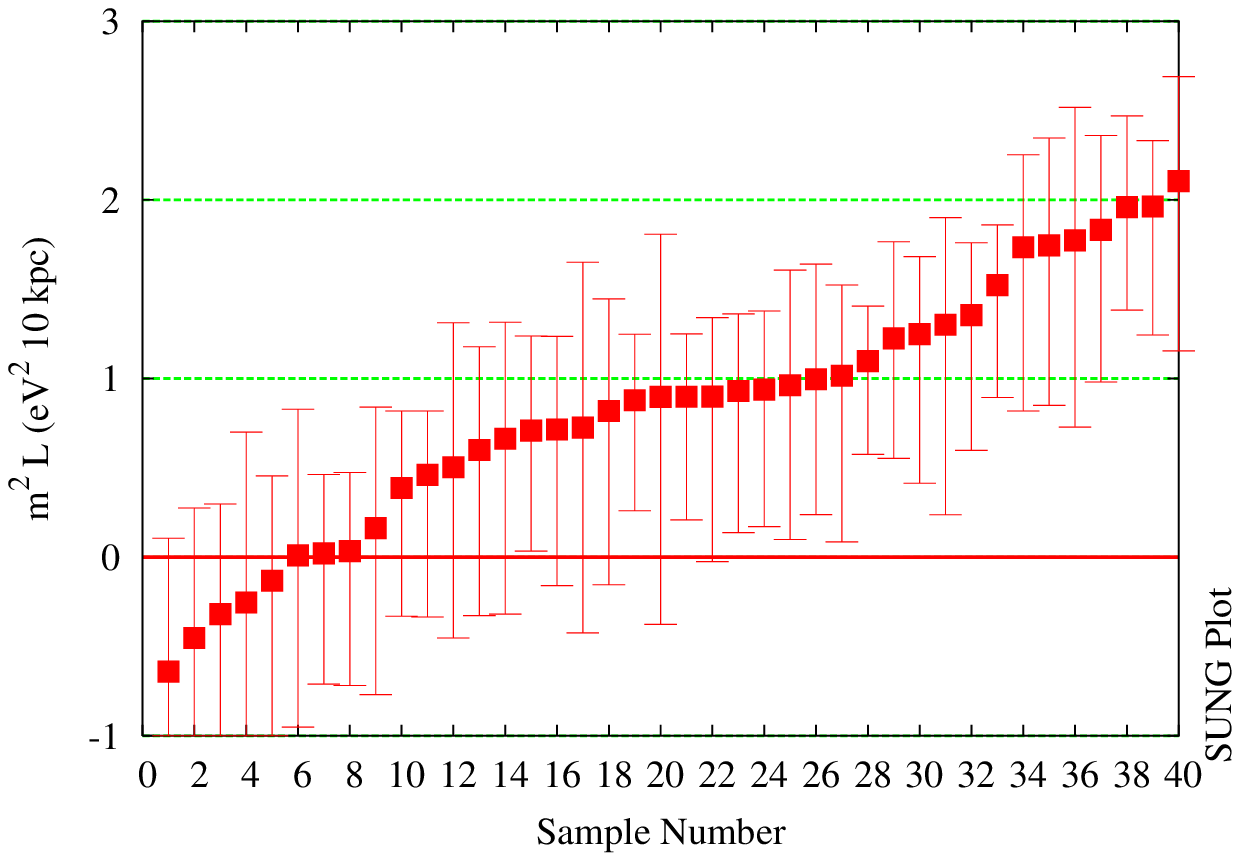}
}
\subfigure[TSNI-7]{
\epsfxsize=65mm
\epsfbox{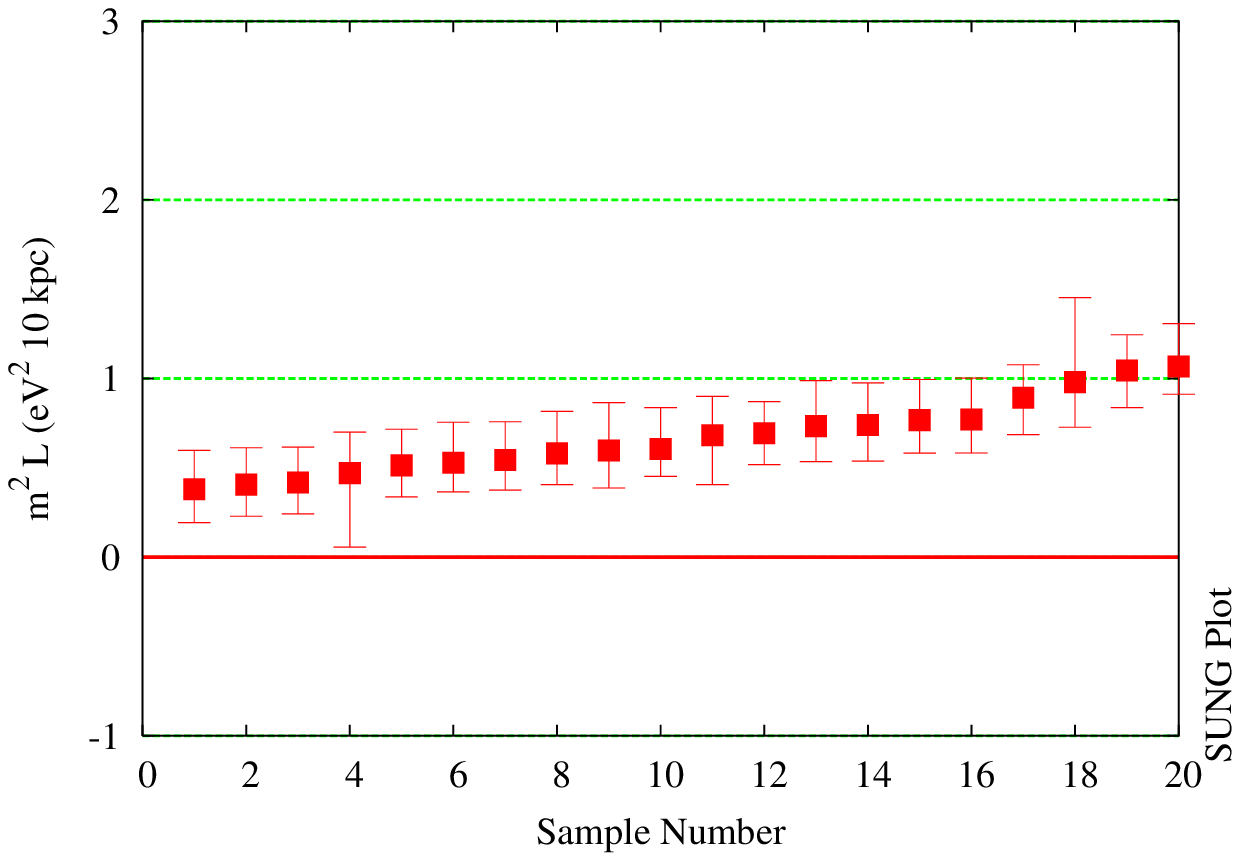}
}
\subfigure[TSNI-8]{
\epsfxsize=65mm
\epsfbox{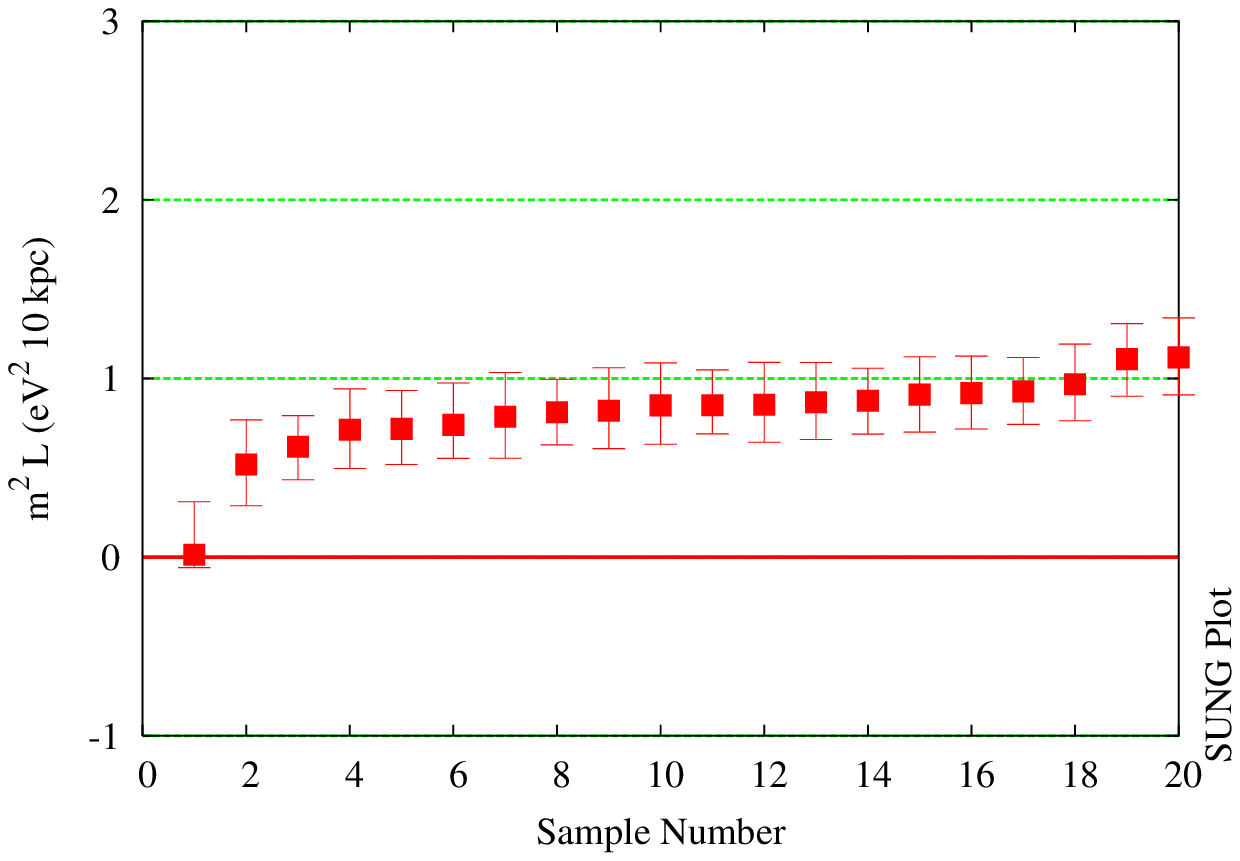}
}
\mycaption{Second sets of band plots for tests performed with supernova model
I}{
Second sets of band plots for tests performed with supernova model I.
}
\label{fig:bandplots.SNI-2}
\end{center}
\end{figure}

\begin{figure}[t]
\begin{center}
\subfigure[TSNII-1]{
\epsfxsize=65mm
\epsfbox{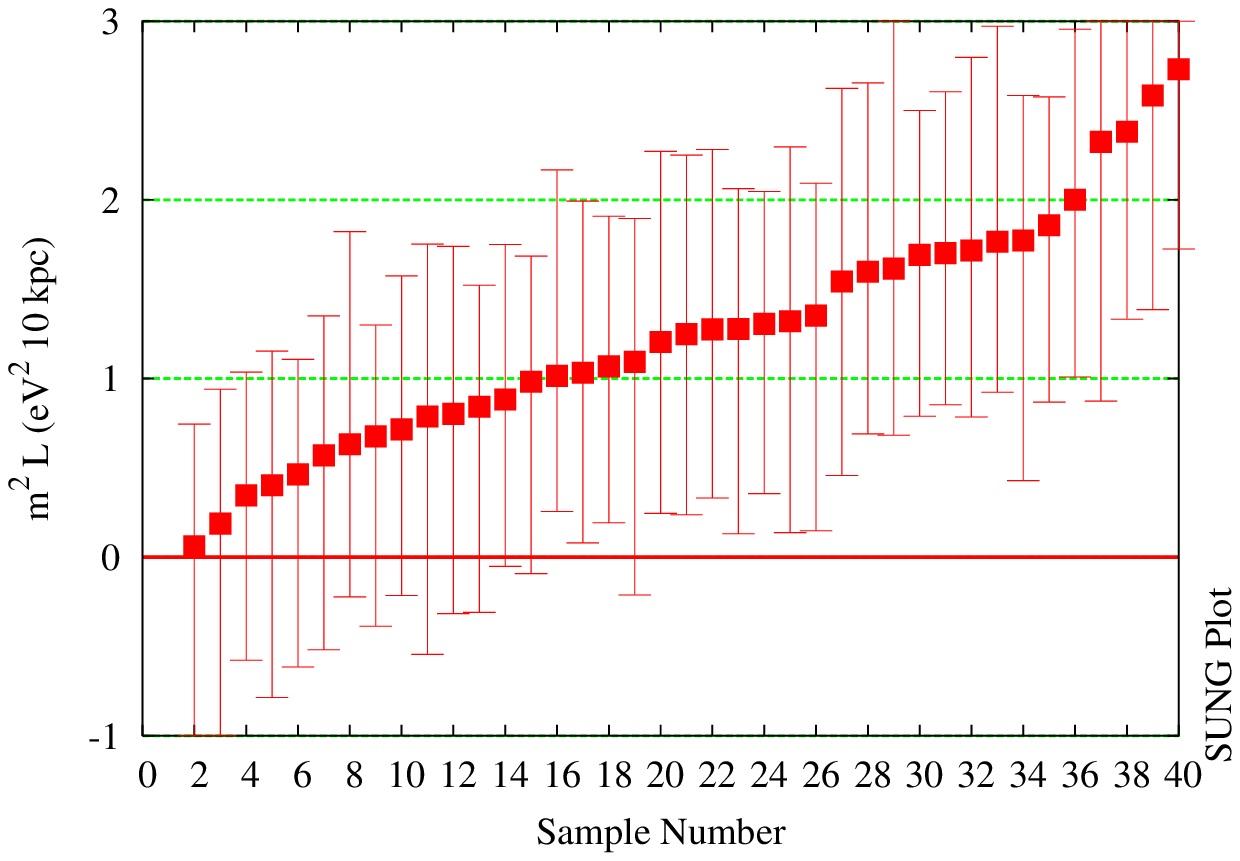}
}
\subfigure[TSNII-2]{
\epsfxsize=65mm
\epsfbox{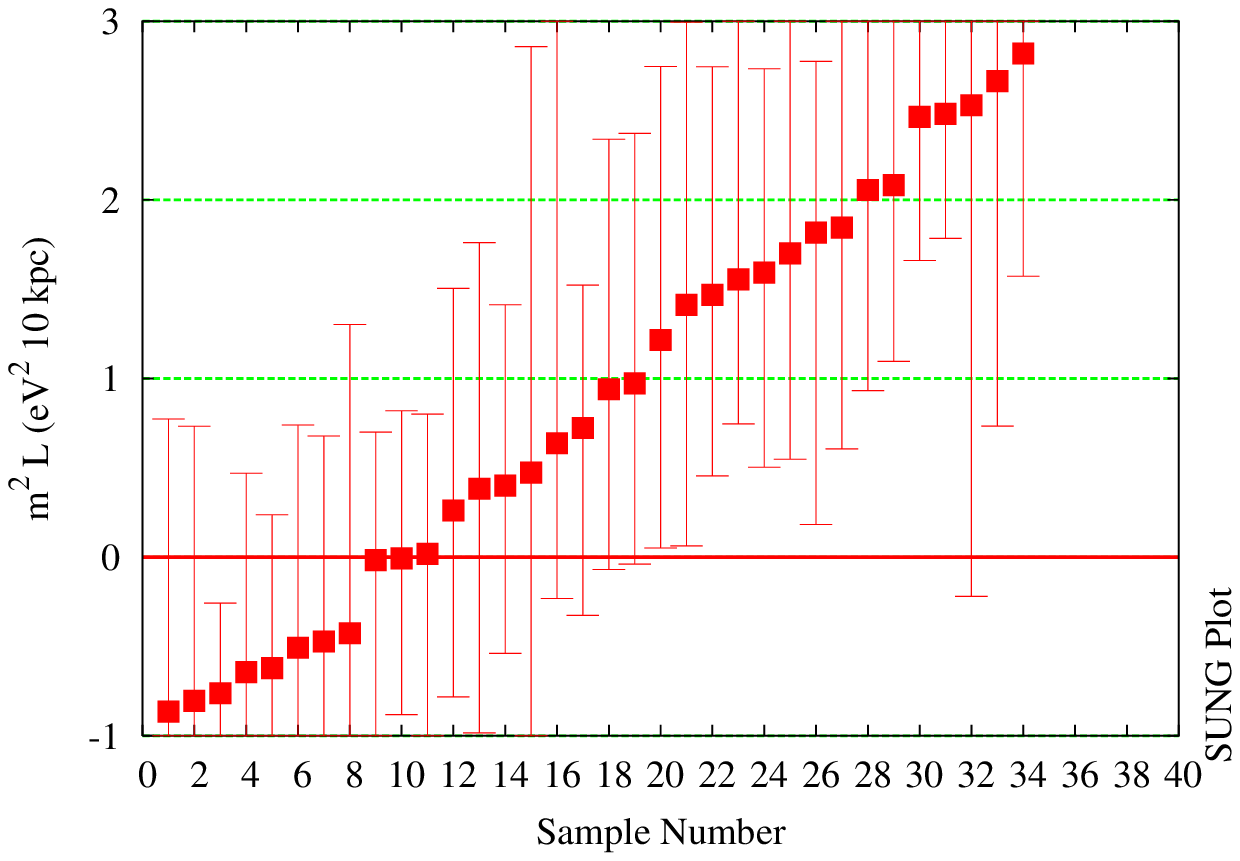}
}
\subfigure[TSNII-3]{
\epsfxsize=65mm
\epsfbox{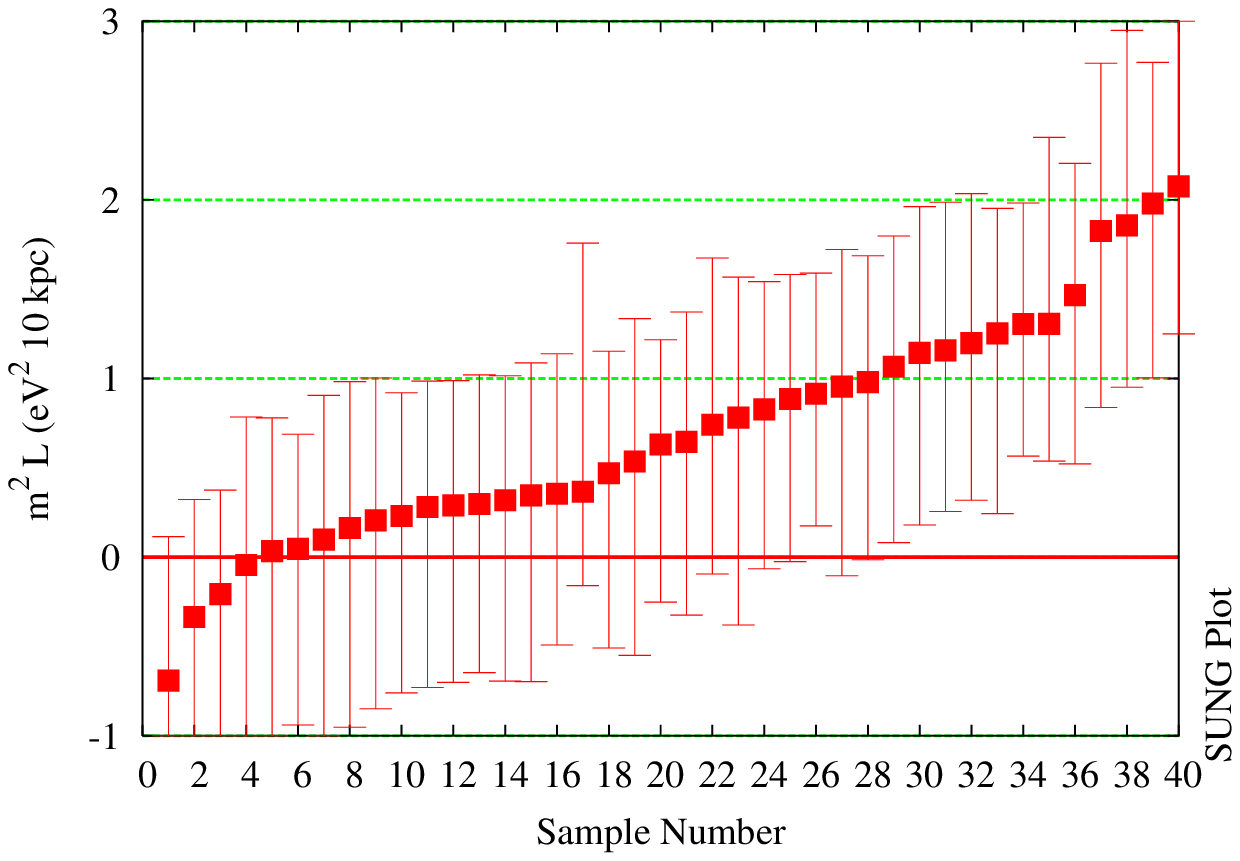}
}
\subfigure[TSNII-4]{
\epsfxsize=65mm
\epsfbox{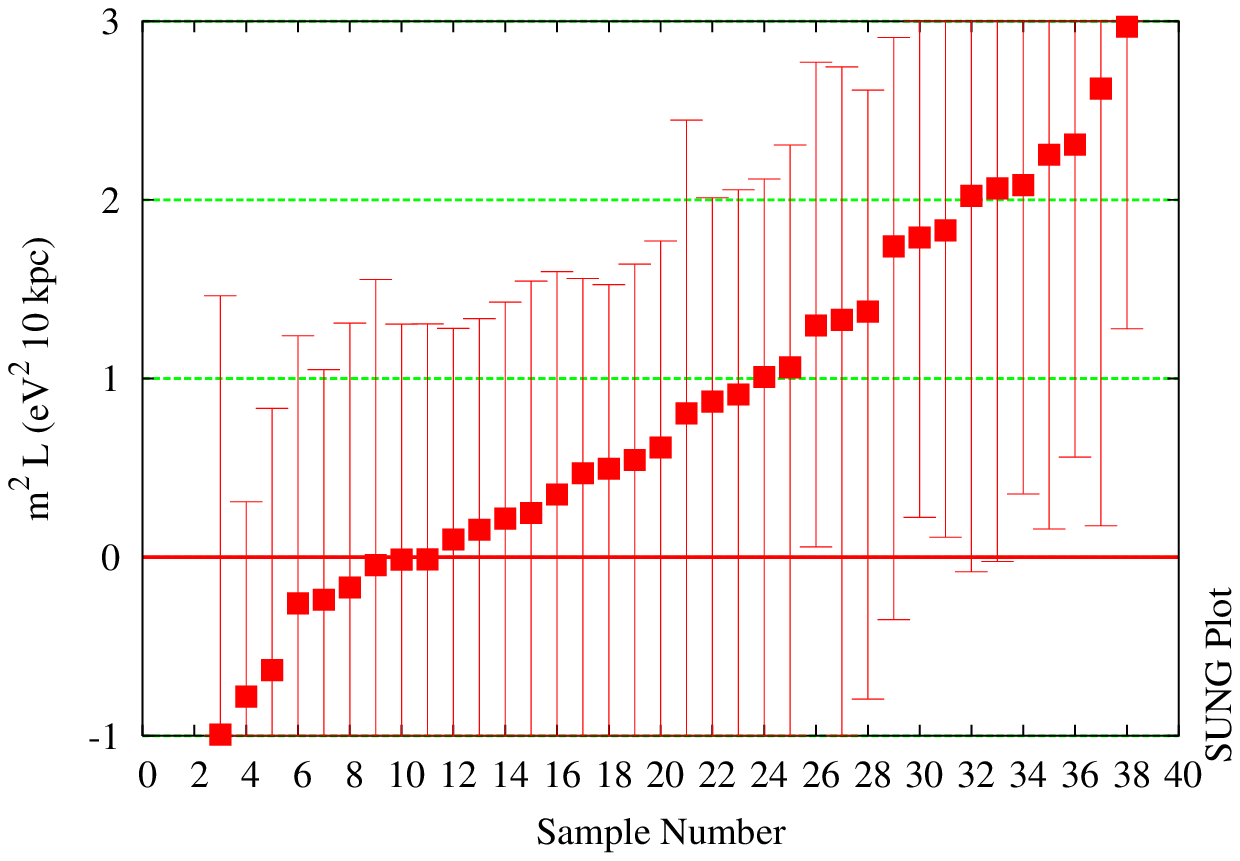}
}
\mycaption{First set of band plots for tests performed with supernova model
II}{
First set of band plots for tests performed with supernova model II.
}
\label{fig:bandplots.SNII-1}
\end{center}
\end{figure}

\begin{figure}[t]
\begin{center}
\subfigure[TSNII-5]{
\epsfxsize=65mm
\epsfbox{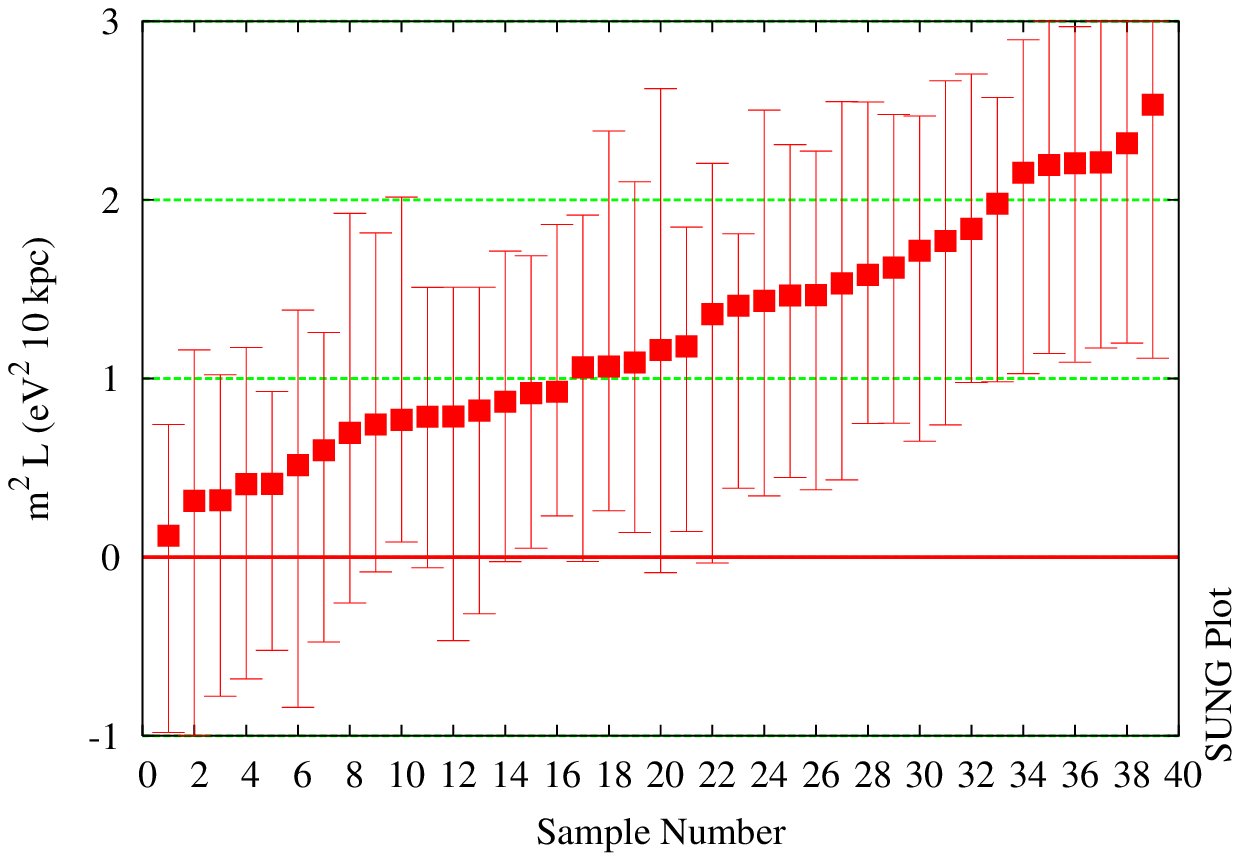}
}
\subfigure[TSNII-6]{
\epsfxsize=65mm
\epsfbox{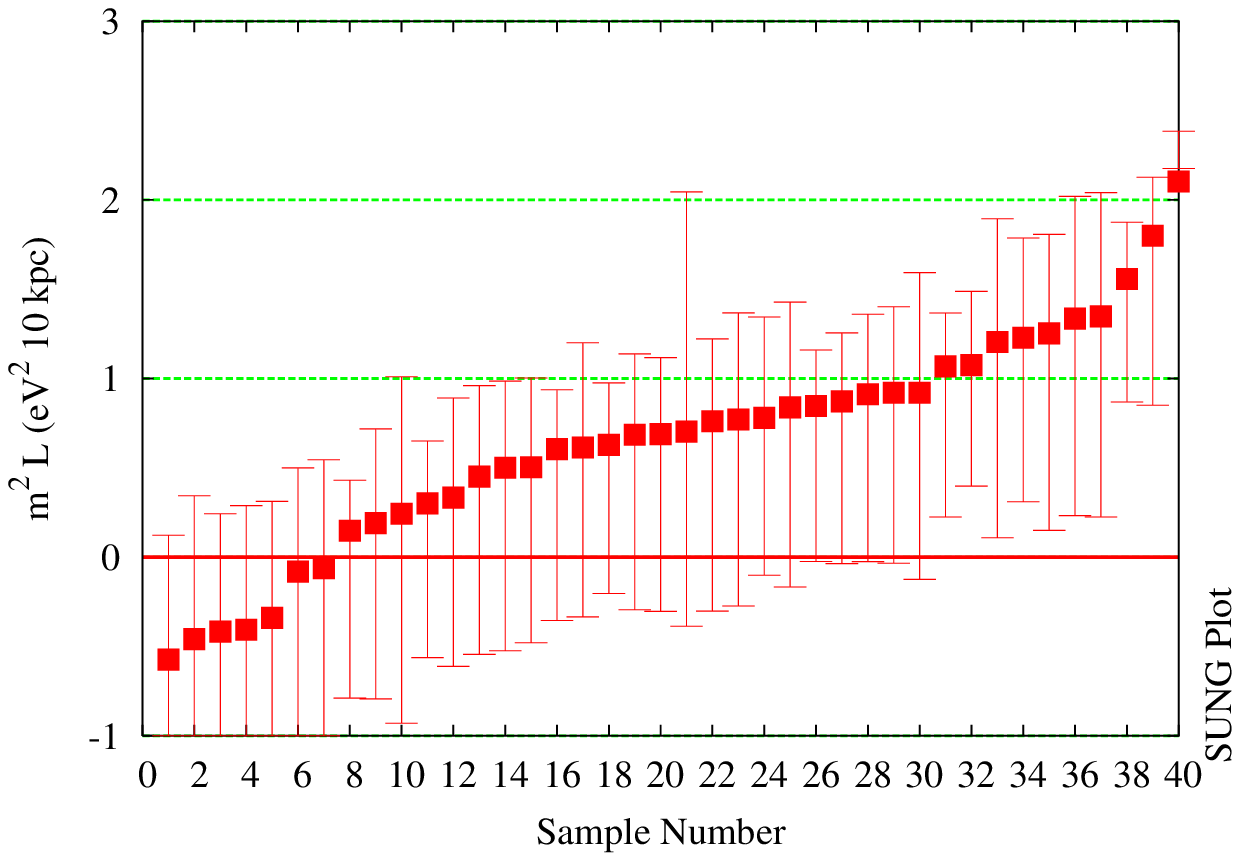}
}
\subfigure[TSNII-7]{
\epsfxsize=65mm
\epsfbox{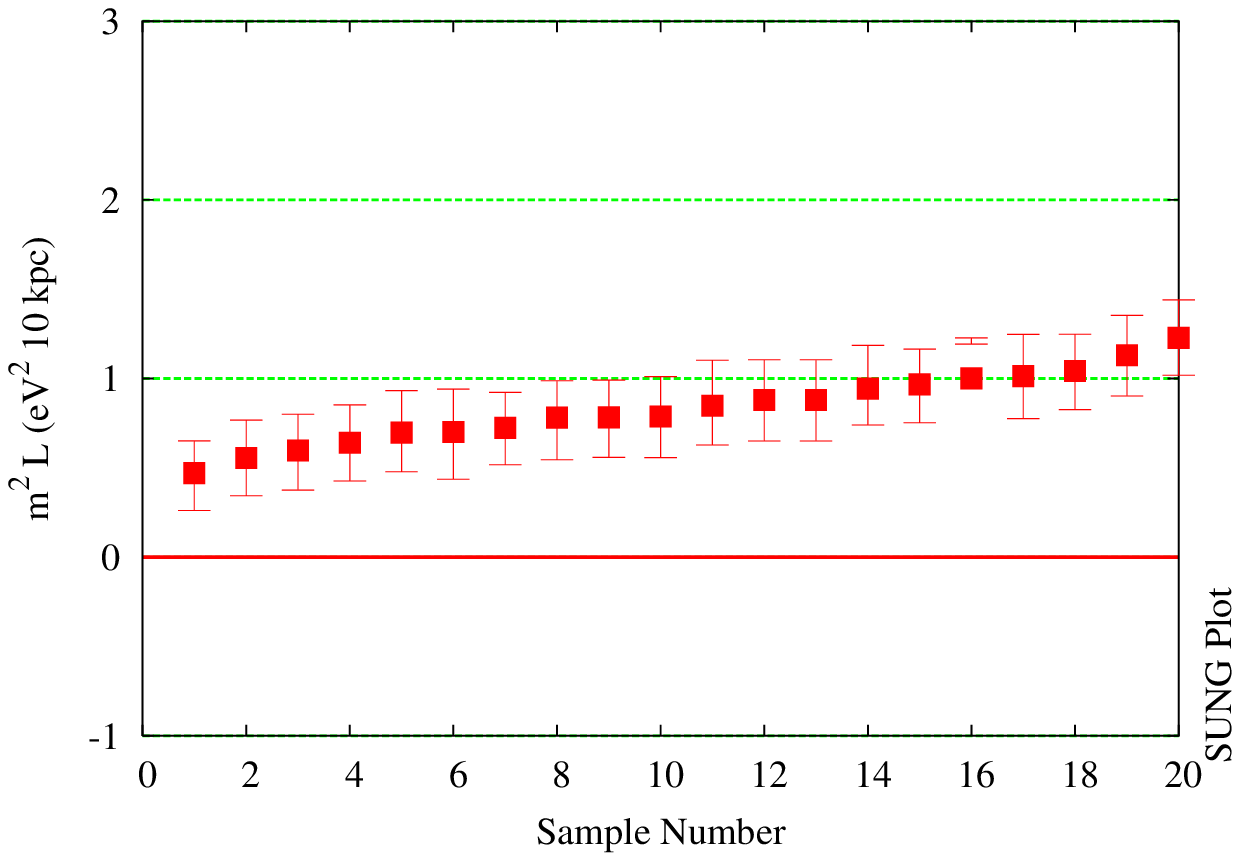}
}
\mycaption{Second set of band plots for tests performed with supernova model
II}{
Second set of band plots for tests performed with supernova model II.
}
\label{fig:bandplots.SNII-2}
\end{center}
\end{figure}

The best-fit values for $\msq$ and its limits are summarized graphically in
the {\it band-plots} presented in
figs.~\ref{fig:bandplots.SNI-1}-\ref{fig:bandplots.SNII-2}.

All the signals on a band plot were generated under the same set of input
conditions.  The squares indicates the position of the best fit value
$m^2_\rm{fit}$ and the ``error bars'' correspond to the lower and upper limits
computed from the posterior pdf $p(\msq|\D)$ using (\ref{eq:mdwlimit}) and
(\ref{eq:muplimit}).

%

\afterpage{\clearpage}
%

\afterpage{\clearpage}
%

\afterpage{\clearpage}
%


Some of the analysis have exceedingly large error bars when compared with the
rest of analysis in the same set. In most of the cases they correspond to
badly behaved analysis, where the extremization procedures used to compute the
profile likelihood found peculiar directions in parameters space corresponding
to large correlations between the mass and the flux parameters.  In a few
cases such ``failures'' can be caused by just one or two neutrinos that
randomly get abnormal values of their energy.  Exclusion of these neutrinos
can generally rescue a well behaved analysis.

The first remarkable fact is that regardless of the wide range of
input conditions (supernova emission model, particular detector, flux
model used to analyze the samples) in almost all cases the average
value of $\msq$ is close to the MC mass used to generate the signals.
This is a clear indication that the signal can act as a self timing
observable and that no particular assumption on the signal time
structure is needed to measure the value of the mass.  This fact
represents a confirmation of our original hypothesis.


The only feature of the analysis sensible to particular input conditions is
the dispersion around the best-fit values of the mass.  As expected, a lower
statistics increase the dispersion and reduces the sensitivity to a neutrino
mass.  This can partially explain the slight difference between the results
for the two supernova models.  Since supernova model I signals have a larger
number of events than supernova model II, the dispersion is smaller for the
first model.  We can conclude that a harder neutrino spectrum would increase
slightly the sensitivity to a a neutrino mass, by increasing the statistics of
the signal.

Results obtained with just SK events (TSNI-,II-1) and SK+KL events
(TSNI-,II-5) do not exhibit any appreciable difference, meaning that
the better energy resolution and lower threshold of KamLAND cannot
compete with the SK much larger statistics.

A qualitative overall analysis of these results shows that sensitivities at
the level of $1$ eV level can be achieved with SK and future LENA detector,
while the sensitivity will be improved by a factor of 2 with a future megaton
water \cerenkov detector.

\subsection{Fit to the flux model}
\label{subsec:fluxmodel.fit}

The ML analysis of the signals described in the previous section provides in
each case also a set of best-fit values of the flux model parameters.  The fit
to the flux model is actually of little interest for the purposes of our work,
but a comparison between the results of such a fit with the input time
profiles and with the histograms of the detected samples, reveals other
interesting aspects of our method.

It must be realized that a direct comparison between the flux model
$\phi(t)$ and the observed time profile $f_t^\rm{obs}(t)$ cannot be
performed directly.  The energy-dependent detection cross-section
depletes the flux at early times when the mean energy of the neutrinos
is lower.  The comparison must be done between the observed profile
and the energy integrated time-profile given by

\beq
f_t^\rm{signal}(t)\equiv\left[N^{-1} \int{dE\,F(E;t)\sigma(E)}\right] \phi(t).
\label{eq:profile.signal.pdf}
\eeq

\begin{figure}[p]
\begin{center}
\subfigure[\it TSNI-1]{
\epsfxsize=65mm
\epsfbox{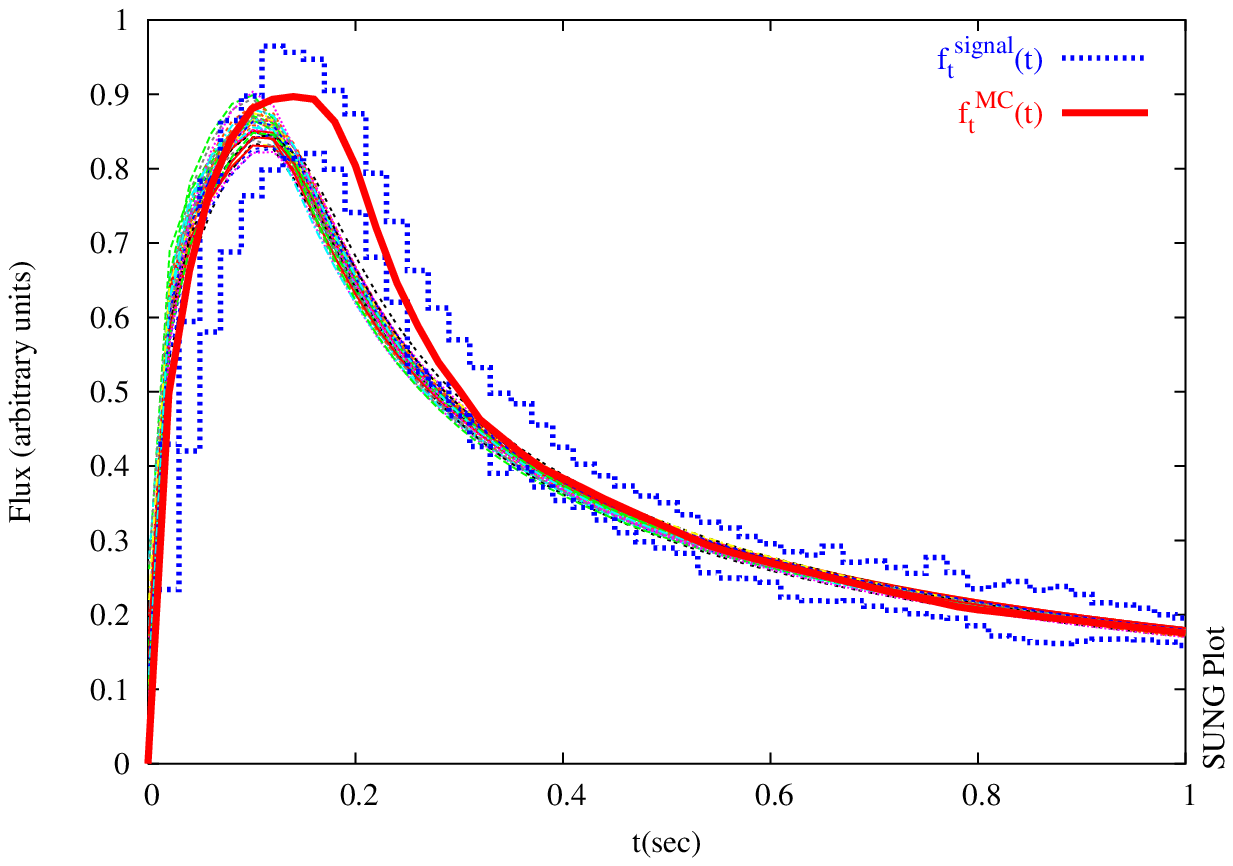}
}
\subfigure[\it TSNI-2]{
\epsfxsize=65mm
\epsfbox{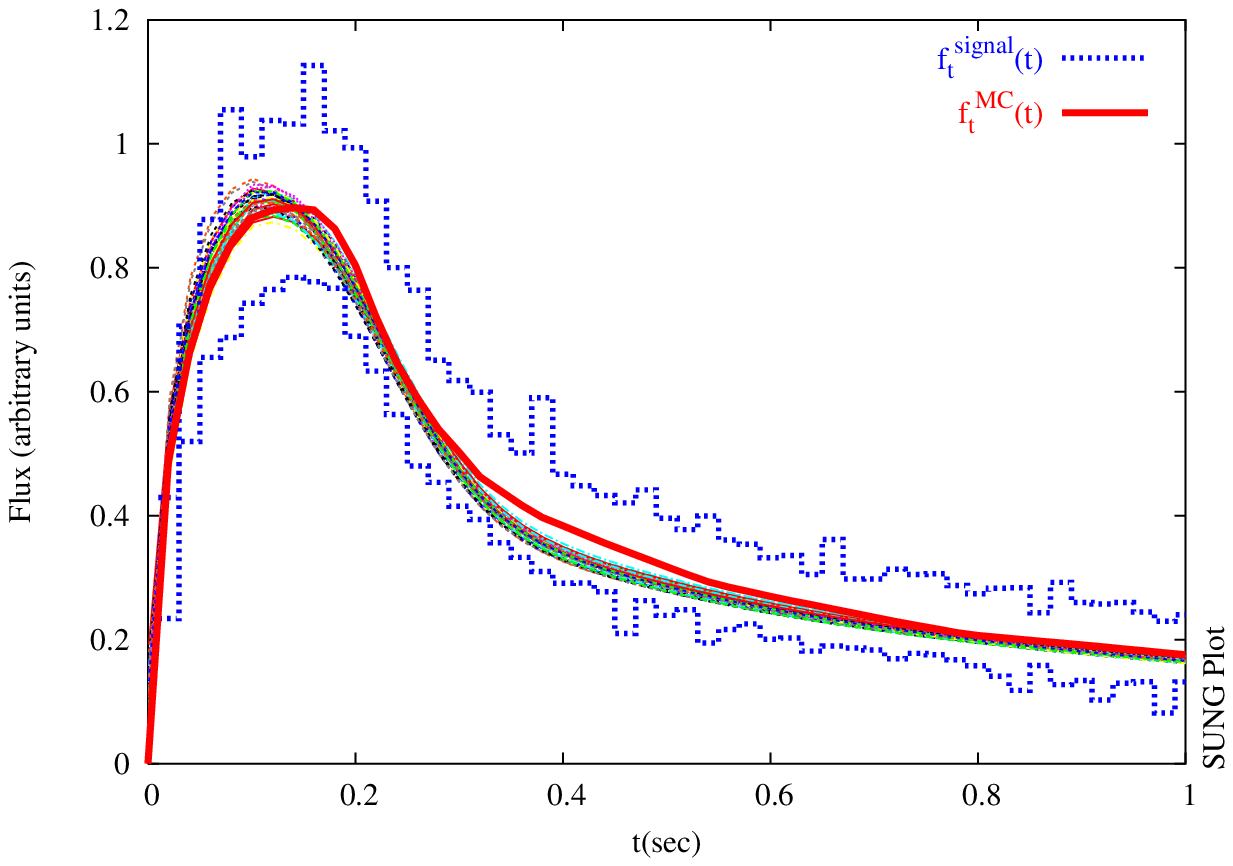}
}
\subfigure[\it TSNI-5]{
  \epsfxsize=65mm
\epsfbox{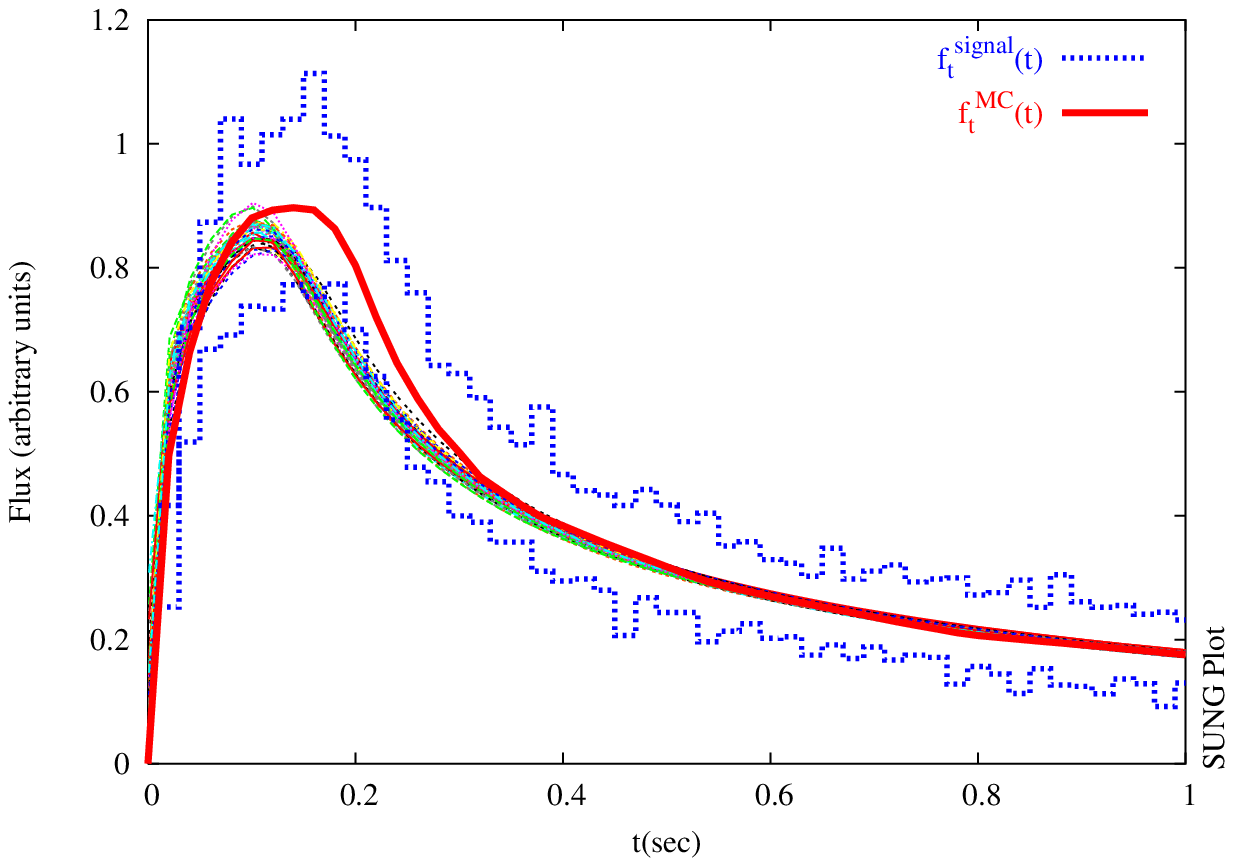}
}
\subfigure[\it TSNI-6]{
\epsfxsize=65mm
\epsfbox{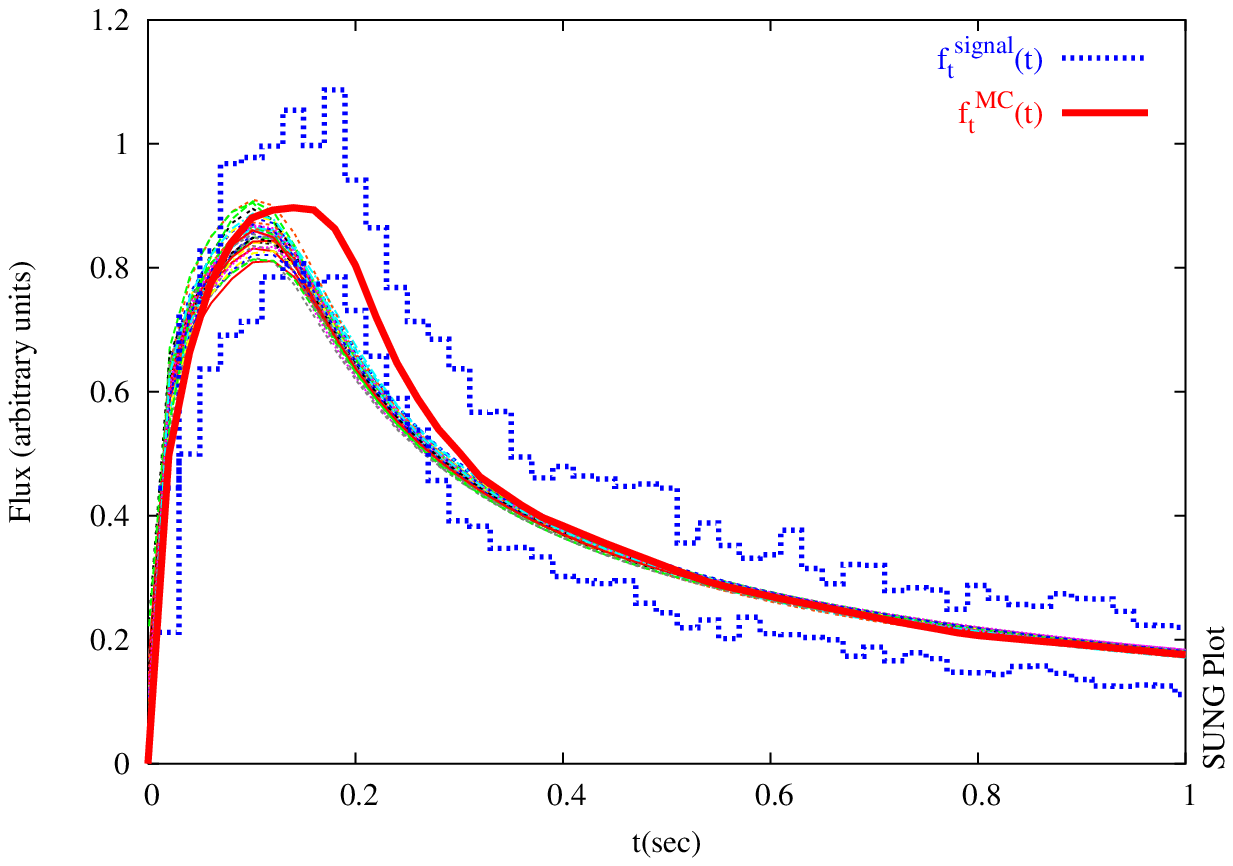}
}
\subfigure[\it TSNI-7]{
\epsfxsize=65mm
\epsfbox{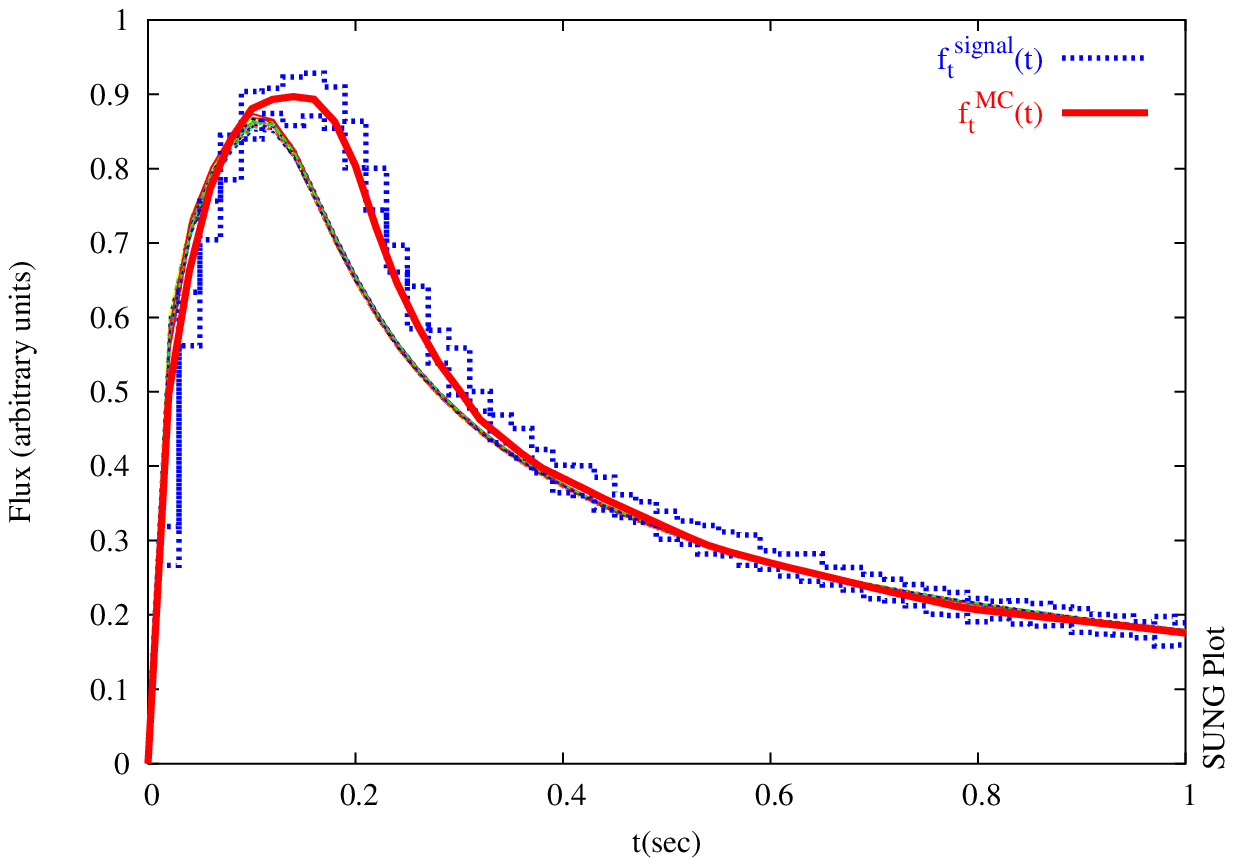}
}
\subfigure[\it TSNI-8]{
\epsfxsize=65mm
\epsfbox{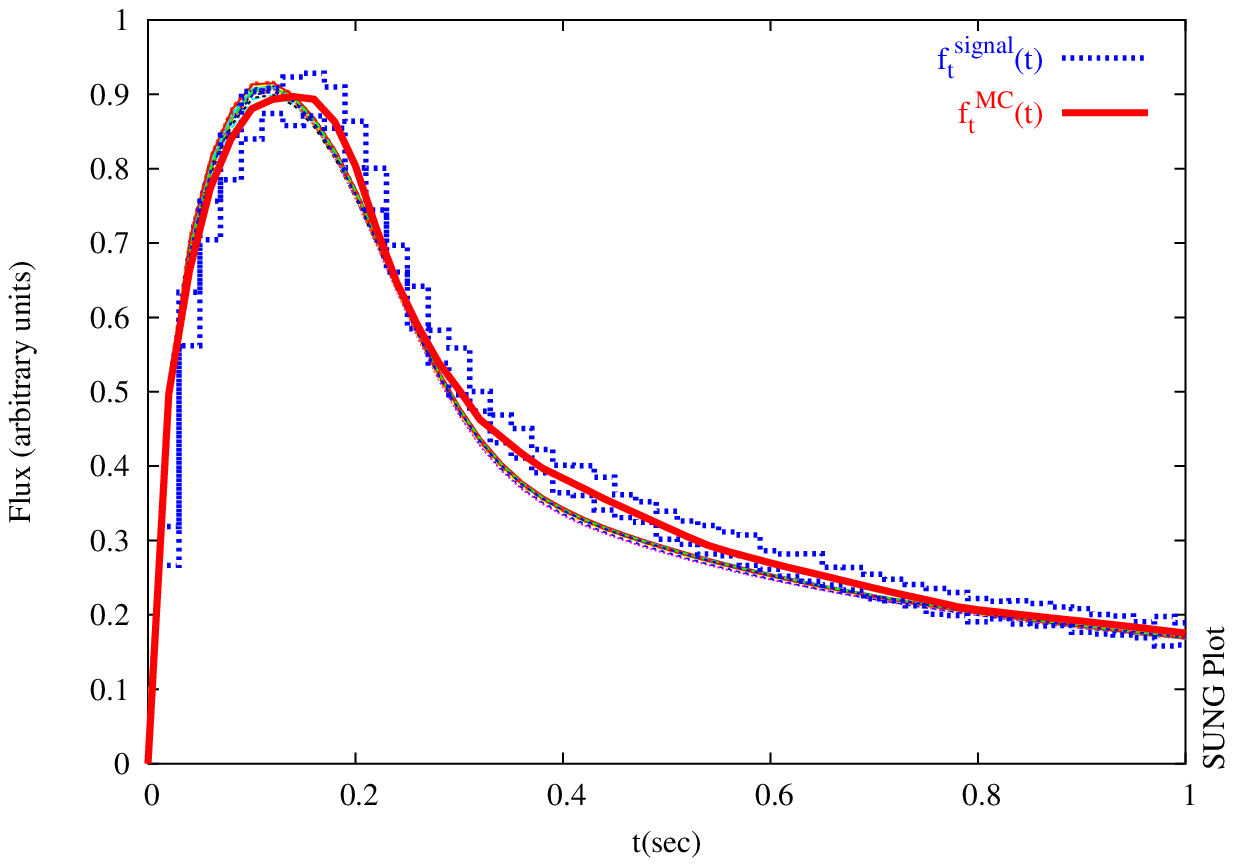}
}
\mycaption{Fits to the flux model for various sets of input conditions using
supernova model I}{
Fits to the flux model for various sets of input conditions using supernova
model I.  In each figure we compare the input time profile $f_t^\rm{MC}(t)$
(\ref{eq:profile.signal.rate}) (thick continuous line), the best-fit flux
model $f_t^\rm{signal}(t)$ for each signal (bunch of curves) and the observed
time profile $f_t^\rm{obs}(t)$ (the outer and inner histograms enclose the
observed profiles for all the signals).
}
\label{fig:flux.fit.SNI}
\end{center}
\end{figure}

\begin{figure}[p]
\begin{center}
\subfigure[\it TSNII-1]{
\epsfxsize=65mm
\epsfbox{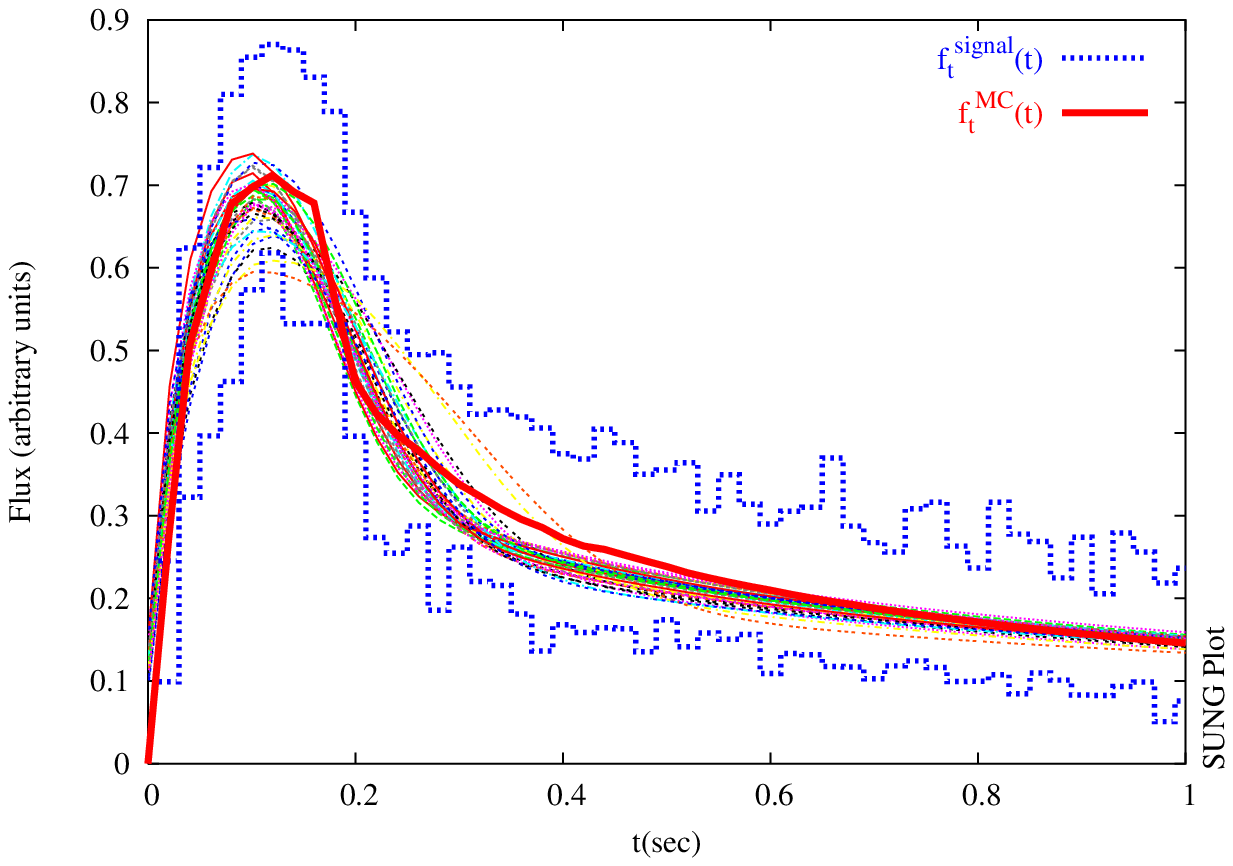}
}
\subfigure[TTSNII-2]{
\epsfxsize=65mm
\epsfbox{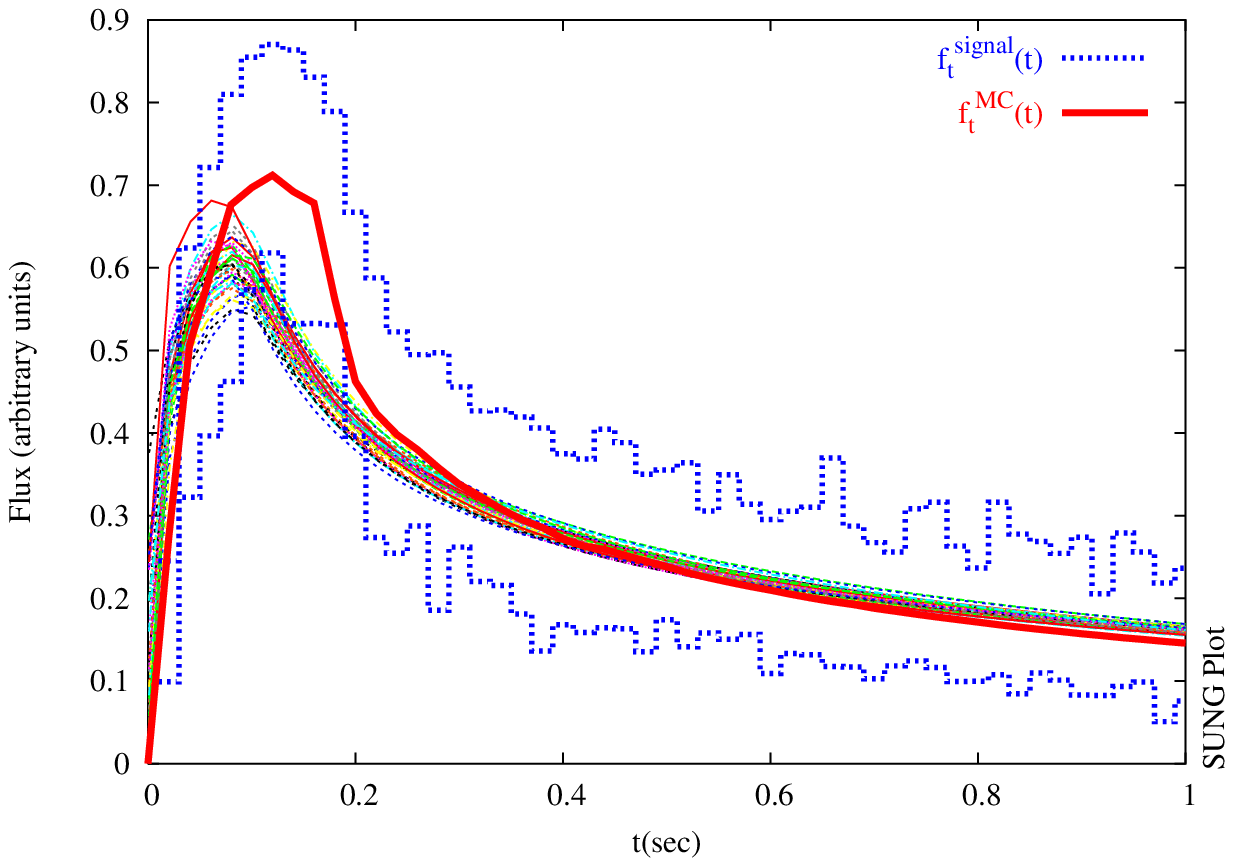}
}
\subfigure[\it TSNII-5]{
\epsfxsize=65mm
\epsfbox{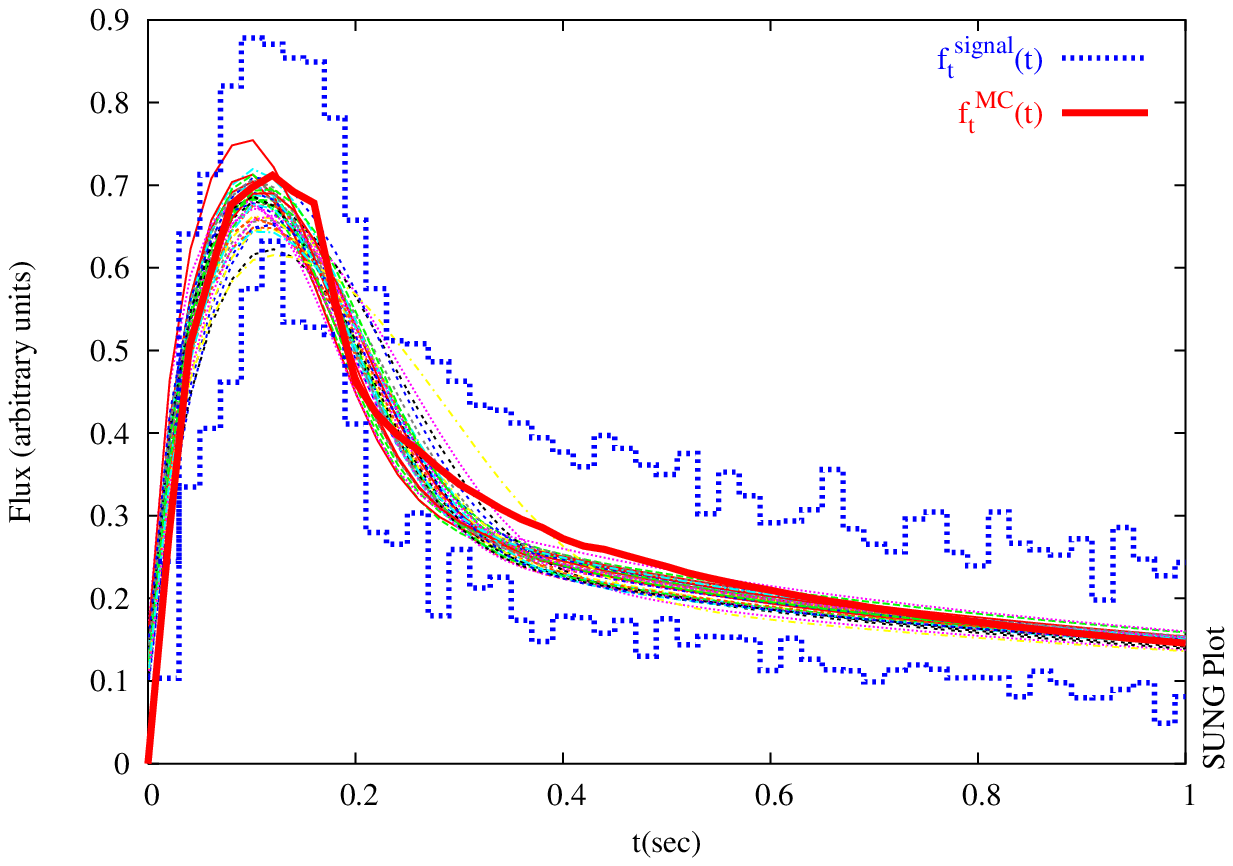}
}
\subfigure[\it TSNII-6]{
\epsfxsize=65mm
\epsfbox{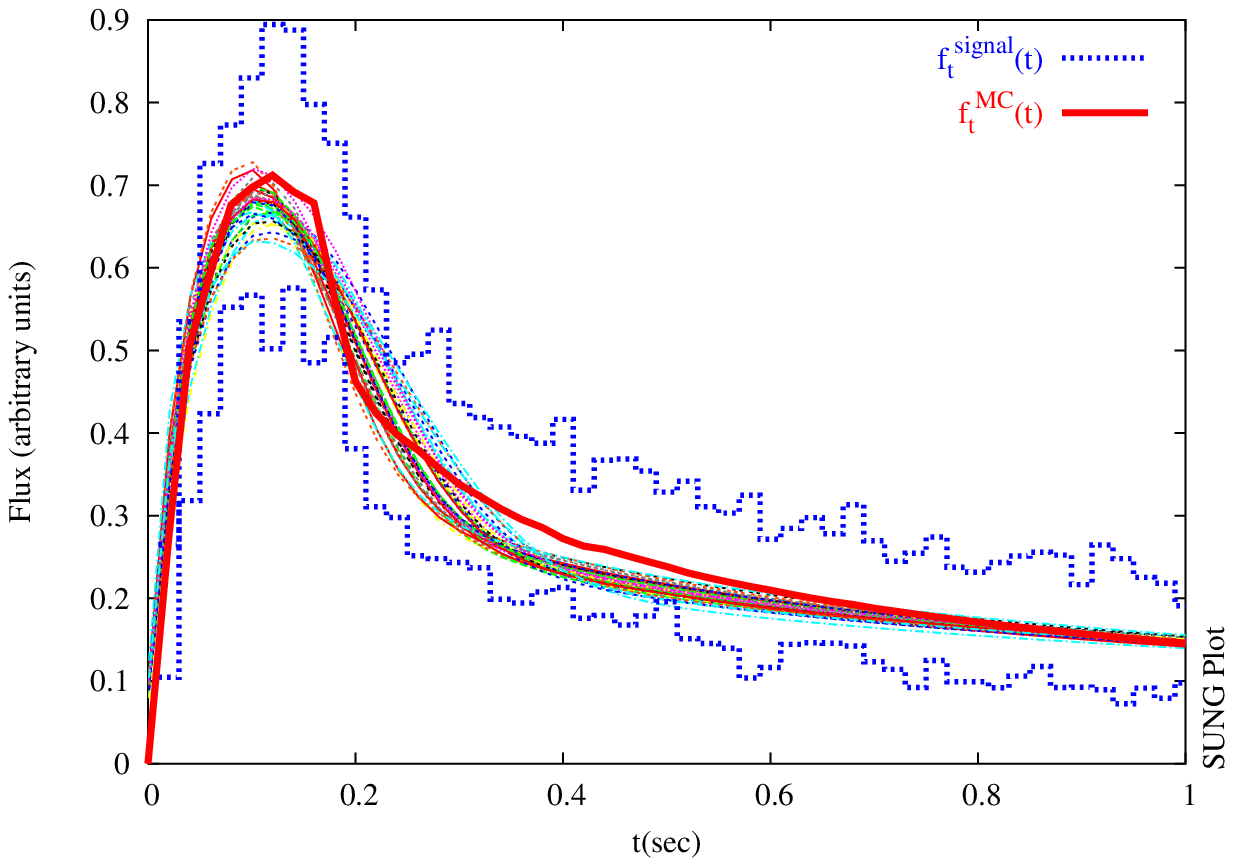}
}
\subfigure[\it TSNII-7]{
\epsfxsize=65mm
\epsfbox{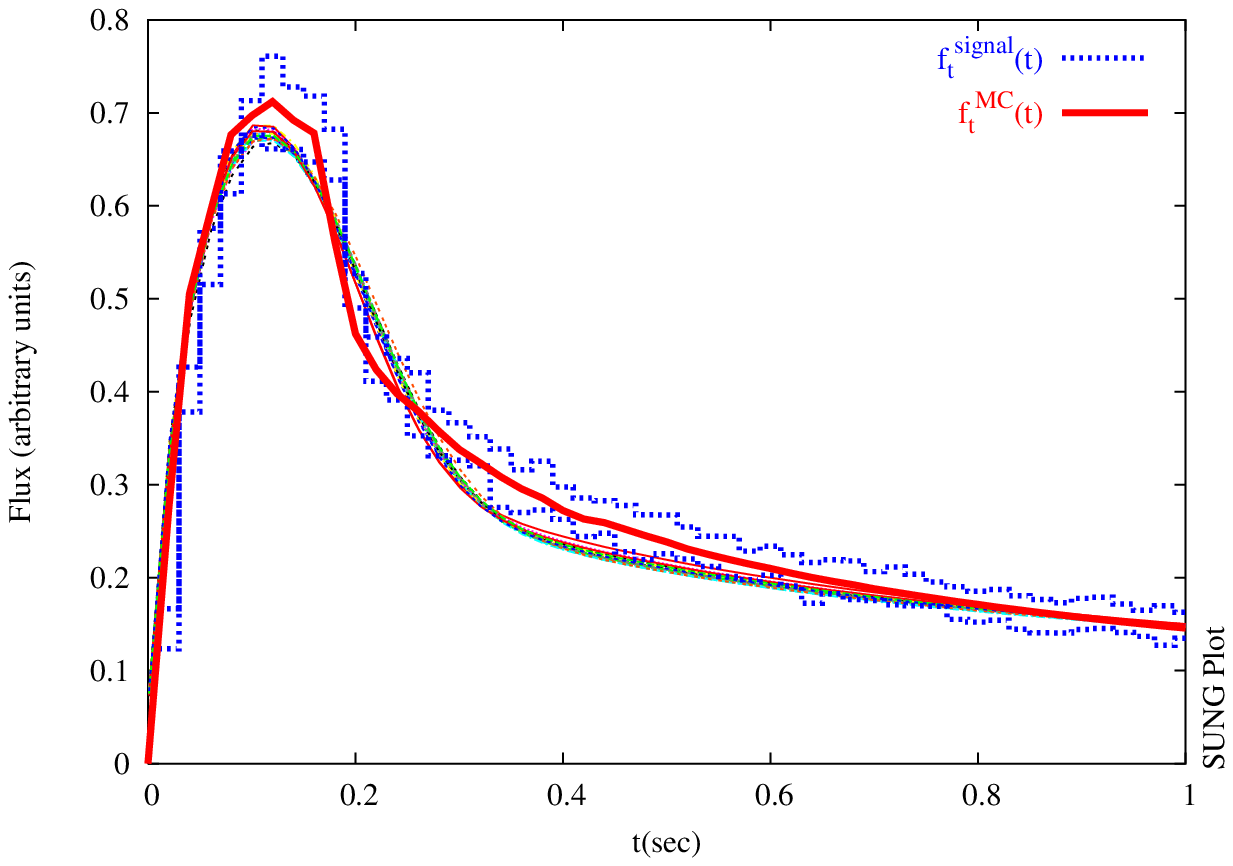}
}
\mycaption{Fits to the flux model for various sets of input conditions using
supernova model II}{
Fits to the flux model for various sets of input conditions using supernova
model II.
}
\label{fig:flux.fit.SNII}
\end{center}
\end{figure}

In figures~\ref{fig:flux.fit.SNI}-\ref{fig:flux.fit.SNII} we depict
comparisons between $f_t^\rm{obs}(t)$ and $f_t^\rm{signal}(t)$ for several of
the tests. Additionally we have included there the expected time-profile
computed from the signal rate (\ref{eq:total.rate}):

\beq
f_t^{MC}(t)\equiv\,n_\rm{tot}^{-1}\int{dE\;\frac{d^2\,n_{\anue}}{dt\,dE}}\,,
\label{eq:profile.signal.rate}
\eeq

The last comparison allows us to understand how well the original emitted flux
can be guessed from the signal analysis.

\afterpage{\clearpage}
%

\afterpage{\clearpage}
%

As can be observed the analytical flux models do not fit perfectly the
signal.  Several fits seem actually rather poor (see for example
TSNII-2).  Nevertheless in almost all the cases in average the right
value of the mass was obtained (see
figs.~\ref{fig:bandplots.SNI-1}-\ref{fig:bandplots.SNII-2}).  A bad
fit to the flux model just increases the dispersion in the mass
values, but does not change too much the best-fit point.  Again, this
is an indication that the right value of the neutrino mass can be
measured even if the fit to the signal detailed shape is only
approximate.

The more flexible flux model II fits better the signals when supernova model
II is used (compare TSNII-1 and TSNII-2).  This flux model allows to describe
more accurately the inflections of the flux for this model of supernova
neutrino emission.  The supernova model I signals are relatively well fitted
by the simpler flux model I, and no significative difference in the best-fit
mass can be observed with respect to flux model II.

The parts of the detected flux that are fitted better are the initial rising
phase and the decay.  In all cases, when the model fits properly these phases,
the fit to the mass is successful.

Altogether, these evidences support the idea that the flux model fit is to a
large extent uncorrelated with the neutrino mass measurement.  Regardless of
the availability of a good astrophysical description of the signal, the method
could extract enough information on the neutrino mass from a high statistics
signal under a wide range of conditions.

\section{Quantifying the sensitivity of the method}
\label{sec:results.sens.quant.}

The results presented in the previous section give us a general idea of the
overall properties of our method.  Now we want to put quantitative limits on
how much information on the neutrino mass could be extracted from a future
supernova signal.

In order to quantify the potential sensitivity of our method we have devised
two different approaches:

{\bf Averaging the upper bounds}.  If the neutrino mass is small and the
sensitivity of the method is not large enough to resolve it, the only
information that we can obtain is what is the largest neutrino mass compatible
with the signal.  In terms of the profile likelihood $\hat{\cal L}$ the mass
upperbound $m_\nu^\rm{up}$ can be computed using the prescription
(\ref{eq:mupper}):

\beq
\nonumber
\int_{0}^{m^\rm{up}_\nu} p(m_\nu|\D)\;dm_\nu \simeq
\int_{-\infty}^{m_\nu^\rm{up}}2 m_\nu \hat{\cal L}(m^2_\nu|\D)\;dm_\nu = CR.
\eeq

We have performed a MC analysis generating $\sim40$ synthetic signals per each
set set of input conditions and assuming a MC neutrino mass
$m_\nu^\rm{MC}\approx0$.  For each signal the mass upperbound with a 90\% of
probability was computed.  We have used a flat prior probability for
$\msq\geqslant0$ and vanishing probability for $\msq\leqslant0$.

\begin{figure}[p]
\begin{center}
\epsfxsize=100mm
\epsfbox{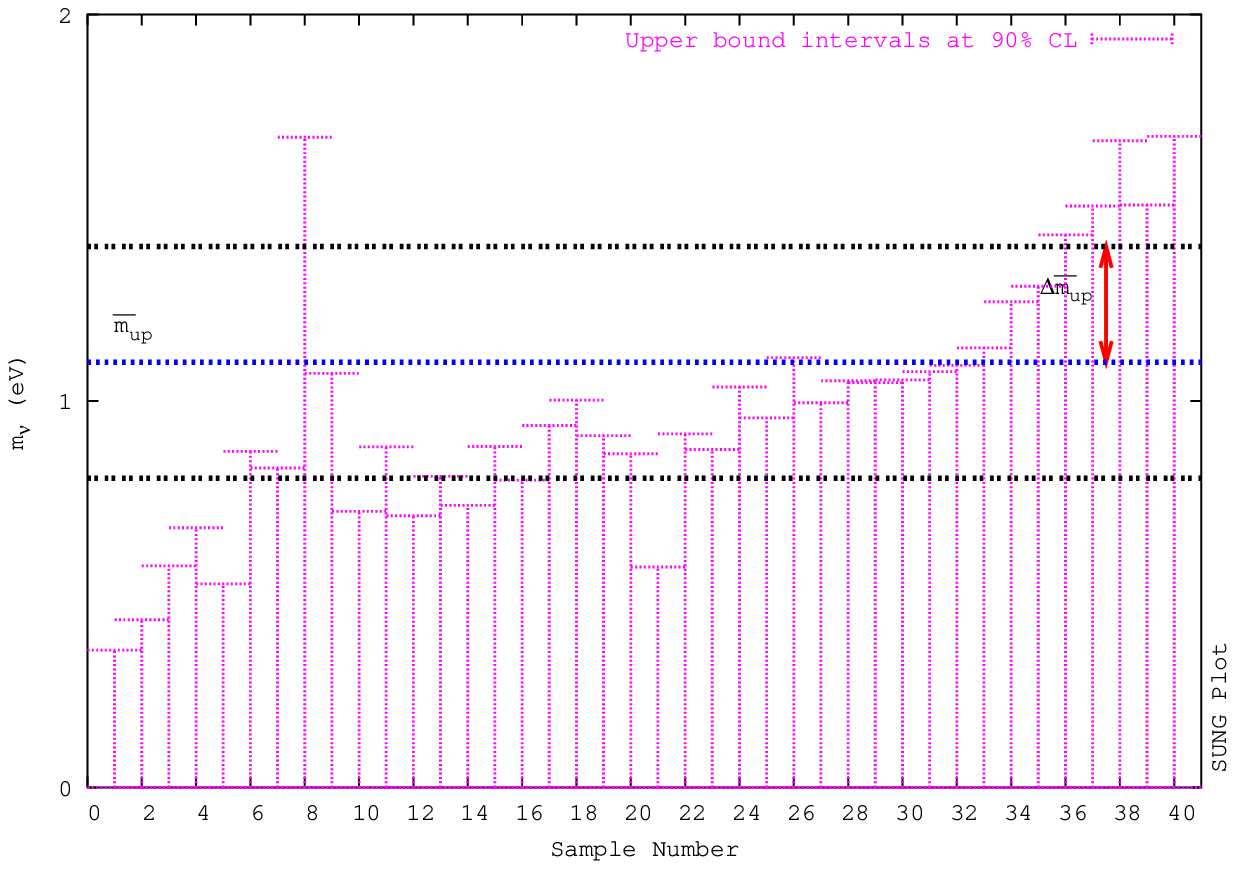} 
\epsfxsize=100mm 
\epsfbox{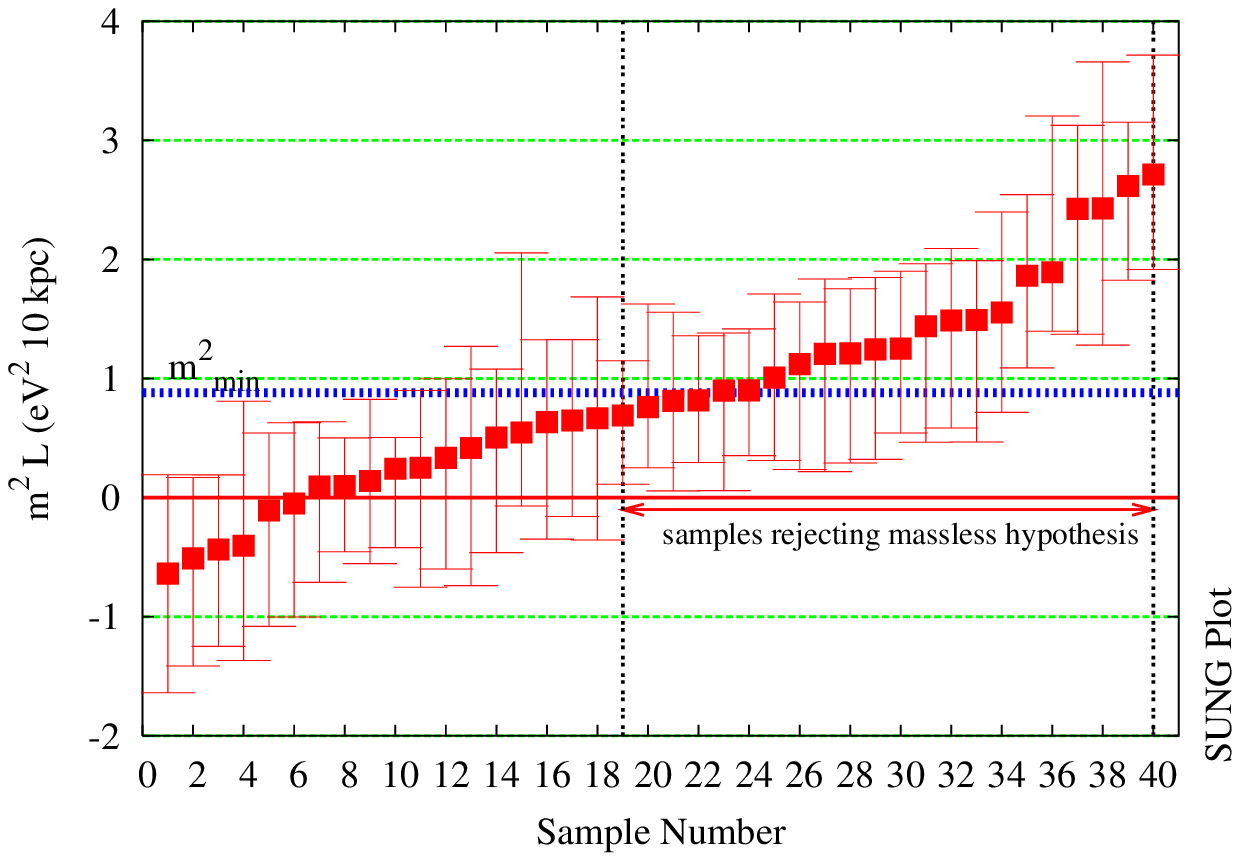} 
%
\mycaption{Two ways to quantify the sensitivity of the method}{
Two ways to quantify the sensitivity of the method. Upper panel: the average
of the mass upper bounds and its dispersion $\Delta\barr{m}_\rm{up}$.  Lower
panel: increasing the MC mass used to generate the samples we find the minimum
value for which more than 50\% of the analysis can reject the zero neutrino
mass hypothesis.
}
\label{fig:quantifiers}
\end{center}
\end{figure}

Using these results we characterize the sensitivity of the method with two
numbers: the average value $\barr{m}_\rm{up}$ of the upper bounds, and the
dispersion $\Delta m_\rm{up}$ around this value (see
fig.~\ref{fig:quantifiers}). Clearly, the only meaning of these two numbers
is that of identifying the possible range for the upper limits that could be
obtained in a real case. In particular, $\Delta m_\rm{up}$ is given only to
indicate to what extent $\barr{m}_\rm{up}$ is a good representative of the
different results of the entire ensemble, and should {\it not} be understood
as the error on the quoted upper limit.

\afterpage{\clearpage}
%

{\bf Rejection of the massless neutrino hypothesis.}  If neutrino mass were
large enough to be resolved, an alternative way to quantify the sensitivity of
the method is to determine for which value of the input mass $m_\nu^\rm{MC}$
the massless hypothesis can be rejected for a significative fraction of the
analyzed signals.

We performed different MC analysis using several values of $m_\nu^{\rm MC}>0$.
In each case for every signal we computed the lower limit using the
prescription (\ref{eq:mdwlimit}):

\beq
\int_{m^2_\rm{low}}^{+\infty}{p(m^2_\nu|D,I)\;d\,m^2_\nu} \simeq
\int_{m^2_\rm{low}}^{+\infty}{ \hat{\cal L}(m^2_\nu|D,I)\;d\,m^2_\nu} = CL
\label{eq:mlowlimit}
\eeq

The minimum value of the input neutrino mass
$m_\nu^\rm{MC}=m_\rm{min}$ for which $m^2_\rm{low}$ was larger than 0
(90\% CR) for more than 50\% of the signals was used as a measure of
the sensitivity (see fig.~\ref{fig:quantifiers}).  In this case
$\hat{\cal L}$ must be normalized over the whole interval
$-\infty<m^2<\infty$ to obtain a physically significant
$m^2_\rm{low}$.

\bigskip

In table~\ref{tab:results} we summarize the results of applying the previous
procedures to quantify the sensitivity of the method.

\begin{table}[t]
\begin{center}
\mycaption{Results for $\barr{m}_\rm{up}$ and $\sqrt{m^2_\rm{min}}$ under
different emission and detection conditions}{
Results for $\barr{m}_\rm{up}$ and $\sqrt{m^2_\rm{min}}$ under
different emission and detection conditions computed inside a 90\%
C.R. (95\% C.R. values are in parenthesis).  The average number of
events of the signals for each set of conditions is included for
reference in the fourth column.  The values reported in columns 5 and
6 are affected by statistical uncertainties at the level of $\sim
5\%$. The typical number of signals per test is $\sim 40$.
}
\begin{tabular}{cp{2cm}l>{\centering}p{2cm}cc}
\multicolumn{6}{c}{\bf supernova model I} \\ \hline\hline
\bf Test & \bf Detector & \bf Distance & \bf N. events ($\times10^3$)&
$\overline{m}_{\rm up} \pm \Delta m_{\rm up}$ (eV) & $\sqrt{m_{\rm min}^2}$
(eV) \\ \hline
TSNI-1 & SK & 10 kpc & 10.0 & $1.0 (1.1)\pm 0.2$  & $1.0 (1.1)$ \\
TSNI-3 & -- & 5 kpc & 40.0 & $1.1 (1.2)\pm 0.2$ & $1.1 (1.2)$  \\ 
TSNI-4 & -- & 15 kpc & 4.4 & $1.3 (1.4)\pm 0.3$ & $1.4 (1.5)$  \\ 
TSNI-5 & SK+KL & 10 kpc & 10.4 & $1.0 (1.1)\pm 0.2$ & $0.9 (1.0)$ \\
TSNI-6 & LENA & 10 kpc & 12.6 & $0.9 (1.0)\pm 0.2$ & $0.9 (1.0)$ \\
TSNI-7 & HK &  10 kpc & 170 & $0.4 (0.5)\pm 0.1$ & $0.5 (0.6)$ \\ 
\blankline{5}
\multicolumn{6}{c}{\bf supernova model II} \\ \hline\hline
\bf Test & \multicolumn{2}{c}{\bf Inputs} & \bf N. events ($\times10^3$)&
$\overline{m}_{\rm up} \pm \Delta m_{\rm up}$ (eV) &
$\sqrt{m_{\rm min}^2}$ (eV) \\ \hline
TSNII-1 & SK & 10 kpc & 5.9 & $1.1 (1.2)\pm 0.3$ & $1.1 (1.2)$ \\ 
TSNII-3 & -- & 5 kpc & 23.7 & $1.1 (1.2)\pm 0.3$ & $1.1 (1.2)$ \\ 
TSNII-4 & -- & 15 kpc & 2.6 & $1.6 (1.7)\pm 0.6$ & $1.6 (1.8)$ \\ 
TSNII-5 & SK+KL &  10 kpc & 6.1 & $1.1 (1.2)\pm 0.3$ & $1.1 (1.2)$ \\
TSNII-6 & LENA &  10 kpc &  7.5 & $0.9 (1.0)\pm 0.3$ & $1.1 (1.2)$\\
TSNII-7 & HK &  10 kpc & 100 & $0.5 (0.6)\pm 0.1$ & $0.5 (0.6)$ \\ \hline\hline
\blankline{5}
\multicolumn{2}{r}{\bf Reference test:}& 10 kpc & 9.6 & $0.8 (0.9)\pm 0.2$ & $0.9 (1.0)$ \\ \hline\hline
\end{tabular}
\label{tab:results}
\end{center}
\afterpage{\clearpage}
\end{table}

\begin{figure}[t]
\begin{center}
\epsfxsize=120mm
\epsfbox{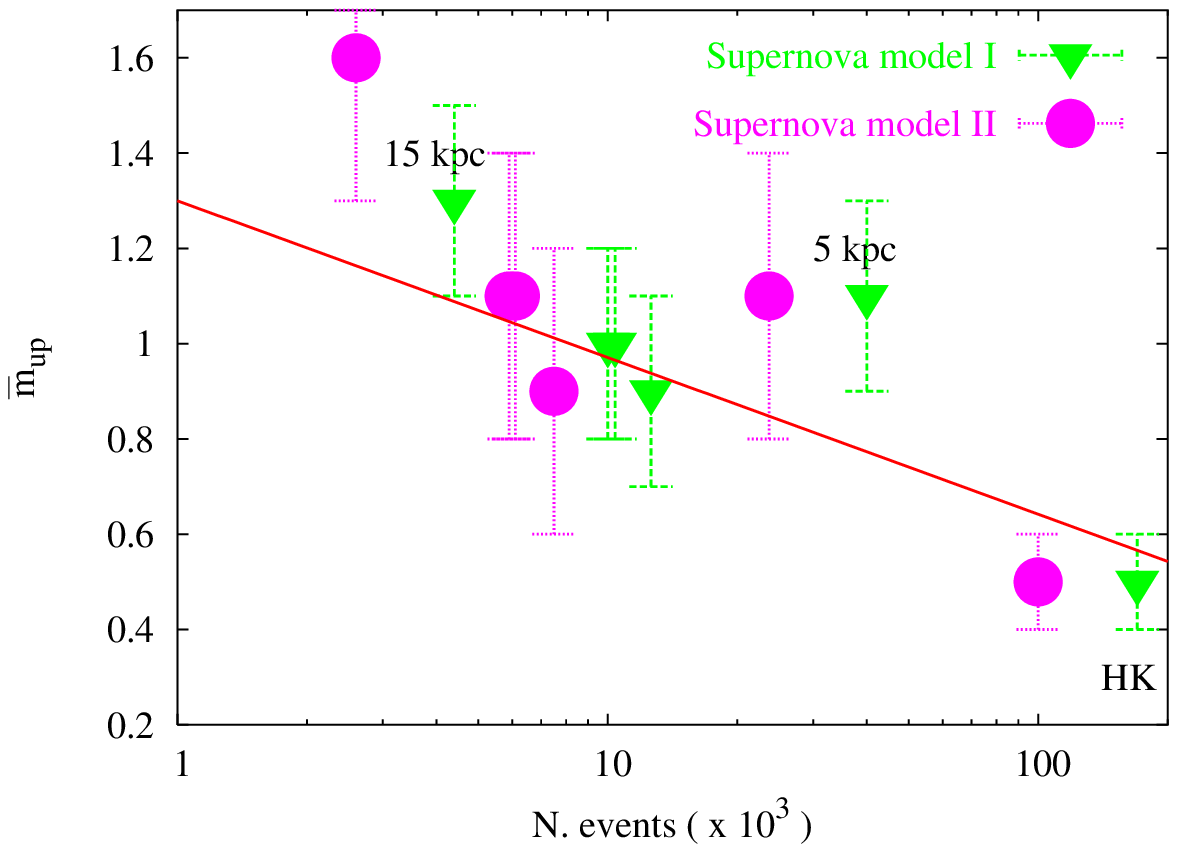}
%
\mycaption{Relation between the number of events in the signal and the
sensitivity of the method}{
Relation between the number of events in the signal and the sensitivity of the
method.  The continuous line highlights the empirical ``logarithmic'' relation
between the sensitivity and the amount of events in the signal.  For different
distances also the different TOF is involved in the determination of the
sensitivity, and the respective results deviate from the linear-log fit.
}
\label{fig:results.graph}
\end{center}
\end{figure}

The first thing we can deduce from the results is that the average number of
events in the signals is the dominant factor. This is clearly illustrated in
fig.~\ref{fig:results.graph} were we have plotted the values of
$\barr{m}_\rm{up}$ against the average number of events.  It must be noticed
that the improvement in the sensitivity due to an increase in the statistics
is not dramatic.  Comparing the sensitivity of the method at HK with the
analogous results of SK we see that an increase by a factor of 10 in the
statistics improves the sensitivity only by a factor of 2.

%
\afterpage{\clearpage}
%

Comparing the results of the combined analysis SK+KL (TSNI-5 and TSNII-5) with
these performed on SK alone (TSNI-1 and TSNII-1) we see that sensitivity is
determined by the larger statistics SK and the better energy resolution and
lower threshold of KamLAND does not impact too much the result.  A similar
conclusion is obtained observing that the planned scintillator detector LENA
presents only a minor improvement in the sensitivity with respect to SK.  Even
if the energy resolution is better, the comparable fiducial volumes that
implies comparable number of events results in quite similar sensitivities for
the two detectors.

Changing the supernova distance two competitive effects arise.  On one hand
decreasing the distance increases the signal statistics, which in principle
implies an increase of the sensitivity.  However, at smaller distances the TOF
delays induced by the mass are also smaller and become harder to identify.  As
can be learned with tests TSNI-3 and TSNII-3 these two effects tend to
compensate each other.

It is natural to ask if anything better could be done to measure neutrino
masses from a supernova neutrino signal.  In the attempt to answer this
question, we have performed the following test: we have produced a sample of
synthetic signals assuming no mixing in the spectrum (no oscillations) and
using as inputs to our MC flux model I (\ref{eq:fluxmodel.simple}) with a suitable
choice of the relevant parameters, together with an analytical
$\alpha$-distribution spectrum corresponding to the average energy profile
given in fig.~\ref{fig:snmodel}a.  We have then performed our usual set of
fits to the neutrino mass (we assume the SK detector) but fixing the value of
the flux parameters to the ones used in the MC, and we also adopt the same
time varying spectrum used to generate the sample. This simulates the ideal
(and unrealistic) situation where the time-energy dependence of the signal at
the source is known, and the only free parameter is the neutrino mass. The
results of this test are given in the last row in table~\ref{tab:results} and
should be compared with the results for tests TSNI-1 and TSNII-1.  We see that
only a mild improvement is achieved with respect to the realistic
situation. This allows us to conclude that the sensitivity to neutrino masses
of the detectors presently in operation is very likely bounded to values not
much below 1$\,$eV.

\section{Additional remarks on the method}
\label{sec:results.otherprop}

To arrive to the results we have just presented, some specific choices about
the overall procedure that was followed had to be made.  Here we will briefly
discuss the impact of some of these choices on the final results.

{\bf Different priors}.  As was stated in
sect.~\ref{subsec:posteriors} the choice of the prior probability for
$\msq$ is a subtle issue.  For all the analysis presented in this work
we have used a $\Theta$ function as the prior probability on $\msq$.
In order to understand the impact that a different prior could have on
the results we have compared the mass limits with the ones obtained
with a flat prior on $m_\nu$.

\begin{figure}[h]
\begin{center}
\epsfxsize=120mm
\epsfbox{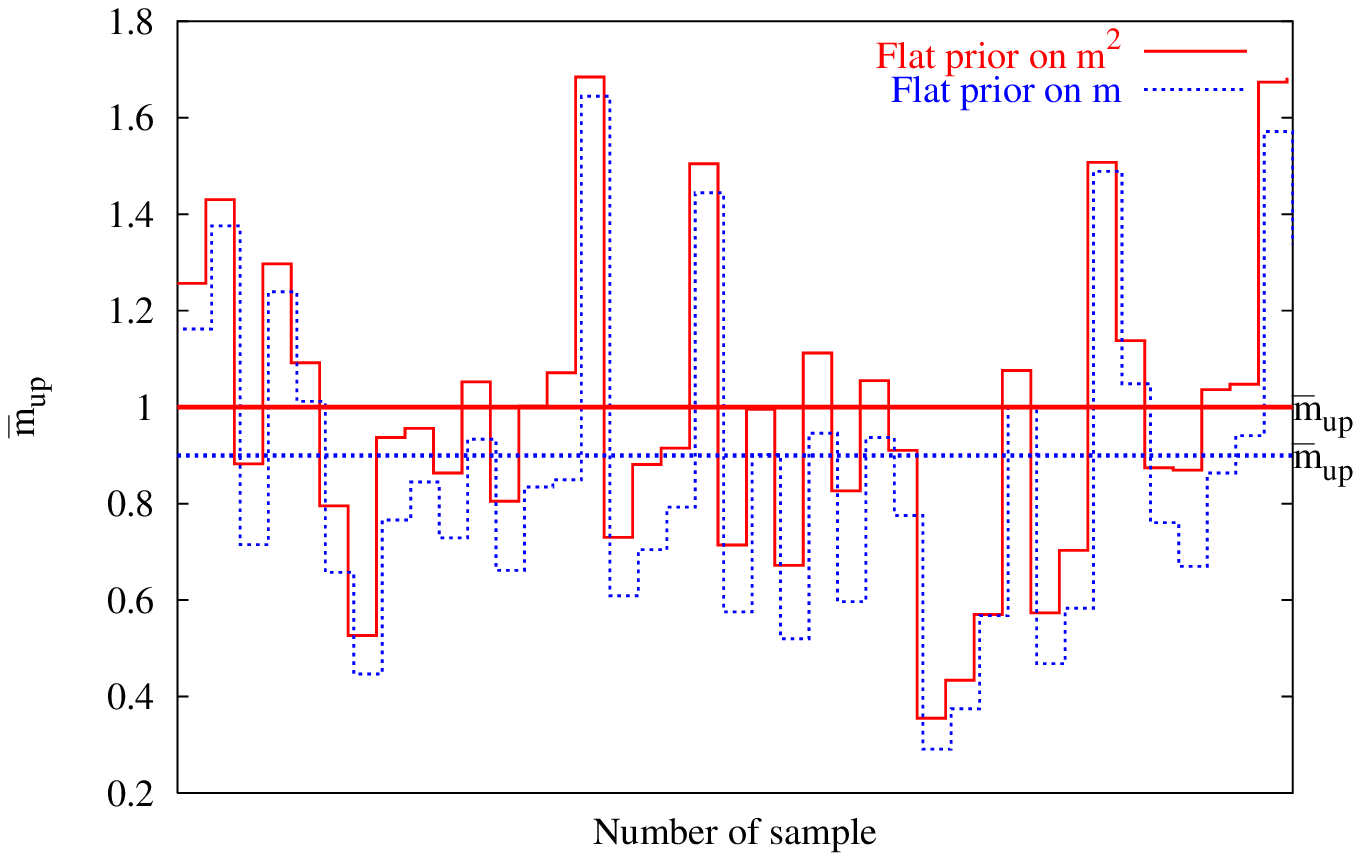}
\mycaption{Upper bounds obtained when two different prior probability
on neutrino mass are used}{
Upper bounds obtained when two different prior probability on neutrino mass
are used.  Continuous line correspond a flat prior on $\msq$ as used for all
the analysis reported in this work.  Dashed line are the upperbounds for the
same signals obtained with a flat prior on $m_\nu$.
}
\label{fig:other.priors}
\end{center}
\end{figure}

In figure~\ref{fig:other.priors} we present the mass upper bounds obtained
with both priors.  We observe that the differences are not large.  In the
context of Bayesian reasoning this means that the signal contains enough
``evidence'' about the mass and therefore a change in the prior does not
affect the results.

%
%

{\bf Estimation of the energy spectrum}. The choice of the spectral
function used to fit the energy distribution of the events represent
an interesting issue in our analysis.  In ref.~\cite{Nardi:2003pr} we
used a Fermi-Dirac distribution.  Comparing these results with the
ones presented here, where the $\alpha$-function was used, it is
evident that the particular spectral shape parametrization has no big
impact on the final results. The main difference is the more direct
way the parameters of the $\alpha$-distribution can be estimated.  In
the Fermi-Dirac case the estimation of the spectral temperature and
effective degeneracy parameter $\eta$ involves a numerical procedure
which mainly slows down the numerical analysis.

{\bf Energy threshold}. As was observed in sect.~\ref{sec:results.sens.quant.}
an increase in the energy resolution of the detector does not affect too much
its sensitivity.  However a too large energy threshold could affect
considerably the method resolution.

In ref.~\cite{Nardi:2003pr} we compared the results for analysis performed on
signals detected in SK assuming two different energy thresholds: 5 MeV, the
threshold used here, and a more conservative threshold of 10 MeV.  In that
work the sensitivity dropped down by a factor of two making apparent that the
low energy neutrinos play a central role in the determination of the
sensitivity.

A new question arises: why the analysis of scintillator detector signals with
a significantly lower threshold does not produce a similar improvement in
sensitivity?  The answer is that the increase in number of events is about
1-2\% (see fig.~\ref{fig:number.zones}) when we go from the 10 MeV threshold
to 5 MeV (and this represents 100-200 additional low energy events in the SK
signal) while lowering the threshold down to $\sim3 MeV$ implies only a
$\sim0.05-0.1$\% increase (5-10 events).

\newpage

\section{Summary and Conclusions}
\label{sec:results.conclusions}

We have described in this work a new method to measure and constrain the
absolute scale of the neutrino mass using a high statistics signal from a
future Galactic supernova.

The method relies on three basic conditions:

\begin{itemize}

\item An almost thermal neutrino spectra at the source (see
  sect.~\ref{sec:spectra}).

\item The distribution of the high energy neutrino events, less affected by
  the TOF delay induced by a non-zero neutrino mass, can be used to
  extrapolate the signal profiles at low energies, since the time scale of the
  spectral evolution is larger than the typical mass induced time delay
  (fig.~\ref{fig:mean.energies}).

\item A theoretical description based on general characteristics of the
  neutrino flux time evolution (an early fast rise followed by a steady decay
  on time scales of several seconds) is sufficiently accurate to construct a
  likelihood for studying the neutrino mass more probable values.

\end{itemize}

We combined a ML estimator and Bayesian techniques to construct the
statistical procedure designed to constrain a neutrino mass, irrespectively of
the particular flux model parameters used to describe the signal.

Different tests corresponding to two different supernova models, different
oscillation schemes and detector were carried out.  For each particular set of
conditions, about 40 complete analysis were performed.

The general results of these tests have been described in detail in
sects.~\ref{sec:results.general}.  In particular, it was shown that regardless
of the fine details of the signal, the measured value of the neutrino mass can
be nailed around the correct value.  The spread in the uncertainty depends
however, on the quality and the amount of details used to describe the signal.

The two analytical models introduced to describe the flux behaved relatively
well when applied to very different numerical neutrino fluxes. Satisfactory
results where obtained with the first more simple flux model
(\ref{eq:fluxmodel.simple}) especially when fitting supernova model I signals.
The more flexible flux model II (\ref{eq:fluxmodel.phen.}) required a tuning of
various parameter that had to be fixed to suitable value to leave just a
reasonable number of free variables in the likelihood multiparameter
extremization. 

The sensitivity of our method was estimated in two different ways,
first, by determining the typical upper bound on $m_\nu$ that could be
obtained in case the neutrino mass is too small to be
resolved. Secondly, by evaluating the minimum mass that could be
distinguished from zero.

\bigskip

We believe that the method that we have proposed represents an
improvement with respect to previous techniques, both in sensitivity
and in the independence from particular astrophysical assumptions.

The sensitivity to neutrino masses of the detectors presently in
operation can reach a level down to 1$\,$eV.  This is sizable better
than present results from tritium $\beta$-decay experiments
\cite{Bonn:2002jw,Lobashev:2001uu}, is competitive with the most
conservative limits from neutrinoless double $\beta$-decay
\cite{Klapdor-Kleingrothaus:2000sn,Aalseth:2002rf,Bilenky:2002aw,Klapdor-Kleingrothaus:2004wj},
and although less precise than cosmological measurements
\cite{Hannestad:2003xv,Elgaroy:2003yh,Crotty:2004gm}, is also
remarkably less dependent from prior assumptions.  Future megaton
water \v{C}erencov detectors as Hyper-Kamiokande will allow for about
a factor of two improvement in the sensitivity.  However, they will
not be competitive with the next generation of tritium $\beta$-decay
\cite{Osipowicz:2001sq,Weinheimer:2002jx} and neutrinoless double
$\beta$-decay experiments (see \cite{Cremonesi:2002is} and references
therein).

\begin{figure}[h]
\begin{center}
\epsfxsize=140mm
\epsfbox{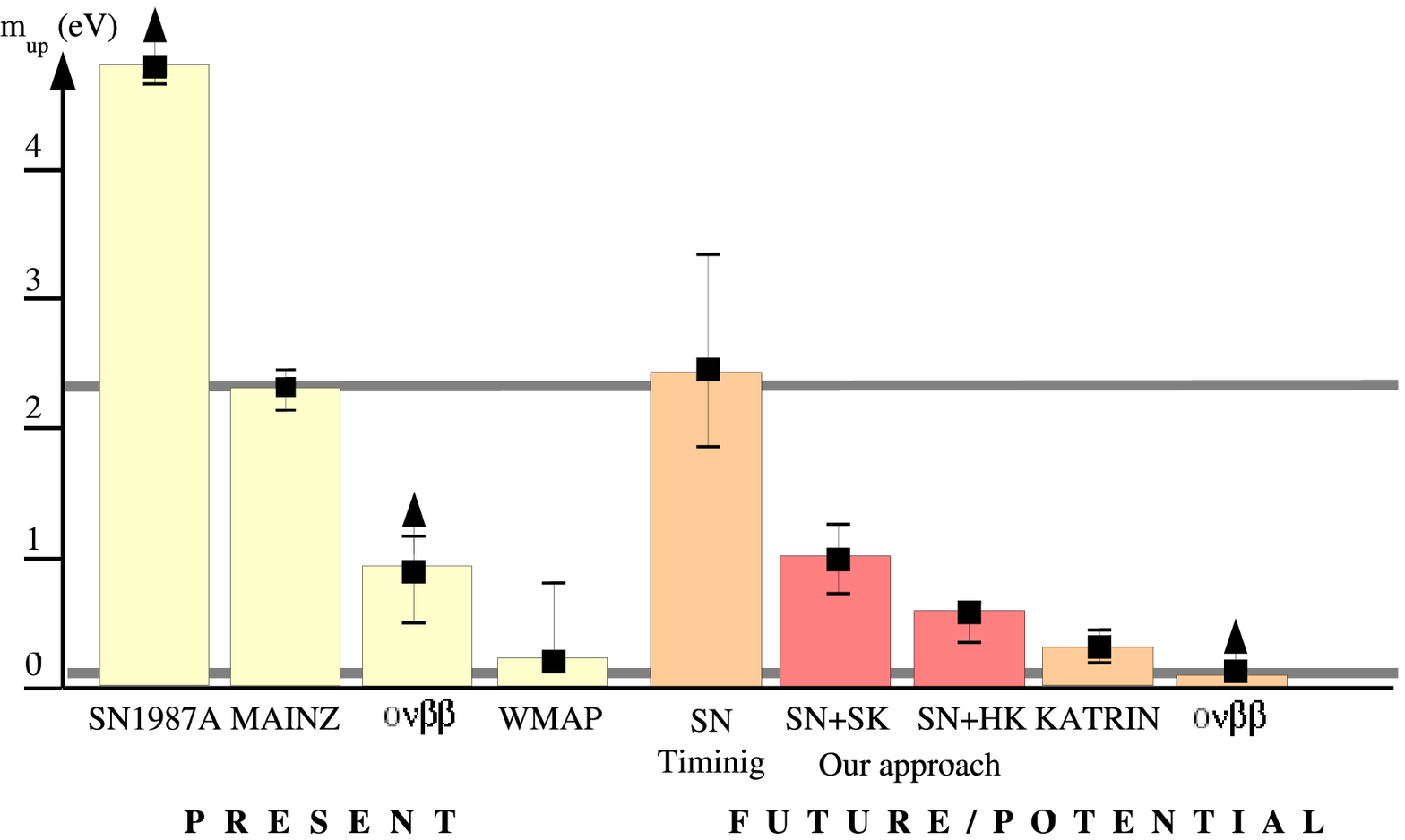}
%
\mycaption{Present and future of mass limits}{
Mass limits from present and future laboratory (MAINZ\cite{Bonn:2002jw},
KATRIN\cite{Weinheimer:2002jx}, \onubb\cite{Cremonesi:2002is}), astrophysical
(SN\cite{Totani:1998nf,Arnaud:2001gt,Beacom:2000qy}) and cosmological studies
(WMAP\cite{Hannestad:2004nb}).  The thick gray lines in the background
indicates the absolute upper limit from Tritium decay $m_\nu<2.3$ eV
\cite{Bonn:2002jw} and the lower limit from oscillation evidences
$m_\nu>\sqrt{\Delta m^2_\rm{atm}}\simeq 0.05$ eV \cite{Bahcall:2004ut}.
}
\label{fig:mass.limits.comparative}
\end{center}
\end{figure}

In figure~\ref{fig:mass.limits.comparative} we compare the limits that could
be obtained with the present method, with other available and foreseen limits
from laboratory, astrophysical and cosmological data.

%
%

We conclude that the occurrence of a Galactic supernova explosion within the
next few years might still provide valuable informations on neutrino masses
that however will not be able to explore a region much below 1 eV. Therefore
as new laboratory experiments and cosmological observations will push the
neutrino mass limits sensibly below 1$\,$eV, the corresponding effects of the
neutrino time of flight delays on a supernova signal will become unmeasurable.


\appendix

\chapter{Bayesian Inference: basic definitions}
\label{ap:bayes}

The method and techniques used in this work to obtain informations
about a neutrino mass from a synthetic supernova signal are based on
Bayesian principles of inference.  Bayesian reasoning in data analysis
has recently gained more and more relevance in a wide range of
scientific disciplines, including frontier physics
\cite{DAgostini:2003qr}. This appendix is aimed to introduce the main
definitions and results of Bayesian inference in data analysis on
which the method proposed in this work is based.  A self contained and
physics oriented introduction to Bayesian reasoning can be found in
\cite{Loredo:2001rx} while a more complete review of Bayesian
techniques and their applications in physics data analysis is given in
\cite{DAgostini:2003qr}.

\section{Basic principles}
\label{sec:bayes.basic}

Bayesian inference is founded on two basic ideas.  The first is the
conception of probability as the {\it degree of belief} that a given
proposition is true as opposite to the {\it conventional statistical}
definition of probability as the long-run relative frequency with
which the event defined by the proposition occurred on many repeated
experiments \cite{Sivia:1998}.  Curiously (or indeed not) the Bayesian
idea of probability correspond to the original concept by Laplace (as
the father of inference reasoning in science) two centuries
ago\cite{Sivia:1998}.  The second idea is that every probabilistic
statement that we can make on any proposition must be a {\it
conditional probability}, i.e. it must be conditioned to the available
relevant information related with the proposition.

Formally, Bayesian inference is based on two central rules that can be
derived from the requirement of logical consistency of the probability
theory\cite{Cox:1946}.  By denoting with $p(A|B)$ the probability that
$A$ is true, given that $B$ is true, it is straightforward to write
down the {\it product rule}:

\beqas
p(M,D|I)&=&p(M|D,I)\times p(D|I)\\
        &=&p(D|M,I)\times p(M|I).
\label{eq:prod.rule}
\eeqas

which states that the probability that propositions $M$ {\it and} $D$ are true
{\it given} the background information $I$ is equal to the probability of $M$,
given $D$ and $I$, times the probability of $D$ given $I$.  On the other hand
we have the {\it sum rule}:

\beq
\sum_i p(M_i|I) = 1\,,
\label{eq:sum.rule}
\eeq

where $M_i$ is a set of mutually {\it exclusive} and {\it exhaustive} set of
propositions and 1 refers to the convention to assign this value to the
probability of a {\it tautological} proposition.

\section{Bayes theorem}
\label{sec:bayes.theorem}

Starting from the two rules introduced in the previous section two central
results can be obtained: the Bayes theorem and the marginalization procedure.

Let us suppose that in a given measurement we obtain the data $D$.  Based on
some background information $I$ we assume that model $M$ can describe the
data.  Using the previous knowledge (product of our expertise) on the
phenomenon we assign to model $M$ a {\it prior probability} $p(M|I)$ which
measure the {\it plausibility} of that model given the background information
$I$.  The {\it posterior probability} of $M$ given the data and the background
information can be computed from $p(M|I)$ using the product rule
(\ref{eq:prod.rule}):

\beq
p(M|D,I) = p(M|I)\,\frac{p(D|M,I)}{p(D|I)}
\label{eq:bayes.theo.basic}
\eeq

This result is called the {Bayes theorem}. Here $p(D|M,I)$ is the probability
that the data $D$ be described by model $M$, and it is called the {\it
sampling probability} for $D$ or the {\it Likelihood} for model $M$.  $p(D|I)$
is called the {\it evidence} for $D$ and represents the probability that the
measurement produce the data $D$ for the entire class of hypotheses (models).

When $M$ is described by a (continuous) set of parameters collectively denoted
as $\{\theta\}$, posterior $p(\{\theta\}|D,I)$ becomes multivariate {\it
probability distribution functions} (pdf) of the parameters, while the
Likelihood $p(D|\{\theta\} ,I)$, that we will denote by the symbol ${\cal
L}(D;\{\theta\})$ in spite of its explicit dependence is not by itself a pdf
for the parameters.  The evidence $p(D|I)$ is independent of $\{\theta\}$ and
plays simply the role of the posterior pdf normalization constant $N\equiv
p(D|I)$.  Bayes theorem now reads:

\beq
p(\{\theta\}|D) = N^{-1}\,p(\{\theta\}|I)\,{\cal L}(D;\{\theta\})\,,
\label{eq:bayes.theo.param.}
\eeq

where for simplicity we have dropped out the reference to the background
information $I$.

An interesting way to see what Bayes theorem establishes is to think
that our knowledge on how a model describes a given phenomenon grows
when new evidence is accumulated.  In this sense we could thing at
Bayes theorem as a recipe for learning \cite{Dose:2002}.

\section{The marginalization procedure}
\label{sec:bayes.margin}

Often some of the parameters that are used to describe a model, thought
important for the computation of the Likelihood, are uninteresting for the
final conclusions that one wants to extract from the experimental evidence.
The principles of Bayesian reasoning provide a natural way to {\it
marginalize} these {\it nuisance parameters}. Using the continuous limit of
the sum rule (\ref{eq:sum.rule}):

\beq
\int p(\theta|D)\,d\theta = 1\,,
\label{eq:sum.rule.continuous}
\eeq

the {\it marginal posterior probability} for parameter $\theta_0$ can be
written as:

\beqa
\nonumber
p(\theta_0|D)&=&\int{d\{\theta\}\,p(\theta_0,\{\theta\}|D)}\\
             &=&N^{-1}\,p(\theta_0)\int{d\{\theta\}\,{\cal
             L}(D;\theta_0,\{\theta\})\,p(\{\theta\})}
\label{eq:marginalization}
\eeqa

Marginalization is one of the most important features of Bayesian
inference when compared with conventional approaches.  However,
performing the multidimensional integrals required for finding the
marginal distribution can be a computational challenge in terms of the
large times necessary to carry out such integrations.  Several
solutions have been devised and are used in the most complicated
problems where Bayesian inference is applied (see section 2 of
ref.\cite{Dose:2002}).  As was described in
sect.~\ref{sec:method.margin.}, we have used the simple approximation
of taking the {\it profile likelihood} (PL) as the marginal
distribution, under the assumption that the likelihood can be
approximated by a multivariate Gaussian (normal) pdf.  In the
following section we will prove that the profile of a normal
likelihood coincides with its marginal distribution.

\subsection{Marginalization of a normal pdf}
\label{subsec:margin.normal}

The general form of a multivariate normal pdf for parameters
$\vvec{\theta}\equiv(\theta_0 ... \theta_{n-1})^\rm{T}$ is:

\beq
{\cal G}_n(\vvec{\theta};\vvec{\mu},\Matrix{C}) =
\frac{1}{(2\pi)^{n/2}\sqrt{\rm{det}(\Matrix{C})}} 
\exp\left[-\frac{1}{2}(\vvec{\theta}-\vvec{\mu})^\rm{T}
\Matrix{C}^{-1}(\vvec{\theta}-\vvec{\mu})\right]\,,
\label{eq:general.normal}
\eeq

where $\vvec{\mu}\equiv(\mu_0 ... \mu_{n-1})^\rm{T}$ are the mean values of
the parameters (here for simplicity we will assume $\mu_i=0$) and $\Matrix{C}$
is the covariance matrix:

\beq
\Matrix{C}=
\left(
\begin{array}{ccc}
\sigma_0^2 & \rho_{01}\sigma_0\sigma_1 & ... \\
\rho_{01}\sigma_0\sigma_1 & \sigma_1^2 & ... \\
... & ... & ...
\end{array}
\right)
\eeq

with $\rho_{ij}$ the correlation coefficient between parameters $i$
and $j$.  

To simplify let's take the particular case of a bivariate normal pdf.  In this
situation eq. (\ref{eq:general.normal}) is given by:

\beqa
\nonumber
{\cal G}_2(\theta_0,\theta_1;\sigma_0,\sigma_1)&=&
\frac{1}{2\pi\sigma_0\sigma_1\sqrt{1-\rho^2}}\exp\left[
-\frac{1}{2(1-\rho^2)}\left(\frac{\theta_0^2}{\sigma_0^2}+\frac{\theta_1^2}{\sigma_1^2}
-\frac{2\rho\theta_0\theta_1}{\sigma_0\sigma_1}\right)\right]\\
	&=& Z^{-1}\exp
	\left[-\frac{1}{2}
	\left(a\,\theta_0^2+b\,\theta_1^2-2c\,\theta_0\theta_1\right)\right]\,,
\label{eq:bivariate.normal}
\eeqa

with $\rho\equiv\rho_{12}$ and in the last equation we have defined
$Z\equiv2\pi\sqrt{1-\rho^2}\sigma_0\sigma_1$,
$a\equiv1/\sigma_0^2(1-\rho^2)$, $b\equiv1/\sigma_1^2(1-\rho^2)$ and
$c\equiv\rho/\sigma_0\sigma_1(1-\rho^2)$.

Using (\ref{eq:bivariate.normal}) the marginal pdf on $\theta_0$,
$p(\theta_0)$ reads,

\beqa
\nonumber
p(\theta_0)&=&\int{d\theta_1\,{\cal
	   G}_2(\theta_0,\theta_1;\sigma_0,\sigma_1)}\\
\nonumber
	&=& Z^{-1}
	\exp\left(-\frac{a\theta_0^2}{2}+\frac{c^2\theta_0^2}{2b}\right)
	\int{d\theta_1 \exp\left[-\frac{1}{2}b(\theta_1+d)^2\right]}\\
\nonumber
	&=& Z^{-1}
	\sqrt{2\pi/b}\,\exp\left(-\frac{1}{2\sigma_0^2}\theta_0^2\right)\\
\nonumber
	&=& \frac{1}{\sqrt{2\pi}\sigma_0}
	\exp\left(-\frac{1}{2\sigma_0^2}\theta_0^2\right)\,,
\eeqa

where after completing the square in the exponential of the second equation
($d=c\theta_0/\sqrt{b}$) the resulting integration is performed using
$\int{\exp(-x^2/2\sigma)}=\sqrt{2\pi}\sigma$.  As can be see from the last
equation, the marginal distribution in $\theta_0$ is simply a normal
distribution with dispersion $\sigma_0$.

Now we will proceed to compute the {\it profile likelihood} (PL)
$\hat{p}(\theta_0)$ which is defined by:

\beq
\hat{p}(\theta_0)\propto{\cal
G}(\theta_0,\theta_{1\rm{max}};\sigma_0,\sigma_1)
\eeq

where $\theta_{1\rm{max}}$ refers to the value of $\theta_1$ for which
${\cal G}$ is maximum at the given value of $\theta_0$ (see
figure~\ref{fig:illust.margin.}).  The proportional symbol '$\propto$'
means that $\hat{p}(\theta_0)$ is not necessarily normalized.

The value of $\theta_{1\rm{max}}$ is found maximizing the denominator in the
exponential of (\ref{eq:bivariate.normal}) which gives
$\theta_{1\rm{max}}=c\theta_0/b$.  This results in,

\beqa
\hat{p}(\theta_0)&\sim&\exp
	\left[-\frac{1}{2}
	\left(a\theta_0^2+\frac{b c^2}{b^2}\theta_0^2-2\frac{c^2}{b}\theta_0^2\right)
	\right]\\
	&\sim& \exp\left(-\frac{1}{2\sigma_0^2}\theta_0^2\right)
\eeqa

Therefore the profile likelihood is also a normal distribution in
$\theta_0$ with dispersion equal to $\sigma_0$ and coincides with the
marginal distribution $p(\theta_0)$.  

This result can be generalized to more than two variables.  {\it When
the likelihood is a multivariate normal distribution in the parameters
the profile likelihood for a given parameter coincides with the
marginal distribution}.

\section{Credible regions}
\label{sec:bayes.cr}

(Marginal) posterior pdf are the general outcome of inference analysis based
on Bayesian reasoning.  It is common to summarize the properties of these
distributions using for example the {\it mode} (the values of parameters where
the pdf is maximum) or the mean value and dispersion of the parameters as
obtained from the posterior using the classical prescriptions.

It is also common to construct {\it credible regions} (CR), defined as the
interval(region) where the parameter value(s) is found with a probability
$CR$,

\beq
\int_{CR}{d\theta\,p(\theta|D)}\equiv CR
\label{eq:credible.region}
\eeq

The credible regions are the analog of the {\it confidence intervals}
in {\it frequentist} statistics that however are computed and
interpreted in a very different way.

\chapter{SUNG: SUpernova Neutrino Generation tool}
\label{ap:sung}

In chapter~\ref{ch:signal} we describe how the properties of a
supernova signal are computed starting with the characteristics of the
emitted neutrino signal as provided by self-consistent simulation of
the core collapse and using the proper description of neutrinos
oscillations in the supernova mantle and the Earth interior and
simulating the neutrino detection process.  Numerically this procedure
involves a particular manipulation of the supernova simulation results
that are used as the inputs of the general description of the signal
and the development of specially designed routines to model neutrino
oscillations under general circumstances in the supernova mantle and
the Earth interior.

Other numerical challenges arise when the purpose is to generate full
statistics signals.  In this case it is necessary on one hand to use
especialized algorithms and techniques to sample the complex observed
rate but in the other hand to design mechanisms to perform a
cross-checking to the generated signal to ensure that its events
already follows the original distribution.

Using the experience and the computer codes created in the frame of
this project we have created a computer package, SUNG (SUpernova
Neutrino Generation tool) intended to offer a simple and acsequible
computer solution to reproduce the results published in this work and
to perform similar aproaches to the analysis of supernova neutrino
signals.

For new or experiencied researchers in the field SUNG is a solution to
fastly obtain results that can require non-trivial programming
efforts.  Certainly it is not an ultimate solution for the computation
and generation of supernova neutrino signals under rather arbitrary
conditions.  The main philosophy behind this effort is that of promote
the creation of computational solutions in the field of neutrino
astrophysics as has been done in other fields (e.g. CMBFAST for CMBR
simulation and analysis~\cite{Seljak:1996is}).

In this appendix we describe the general features of SUNG and some of
the methods and algorithms used to perform the most important tasks
that the package can perform.

For a more complete description of the package including simple
reference manuals for the final user and the programmer please refer
to the web site {\tt http://www.sungweb.tk}.

\section{Package structure}
\label{sec:sung.structure}

SUNG is more than a simple single program to generate supernova
neutrino signals.  It is also a especialized programming framework
that can be used for a wide variety of different applications.

SUNG has three main components:

\begin{itemize}

\item A library of ({\tt ANSI C}) routines for data analysis originally
  designed or adapted from other libraries, and other routines created
  to simulate neutrino oscillations in the Supernova mantle and inside
  the Earth.

  The routines has been designed and organized to offer a coherent
  programming framework.  Even the especialized numerical routines of
  other scientific libraries are inserted into routines especially
  formated for this library~\cite{Zuluaga:SUNGWEB2004}.  

  In the core of the numerical integration, special functions,
  interpolation and random number generation routines SUNG uses
  routines of the GNU scientific library (GSL)\cite{GSL}.  Function
  minimization routines used in fit procedures use the MINUIT package
  routines from the CERN library \cite{CERNLIB}.  

  The SUNG library also includes a series of routines to generate data
  plots in eps and ps format using the plotting package GNUPLOT
  \cite{GNUPLOT}.

  In this sense the SUNG library can be seen as a special purpose
  common interface to those scientific libraries and packages.

\item A set of {\tt C} programs, written down using the routines in
  the SUNG library and {\tt perl/tcl} scripts designed to perform
  mainly four tasks: manipulate supernova simulation results, compute
  the properties of supernova neutrino signals under different input
  conditions, generate full-statistics realizations of those signals
  and analyze high-statistics supernova neutrino signals using a
  suitable analytical model.

  All the programs are designed to be ran from the command-line.  That
  programs are the best example on how complex tasks can be programmed
  using the routines of the package library.

\item The package have a special Graphical User Interface ({\tt GUI}) which
  has been designed to run on a web browser.  This {\em\tt webGUI}
  consist of an elaborated set of php scripts and works almost as a
  regular off-line GUI but with the advantage of not depend on any
  particular GUI library.  Once installed in a server (with an Apache
  and Php server) it can be run locally or remotely on a platform
  independent basis.

  A running version of the SUNG web interface can be found in the web
  site of the project {\tt http://www.sungweb.tk}.

\end{itemize}

\section{Methods and Algorithms}
\label{sec:sung.algorithms}

In the following paragraphs we describe the methods and algorithms
used by the package to perform the most important general tasks for
which it was designed.  

It is important to recall that this work was performed using the same
computer codes of the package and therefore the algorithms described
below are the same ones involved in the production of the results
presented here.

\subsection{Manipulation of Supernova simulations}
\label{subsec:sung.manipulation}

In order to describe the properties of an observed supernova neutrino
signal it is necessary to know the properties of the emitted
neutrinos.

There are three main pieces of information that describe almost
completely the neutrino emission from a supernova at a given time: the
luminosity, the average energy and the non-thermal distorsion
(quantified with an $\alpha$-parameter for example).  In the ideal
case the non-thermal distortions information is replaced by the
detailed knowledge of the spectral shape at each time.

Most of the supernova simulations results produce as an output that
informations and in our case they are the input of the computation of
the observed signal properties.

In the manipulations of results from supernova simulations several
problems must be faced.  One of the most common problems arise when a
given simulation result is limited to a shorter interval than the
total assumed duration of the signal (20 sec in this work).  This is
the case for Supernova model II \cite{Buras:Private2004} where
neutrino emission properties are just reliable for times less than 300
msec.

The routines included in the package are not prepared to face this
kind of limitations and an external extrapolation of the results is
required (see sec.~\ref{sec:results.tests}).

All the data files describing the evolution and the luminosity of the
spectral properties are treated by the SUNG library as continuous
functions using linear interpolation.

\subsection{Expected signal properties}
\label{subsec:sung.sigproperties}

The signal properties computed by SUNG are the same described in
sec.~\ref{sec:sign.char}:

\begin{itemize}

\item {\bf Number of events}.  This property is computed using the
definition (\ref{eq:signal.number}):

\beq
N(\Delta E_{12},\Delta
t_{12})=\int_{E_1}^{E_2}{dE\int_{t_1}^{t_2}{dt\,\frac{d^2n\,(E,t)}{dE\,dt}}}.
\eeq

Where the detected signal rate is given by (\ref{eq:total.rate}) and
(\ref{eq:detected.flux}):

\beqa
\nonumber
\frac{d^2n_\anue(E,t)}{dE\,dt} & = & N_T\, \int{dE'\,
S_{\barr{e}}^\rm{det}(E',t)\, \s(E')\, \epsilon(E')\, {\cal R}(E,E')}\,,\\
\nonumber
L^2 S^\rm{det}_{\barr{e}}(E,t) & = &\barr{p}\,S_{\barr{e}} +
(1-\barr{p})\,S_{\barr{x}}.
\eeqa

The three dimensional integration involved in the computation of $N$
is performed using a recursive procedure \cite{NRC:1992} supported by
one-dimensional adaptative Gaussian-Kronrod integrations.  

\item {\bf Time-profiles}.  There are two kind of time-profiles
computed by SUNG. The energy integrated time-profile as defined in
sec.~\ref{subsec:energy.integrated.profile} also called in the package
context the {\it marginal time-profile}
(\ref{eq:energy.int.timeprofile}):

\beq
\nonumber
f_t^\rm{det}(t)=\int{dE\,\frac{d^2n\,(E,t)}{dE\,dt}}.
\eeq

The other profile is given by the histogram that describes the number
of events inside narrow time windows of regular width $\Delta t$.
This histogram is given by:

\beq
\nonumber
h^t_i=\int_{t_i-\Delta t/2}^{t_i+\Delta t/2}{dt\,\int{dE\,\frac{d^2n\,(E,t)}{dE\,dt}}}.
\eeq

\item {\bf Energy-profiles}.  The energy-profiles (time integrated
  spectrum and energy-profile histogram) are computed similarly using
  the definitions given in sec.~\ref{subsec:time.integrated.spectrum}.

\end{itemize}

\subsection{Signal generation}
\label{subsec:sung.generation}

The generation of full statistics signals is the central problem
solved by SUNG.

As explained in sec.~\ref{sec:sign.generation} we can conceive several
alternative algorithms to generate the events in a supernova neutrino
signal including all the effects of the neutrino propagation,
oscillation and detection.  

SUNG uses the approach of generate the signal by sampling directly the
detected rate.

The algorithm used by SUNG to perform this task is described in the
following lines:

\begin{quotation}

\medskip
{\bf Signal generation Algorithm}
\medskip

\begin{enumerate}

\item Compute the number of events that different neutrinos emitted as
  a given flavor will produce in the detector:

\beqa
\nonumber
N_{\anue\rightarrow\anue} & = & \int{dE\,\int{dt\, N_T
    \barr{p}\,\frac{S_{\barr{e}}}{L^2}}}\\
\nonumber
N_{\anux\rightarrow\anue} & = & \int{dE\,\int{dt\, N_T
    (1-\barr{p})\,\frac{S_{\barr{x}}}{L^2}}}
\eeqa

 where $N_{\anue\rightarrow\anue}$ ($N_{\anux\rightarrow\anue}$) are the
 number of $\anue$ events produced by neutrinos emitted from the
 supernova core as $\anue$ ($\anux$).

\item Reshuffle the total number of neutrinos computed in the previous
  step according to a Poisson distribution with average equal to
  $N_\rm{tot}=N_{\anue\rightarrow\anue}+N_{\anux\rightarrow\anue}$.
  Acoording to the reshuffled value ${N'}_\rm{tot}$ compute the finally detected
  number of neutrinos of each flavor $N'_{\anue,\anux\rightarrow\anue}$
  obeying the original proportions.

\item Generate $N'_{\anue\rightarrow\anue}$
  ($N'_{\anux\rightarrow\anue}$) pairs of neutrino energies and times
  $\{E_i,t_i\}$ according to the respective neutrino detected flux
  (\ref{eq:emitted.signal}), $\barr{p}\,S_{\barr{e}}/L^2$
  ($(1-\barr{p})\,S_{\barr{x}}/L^2$) with $S_{\barr{e},\barr{x}}$
  given by (\ref{eq:emitted.signal}):

\beq
\nonumber
S_\a(E,t)=\frac{L_\a(t)}{\barr{E_\a}(t)}\,F^\rm{em}_\a(E;t).
\eeq

\item Shift the neutrino time according to the mass induced delay
(\ref{eq:tof.delay}):

\beq
\nonumber
t_i=t_i+\Delta t_\rm{tof} (\msq,L,E_i)
\eeq

\item Reset the neutrino times in order to place the first neutrino
  detected at $t_1=0$.

\item Compute the enegry of the secondary produced by the detection of
  each neutrino: $E^\rm{p}_i=E_i-Q_{pn}$.

\item Reshuffle the positron energy according to the energy
  measurement uncertainty (\ref{eq:Euncertainty}):

\beq
\nonumber
{E'}^\rm{p}_i\rightarrow \cal{G}(\barr{x}=E^\rm{p}_i,\sigma=\Delta E)
\eeq

\end{enumerate}

\end{quotation}

The result of this procedure is the set $\{{E'}^\rm{p}_i,t_i\}$ with
$i=1,\ldots,{N'}_\rm{tot}$.

\chapter{List of symbols and abbreviations}
\label{ap:abbrev}

\section{List of symbols}
\label{sec:app.symbols.}

\setlength{\tabcolsep}{1mm}
\begin{center}
\tablehead{%
\hline\hline
\bf Symbol & \bf Definition \\\hline\hline}
\tabletail{\hline\hline}
\begin{supertabular}{l>{}p{12cm}}
$m_\nu$ & Absolute scale of neutrino masses.  In the case of
  non-degenerate masses it refers to the electron neutrino mass in the
  context of this work.\\ \hline\blankline{2}
$\nux$ & Non-electron neutrinos and antineutrinos $\equiv\nu_{\mu,\tau},
  \barr{\nu}_{\mu,\tau}$.\\ \hline\blankline{2}
$L$ & Supernova distance.\\ \hline\blankline{2}
$R_\rm{NS}, M_\rm{NS}$ & Radius and mass of a neutron star.\\
\hline\blankline{2}
$\Delta t_\rm{tof}$ & Mass induced time-of-flight delay.\\
\hline\blankline{2}
$dn^\rm{em}/dt$ & Rate of neutrino emission, i.e. number of neutrinos
of any energy emitted per unit of time.\\ \hline\blankline{2}
$F(E)$ & Neutrino spectrum at the source.\\ \hline\blankline{2}
$p$ & Pinching parameter, $p\propto\aver{E^2}/\aver{E}$.\\
\hline\blankline{2}
$\vvec{\nu_W}$ & Neutrino flavor eigenstates.\\ \hline\blankline{2}
$\vvec{\nu_M}$ ($\,\vvec{\nu_M^m}\,$) & Neutrino mass eigenstates in
vacuum (in matter).\\ \hline\blankline{2}
$U_{\a i}$ & entries of the MNS matrix.\\
\hline\blankline{2}
$\mu_{i}$ & In matter effective neutrino masses.\\
\hline\blankline{2}
$\gamma$ & Adiabaticity parameter.\\
\hline\blankline{2}
$P_{H,L}$ ($\barr{P}_{H,L}$) & Supernova mantle jumping
probabilities.\\ \hline\blankline{2}
$p_{\a i}$ ($\barr{p}_{\a i}$) & Supernova mantle conversion
probabilities $\equiv P(\nu_\a\rightarrow\nu_i)$.\\
\hline\blankline{2}
$P_{\a\b}$ ($P_{\barr{\a}\barr{\b}}$) & Conversion probabilities $\equiv
P(\nu_\a\rightarrow\nu_\b)$.\\ \hline\blankline{2}
$p_{\oplus}$ ($\barr{p}_{\oplus}$) & Conversion probabilities in the
Earth.\\ \hline\blankline{2}
$S_\a(E,t)$ & Total flux of neutrino flavor $\a$ at the source,
i.e. number of neutrinos $\a$ emitted per unit of time and per unit of
energy.\\ \hline\blankline{2}
$S_\a^\rm{det}(E,t)$ & Total flux of neutrino flavor $\a$ arriving to
the detector, i.e. number of neutrinos $\a$ arriving per unit of time
and per unit of energy.\\ \hline\blankline{2}
$\barr{E}_\a$ & Average energy of neutrinos $\a$ at the source.\\
\hline\blankline{2}
$d^2n_\a/dE\,dt$ & Total rate of neutrino flavor $\a$ in the detector,
i.e. number of neutrinos $\a$ detected per unit of time and per unit
of energy.\\ \hline\blankline{2}
$f_E(E)$ & Time-integrated energy spectrum.\\
\hline\blankline{2}
$f_t(t)$ & Energy-integrated time profile.\\
\hline\blankline{2}
$\M$ & A supernova emission model.  If the model can be parametrized
  with a set of parameters $\{\theta\}$ the model is represented by
  $\M(\t)$.\\ \hline\blankline{2}
$\D$ & Data set, i.e. set of energy and time pairs $\{E_i,t_i\}$.\\
  \hline\blankline{2}
$f(E,t)$ & Probability distribution function used to describe the
  signal (model density probability).\\ \hline\blankline{2}
$\cal L$ & Likelihood function.\\ \hline\blankline{2}
$\phi(t)$ & Flux model.\\ \hline\blankline{2}
${\cal G(x;\mu,\s)}$ & Normal distribution with mean $\mu$ and
  standard deviation $\s$.\\ \hline\blankline{2}
$\tilde{f}(E,t)$ & Regularized model density probability.\\
\hline\blankline{2}
$\hat{\cal L}$ & Profile Likelihood.\\
\end{supertabular}
\end{center}

\newpage

\section{List of abbreviations}
\label{sec:app.abbrev.}

\begin{center}
\tablehead{%
\hline\hline
\bf Abrev. & \bf Definition \\\hline\hline
\blankline{2}}
\tabletail{\hline\hline}
\begin{supertabular}{ll}
SN & Supernova/Supernovae.\\ \hline\blankline{2}
TOF & Time-of-flight.\\ \hline\blankline{2}
LTE & Local Thermodynamic Equilibrium.\\ \hline\blankline{2}
MNS & Maki-Nakagawa-Sakata matrix.\\ \hline\blankline{2}
NH & Normal hierarchy.\\ \hline\blankline{2}
IH & Inverted hierarchy.\\ \hline\blankline{2}
LF & Likelihood function.\\ \hline\blankline{2}
log-LF & Logarithm of Likelihood Function.\\ \hline\blankline{2}
ML & Maximum Likelihood analysis.\\ \hline\blankline{2}
MC & Monte Carlo.\\ \hline\blankline{2}
CR & Credible regions (credible region probability).\\
\hline\blankline{2}
PL & Profile likelihood.\\ \hline\blankline{2}
SK & Super-Kamiokande.\\ \hline\blankline{2}
KL & KamLAND.\\ \hline\blankline{2}
HK & Hyper-Kamiokande.\\
\end{supertabular}
\end{center}



\bibliographystyle{Biblio/utphys} %

\bibliography{jzphd-thesis}

\providecommand{\href}[2]{#2}\begingroup\raggedright\begin{thebibliography}{10%
0}

\bibitem{Gamow:1940aa}
G.~{Gamow} and M.~Schoenberg, ``The possible role of neutrinos in stellar
  evolution,'' {\em Phys.\ Rev.} {\bf 58} (1940) 1117.

\bibitem{1935MNRAS..95..207C}
S.~{Chandrasekhar}, ``The highly collapsed configurations of a stellar mass
  (second paper),'' {\em \mnras} {\bf 95} (Jan., 1935) 207--225.

\bibitem{Raffelt:2003en}
G.~G. Raffelt, M.~T. Keil, R.~Buras, H.-T. Janka, and M.~Rampp, ``Supernova
  neutrinos: Flavor-dependent fluxes and spectra,''
\href{http://www.arXiv.org/abs/astro-ph/0303226}{{\tt astro-ph/0303226}}.

\bibitem{Janka:2002ei}
H.-T. Janka, R.~Buras, K.~Kifonidis, T.~Plewa, and M.~Rampp, ``Core collapse
  and then? the route to massive star explosions,''
\href{http://www.arXiv.org/abs/astro-ph/0212316}{{\tt astro-ph/0212316}}.

\bibitem{Buras:2003sn}
R.~Buras, M.~Rampp, H.~T. Janka, and K.~Kifonidis, ``Improved models of stellar
  core collapse and still no explosions: What is missing?,'' {\em Phys. Rev.
  Lett.} {\bf 90} (2003) 241101,
\href{http://www.arXiv.org/abs/astro-ph/0303171}{{\tt astro-ph/0303171}}.

\bibitem{Janka:2004jb}
H.-T. Janka, R.~Buras, K.~Kifonidis, A.~Marek, and M.~Rampp, ``Core-collapse
  supernovae at the threshold,''
\href{http://www.arXiv.org/abs/astro-ph/0401461}{{\tt astro-ph/0401461}}.

\bibitem{Bethe:1984ux}
H.~A. Bethe and R.~Wilson, James, ``Revival of a stalled supernova shock by
  neutrino heating,'' {\em Astrophys. J.} {\bf 295} (1985)
14--23.

\bibitem{Fukuda:1998mi}
{\bf Super-Kamiokande} Collaboration, Y.~Fukuda {\em et al.}, ``Evidence for
  oscillation of atmospheric neutrinos,'' {\em Phys. Rev. Lett.} {\bf 81}
  (1998) 1562--1567,
\href{http://www.arXiv.org/abs/hep-ex/9807003}{{\tt hep-ex/9807003}}.

\bibitem{Fukuda:1998ah}
{\bf Super-Kamiokande} Collaboration, Y.~Fukuda {\em et al.}, ``Measurement of
  the flux and zenith-angle distribution of upward through-going muons by
  super-kamiokande,'' {\em Phys. Rev. Lett.} {\bf 82} (1999) 2644--2648,
\href{http://www.arXiv.org/abs/hep-ex/9812014}{{\tt hep-ex/9812014}}.

\bibitem{Ambrosio:1998wu}
{\bf MACRO} Collaboration, M.~Ambrosio {\em et al.}, ``Measurement of the
  atmospheric neutrino-induced upgoing muon flux using macro,'' {\em Phys.
  Lett.} {\bf B434} (1998) 451--457,
\href{http://www.arXiv.org/abs/hep-ex/9807005}{{\tt hep-ex/9807005}}.

\bibitem{Ambrosio:2000ja}
{\bf MACRO} Collaboration, M.~Ambrosio {\em et al.}, ``Low energy atmospheric
  muon neutrinos in macro,'' {\em Phys. Lett.} {\bf B478} (2000) 5--13,
\href{http://www.arXiv.org/abs/hep-ex/0001044}{{\tt hep-ex/0001044}}.

\bibitem{Sanchez:2003rb}
{\bf Soudan 2} Collaboration, M.~Sanchez {\em et al.}, ``Observation of
  atmospheric neutrino oscillations in soudan 2,'' {\em Phys. Rev.} {\bf D68}
  (2003) 113004,
\href{http://www.arXiv.org/abs/hep-ex/0307069}{{\tt hep-ex/0307069}}.

\bibitem{Ahmad:2001an}
{\bf SNO} Collaboration, Q.~R. Ahmad {\em et al.}, ``Measurement of the charged
  current interactions produced by b-8 solar neutrinos at the sudbury neutrino
  observatory,'' {\em Phys. Rev. Lett.} {\bf 87} (2001) 071301,
\href{http://www.arXiv.org/abs/nucl-ex/0106015}{{\tt nucl-ex/0106015}}.

\bibitem{Fukuda:1996sz}
{\bf Kamiokande} Collaboration, Y.~Fukuda {\em et al.}, ``Solar neutrino data
  covering solar cycle 22,'' {\em Phys. Rev. Lett.} {\bf 77} (1996)
1683--1686.

\bibitem{Cleveland:1998nv}
B.~T. Cleveland {\em et al.}, ``Measurement of the solar electron neutrino flux
  with the homestake chlorine detector,'' {\em Astrophys. J.} {\bf 496} (1998)
505--526.

\bibitem{Hampel:1998xg}
{\bf GALLEX} Collaboration, W.~Hampel {\em et al.}, ``Gallex solar neutrino
  observations: Results for gallex iv,'' {\em Phys. Lett.} {\bf B447} (1999)
127--133.

\bibitem{Altmann:2000ft}
{\bf GNO} Collaboration, M.~Altmann {\em et al.}, ``Gno solar neutrino
  observations: Results for gno i,'' {\em Phys. Lett.} {\bf B490} (2000)
  16--26,
\href{http://www.arXiv.org/abs/hep-ex/0006034}{{\tt hep-ex/0006034}}.

\bibitem{Fukuda:2001nj}
{\bf Super-Kamiokande} Collaboration, S.~Fukuda {\em et al.}, ``Solar b-8 and
  he p neutrino measurements from 1258 days of super-kamiokande data,'' {\em
  Phys. Rev. Lett.} {\bf 86} (2001) 5651--5655,
\href{http://www.arXiv.org/abs/hep-ex/0103032}{{\tt hep-ex/0103032}}.

\bibitem{Abdurashitov:2002nt}
{\bf SAGE} Collaboration, J.~N. Abdurashitov {\em et al.}, ``Measurement of the
  solar neutrino capture rate by the russian-american gallium solar neutrino
  experiment during one half of the 22-year cycle of solar activity,'' {\em J.
  Exp. Theor. Phys.} {\bf 95} (2002) 181--193,
\href{http://www.arXiv.org/abs/astro-ph/0204245}{{\tt astro-ph/0204245}}.

\bibitem{Smy:2003jf}
{\bf Super-Kamiokande} Collaboration, M.~B. Smy {\em et al.}, ``Precise
  measurement of the solar neutrino day/night and seasonal variation in
  super-kamiokande-i,'' {\em Phys. Rev.} {\bf D69} (2004) 011104,
\href{http://www.arXiv.org/abs/hep-ex/0309011}{{\tt hep-ex/0309011}}.

\bibitem{Ahmed:2003kj}
{\bf SNO} Collaboration, S.~N. Ahmed {\em et al.}, ``Measurement of the total
  active b-8 solar neutrino flux at the sudbury neutrino observatory with
  enhanced neutral current sensitivity,'' {\em Phys. Rev. Lett.} {\bf 92}
  (2004) 181301,
\href{http://www.arXiv.org/abs/nucl-ex/0309004}{{\tt nucl-ex/0309004}}.

\bibitem{Eguchi:2002dm}
{\bf KamLAND} Collaboration, K.~Eguchi {\em et al.}, ``First results from
  kamland: Evidence for reactor anti- neutrino disappearance,'' {\em Phys. Rev.
  Lett.} {\bf 90} (2003) 021802,
\href{http://www.arXiv.org/abs/hep-ex/0212021}{{\tt hep-ex/0212021}}.

\bibitem{Araki:2004mb}
{\bf KamLAND} Collaboration, T.~Araki {\em et al.}, ``Measurement of neutrino
  oscillation with kamland: Evidence of spectral distortion,''
\href{http://www.arXiv.org/abs/hep-ex/0406035}{{\tt hep-ex/0406035}}.

\bibitem{Oyama:1998bd}
{\bf K2K} Collaboration, Y.~Oyama, ``K2k (kek to kamioka) neutrino oscillation
  experiment at kek-ps,''
\href{http://www.arXiv.org/abs/hep-ex/9803014}{{\tt hep-ex/9803014}}.

\bibitem{Paes:2001nd}
H.~Paes and T.~J. Weiler, ``Absolute neutrino mass determination,'' {\em Phys.
  Rev.} {\bf D63} (2001) 113015,
\href{http://www.arXiv.org/abs/hep-ph/0101091}{{\tt hep-ph/0101091}}.

\bibitem{Bilenky:2002aw}
S.~M. Bilenky, C.~Giunti, J.~A. Grifols, and E.~Masso, ``Absolute values of
  neutrino masses: Status and prospects,'' {\em Phys. Rept.} {\bf 379} (2003)
  69--148,
\href{http://www.arXiv.org/abs/hep-ph/0211462}{{\tt hep-ph/0211462}}.

\bibitem{Bonn:2002jw}
J.~Bonn {\em et al.}, ``Results from the mainz neutrino mass experiment,'' {\em
  Prog. Part. Nucl. Phys.} {\bf 48} (2002)
133--139.

\bibitem{Lobashev:2001uu}
V.~M. Lobashev {\em et al.}, ``Direct search for neutrino mass and anomaly in
  the tritium beta-spectrum: Status of \'troitsk neutrino mass\' experiment,''
  {\em Nucl. Phys. Proc. Suppl.} {\bf 91} (2001)
280--286.

\bibitem{Klapdor-Kleingrothaus:2000sn}
H.~V. Klapdor-Kleingrothaus {\em et al.}, ``Latest results from the
  heidelberg-moscow double-beta-decay experiment,'' {\em Eur. Phys. J.} {\bf
  A12} (2001) 147--154,
\href{http://www.arXiv.org/abs/hep-ph/0103062}{{\tt hep-ph/0103062}}.

\bibitem{Klapdor-Kleingrothaus:2004wj}
H.~V. Klapdor-Kleingrothaus, I.~V. Krivosheina, A.~Dietz, and O.~Chkvorets,
  ``Search for neutrinoless double beta decay with enriched ge- 76 in gran
  sasso 1990-2003,'' {\em Phys. Lett.} {\bf B586} (2004) 198--212,
\href{http://www.arXiv.org/abs/hep-ph/0404088}{{\tt hep-ph/0404088}}.

\bibitem{Aalseth:2002rf}
{\bf 16EX} Collaboration, C.~E. Aalseth {\em et al.}, ``The igex ge-76
  neutrinoless double-beta decay experiment: Prospects for next generation
  experiments,'' {\em Phys. Rev.} {\bf D65} (2002) 092007,
\href{http://www.arXiv.org/abs/hep-ex/0202026}{{\tt hep-ex/0202026}}.

\bibitem{Hannestad:2004nb}
S.~Hannestad, ``Neutrinos in cosmology,'' {\em New J. Phys.} {\bf 6} (2004)
  108,
\href{http://www.arXiv.org/abs/hep-ph/0404239}{{\tt hep-ph/0404239}}.

\bibitem{Hannestad:2003xv}
S.~Hannestad, ``Neutrino masses and the number of neutrino species from wmap
  and 2dfgrs,'' {\em JCAP} {\bf 0305} (2003) 004,
\href{http://www.arXiv.org/abs/astro-ph/0303076}{{\tt astro-ph/0303076}}.

\bibitem{Elgaroy:2003yh}
O.~Elgaroy and O.~Lahav, ``The role of priors in deriving upper limits on
  neutrino masses from the 2dfgrs and wmap,'' {\em JCAP} {\bf 0304} (2003) 004,
\href{http://www.arXiv.org/abs/astro-ph/0303089}{{\tt astro-ph/0303089}}.

\bibitem{Crotty:2004gm}
P.~Crotty, J.~Lesgourgues, and S.~Pastor, ``Current cosmological bounds on
  neutrino masses and relativistic relics,'' {\em Phys. Rev.} {\bf D69} (2004)
  123007,
\href{http://www.arXiv.org/abs/hep-ph/0402049}{{\tt hep-ph/0402049}}.

\bibitem{Beacom:2004yd}
J.~F. Beacom, N.~F. Bell, and S.~Dodelson, ``Neutrinoless universe,'' {\em
  Phys. Rev. Lett.} {\bf 93} (2004) 121302,
\href{http://www.arXiv.org/abs/astro-ph/0404585}{{\tt astro-ph/0404585}}.

\bibitem{Zatsepin:1968aa}
G.~T. Zatsepin, ``Possibility of determining the upper limit of the neutrino
  mass from the time of flight,'' {\em JETP Lett.} {\bf 8} (1968) 205.

\bibitem{Pakvasa:1972gz}
S.~Pakvasa and K.~Tennakone, ``Neutrinos of non-zero rest mass,'' {\em Phys.
  Rev. Lett.} {\bf 28} (1972)
1415.

\bibitem{Piran:1981zz}
T.~Piran, ``Neutrino mass and detection of neutrino supernova bursts,'' {\em
  Phys. Lett.} {\bf B102} (1981)
299--302.

\bibitem{1982Ap&SS..81..483S}
Z.~F. {Seidov}, ``The supernova neutrino pulse shape in the scintillation
  detector,'' {\em \apss} {\bf 81} (Jan., 1982) 483--488.

\bibitem{Hirata:1988ad}
K.~S. Hirata {\em et al.}, ``Observation in the kamiokande-ii detector of the
  neutrino burst from supernova sn1987a,'' {\em Phys. Rev.} {\bf D38} (1988)
448--458.

\bibitem{VanDerVelde:1987hh}
{\bf IMB} Collaboration, J.~C. Van Der~Velde {\em et al.}, ``Neutrinos from
  sn1987a in the imb detector,'' {\em Nucl. Instrum. Meth.} {\bf A264} (1988)
28--31.

\bibitem{Alekseev:1988gp}
E.~N. Alekseev, L.~N. Alekseeva, I.~V. Krivosheina, and V.~I. Volchenko,
  ``Detection of the neutrino signal from sn1987a in the lmc using the inr
  baksan underground scintillation telescope,'' {\em Phys. Lett.} {\bf B205}
  (1988)
209--214.

\bibitem{Schramm:1987ra}
D.~N. Schramm, ``Neutrinos from supernova sn1987a,'' {\em Comments Nucl. Part.
  Phys.} {\bf 17} (1987)
239.

\bibitem{Arnett:1987iz}
W.~D. Arnett and J.~L. Rosner, ``Neutrino mass limits from sn1987a,'' {\em
  Phys. Rev. Lett.} {\bf 58} (1987)
1906.

\bibitem{Bahcall:1987nx}
J.~N. Bahcall and S.~L. Glashow, ``Upper limit on the mass of the
  electron-neutrino,'' {\em Nature} {\bf 326} (1987)
476.

\bibitem{Spergel:1987ex}
D.~N. Spergel and J.~N. Bahcall, ``The mass of the electron-neutrino: Monte
  carlo studies of sn1987a observations,'' {\em Phys. Lett.} {\bf B200} (1988)
366.

\bibitem{Abbott:1987bm}
L.~F. Abbott, A.~De~Rujula, and T.~P. Walker, ``Constraints on the neutrino
  mass from the supernova data: A systematic analysis,'' {\em Nucl. Phys.} {\bf
  B299} (1988)
734.

\bibitem{Loredo:2001rx}
T.~J. Loredo and D.~Q. Lamb, ``Bayesian analysis of neutrinos observed from
  supernova sn 1987a,'' {\em Phys. Rev.} {\bf D65} (2002) 063002,
\href{http://www.arXiv.org/abs/astro-ph/0107260}{{\tt astro-ph/0107260}}.

\bibitem{Fargion:1981gg}
D.~Fargion, ``Time delay between gravitational waves and neutrino burst from a
  supernova explosion: a test for the neutrino mass,'' {\em Nuovo Cim. Lett.}
  {\bf 31} (1981) 499--500,
\href{http://www.arXiv.org/abs/hep-ph/0110061}{{\tt hep-ph/0110061}}.

\bibitem{Arnaud:2001gt}
N.~Arnaud {\em et al.}, ``Gravity wave and neutrino bursts from stellar
  collapse: A sensitive test of neutrino masses,'' {\em Phys. Rev.} {\bf D65}
  (2002) 033010,
\href{http://www.arXiv.org/abs/hep-ph/0109027}{{\tt hep-ph/0109027}}.

\bibitem{Totani:1998nf}
T.~Totani, ``Electron neutrino mass measurement by supernova neutrino bursts
  and implications on hot dark matter,'' {\em Phys. Rev. Lett.} {\bf 80} (1998)
  2039--2042,
\href{http://www.arXiv.org/abs/astro-ph/9801104}{{\tt astro-ph/9801104}}.

\bibitem{Beacom:2000ng}
J.~F. Beacom, R.~N. Boyd, and A.~Mezzacappa, ``Technique for direct-ev scale
  measurements of the mu and tau neutrino masses using supernova neutrinos,''
  {\em Phys. Rev. Lett.} {\bf 85} (2000) 3568--3571,
\href{http://www.arXiv.org/abs/hep-ph/0006015}{{\tt hep-ph/0006015}}.

\bibitem{Beacom:2000qy}
J.~F. Beacom, R.~N. Boyd, and A.~Mezzacappa, ``Black hole formation in core
  collapse supernovae and time- of-flight measurements of the neutrino
  masses,'' {\em Phys. Rev.} {\bf D63} (2001) 073011,
\href{http://www.arXiv.org/abs/astro-ph/0010398}{{\tt astro-ph/0010398}}.

\bibitem{Mayle:1986ic}
R.~Mayle, J.~R. Wilson, and D.~N. Schramm, ``Neutrinos from gravitational
  collapse,'' {\em Astrophys. J.} {\bf 318} (1987)
288--306.

\bibitem{Burrows:1991kf}
A.~Burrows, D.~Klein, and R.~Gandhi, ``The future of supernova neutrino
  detection,'' {\em Phys. Rev.} {\bf D45} (1992)
3361--3385.

\bibitem{Keil:2003sw}
M.~T. Keil, {\em Supernova neutrino spectra and applications to flavor
  oscillations}.
\newblock PhD thesis, 2003.
\newblock
\href{http://www.arXiv.org/abs/astro-ph/0308228}{{\tt astro-ph/0308228}}.
\newblock

\bibitem{Spergel:1987ch}
D.~N. Spergel, T.~Piran, A.~Loeb, J.~Goodman, and J.~N. Bahcall, ``A simple
  model for neutrino cooling of the lmc supernova,'' {\em Science} {\bf 237}
  (1987)
1471.

\bibitem{Janka:1995cu}
H.~T. Janka, ``When do supernova neutrinos of different flavors have similar
  luminosities but different spectra?,'' {\em Astropart. Phys.} {\bf 3} (1995)
  377--384,
\href{http://www.arXiv.org/abs/astro-ph/9503068}{{\tt astro-ph/9503068}}.

\bibitem{Woosley:1994ux}
S.~E. Woosley, J.~R. Wilson, G.~J. Mathews, R.~D. Hoffman, and B.~S. Meyer,
  ``The r process and neutrino heated supernova ejecta,'' {\em Astrophys. J.}
  {\bf 433} (1994)
229--246.

\bibitem{Buras:Private2004}
R.~Buras, ``Private communication,''.

\bibitem{Raffelt:2001kv}
G.~G. Raffelt, ``Mu- and tau-neutrino spectra formation in supernovae,'' {\em
  Astrophys. J.} {\bf 561} (2001) 890--914,
\href{http://www.arXiv.org/abs/astro-ph/0105250}{{\tt astro-ph/0105250}}.

\bibitem{Keil:2002in}
M.~T. Keil, G.~G. Raffelt, and H.-T. Janka, ``Monte carlo study of supernova
  neutrino spectra formation,'' {\em Astrophys. J.} {\bf 590} (2003) 971--991,
\href{http://www.arXiv.org/abs/astro-ph/0208035}{{\tt astro-ph/0208035}}.

\bibitem{Raffelt:1996wa}
G.~G. Raffelt, {\em Stars as laboratories for fundamental physics: The
  astrophysics of neutrinos, axions, and other weakly interacting particles}.
\newblock Chicago, USA: Univ. Pr. (1996) 664 p.

\bibitem{Minakata:2000rx}
H.~Minakata and H.~Nunokawa, ``Inverted hierarchy of neutrino masses disfavored
  by supernova 1987a,'' {\em Phys. Lett.} {\bf B504} (2001) 301--308,
\href{http://www.arXiv.org/abs/hep-ph/0010240}{{\tt hep-ph/0010240}}.

\bibitem{Lunardini:2001pb}
C.~Lunardini and A.~Y. Smirnov, ``Supernova neutrinos: Earth matter effects and
  neutrino mass spectrum,'' {\em Nucl. Phys.} {\bf B616} (2001) 307--348,
\href{http://www.arXiv.org/abs/hep-ph/0106149}{{\tt hep-ph/0106149}}.

\bibitem{Dighe:1999bi}
A.~S. Dighe and A.~Y. Smirnov, ``Identifying the neutrino mass spectrum from
  the neutrino burst from a supernova,'' {\em Phys. Rev.} {\bf D62} (2000)
  033007,
\href{http://www.arXiv.org/abs/hep-ph/9907423}{{\tt hep-ph/9907423}}.

\bibitem{Aglietta:2001jf}
M.~Aglietta {\em et al.}, ``Effects of neutrino oscillations on the supernova
  signal in lvd,'' {\em Nucl. Phys. Proc. Suppl.} {\bf 110} (2002) 410--413,
\href{http://www.arXiv.org/abs/astro-ph/0112312}{{\tt astro-ph/0112312}}.

\bibitem{Takahashi:2001dc}
K.~Takahashi and K.~Sato, ``Earth effects on supernova neutrinos and their
  implications for neutrino parameters,'' {\em Phys. Rev.} {\bf D66} (2002)
  033006,
\href{http://www.arXiv.org/abs/hep-ph/0110105}{{\tt hep-ph/0110105}}.

\bibitem{Lunardini:2003eh}
C.~Lunardini and A.~Y. Smirnov, ``Probing the neutrino mass hierarchy and the
  13-mixing with supernovae,'' {\em JCAP} {\bf 0306} (2003) 009,
\href{http://www.arXiv.org/abs/hep-ph/0302033}{{\tt hep-ph/0302033}}.

\bibitem{Dighe:2003jg}
A.~S. Dighe, M.~T. Keil, and G.~G. Raffelt, ``Identifying earth matter effects
  on supernova neutrinos at a single detector,'' {\em JCAP} {\bf 0306} (2003)
  006,
\href{http://www.arXiv.org/abs/hep-ph/0304150}{{\tt hep-ph/0304150}}.

\bibitem{Maki:1962mu}
Z.~Maki, M.~Nakagawa, and S.~Sakata, ``Remarks on the unified model of
  elementary particles,'' {\em Prog. Theor. Phys.} {\bf 28} (1962)
870.

\bibitem{Eidelman:2004wy}
{\bf Particle Data Group} Collaboration, S.~Eidelman {\em et al.}, ``Review of
  particle physics,'' {\em Phys. Lett.} {\bf B592} (2004)
1.

\bibitem{Kuo:1989qe}
T.~K. Kuo and J.~Pantaleone, ``Neutrino oscillations in matter,'' {\em Rev.
  Mod. Phys.} {\bf 61} (1989)
937.

\bibitem{Fogli:2001pm}
G.~L. Fogli, E.~Lisi, D.~Montanino, and A.~Palazzo, ``Supernova neutrino
  oscillations: A simple analytical approach,'' {\em Phys. Rev.} {\bf D65}
  (2002) 073008,
\href{http://www.arXiv.org/abs/hep-ph/0111199}{{\tt hep-ph/0111199}}.

\bibitem{Wolfenstein:1977ue}
L.~Wolfenstein, ``Neutrino oscillations in matter,'' {\em Phys. Rev.} {\bf D17}
  (1978)
2369.

\bibitem{Mikheev:1986wj}
S.~P. Mikheev and A.~Y. Smirnov, ``Resonant amplification of neutrino
  oscillations in matter and solar neutrino spectroscopy,'' {\em Nuovo Cim.}
  {\bf C9} (1986)
17--26.

\bibitem{Bahcall:2004ut}
J.~N. Bahcall, M.~C. Gonzalez-Garcia, and C.~Pena-Garay, ``Solar neutrinos
  before and after neutrino 2004,'' {\em JHEP} {\bf 08} (2004) 016,
\href{http://www.arXiv.org/abs/hep-ph/0406294}{{\tt hep-ph/0406294}}.

\bibitem{Kuo:1987qu}
T.~K. Kuo and J.~Pantaleone, ``Supernova neutrinos and their oscillations,''
  {\em Phys. Rev.} {\bf D37} (1988)
298.

\bibitem{Mikheev:1988in}
S.~P. Mikheev and A.~Y. Smirnov, ``3 nu oscillations in matter and solar
  neutrino data,'' {\em Phys. Lett.} {\bf B200} (1988)
560--564.

\bibitem{Akhmedov:1998ui}
E.~K. Akhmedov, ``Parametric resonance of neutrino oscillations and passage of
  solar and atmospheric neutrinos through the earth,'' {\em Nucl. Phys.} {\bf
  B538} (1999) 25--51,
\href{http://www.arXiv.org/abs/hep-ph/9805272}{{\tt hep-ph/9805272}}.

\bibitem{Strumia:2003zx}
A.~Strumia and F.~Vissani, ``Precise quasielastic neutrino nucleon cross
  section,'' {\em Phys. Lett.} {\bf B564} (2003) 42--54,
\href{http://www.arXiv.org/abs/astro-ph/0302055}{{\tt astro-ph/0302055}}.

\bibitem{Nardi:2003pr}
E.~Nardi and J.~I. Zuluaga, ``Exploring the sub-ev neutrino mass range with
  supernova neutrinos,'' {\em Phys. Rev.} {\bf D69} (2004) 103002,
\href{http://www.arXiv.org/abs/astro-ph/0306384}{{\tt astro-ph/0306384}}.

\bibitem{Totani:1997vj}
T.~Totani, K.~Sato, H.~E. Dalhed, and J.~R. Wilson, ``Future detection of
  supernova neutrino burst and explosion mechanism,'' {\em Astrophys. J.} {\bf
  496} (1998) 216--225,
\href{http://www.arXiv.org/abs/astro-ph/9710203}{{\tt astro-ph/9710203}}.

\bibitem{Nardi:2004zg}
E.~Nardi and J.~I. Zuluaga, ``Constraints on neutrino masses from a galactic
  supernova neutrino signal at present and future detectors,'' {\em Nucl.
  Phys.} {\bf B731} (2005) 140--163,
\href{http://www.arXiv.org/abs/hep-ph/0412104}{{\tt hep-ph/0412104}}.

\bibitem{Mayle:1993uj}
R.~W. Mayle, J.~R. Wilson, and M.~Tavani, ``Pions, supernovae, and the
  supranuclear matter density equation of state,'' {\em Astrophys. J.} {\bf
  418} (1993)
398--404.

\bibitem{1989A&A...224...49J}
H.-T. {Janka} and W.~{Hillebrandt}, ``{Neutrino emission from type II
  supernovae - an analysis of the spectra},'' {\em \aap} {\bf 224} (Oct., 1989)
  49--56.

\bibitem{Rampp:2000a}
M.~Rampp, {\em Radiation Hydrodynamics with Neutrinos: Stellar Core Collapse
  and the Explosion Mechanism of Type II Supernovae}.
\newblock PhD thesis, 2000.

\bibitem{Rampp:2000ws}
M.~Rampp and H.~T. Janka, ``Spherically symmetric simulation with boltzmann
  neutrino transport of core collapse and post-bounce evolution of a 15 solar
  mass star,'' {\em Astrophys. J.} {\bf 539} (2000) L33--L36,
\href{http://www.arXiv.org/abs/astro-ph/0005438}{{\tt astro-ph/0005438}}.

\bibitem{Rampp:2002bq}
M.~Rampp and H.~T. Janka, ``Radiation hydrodynamics with neutrinos: Variable
  eddington factor method for core-collapse supernova simulations,'' {\em
  Astron. Astrophys.} {\bf 396} (2002) 361,
\href{http://www.arXiv.org/abs/astro-ph/0203101}{{\tt astro-ph/0203101}}.

\bibitem{Buras:2002wt}
R.~Buras, H.-T. Janka, M.~T. Keil, G.~G. Raffelt, and M.~Rampp,
  ``Electron-neutrino pair annihilation: A new source for muon and tau
  neutrinos in supernovae,'' {\em Astrophys. J.} {\bf 587} (2003) 320--326,
\href{http://www.arXiv.org/abs/astro-ph/0205006}{{\tt astro-ph/0205006}}.

\bibitem{Nakamura:2003hk}
K.~Nakamura, ``Hyper-kamiokande: A next generation water cherenkov detector,''
  {\em Int. J. Mod. Phys.} {\bf A18} (2003)
4053--4063.

\bibitem{Oberauer:2004ji}
L.~Oberauer, ``Low energy neutrino physics after sno and kamland,'' {\em Mod.
  Phys. Lett.} {\bf A19} (2004) 337--348,
\href{http://www.arXiv.org/abs/hep-ph/0402162}{{\tt hep-ph/0402162}}.

\bibitem{Osipowicz:2001sq}
{\bf KATRIN} Collaboration, A.~Osipowicz {\em et al.}, ``Katrin: A next
  generation tritium beta decay experiment with sub-ev sensitivity for the
  electron neutrino mass,''
\href{http://www.arXiv.org/abs/hep-ex/0109033}{{\tt hep-ex/0109033}}.

\bibitem{Weinheimer:2002jx}
{\bf KATRIN} Collaboration, C.~Weinheimer, ``Katrin, a next generation tritium
  beta decay experiment in search for the absolute neutrino mass scale,'' {\em
  Prog. Part. Nucl. Phys.} {\bf 48} (2002)
141--150.

\bibitem{Cremonesi:2002is}
O.~Cremonesi, ``Neutrinoless double beta decay: Present and future,'' {\em
  Nucl. Phys. Proc. Suppl.} {\bf 118} (2003) 287--296,
\href{http://www.arXiv.org/abs/hep-ex/0210007}{{\tt hep-ex/0210007}}.

\bibitem{DAgostini:2003qr}
G.~D'Agostini, ``Bayesian inference in processing experimental data: Principles
  and basic applications,'' {\em Rept. Prog. Phys.} {\bf 66} (2003) 1383--1420,
\href{http://www.arXiv.org/abs/physics/0304102}{{\tt physics/0304102}}.

\bibitem{Sivia:1998}
D.~Sivia, {\em Data analysis: A Bayesian tutorial}.
\newblock Oxford, UK: Oxford Science Publications (1998).

\bibitem{Cox:1946}
R.~Cox, ``Probability, frequency and reasonable expectation,'' {\em American
  Journal of Physics} {\bf 14} (1946) 1--13.

\bibitem{Dose:2002}
V.~Dose, ``Bayes in five days.''
\newblock CIPS, MPI für Plasmaphysik, Garching, Germany, Reprint 83, May 2002.

\bibitem{Seljak:1996is}
U.~Seljak and M.~Zaldarriaga, ``A line of sight approach to cosmic microwave
  background anisotropies,'' {\em Astrophys. J.} {\bf 469} (1996) 437--444,
\href{http://www.arXiv.org/abs/astro-ph/9603033}{{\tt astro-ph/9603033}}.

\bibitem{Zuluaga:SUNGWEB2004}
J.~Zuluaga, ``Programming with sung: reference manual,'' 2004.
\newblock {\tt http://www.sungweb.tk}.

\bibitem{GSL}
``Gnu scientific library (gsl).''
\newblock GNU Project, {\tt http://www.gnu.org/software/gsl}.

\bibitem{CERNLIB}
``{CERNLIB libraries}.''
\newblock Computing and Network division, CERN. {\tt
  http://cernlib.web.cern.ch/cernlib/libraries.html}.

\bibitem{GNUPLOT}
T.~William and C.~Kelley, ``Gnuplot,''. {\tt http://www.gnuplot.info}.

\bibitem{NRC:1992}
W.~H. Press, B.~P. Flannery, S.~A. Teukolsky, and W.~T. Vetterling, {\em
  Numerical Recipes: The Art of Scientific Computing}.
\newblock Cambridge University Press, Cambridge (UK) and New York, 2nd~ed.,
  1992.

\end{thebibliography}\endgroup


\end{document}